\documentclass[final,5p,times,preprint]{elsarticle}

\usepackage{setspace}
\usepackage{amssymb}
\usepackage{amsmath}
\usepackage{graphicx}
\usepackage{amsfonts}
\usepackage{epsfig}
\usepackage{braket}
\usepackage{xcolor}
\usepackage{bm}
\usepackage[normalem]{ulem}

\newcommand{\ivec}{\vec}
\def\ss{\mbox{\boldmath $\sigma$}}
\newcommand{\be}{\begin{equation}}
\newcommand{\ee}{\end{equation}}
\newcommand{\bea}{\begin{eqnarray}}
\newcommand{\eea}{\end{eqnarray}}

\newcommand{\svec}[1]{{\mbox{\boldmath${ #1}$}}}
\newcommand{\ben}{\begin{enumerate}}
\newcommand{\een}{\end{enumerate}}

\newcommand{\beq}{\begin{equation}}
\newcommand{\eeq}{\end{equation}}
\newcommand{\beqn}{\begin{eqnarray}}
\newcommand{\eeqn}{\end{eqnarray}}

\newcommand{\beqd}{\begin{eqnarray*}}
\newcommand{\eeqd}{\end{eqnarray*}}
\newcommand{\bcen}{\begin{center}}
\newcommand{\ecen}{\end{center}}
\newcommand{\btab}{\begin{tabular}}
\newcommand{\etab}{\end{tabular}}
\newcommand{\bsub}{\begin{subequations}}
\newcommand{\esub}{\end{subequations}}
\newcommand{\bit}{\begin{itemize}}
\newcommand{\eit}{\end{itemize}}
\newcommand{\brule}{\begin{ruledtabular}}
\newcommand{\erule}{\end{ruledtabular}}
\newcommand{\bpm}{\begin{pmatrix}}
\newcommand{\epm}{\end{pmatrix}}

\newcommand{\ls}{\left[}
\newcommand{\rs}{\right]}

\renewcommand{\emph}[1]{\textit{\textbf{#1}}}
\newcommand{\re}{\nonumber\\}

\begin{document}


\title{Indirect methods in nuclear astrophysics with relativistic radioactive beams}
\author[TU]{Thomas Aumann} 
\ead{taumann@ikp.tu-darmstadt.de}
\author[Texas,TU]{Carlos A. Bertulani}
\ead{carlos.bertulani@tamuc.edu}
\address[TU]{Institut f\"ur Kernphysik, Technische Universit\"at Darmstadt, Germany}
\address[Texas]{Department of Physics and Astronomy, Texas A \& M University-Commerce, Commerce, Texas  75429, USA}

\begin{abstract}
Reactions with radioactive nuclear beams at relativistic energies have opened new doors to clarify the mechanisms of stellar evolution and cataclysmic events involving stars and during the big bang epoch. Numerous nuclear reactions of astrophysical interest cannot be assessed directly in laboratory experiments. Ironically, some of the information needed to describe such reactions, at extremely low energies (e.g., keVs), can only be studied on Earth by using relativistic collisions between heavy ions at GeV energies. In this contribution, we make a short review of experiments with relativistic radioactive beams and of the  theoretical methods needed to understand the physics of stars, adding to the knowledge inferred from astronomical observations. We continue by introducing a more detailed description of how the use of relativistic radioactive beams can help to solve astrophysical puzzles and  several successful experimental methods. State-of-the-art theories are discussed at some length with the purpose of helping us understand the experimental results reported. The review is not complete and  we have focused most of it to traditional methods aiming at the determination of the equation of state of symmetric and asymmetric nuclear matter and the role of the symmetry energy. Whenever possible, under the limitations of our present understanding of experimental data and theory, we try to pinpoint the information still missing to further understand how stars evolve, explode, and how their internal structure might be. We try to convey the idea that in order to improve microscopic theories for many-body calculations, nuclear structure, nuclear reactions, and astrophysics, and in order to constrain and allow for  convergence of our understanding of stars, we still need considerable improvements in terms of accuracy of experiments and the development of new and dedicated nuclear facilities to study relativistic reactions with radioactive beams.
\end{abstract}


\maketitle
\tableofcontents

\section{Introduction}


About 8 decades have passed by since the first nuclear induced reaction was carried out in a terrestrial laboratory. This happened  in 1932 at Cambridge University, in an experiment lead by  J. Cockcroft and E. Walton 
\cite{Cockcroft1932,Cockcroft1932-2}, who used a new technique at the time to accelerate and collide protons with a $^7$Li target, yielding two alpha particles as reaction byproducts, namely
\begin{equation}
{\rm p} + \ ^7{\rm Li} \longrightarrow \alpha + \alpha.
\label{p7Liaa}
\end{equation}

Till now reactions involving lithium are of great interest for mankind for worse or for better. For example, $^6$Li deuterides, i.e., a chemical combination in the form $^6$LiD, can be used as  fuel in thermonuclear weapons. Nuclear fission triggers explosion in such weapons to first induce heat and compress the $^6$LiD deuteride, and to bombard the $^6$LiD with neutrons. The ensuing reaction 
\begin{equation}
^6{\rm Li} + {\rm n}  \longrightarrow \alpha + \ ^3{\rm H}
\label{n6Lia3H}
\end{equation}
is followed by ${\rm D}+\, ^3{\rm H} \rightarrow \alpha + {\rm n}$ which liberates about 17.6 MeV of energy. In contrast, $^7$Li hydrides, in the chemical form of $^7$LiH,  is  a good moderator in nuclear reactors, because it reacts with less probability with neutrons, therefore forming less tritium, $^3$H. 

After so many decades since the Cockcroft and Walton experiment, both $^{6}$Li and $^{7}$Li isotopes are still of large interest for nuclear physics and nuclear astrophysics and other areas of science. For example, in cosmology both isotopes are at the center of the so-called ``lithium puzzle" in the Big Bang Nucleosynthesis (BBN) theory \cite{BERTULANI201656,Hou_2017,PITROU20181} which is an important part of our understanding of how the universe evolved from a primordial soup of fundamental particles to the present day universe with planets, stars and galaxies. 

The reaction in Eq. \eqref{p7Liaa} was made possible due to the use of electric and magnetic (EM) fields to accelerate charged particles. In the decades following the Cockcroft and Walton experiment, advances in using EM fields in nuclear accelerators developed enormously and almost all stable nuclear isotopes are now amenable to accelerate with different charge states, in some cases even with fully electron-stripped nuclei. In the last few decades a new era in nuclear physics emerged with a large investment on nuclear accelerators using short-lived unstable nuclei. This enabled the nuclear science community to study the structure and dynamics of unstable nuclei used as projectiles incident on stable nuclear targets. It also provided the access to key information on astrophysical nuclear reactions by means of indirect techniques as we will discuss in this review.

Assuming that the extraction of any nuclear isotope from ion sources would be possible, a back-of-the-envelope estimate of the lifetime limit for a nucleus down the beam line can be done. Taking as an example 100 m along the accelerator and projectiles moving close to the speed of light, the lowest lifetime admissible for a nuclear beam, before it decays, would be $\tau = 100/3\times 10^8 \sim 0.3\times 10^{-6}$ s, or about one $\mu$s\footnote{In fact, it would be longer due to Lorentz contraction at very high bombarding energies.}. However, ion production and release times from ion sources increases this number, down to a few ms. Accelerated unstable nuclei with such short lifetimes have only been possible with the developments in the last decades, with better techniques for production and extraction of nuclei from ion sources and the accomplishment of better accelerator technologies. Target manufacturing and new detector construction ideas have also been crucial to study reactions induced with short-lived nuclei. Reactions involving short-lived nuclear isotopes such as $^{11}$Li ($T_{1/2}$=$1.5$ ms) or $^{100}$Sn ($T_{1/2}$=$1.6$ s) are common in radioactive nuclear beam facilities. By relativistic we mean beam energies in the range $100-2000$ MeV/nucleon, typical bombarding energies available in a few radioactive beam facilities in the world such as the RIKEN/Japan or the GSI/Germany facility. These laboratories have the advantage of  probing properties of neutron-rich matter, with large beam luminosities, and newly developed detectors using inverse kinematic techniques.    

Among many subjects, nuclear astrophysics deals with the synthesis of nuclei in high temperature and pressure conditions existing within stars. Using the big bang theory, astrophysicists were able to explain how thermonuclear reactions lead to the production of 75\% of hydrogen and 25\% of helium observed in the universe.  The small traces of other elements have also been explained in terms of stellar processing of nuclear fuel.  Population III stars (oldest ones) were formed from light primordial nuclei originated from big bang nucleosynthesis. Population III stars with large masses  ejected heavier elements  during energetic explosions. Population II stars, very poor in metals, are the next generation of stars and are found in the halo of galaxies.  They can also form carbon, oxygen, calcium and iron which can be ejected by stellar winds to the interstellar medium.  The stardust generated by stellar explosions and stellar winds generated the population I stars found in the disk of galaxies containing large metallicities. The cores left over during supernova explosions can either form a neutron star or a black hole. To understand all these physical processes, one needs to develop a large number of theoretical models,  pursue dedicated astronomical observations and, if possible, perform nuclear experiments on earth.  

The nuclear physics contribution to the synthesis of light elements during the big bang, the formation of medium-heavy elements in stellar cores and of heavier elements, up to uranium, in supernova explosions and in neutron star mergers, involves a very large number of unsettled puzzles. Some of these puzzles can be tackled by studying nuclear reactions in nuclear physics laboratories with either extremely low energy beams or extremely large energies. In particular, reactions involving rare nuclear species often require the use of fast nuclear beams. It has been realized in the last few decades that relativistic reactions with radioactive beams can fill the gap of our knowledge in many aspects of nuclear astrophysics related to stellar evolution, stellar structure and the synthesis of the elements in the universe. 

Neutron stars are interesting objects because they consist of nuclear matter compressed to incredibly high densities. The behavior of such dense matter under compression is determined by the so-called neutron star equation of state (EoS) which also determines their basic properties, such as their masses and radii  \cite{Lattimer:2001}. Many theoretical predictions exist for the EoS and learning about it  under such extreme conditions is important to advance our knowledge of nuclear physics. Recently, gravitational waves have been detected by the LIGO-Virgo collaboration \cite{LIGOPRL116.061102}. The gravitational waves are thought to be caused by merging  black holes or neutron stars. The event named GW170817 \cite{Abbott_2017} is likely caused by the  orbiting of two neutron stars during the coalescence phase.  The shape of the gravitational-wave signal depends not only on the masses of the neutron stars but also in their so-called tidal deformabilities which describe how much they are deformed by tidal forces during the merging phase. The  effect of tidal deformability could modify the orbital decay caused by the emission of gravitational waves. Such astronomical observations are important to determine the neutron star EoS. Nuclear physics experiments, in particular involving neutron-rich nuclei, are also crucial for a consistent determination of the EoS of both symmetric and asymmetric nuclear matter  

In this review, we will focus on a few of the indirect techniques that have emerged in the last decades and that are a main part of our own research agenda.  Previous reviews covering other parts of this vast subject have been published, e.g., in Refs. \cite{BERTULANI201656,BERTULANI2010195}. In section 2 we make a short description of the mechanisms involved in nucleosynthesis, the challenges in performing measurements at very low energies, the physics of neutron stars and their equation of state. Some microscopic theories used  for the purpose are also discussed. In particular, we focus the discussion on constraining  the slope parameter of the equation of state. In section 3 we discuss some of the indirect techniques used to tackle the astrophysical  problems we raised in section 2. These include electron scattering off exotic nuclei, elastic and inelastic hadronic scattering,  total nuclear reaction cross sections, Coulomb excitation, pygmy resonances, dipole polarizability and electromagnetic response, charge exchange reactions and central collisions. In section 4 we present our conclusions. 

We  apologize to the authors and their publications that we might inadvertently missed to cite properly in this review. 

\section{Nuclear physics in astrophysics}

\subsection{Nucleosynthesis}

Cosmology and stellar evolution involve many aspects of nuclear physics. These areas of science deal with a dynamical scenario involving matter and radiation densities, thermodynamics, chemical composition, hydrodynamics, and other physical quantities and processes. Nuclear physics enters in most of the dynamical parts of the relevant scenarios through the determination  of rates of nuclear transmutations, such as particle, electromagnetic and weak-decay processes, as well as fusion and rearrangement reactions. For example, in the hot stellar plasmas the two-body reaction rate, or number of reactions per unit volume and per unit time involving particles $i$ and $k$, is given by 
\begin{equation}
\Gamma_{ik}={n_{i}n_{k}\left<\sigma v\right>\over 1+\delta_{ik}}, \label{Gamik}
\end{equation} 
where $n_{i(k)}$ is the number density of particle $i(k)$,  $v$ is the relative velocity between particle $i$ and $k$, and $\left<\sigma v\right>$ is the average of the cross section $\sigma$ for the reaction over the Maxwell-Boltzmann relative motion distribution of the particles. The Kronecker-delta factor $\delta_{ik}$ prevents double counting for the case $i=k$. Thus, the reaction rate is given by
\begin{align}
 \Gamma_{j,k}&={n_{i}n_{k}\over 1+\delta_{ik}}\left({\frac{8}{\pi{m_{ik}k_B^3T^3}}}\right)^{1/2} \int_{0}^{\infty}\sigma(E)\exp\left(-{E\over k_BT}\right)EdE, \label{astrophys5}
\end{align}
where $m_{ik}$ is the reduced mass of the $i$-$k$ pair. 

In rare cases, stellar modeling also needs information of reaction rates involving three particles, such as in the triple-alpha capture process, $\alpha+\alpha+\alpha$$\longrightarrow$ $ ^{12}{\rm C}+\gamma  $. This case can be treated as a sequential process in which two $\alpha$-particles are radiatively captured to form the unbound $^8$Be nucleus, followed by another $\alpha$ capture to create $^{12}$C. The reaction rate for this reaction is given by
\begin{eqnarray}
\Gamma_{\alpha\alpha\alpha}&=&3\left({\frac{8\pi \hbar}{ {m_{\alpha\alpha}^2}}}\right) \left( {m_{\alpha\alpha}\over 2\pi k_BT} \right)^{3/2} \\ \label{astrophys1}
 &\times& \int_{0}^{\infty}{\sigma_{\alpha\alpha}(E)\over \Gamma_\alpha(^8{\rm Be};E)}\exp\left(-{E\over k_BT}\right)\left<\sigma v\right>_{\alpha ^8{\rm Be}} E dE, \nonumber 
\end{eqnarray}
where $E$ is the $\alpha\alpha$ energy relative to the $\alpha\alpha$ threshold, $\sigma_{\alpha\alpha}$ is the $\alpha\alpha$ elastic cross section described by a narrow Breit-Wigner function, and $\Gamma_\alpha(^8{\rm Be};E)$ is the $\alpha$-decay width of the $^8$Be ground state treated as energy dependent. In $^8$Be, the width of this resonance is about $\Gamma_\alpha=6$ eV. The reaction rate $\left<\sigma v\right>_{\alpha- ^8{\rm Be}}$ in Eq. \eqref{astrophys1} is obtained in the same way as in Eq. \eqref{astrophys5}, but with the integral running over $E'$, where $E'$ is now the energy with respect to the $\alpha$-$^8{\rm Be}$  threshold (which varies with the $^8{\rm Be}$ formation energy $E$) and the integrand is replaced by $\sigma_{\alpha- ^8Be}(E';E)\exp(-E'/k_B T)E'$. 

The reaction rates for nuclear fusion are crucial for nuclear astrophysics, namely the study of stellar formation and evolution. To determine the chemical evolution, stellar modelers need to solve a set of coupled equations for the number densities in the form
\begin{equation}
\frac{{\partial n_{i}}}{{\partial t}} =\sum_{j} m_{j}^{i} \Gamma_{j} +\sum_{j,k} m_{j,k}^{i} \Gamma_{j,k} +\sum_{j,k,l}  m_{j,k,l}^{i} \Gamma_{j,k,l},\label{astrophys2}
\end{equation}
where $m^{i}_Y$ are positive or negative integers  specifying how many particles of species $i$ are created or destroyed in a reaction $Y$ out of the combination of the particles forming them. Such reactions  can be due to (a) decays ($Y=i$), electron/positron capture, neutrino-induced reactions, and photodisintegrations,  in which case $\Gamma_{i}=\lambda_{i}n_{i}$, where $\lambda_i$ is the decay-constant,  (b)  two-particle reactions ($Y=\{i,j\}$),  and (c) three-particle reactions ($Y=\{i,j,k\}$), as described above.

Using the concept of nuclear abundances defined as $Y_{i}=n_{i}/(\rho N_{A})$, where $\rho$ is the density and $N_A$ the Avogadro number, for a nucleus with atomic weight $A_{i}$, then $A_{i}Y_{i}$\ is the mass fraction of this nucleus in the environment and  $\sum
A_{i}Y_{i}=1$ \footnote{This only works in cgs units.}. The reaction network equations (\ref{astrophys2}) can be re-written as
\begin{eqnarray}
{d {Y}_{i}\over dt}&=&\sum_{j}N_{j}^{i}\lambda_{j}Y_{j}+\sum_{j,k}{N_{i}\over N_{j}! N_{k}!}\rho
N_{A}\left<\sigma\mathrm{v}\right>_{j,k}Y_{j}Y_{k} \nonumber \\
&+&\sum_{j,k,l}{N_{i}\over N_{j}! N_{k}!N_{l}!}\rho^{2}N_{A}^{2}\left<\sigma\mathrm{v}\right>_{j,l,k}Y_{j}
Y_{k}Y_{l}.\label{astrophys13}
\end{eqnarray}
Here $N_{i}$ denotes the number of nuclides  $i$ produced in the reaction,  being a negative number for destructive processes.

The energy generation in stars per unit volume per unit time is obtained by adding  the mass excess $\Delta M_{i}c^{2}$ of all nucleus $i$ created during the time step, i.e.,
\begin{equation}
{d{\epsilon}\over dt} =-\sum_{i} {d{Y}_{i}\over dt}N_{A} \Delta M_{i} c 
^{2} .\label{astrophys15}
\end{equation}

Upon including reaction networks in stellar modeling one obtains the chemical evolution and energy generation in diverse stellar and cosmological scenarios. For example, to simulate the BBN one needs information about a chain of nuclear reactions involving light nuclei ($A_i <7$), up to the formation of $^7$B and $^7$Li.  For solar physics one needs to know features of the reactions in the pp-chain which convert hydrogen to helium and dominates the energy generation and nucleosynthesis in stars with masses less than or equal to that of the Sun. In heavier stars, with masses larger than the Sun, the energy and nucleosynthesis process  are dominated by the Carbon-Nitrogen-Oxygen (CNO) cycle, fusing hydrogen into helium via a six-stage sequence of reactions. To describe  neutron star mergers one needs to know cross sections for rapid neutron capture (r-process) reactions,  a rapid sequence of neutron capture followed by beta-decay occurring in neutron-rich environments \cite{Burbidge.29.547,Fowler.56.149,Kratz.1993ApJ,WallersteinRMP.69.995,kaeppeler.48.1.175,Arnould_1999,Langanke_2001,THIELEMANN2011346,THIELEMANNannurev101916}.  

The two mechanisms involving most neutron-capture elements are the r-process and the s-process (slow neutron capture). The r-process is thought to generate about half of the nuclides with $A\gtrsim 100$ \cite{Burbidge.29.547}.  The high abundance patter around $Z \sim 56-76$, known as the main r-process, is very consistent with halo star and the solar system r-process abundances \cite{ARNOULD200797}. A second ($A\sim 130$) and third ($A \sim 195$) r-process peaks are also observed. Abundances around the second and third r-process peaks, as well as for the intermediate nuclei between them, are dependent on  nuclear properties of the r-process under steady beta-decay flow and fission cycling \cite{Seger1965ApJS}. Fission cycling occurs under sufficiently high neutron-rich conditions where the r-process extends to nuclei decaying through fission channels. Fission has an impact on the r-process,  terminating its path near the transuranium region leading to material returning to the $A \sim 130$ peak. The fission products then become new seed nuclei for the r-process, facilitating steady beta-decay flow.

Many stellar nucleosynthesis scenarios, require knowledge of reactions  involving short-lived nuclei. These are available now in large quantities in radioactive-beam facilities. They are usually produced in flight and therefore the extraction of information of interest for astrophysics are done indirectly using experimental and theoretical techniques, many of which developed in the last few decades. 

\subsection{Some specific reactions}

\subsubsection{CNO cycle}

\begin{figure}[t]
\begin{center}
\includegraphics[height=1.9in]
{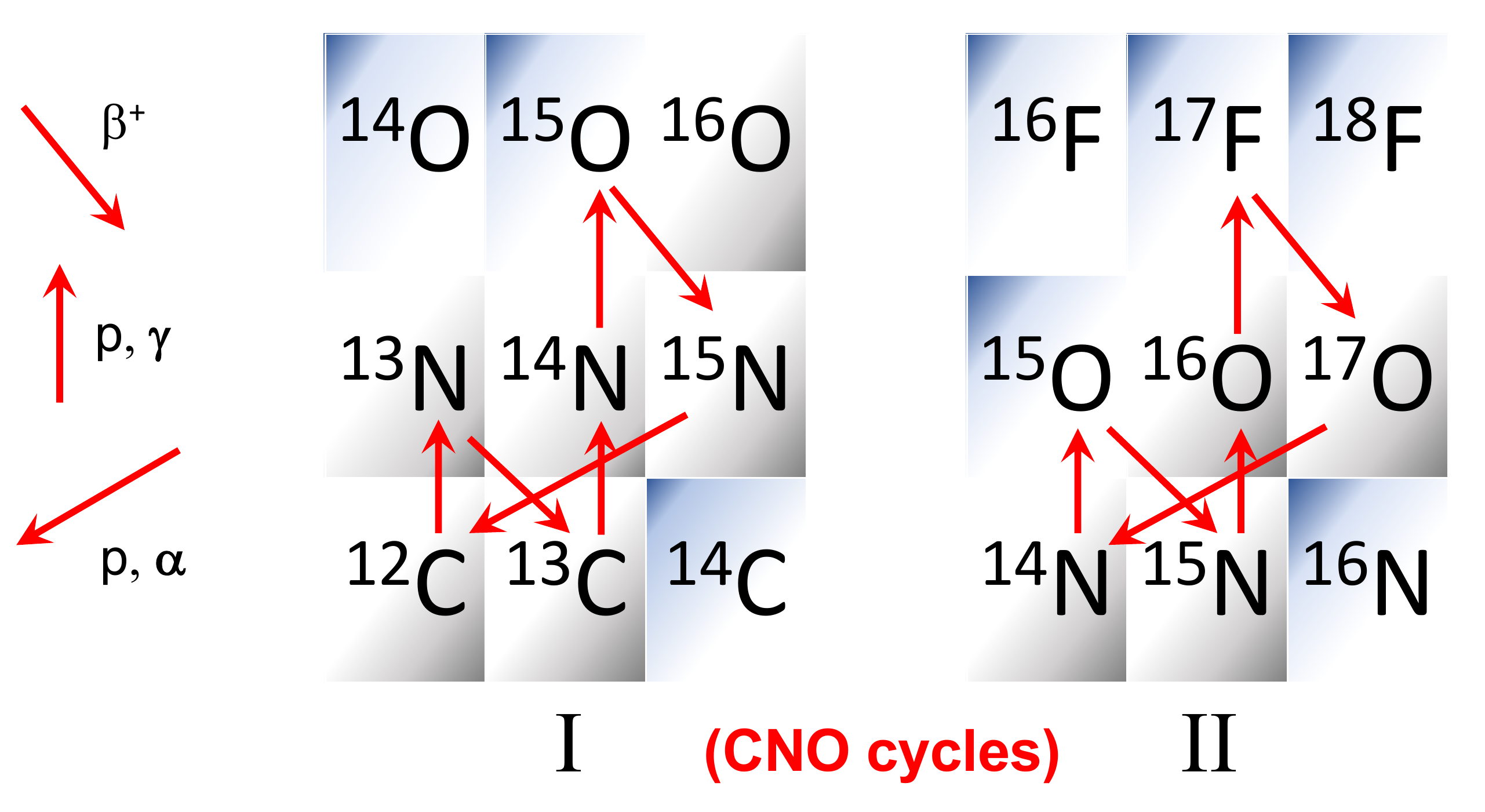} 
\caption{
The CNO cycles I and II.  Cycle I, also known as CN cycle \cite{Bethe1937}.Cycle II is a breakout of the CN cycle and is sometimes called by ON cycle. The stable nuclei are represented in boxes with bottom-right gray shaded areas.
}
\label{CNOcycle}
\end{center}
\end{figure}

An example of a reaction network of interest for nuclear astrophysics is the CNO cycle shown schematically in Figure \ref{CNOcycle}.
It has a cycle I, also known as CN cycle, involving the reactions 
\begin{equation}
^{12}{\rm C(p},\gamma)^{13}{\rm N(e}^+\nu_e)^{13}
{\rm C(p},\gamma)^{14}{\rm N(p},\gamma)^{15}{\rm O(e}^+\nu)^{15}{\rm
N(p},\alpha)^{12} {\rm C},
\end{equation}
where  an $\alpha$-particle is synthesized  out of four protons, effectively as $4 p \rightarrow ^4{\rm He} + 2 {\rm e}^+ + 2\nu_e$, and an energy of $Q=26.7$ MeV is released.   The  $^{15}{\rm N (p}, \gamma)^{16}$O reaction leads to a breakout from the CN cycle,  returning to the CN cycle  via a number of reactions within the  CNO cycle II, also known as ON cycle,
\begin{equation}^{15}{\rm N (p}, \gamma)^{16}{\rm O(p}, \gamma) ^{17}{\rm F(e}^+\nu_e)^{17}{\rm
O(p},\alpha)^{14}{\rm N}\ .\end{equation}
Two low-energy neutrinos are produced in the beta decays of $^{13}$N ($t_{1/2} = 10 \ \rm min$) and $^{15}$O ($t_{1/2} = 122 \ \rm s)$. The ON cycle is slower by a factor of 1000 than the CN cycle because the S-factor for $^{15}{\rm N(p}, \alpha)^{12}$C  is about 1000 times larger than the S-factor  for the $^{15}{\rm N(p}, \gamma)^{16}$O reaction \cite{Calcioli:2011}. 

The determination of the cross sections for the reactions in the CNO cycle and its breakout cycles are still a matter of intense investigation. For example,  asymptotic giant branch (AGB) stars are believed to be one of the major source of interstellar medium (ISM) in the galaxies. The composition of the AGB ejecta, which constitute the late phase in the evolution of stars with masses in the range $1M_\odot \lesssim M  \lesssim 8M_\odot$,   is strongly dependent on the mixing processes occurring during the stellar lifetime. A telltale of these processes are the isotopic ratios of carbon ($^{12,13}$C), nitrogen ($^{14,15}$N), and oxygen ($^{16,17,18}$O). Systematic astrophysical  modeling of stars in the red giant branch (RGB) show that they undergo the so-called dredge-up mechanism, a convective mixing process that carries nuclei from internal layers to the surface. Some calculations have shown that the observed ratios cannot be explained on the basis of the hydrostatic H-burning through the CNO cycle. An accurate experimental determination and theoretical description of the reactions involved in these cycles are still being pursued. For more discussion on this subject, see Refs. \cite{Wiescher_1999,Abia2017}.

\subsubsection{Radiative capture, beta-decay and electron capture}

To model a stellar environment, one needs to input densities and temperatures in the reaction networks, Eq. (\ref{astrophys13}), and to acquire knowledge on the decay constants $\lambda$, both in direct and inverse $\beta$-decay (electron capture), particle and photon emission,  cross sections for the elastic scattering, and the reaction rates for disintegration and formation of nuclei. 

Radiative capture reactions, involving the emission of electromagnetic radiation, are involved in the pp chain and CNO cycles. They are also prominent in the explosive conditions in novae, x-ray bursts, and supernovae. Often, radiative captures are the only proton- or $\alpha$-induced reactions possible with positive $Q$ values. They proceed slowly compared to strong interactions, and frequently control the reaction flow and nucleosynthesis rates. Measuring the radiative capture cross sections at the relevant energies is very difficult due to the vanishingly small reaction probability.  Moreover, when one of the participating nuclei is radioactive, it is hard to achieve the necessary beam intensity to enable an accurate measurement at low energies. These difficulties has induced the use of indirect experimental techniques often involving the use of relativistic radioactive beams leading to much higher reaction yields. But, using the measured cross sections and indirectly determining the radiative capture cross sections of astrophysical interest can only be done reliably with the help of nuclear reaction theory. 

Radiative capture cross sections are related to photo-decay constants through the detailed balance theorem, leading to  the  decay rate for the reaction $i+j\rightarrow k+\gamma $, given by
\begin{equation}
\lambda_{k,\gamma}(T)={\frac{{\omega_{i}\omega_{j}}}{\omega_{k}}}\left({\frac{{A_{i}A_{j}}
}{A_{k}}}\right)^{3/2}\left({\frac{{m_{u}kT}}{{2\pi\hbar^{2}}}}\right)^{3/2}
 \left<\sigma\mathrm{v}\right>_{i,j}
\mathrm{exp}%
\left(-{Q\over kT}\right),\label{astrophys8}%
\end{equation}
where $Q$ is the energy released, $m_u$ is the mass unit,  $T$ is the temperature, $A_i$ are the nuclear mass numbers, and  $\omega_i (T)=\sum_{m}(2J_{im}+1)\exp(-E_{im}/kT)$ are partition functions. 

An example of actual relevance of radiative capture reactions is the pp-chain in the sun which begins with the p(p,e$^+ \nu_e$)d reaction, followed by the production of $^4$He via three possible reaction pathways involving nuclei with $A \leqslant 8$. The radiative capture reactions d(p,$\gamma$)$^3$He, $^3$He($\alpha,\gamma$)$^7$Be, and $^7$Be(p,$\gamma$)$^8$B are important parts of the pp chain and for the production of solar neutrinos from $^7$Be and $^8$B decays. The precise determination of the reaction rates of the last two reactions still remain a goal of contemporary nuclear astrophysics \cite{AdelbergerRMP.83.195}.

The late evolution stages of massive stars are strongly influenced by weak interactions, which  determine the core entropy and electron-to-baryon ratio $Y_e$ in
pre-supernovae and influences its core mass, driving the stellar matter neutron richer. Electron capture reduces pressure support by the remaining electrons, while $\beta$-decay acts in the opposite direction. Both electron capture and $\beta$-decay generate neutrinos, which escape the star and carry away energy and entropy from the core  when densities are less than $10^{11}$ g/cm$^3$. Electron captures and beta decays occur during hydrostatic burning stages in the very dense stellar core where the Fermi energy, or chemical potential, of the degenerate electron gas is sufficiently large to overcome the negative $Q$ values for the capture reactions. The capture rates are  dominated by Fermi (F) and Gamow-Teller (GT) transitions in the nuclei. These rates depend on the temperature and electron-number density, which can be related to the electron capture cross sections by
\begin{equation}
\lambda_{ec}(T)={1\over \pi^{2}\hbar^{3}}\sum_{{i,f}}\int_{\epsilon_e^{0}}^{\infty} p_{e}^{2}\sigma_{{ec}}(\epsilon_{e},\epsilon_{i},\epsilon_{f})f(\epsilon_{e},\mu_{e},T)d\epsilon_{e},\label{astrophys9}%
\end{equation}
where $m_{e}$ is the electron rest mass, $\epsilon_e^{0}={\rm max}(Q_{if},m_{e}c^{2})$, and $p_{e} = (\epsilon^2_{e} - m^{2}_{e}c^{4})^{1/2}/c$ is the momentum of the captured electron with energy $\epsilon_{e}$. In a supernova collapsing core the conditions are such that the electrons obey the Fermi-Dirac distribution 
\be
f(\epsilon_{e},\mu_{e},T)=[1+\exp{(\epsilon_{e}-\mu_{e})/k_{B}T}]^{-1},
\ee
with the electron chemical potential, depending on the electron density, given by $\mu_{e}$. 
The cross section for electron capture with an energy $\epsilon_e$, leading to a transition from a proton single-particle state with energy $\epsilon_{i}$ to a neutron single-particle state $\epsilon_{f}$, is denoted by $\sigma_{ec}(\epsilon_{e}, \epsilon_{i}, \epsilon_{f})$. The spectrum of emitted neutrinos produced by electron captures on a particular nucleus is given by (here $\hbar=c=1$)
\begin{equation}
\phi_{\nu}(\epsilon_{\nu})={1\over \lambda_{ec}}{1\over \pi^{2}\hbar^{3}}\sum_{{i,f}} p_{e}^{2}\sigma_{{ec}}(\epsilon_{e},\epsilon_{i},\epsilon_{f})f(\epsilon_{e},\mu_{e},T),\label{astrophys10}
\end{equation}
A similar method is used to obtain positron capture rates.

\subsubsection{Neutrino induced reactions}

At high densities ($\rho>10^{12}$ g/cm$^{3}$) neutrino scattering cross sections on nuclei and electrons also become important. Such densities occur in  core-collapse supernovae.  Densities of order $10^{11} - 10^{15}$ g cm$^{-3}$ and temperatures ranging from 1 to 50 MeV are reached. Neutrinos are produced in large numbers via electron-positron annihilation ($e^+e^- \leftrightarrow \nu \bar{\nu}$), nucleon-nucleon bremsstrahlung, etc. Among these, the charged current absorption and emission processes $\nu n \leftrightarrow pe^-$ and $\bar{\nu} p \leftrightarrow  ne^+$ dominate. These and other processes  couple the neutrinos to dense nuclear matter, influencing the diffusive energy transport in the core  to the less dense outer layers where the neutrinos stream freely. The neutrino heating in the tenuous layers behind the shock is thought ignite the supernova explosion \cite{Colgate1966ApJ143.626C,BetheWilson1985ApJ.295.14B,Burrows1995ApJ450.830B}. The neutrino wind is also thought to be  a site for r-process nucleosynthesis. Numerous neutrino induced cross sections are needed to understand the whole mechanism of supernova explosions.

We use the notation $p_\ell\equiv\{{\bf p}_\ell, E_\ell\}$ and $q_\nu\equiv\{{\bf q},E_{\nu}\}$ for the lepton and the neutrino momenta, and $ k = P_i-P_f\equiv \{{\bf k},k_\emptyset \},$ for  the momentum transfer,  where $P_i$ ($P_f$) is the momentum of the initial (final) nucleus, ${\rm M}$ is the nucleon
 mass, ${\rm m}_\ell$ is the charged lepton mass, and $g_{\scriptscriptstyle V}$, $g_{\scriptscriptstyle A}$, $g_{\scriptscriptstyle M}$ and $g_{\scriptscriptstyle P}$ are the dimensionless effective vector, axial-vector, weak-magnetism and pseudoscalar coupling constants, respectively. Their
numerical values are 
$ g_{\scriptscriptstyle V}=1$, $g_{\scriptscriptstyle A}=1.26$,
$g_{\scriptscriptstyle M}=\kappa_p-\kappa_n=3.70$, and $g_{\scriptscriptstyle P}= g_{\scriptscriptstyle
A}(2\mathrm{M} \mathrm{m}_\ell )/(k^{2}+\mathrm{m}_\pi^2)$. The cross section within first-order perturbation theory  for the process $\nu_{e}+(Z,A)\rightarrow (Z+1,A)+e^{-}$, with momentum $k=p_\ell-q_\nu$, is given by
\begin{equation}
\sigma
(E_\ell,J_f) = \frac{|{\bf p}_\ell| E_\ell}{2\pi} F(Z\pm1,E_\ell)
\int_{-1}^1
d(\cos\theta){\mathcal T}_{\sigma}({\rm q},J_f),\label{astrophys11}
\end{equation}
where $$F(Z\pm1,E_\ell)=  {2\pi \eta \over  \exp(2 \pi \eta) -1}, \ \ \ {\rm with} \ \ \ \eta = {Z_\pm Z_e \alpha \over v_\ell},$$ is  the Fermi function
($Z_\pm=Z+1$, for neutrino, and $Z-1$, for antineutrino), $\theta\equiv
\hat{\bf q}\cdot\hat{\bf p}$ is the angle between  the emerging electron and the incident neutrino, and the transition amplitude for initial (final) angular momentum $J_i$ ($J_f$) is
\[
{\mathcal T}_{\sigma}(\kappa,J_f)= \frac{1}{2J_i+1} \sum_{ s_\ell,s_\nu }\sum_{M_i, M_f }
\left|\left<{J_fM_f}\left|H_{{ {W}}}\right|{J_iM_i}\right>\right|^{2},
\label{24}\]
where $H_W$ is the weak interaction Hamiltonian and $J_i$ ($J_f$) is the initial (final) angular momentum. ${\mathcal T}_{\sigma}(\kappa,J_f)$ depends on the neutrino leptonic traces and on the nuclear matrix elements \cite{FullerMeyer:1995,SAMANA20101123,SamanaBertulani.PRC.78.024312} \footnote{Indices $\emptyset$ and $z$ are used for 
time- the third-component of four-vectors, respectively.}:
\begin{eqnarray}
{\mathcal T}_{\sigma}(\kappa,J_f)&=&\frac{4\pi
G^2}{2J_i+1}\sum_{{\sf J}}\Bigg[ |\left<{J_f}||{\sf O}_{\emptyset{\sf J}}||{J_i}\right>|^2{\mathcal L}_{\emptyset} \nonumber \\
&+&\sum_{\sf {M}=0\pm 1}|\left<{J_f}||{\sf O}_{{\sf M J}}||{J_i}\right>|^2{\mathcal L}_{\sf M} \label{25}\nonumber \\
&-&2\Re\left(|\left<{J_f}||{\sf O}_{\emptyset{\sf J}}||{J_i}\right> \left<{J_f}||{\sf O}_{0{\sf J}}||{J_i}\right>\right){\mathcal L}_{\emptyset z}\Bigg],
\end{eqnarray}
with $G=(3.04545\pm 0.00006){\times} 10^{-12}$  in natural units, and
\begin{eqnarray}
{\mathcal L}_{\emptyset}&=&1+\frac{|{\bf p}|\cos\theta}{E_\ell},
\ \ \ 
{\mathcal L}_{\emptyset
z}=\left(\frac{q_z}{E_\nu}+\frac{p_z}{E_\ell}\right),
\nonumber \\
{\mathcal L}_{0}\equiv {\mathcal L}_z&=&1+\frac{2q_zp_z}{E_\ell
E_\nu}-\frac{|{\bf p}|\cos\theta}{E_\ell},
\nonumber \\
{\mathcal L}_{\pm1}&=&1-\frac{q_zp_z}{E_\ell E_\nu}\pm
\left(\frac{q_z}{E_\nu}-\frac{p_z}{E_\ell}\right) S_1,
\label{26}
\end{eqnarray}
where 
\begin{equation} q_z={\hat k}\cdot
{\bf q}=\frac{E_\nu(|{\bf p}|\cos\theta-E_\nu)}{\kappa},
\ \ \ \ 
p_z={\hat k}\cdot {\bf p}=\frac{|{\bf p}|(|{\bf p}|-E_\nu\cos\theta)}{\kappa},
\label{27} 
\end{equation}
with $S_1=+ 1$ $(-1)$ for neutrino (antineutrino) scattering. For simplicity, we omitted the  isospin operators $\tau_\pm$ that are responsible for one unit isospin change in the matrix elements written above.

The operators in \eqref{25} are
\begin{eqnarray}
&{\sf O}_{\emptyset{\sf J}}=g_{\scriptscriptstyle{V}}{\mathcal M}_{\sf J}^{\scriptscriptstyle V}
+2i\overline{g}_{\mbox{\tiny A}}{\mathcal M}^{\scriptscriptstyle A}_{\sf J}
+i(\overline{g}_{\mbox{\tiny A}}+\overline{g}_{\mbox{\tiny P1}}){\mathcal M}^{\scriptscriptstyle A}_{z{\sf J}},
\nonumber \\
&{\sf O}_{{\sf M}{\sf J}} =i(\delta_{{\sf M}z}\overline{g}_{\mbox{\tiny P2}}-g_{\mbox{\tiny A}} +
{\sf M} \overline{g}_{\mbox{\tiny W}}){\mathcal M}^{\scriptscriptstyle A}_{{\sf M}{\sf J}}
+2\overline{g}_{\mbox{\tiny V}}{\mathcal M}^{\scriptscriptstyle V}_{{\sf M}{\sf J}}-\delta_{{\sf M} z}\overline{g}_{\mbox{\tiny V}}{\mathcal M}_{\sf J}^{\scriptscriptstyle V}, 
\nonumber \\
\end{eqnarray}
with the notation $\hat{\bf k}={\bf k}/\kappa$, $\kappa \equiv|{\bf k}|$, the coupling constants defined as
\begin{eqnarray}
&&\overline{g}_{\mbox{\tiny V}}={g}_{\mbox{\tiny V}}\frac{\kappa}{2{\rm M}};~
\overline{g}_{\mbox{\tiny A}}={g}_{\mbox{\tiny A}}\frac{\kappa}{2{\rm M}};~ \overline{g}_{\mbox{\tiny W}}=({g}_{\mbox{\tiny V}}
+{g}_{\mbox{\tiny M}})\frac{\kappa}{2{\rm M}}; \nonumber \\
&&\overline{g}_{\mbox{\tiny P1}}={g}_{\mbox{\tiny P}}\frac{\kappa}{2{\rm M}}\frac{q_\emptyset}{{\rm m}_\ell};~
\overline{g}_{\mbox{\tiny P2}}={g}_{\mbox{\tiny P}}\frac{\kappa}{2{\rm M}}\frac{\kappa}{{\rm m}_\ell},
\label{8}
\end{eqnarray}
and
\begin{eqnarray}
{\mathcal M}^{\scriptscriptstyle V}_{\sf J}&=&j_{\sf J}(\rho) Y_{{\sf J}}(\hat{\bf r});
\ \ \
{\mathcal M}^{\scriptscriptstyle A}_{\sf J}=
{\kappa}^{-1}j_{\sf J}(\rho)Y_{\sf J}(\hat{\bf r})(\mbox{\boldmath$\sigma$}\cdot\mbox{\boldmath$\nabla$});
\nonumber\\
{\mathcal M}^{\scriptscriptstyle A}_{{\sf MJ}}&=&\sum_{{\sf L}}i^{\sf  J-L-1}\ F_{{\sf MLJ}}j_{\sf L}(\rho)
\left[Y_{{\sf L}}(\hat{\bf r})\otimes{\mbox{\boldmath$\sigma$}}\right]_{{\sf J}};
\label{16}\nonumber \\
{\mathcal M}^{\scriptscriptstyle V}_{{\sf MJ}}&=&{\kappa}^{-1}\sum_{{\sf L}}i^{\sf  J-L-1}
F_{{\sf MLJ}}j_{\sf L}(\rho) [Y_{\sf L}(\hat{\bf r})\otimes\mbox{\boldmath$\nabla$}]_{{\sf J}},\nonumber \\
\end{eqnarray}
where $\rho=\kappa r$.

 Neutrino scattering cross sections are nearly impossible to measure on earth, especially for all neutrino reactions occurring in stellar environments. Therefore, neutrino scattering cross sections are obtained by calculation of Eq.  (\ref{astrophys11}) in a theoretical framework such as the Random Phase Approximation (RPA) and its variations \cite{LangankePinedoRMP.75.819}. But,  the reactions occurring in typical stellar scenarios have small momentum transfer, so that $\rho \ll 1$ and the angular dependence can be neglected. One obtains for example, for charged-current,
\begin{eqnarray}
{\mathcal T}_{\sigma}(\kappa,J_f)&\sim& {\mathcal C} \Big[ |\left<{J_f}||\Sigma_{k=1}^A 
\tau_\pm(k)||{J_i}\right>|^2 \nonumber \\
&+&g_{\tiny A}^2|\left<{J_f}||\Sigma_{k=1}^A \sigma(k)
\tau_\pm(k)||{J_i}\right>|^2\Big]
 \label{GT},
\end{eqnarray}
where ${\mathcal C}$ depends on the particle energies. The operator $\tau_+$ changes a neutron into a proton, $\tau_+$ $\mid$ n$\rangle$ = $\mid$ p$\rangle$, and $\tau_-$ changes a proton into a neutron, $\tau_-$$\mid$ p$\rangle$ = $\mid$ n$\rangle$.
The matrix elements are known as  Fermi and Gamow-Teller matrix elements, respectively.  

Next-to-leading-order (NLO) terms for the expansion of the matrix elements in powers of $\rho$  probe the nucleus  at shorter scales and for small $\bf{k}$. They are a consequence of the expansion $\exp(i {\bf k} \cdot {\bf r}) \sim 1 + i {\bf k} \cdot {\bf r} $, leading to  ``first forbidden'' terms $ \sum_{i=1}^A {\bf r}_i \tau_3(i) $ and $ \sum_{i=1}^A [{\bf r}_i \otimes \mbox{\boldmath$\sigma$}(i)]_{J=0,1,2} \tau_3(i) $ within the matrix elements. The NLO terms  generate collective radial excitations, such as giant resonances. They  to dominate the cross sections for high energy neutrinos in supernova environments.

Neutrinos detected on earth originating from stellar events give rise to a number of events in the neutrino detectors given by\
\[
N_{\alpha}=N_t \int_0^\infty  {\mathcal F}_\alpha(E_\nu) \cdot \sigma(E_\nu)
\cdot \epsilon(E_\nu) dE_\nu,
\label{4}\]
where  $\alpha=\nu_e,{\bar \nu}_e,\nu_x$ and
$(\nu_x=\nu_\tau,\nu_\mu,{\bar \nu}_\mu,{\bar \nu}_\tau)$ stands for neutrino and antineutrino types, $N_t$ is the number of target nuclei used in the detector, ${\mathcal F}_\alpha(E_\nu)$ is the  neutrino flux arriving the detector, $\sigma(E_\nu)$ is the neutrino-target nucleus cross section,
$\epsilon(E_\nu)$ is the detector efficiency for the neutrino energy $E_\nu$.  Dark matter detection experiments are often based on neutrino detectors and vice-versa. Very sensitive detectors are being built to detect neutrinos from  the Sun, the atmosphere, and from supernova. A new window in neutrino physics and astrophysics has been opened with the development of such new detectors (see, e.g., Ref. \cite{Dutta_2019}).

In Ref. \cite{SamanaBertulani.PRC.78.024312} it was shown that while microscopic calculations using different models for neutrino induced cross sections  might  agree reasonably well at low neutrino energies, they substantially deviate from each other at larger energies, e.g., $E_\nu \gtrsim 5$ MeV. This has a large impact on the simulation of  detector efficiencies of high energy neutrinos streaming off supernova explosions.  

\subsection{Challenges in Measurements of Low Energy Astrophysical Reactions}

\subsubsection{Reaction rates in astrophysical environments}
\begin{figure}[t]
\begin{center}
\includegraphics[height=3.9in]
{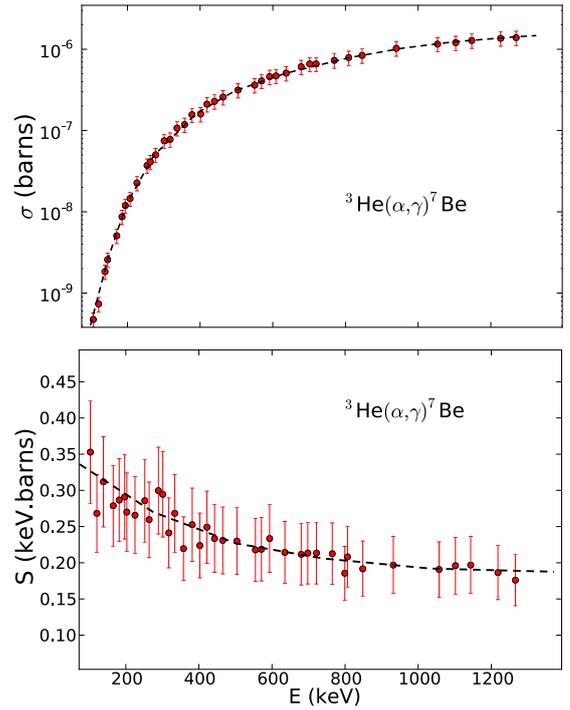} 
\caption{{\it Upper figure:} Cross section data for the reaction $^{3}$He($\alpha,\gamma$)$^{7}$Be as a function of the relative energy. {\it Bottom figure:} Same data expressed in terms of the astrophysical S-factor defined in Eq. (\ref{astrophys13a}). Data are from Ref. \cite{Kraewinkel1982}.}
\label{3He4He}
\end{center}
\end{figure}

Many fusion reactions involve charged nuclei that need to tunnel through a large Coulomb barrier  at  the very low relative energies in stars. This renders cross sections that drop many orders of magnitude as the relative energy decreases. See for example, Fig. \ref{3He4He} (upper panel)  for the cross section of the $^{3}$He$(\alpha,\gamma)^{7}$Be reaction relevant for solar physics. To avoid dealing with the trivial exponential fall of the capture cross section due to decreasing tunneling probability as the relative energy decreases, it is convenient to define the astrophysical $S$\ factor 
\begin{equation}
S(E)=E\sigma(E)\exp(2\pi\eta),\label{astrophys13a}
\end{equation}
where $\eta=Z_{j}Z_{k}e^{2}/\hbar v$ is the Sommerfeld parameter, and  $E$ is the relative energy, and $v$  is the velocity of the ions $j$ and $k$. The ``quantum area'' is $\lambda^{2} \propto 1/E$, where $\lambda$ is the wavelength for the relative motion. Then $\exp(-2\pi\eta)$ is a rough estimate for the quantum tunneling probability for s-wave scattering. Therefore, the S-factor definition in Eq. (\ref{astrophys13a}) removes the steep decrease of the cross section with the decrease of the relative energy, as we see in Fig. \ref{3He4He} (lower panel). In terms of the S-factor, the reaction rate, Eq. (\ref{Gamik}),  for the pair $jk$ becomes
\begin{equation}
\Gamma_{jk}={n_{j}n_{k}\over (1+\delta_{jk})}\left(  \frac{8}{\pi m_{jk}}\right)  ^{1/2}\frac{1}{\left(kT\right)  ^{3/2}}\int_{0}^{\infty}S(E)\exp\left[  -\frac{E}{kT}-\frac
{b}{E^{1/2}}\right]  ,\label{astrophys13b}
\end{equation}
where $b=2\pi\eta E^{1/2}=(2m_{jk})^{1/2}\pi e^{2}Z_{j}Z_{k}/\hbar $ and $m_{jk}$ the reduced mass in units of $m_u$.
 
For reactions induced by neutrons no Coulomb barrier exists, but quantum mechanics yields a non-zero reflection probability. The transmission probability of a neutron to an attractive potential region is proportional to its velocity v. This, combined with the proportionality of the cross section to the ``quantum area'', $1/E$, implies that the cross sections for neutron capture are better  rewritten as
\begin{equation}
\sigma(E)= {R(E)\over v},\label{astrophys14}
\end{equation}
where $R(E)$ has now also a flatter dependence on $E$.

Short-lived nuclei are very common in explosive stellar environments. They are the seeds of other reactions leading to stable nuclear species, which might pass by a large variety of unstable and stable nuclei. Cross sections for reaction involving short-lived nuclei are not well know in many cases.  During stellar hydrostatic burning stages, reactions with charged-particles at very low energies are very hard to study in direct measurements and frequently impossible to be done. But indirect techniques using radioactive beams have been developed to access partial or full information on decay constants and cross sections of interest for nuclear astrophysics. These techniques also allow  the extraction of information for stellar reactions involving stable nuclei  that were not possible prior to the advent of radioactive beam facilities.  

 \begin{figure}[t]
\begin{center}
\includegraphics[height=2.1in]
{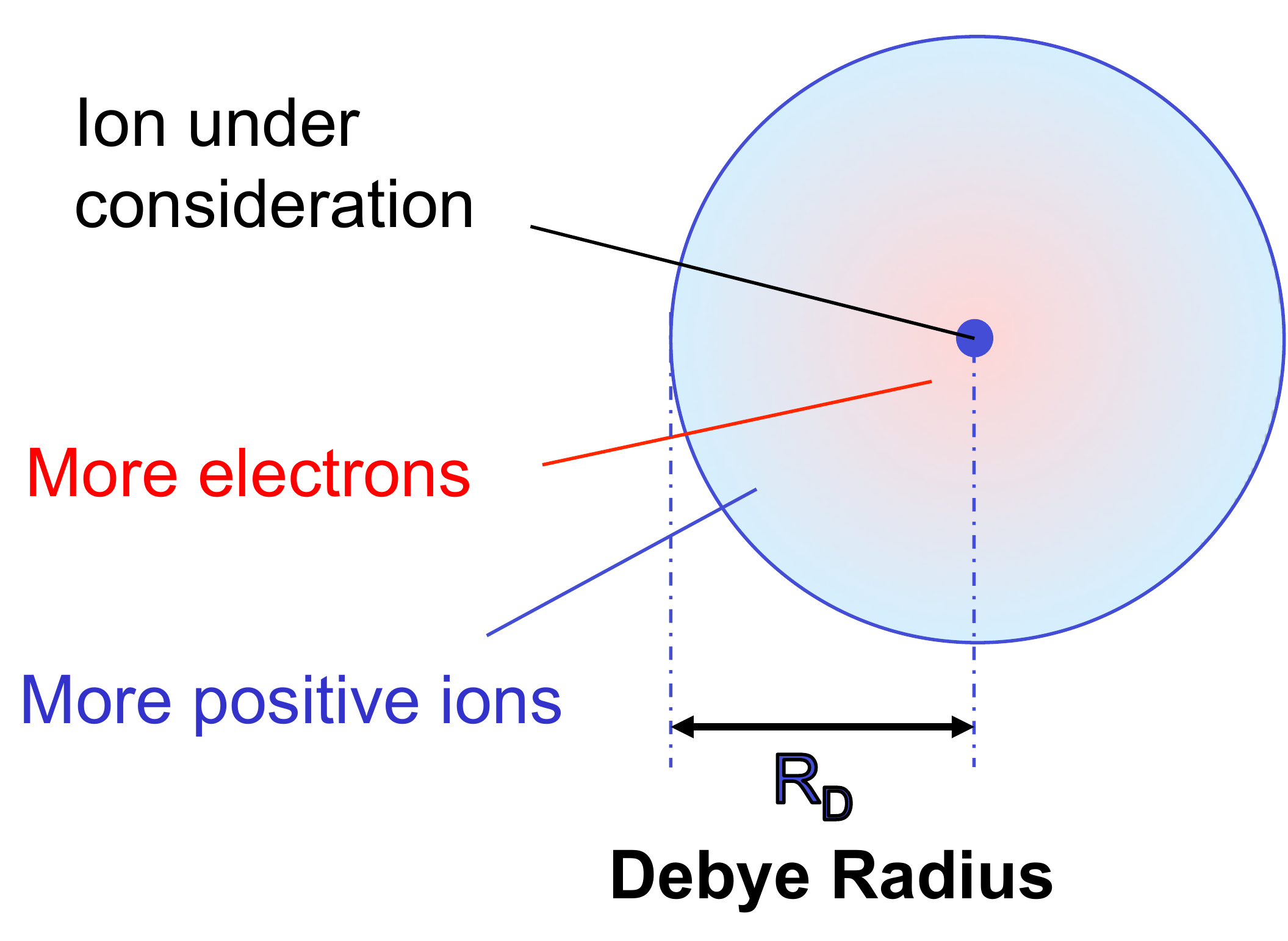} 
\caption{Schematic view of the Debye-H\"uckel sphere approximation used to describe electron screening in plasmas.}
\label{Debye}
\end{center}
\end{figure}

\subsubsection{Environment electrons in stars and on earth}
In stellar plasmas the free electrons shield the nuclear charges and reduce the nuclear reaction rates. When the potential energy  is smaller than the kinetic energy of the particles, one can account for electron screening  effects using the Debye-H\"uckel approximation, so that the reaction-rate in Eq.  \eqref{Gamik} is modified with
\begin{equation} \langle \sigma v
\rangle_{screened} = f_{ij}(\rho,T) \langle \sigma v \rangle_{bare},
\end{equation}
where
\begin{equation}
f_{ij}(\rho,T)=1+0.188{Z_iZ_j\rho^{1/2}\xi^{1/2}\over T_6^{3/2}},  \ \ \ {\rm with} \ \ \xi =\sum_k (Z_k^2 +Z_k)^2 Y_k .
\label{screened}
\end{equation}
Here, $T_{6}$ is the plasma temperature in units of billions  of degree Kelvin.  The Debye-H\"uckel approximation, shown schematically in Figure  \ref{Debye} is valid for electron densities $n_{e}$ such that within a radius $R_{D}$ a mean field approximation is valid, $n_{e}R_{D}\gg 1$. Based on these assumptions, the agglomeration of more negative electrons than positive ions within the sphere leads to an enhancement factor, $f(E)$, in the plasma so that the reaction rate in the plasma is reduced compared to the reaction rate in a charge-free environment,  $\langle \sigma v \rangle_{plasma} = f(E) \langle \sigma v \rangle_{bare}$. The screening causes a change in the Coulomb potential between two ions in the form $V(r) =Z_{I}Z_{j}e^{2}\exp(-r/R_{D})/r$ and one finds that $f=\exp(U_{e}/k_{B}T)$, where $U_{e}$, with energy dimensions is know as the screening potential. In the Sun, $R_{D} \sim \sqrt{kT/n}\sim 0.218$ ${\buildrel _{\circ} \over {\mathrm{A}}}$, and $f\sim 1.2$, a 20\% effect for the reaction $^{7}$Be(p,$\gamma)^{8}$B, important for the high-energy  neutrino production in the Sun. In Ref. \cite{Carraro1988} the authors calculate the enhancement $f(E)$ for weakly screened thermonuclear reactions, taking into account their dependence on the velocity of the colliding ions. They find enhancements are appreciably smaller than those given by the standard adiabatic Debye-H\"uckel approximation if the Gamow velocity is greater than the ion thermal velocity. The mean field approximation following the Debye-H\"uckel picture is not strictly valid under the conditions prevailing in the core of the Sun. A kinetic approach should be implemented.

 \begin{figure}[t]
\begin{center}
\includegraphics[height=1.6in]
{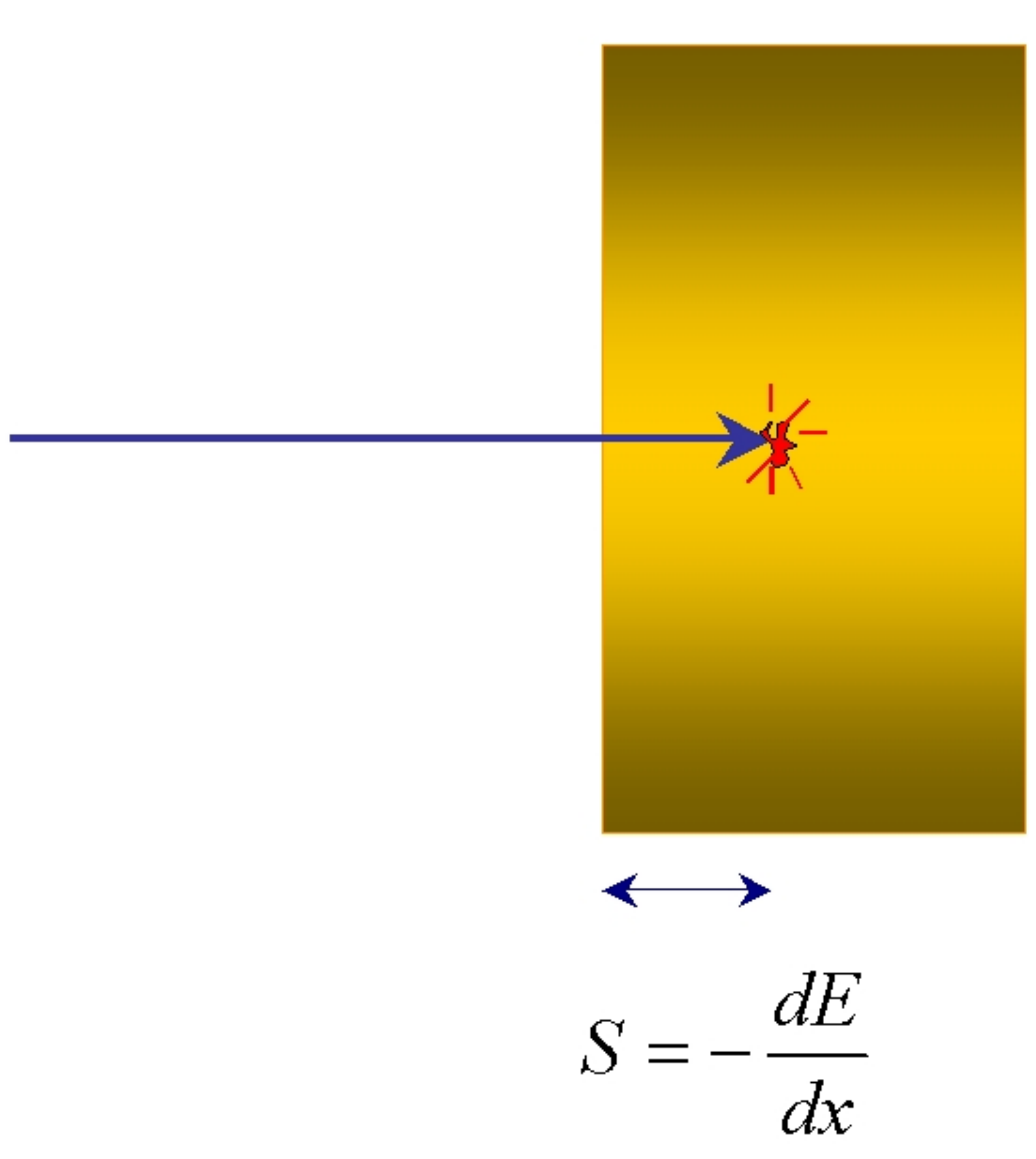} 
\ \ \ 
\includegraphics[height=2.3in]
{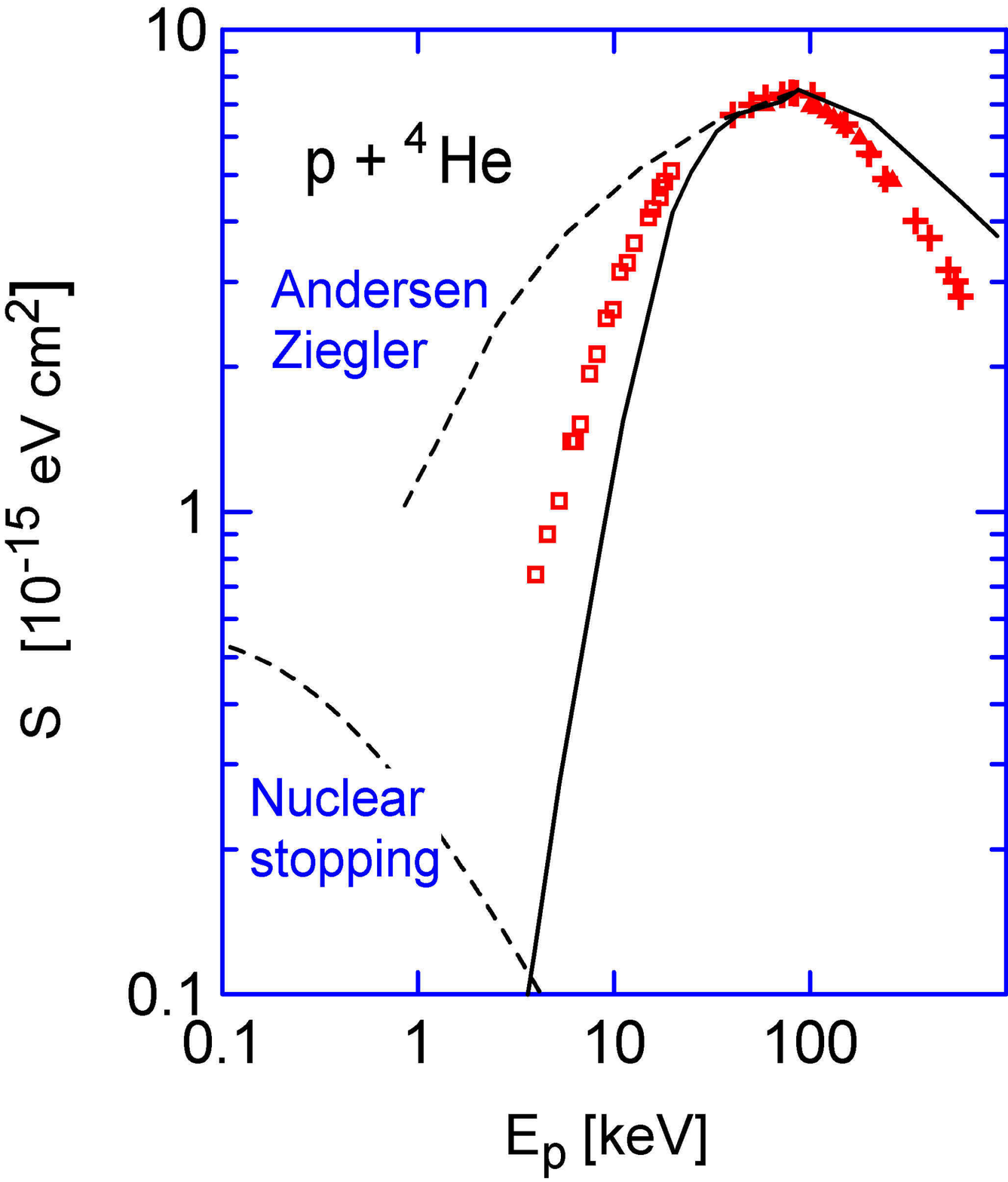} 
\caption{{\it Left:} Schematic representation of the stopping o low energy ions in nuclear targets. {\it Right:} Calculated stopping power in p + $^4$He collisions at energies of astrophysical relevance.}
\label{stopping}
\end{center}
\end{figure}

In Ref. \cite{Shaviv2001} the authors state that the energy exchange between any two scattering ions and the plasma is positive at low relative kinetic energies and negative at high energies. The turnover in a hydrogen plasma occurs at  $E_{kin-rel} \sim 2kT  < E_{Gamow} \sim 6kT$ for the p-p reaction. The net energy exchange, i.e., the sum over all pairs of scattering particles, vanishes in equilibrium.
They claim that fluctuations and non-spherical effects are crucial in affecting the screening. They derive screening corrections, which for the p-p reaction is found to enhance the transition rates, while for higher Z reactions, like $^{7}$Be(p,$\gamma)^{8}$B,  are suppressed relative to the classical Salpeter, or Debye-H\"uckel, theory. A detailed discussion of the screening in stellar plasmas can be found in Ref. \cite{AdelbergerRMP.83.195}, where no conclusion is reached on the apparent contradiction among the several models existing in the literature.  In the  ``strong screening'' regime at low density plasmas other models are more appropriate \cite{AdelbergerRMP.83.195}.  Screening induced in plasmas with intermediate densities obeying the relation $n_eR_D\approx 1$ require more complicated models, raging from the inclusion of  dynamical to quantum field theory contributions \cite{AdelbergerRMP.83.195}. 

Reaction rates of astrophysical interest measured in the laboratory are also increased by the presence of atomic electrons bound in the nuclei  \cite{Assenbaum1987,ROLFS1995297,ROLFS200123}, which reduce the Coulomb barrier. Experimental findings on the incremental factors are at odds with some apparently well founded electron screening theories  \cite{ENGSTLER1988179,Engstler1992,Angulo:1993,Prati1994,Greife1995}.  Due to screening, he fusion cross section measured at  laboratory energy $E$ is equal to that at energy $E+U_e$, with $U_e$ known as the screening potential. That is,
\begin{equation}
\sigma \;(E+U_e)= \exp \left[ \pi \eta (E)\frac{U_e}E\right] \ \sigma (E)\;,  \label{sig1}
\end{equation}
since the factor $S(E)/E$ has a much smaller dependence on the energy  than the term $\exp \left[ -2\pi \eta(E)\right] $ . Dynamical calculations as well as the consideration of several atomic effects have not been able to explain the fact that $U_e$ as measured experimentally is about a factor of two larger than that obtained theoretically  \cite{Shoppa1993,Assenbaum1987,ROLFS1995297,ROLFS200123,BALANTEKIN1997324,Flambaum:1999,HaginoBalantekin2002,Fiorentini2003}).  

In Refs. \cite{LANGANKE1996211,BangPRC1996} one has questioned if the  stopping power corrections used in the  experimental analysis were properly accounted for. As shown in Figure \ref{stopping} (left) the fusion of a low energy ion can occur at any point within the target, and the stopping power, $S$, accounts for the energy loss, $S=-dE/dx$, of the ions as they penetrate the target. The proper reaction energy $E_{eff} = E_{ion}-\langle S.dx\rangle$, in laboratory experiments of fusion reactions, need to account for the average energy loss, $\langle S.dx\rangle$. The stopping power at very low energies was further studied in Ref. \cite{BertulaniPRC.2000,BERTULANI200435} for H$^+$ + H, H$^{+}$ + He, and He$^{+}$ + He collisions. These are the simplest few electron systems that can be treated with a relatively accurate theory, and one has verified that the stopping power is in fact  smaller than the those predicted by the experimental extrapolations of the Andersen-Ziegler tables \cite{Ziegler:1985}. Because at very low ion energies the electrons in the atoms respond nearly adiabatically to the time-dependent interaction, the main cause of stopping are charge exchange, i.e., when an electron jumps from one atom to the other, or by Rutherford scattering, i.e., straggling, in the target (usually denoted as ``nuclear stopping''). Such findings are in agreement with previously determined stopping-power values reported in Ref. \cite{GolserPRL1991}. This is shown in Figure \ref{stopping} (right) \cite{BERTULANI200435}. The same trend was found for  atomic He$^{+}+$He \cite{BERTULANI200435}. A  ``quenching'' of the nuclear recoil  contribution to the stopping power was observed 
experimentally in Ref. \cite{FORMICOLA2003609} and explained in Ref. \cite{BERTULANI200435}.  Several fusion reactions were further studied in deuterated metals and a large increase of the cross sections were found  \cite{Raiola2004,Raiola2006,Czerski2004,Kasagi2004,HukePRC2008,Cvetinovic2015}. No plausible theoretical explanation seems to exist to explain such discrepancies. 

 \begin{figure}[t]
\begin{center}
\includegraphics[height=1.4in]
{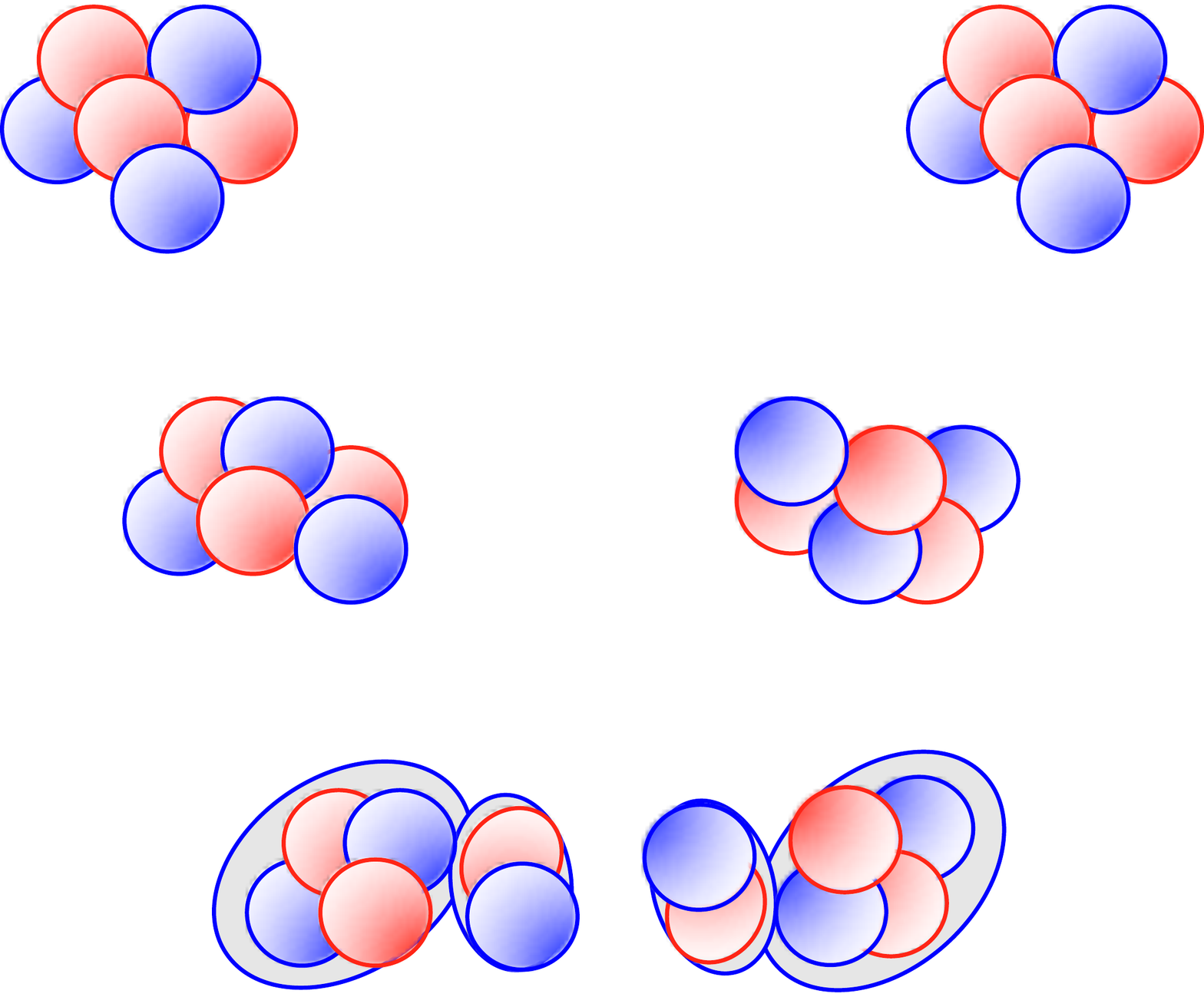} 
\ \ \ \ \ \ 
\includegraphics[height=1.4in]
{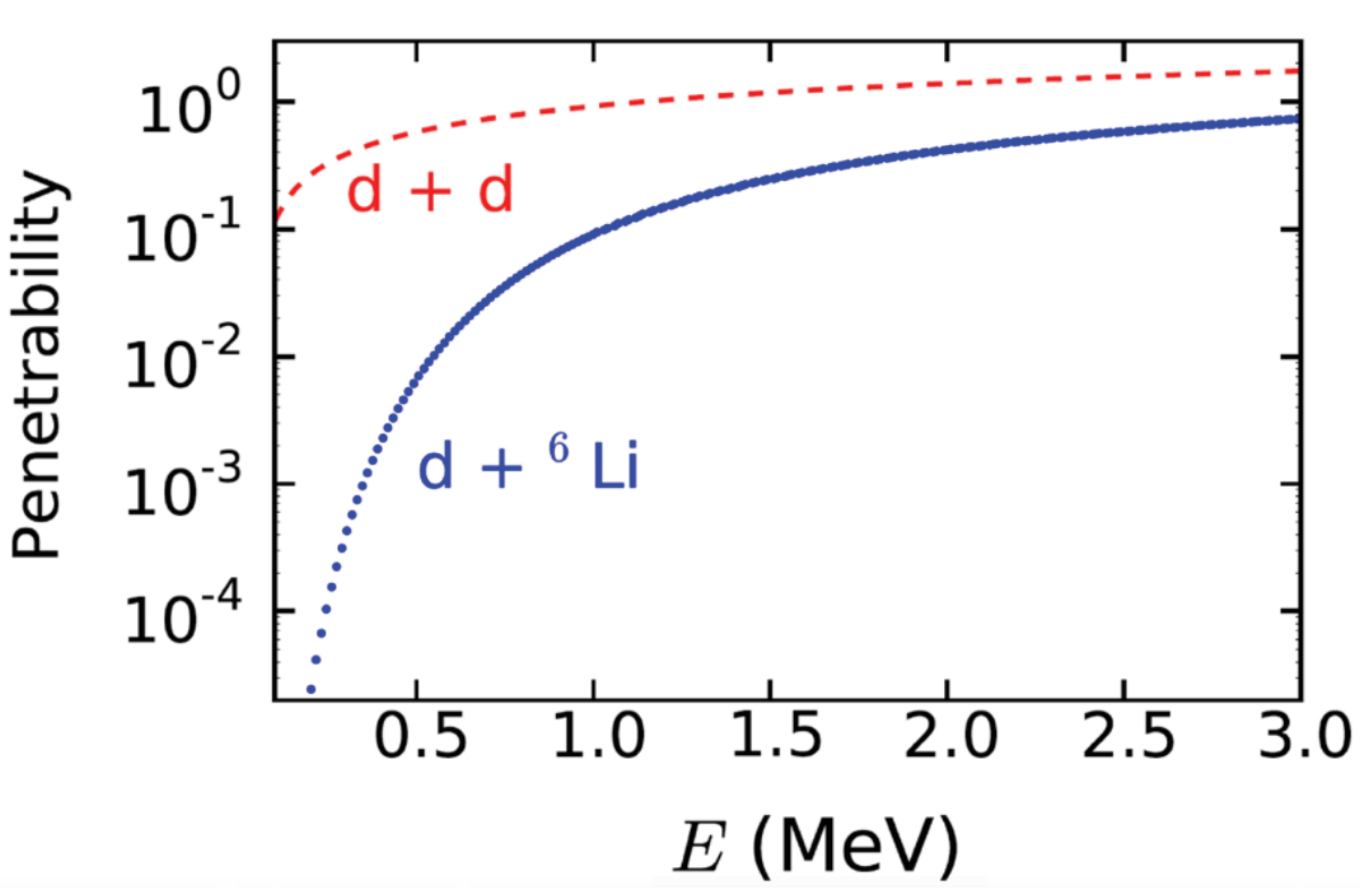} 
\caption{{\it Left:} Schematic view of polarization and orientation as nuclei with large probabilities for cluster-like structures approach each other, shown here for $^{6}$Li + $^{6}$Li. {\it Right:} Barrier penetrabilities for d + d and for d + $^{6}$Li reactions as a function of the relative motion energy.}
\label{clustering}
\end{center}
\end{figure}

\subsubsection{Clusterization in light nuclei}
Recently, it has been proposed that a possible solution of the ``electron screening puzzle'', maybe due to clusterization and polarization effects in nuclear reactions involving light nuclei at very low energies \cite{SPITALERI2016275,BertulaniSpitaleri2017}. Different tunneling distances  for each cluster induce a reduction of the overall tunneling probability. Such clustering effects can also be induced by polarization as the nuclei approach each other, as shown in Figure \ref{clustering} (left). It was shown that this is possibly the only way to explain why the reaction $^6$Li + $^6$Li $\rightarrow \ 3\alpha$ yields the experimentally observed cross sections, much higher in value than those predicted by theory.  In fact, if the Coulomb barrier penetrability used in the $^6$Li + $^6$Li were due to structureless $^6$Li ions, the cross section for $^6$Li + $^6$Li $\rightarrow \ 3\alpha$ would be nearly zero, or at least one could not measure it, but it is observed experimentally at low energies.   Therefore, one expects that it is highly probable that the deuterons within $^6$Li penetrate a smaller barrier and form $\alpha$ particles, thus explaining the puzzle. In Refs. \cite{SPITALERI2016275,BertulaniSpitaleri2017} it was shown that several reactions of astrophysical interest with light nuclei can be explained in this way. This indicates that more precise experiments need to be carried out to allow for a  critical review of theory versus experimental values of the electronic screening potentials $U_e$ and the role of clusterization in astrophysical reactions.

\subsection{Neutron stars}

\subsubsection{General observations}

Neutron-star masses can be deduced from observations of supernova explosions and from binary stellar systems. For example, Newtonian mechanics (i.e., Kepler's third law) relates the mass of a neutron star $M_{NS}$ and its companion mass $M_{C}$ in a binary system though 
\begin{equation}
{(M_{C} \sin \theta)^{3} \over (M_{NS} + M_{C})^{2}} = {T_{NS}v_{\theta}^{3}\over 2\pi G}, \label{nstarm}
\end{equation}
where $T_{NS}$ is the period of the orbit, $v_{\theta}$ is its orbital velocity projected along the line of sight, and $\theta$ is the angle of inclination of the orbit. But this equation is not enough to determine $M_{NS}$ as we need to know the mass of the companion, too. General relativity predicts an advance of the periastron of the orbit, $d\omega /dt$, given by
\begin{equation}
{d\omega \over dt} = 3 \left( {2\pi \over T_{NS}}\right)^{5/3}T_{\odot}^{2/3}{(M_{NS}+M_{C})^{2/3} \over 1-e},
\end{equation}
where and $T_{\odot}=GM/c^{3} = 4.9255\times 10^{-6}$ s. The observation of this quantity and comparison to this equation, together with the range parameter $R=T_{\odot}M_{C}$ associated with the Shapiro time delay of the pulsar signal as it propagates through the gravitational field of the companion star, and the Keplerian equation \ref{nstarm}, allows one to determine the neutron star mass $M_{NS}$. Other so-called post-Keplerian parameters such as the the orbital decay due to the emission of quadrupole gravitational radiation, or the Shapiro delay shaper parameter, can be used together with Eq. \ref{nstarm} to determine the mass of a neutron star \cite{Taylor:1992}. Typical masses obtained  with this procedure range from $M_{NS}=1.44M_{\odot}$ from observations of binary pulsars systems \cite{TaylorHulse:1975} to $M_{NS}=2.01M_{\odot}$ from millisecond pulsars in binary systems formed by a neutron star and a white dwarf \cite{Demorest:2010,Antoniadis:2013}.

 \begin{figure}[t]
\begin{center}
\includegraphics[height=2.2in]
{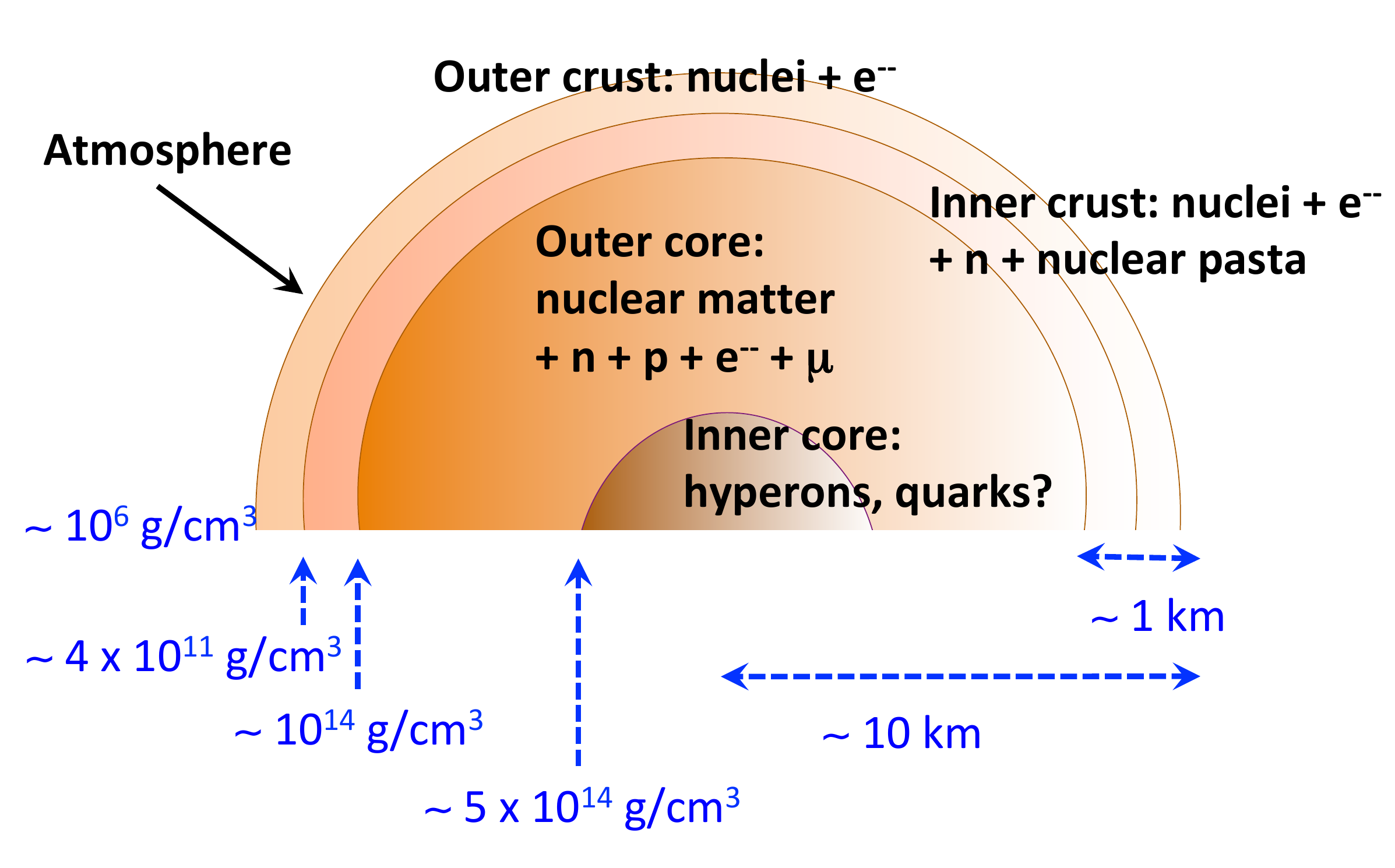} 
 \caption{Schematic view of the structure of a neutron star, showing the approximate sizes and features of the star crust and the core holding the bulk  of its neutron matter.}
\label{NSstruc}
\end{center}
\end{figure}

The radius of a neutron star is more difficult to determine because they are very small. But measurements of the X-ray flux $\Phi$  stemming from a neutron star in a binary system at a distance $d$ from us can be used, assuming that the radiation stems from a blackbody at temperature $T$, leading to the effective radius
\begin{equation}
R_{eff}=\sqrt{\Phi d^{2}\over \sigma T^{4}} \ .
\end{equation}
This can be used together with a relativistic correction to get the neutron star radius $R$ from
\begin{equation}
R=R_{eff}\sqrt{1- {2GM_{NS}\over Rc^{2}}}.
\end{equation}
The radii of neutron stars found with this method range within $9-14$ km \cite{Steiner:2013,Lattimer:2014,Guillot:2013,GuillotRut:2013}.

Neutron stars can rotate very fast  due to the conservation of angular momentum when they were created as leftovers of supernova explosions. In a binary system the matter absorption from the companion star can increase its rotation. The angular speed can reach several hundred times per second and turn it into an oblate form. It slows down because its rotating magnetic field radiates energy into free space and its shape becomes more spherical.  The slow-down rate is nearly constant and extremely small, of the order of $-(d\Omega/dt)/\Omega= 10^{-10} - 10^{-21}$ s/rotation. Sometimes a neutron star suddenly rotates faster, a phenomenon know as a ``glitch'',  thought to be associated with a ``star quake'' when  there is a rupture in their stiff crust. As a consequence, the equatorial radius contracts and angular momentum conservation leads to an increase of rotation. Glitches could also be due to vortices in the superfluid core transiting from a metastable state to a lower-energy state  \cite{AndersonItoh:1975}.

 \begin{figure}[t]
\begin{center}
\includegraphics[height=2.2in]
{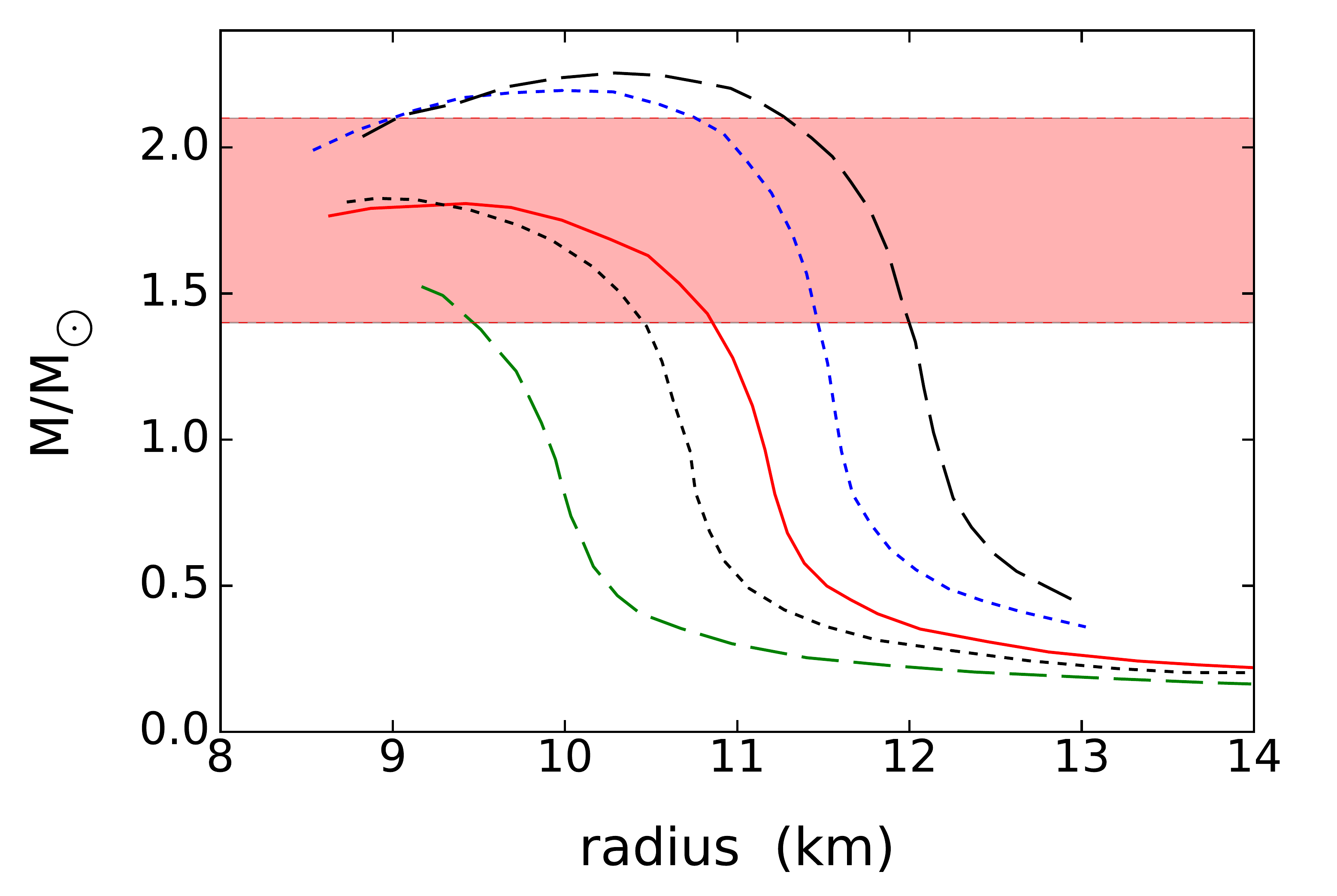} 
 \caption{Mass of a neutron star (in units of solar masses) versus radius in kilometers calculated (solid, dashed and dotted curves) with a few realistic EoS. The horizontal band displays the  limits of the observed masses \cite{Rezzolla_2018}.}
\label{NSs}
\end{center}
\end{figure}

The hitherto compiled knowledge about neutron stars seems to indicate that they contain  a dense core surrounded by a much thinner crust with mass $\lesssim M_\odot/100$ and a thickness of $\lesssim 1$ km  \cite{Lattimer:2001} (see Fig. \ref{NSstruc}). The crust, divided into outer and inner crust, consists of neutron-rich nuclei coexisting  with strongly degenerate electrons and for densities  $\rho \gtrsim 4 \times 10^{11}$ g cm$^{-3}$ free neutrons drip from nuclei. The density increases going from crust to the core and at their interface the density is  $\approx \rho_{0}/2$, where $\rho_0=2.8\times 10^{14}$ g cm$^{-3}$ is the saturation  density of nuclear matter.  The outer core is probably composed of neutrons, protons and electrons, whereas the inner core, found  in massive and more compact neutron stars might contain pion, kaon or hyperon condensates, or quark matter \cite{Lattimer:2012}. The Equation of State (EoS), i.e., how pressure depends on the density in the core of a neutron star, is usually separated in  an EoS for the outer core ($\rho\lesssim 2\rho_0$) and another for the inner core ($\rho\gtrsim 2 \rho_0$).   

\subsubsection{Structure equations}

The equations of hydrostatic equilibrium for a neutron star are modified to account for special and general relativity corrections. The so-called Tolman-Oppenheimer-Volkoff (TOV) equation is the state-of-the-art equation  for the structure of a spherically symmetric neutron star in static equilibrium, given by 
\begin{equation}\frac{dp(r)}{dr}=-\frac{G}{r^2}\left[\rho(r)+\frac{p(r)}{c^2}\right]\left[m(r)+4\pi r^3  \frac{p(r)}{c^2}\right]\left[1-\frac{2Gm(r)}{c^2r}\right]^{-1},
\label{stars:tov}
\end{equation}
where $m(r)$ is the total mass within radius $r$, 
\begin{equation}
m(r)=4\pi \int_{0}^{r} r'^{2}\rho^{2}(r') dr',
\label{stars:tov2}
\end{equation}
and $\rho(r)$ and $p(r)$ is the local density and pressure, respectively.
The second terms within the square brackets in Eq. (\ref{stars:tov}) stem from special and general relativity corrections to order $1/c^2$, and in their absence one has the equations of hydrostatic equilibrium following Newton's gravitation theory.  To obtain how mass increases with the radial distance, the two equations above need to be supplemented by an EoS linking the pressure to density, $p(\rho)$. The relativistic corrections factors have all  the effect of enhancing the gravity at a distance $r$ from the center.

The TOV equation is integrated enforcing the boundary conditions $M(0)=0$, $\rho(R)=0$ and $p(R)=0$, where $R$ denotes the neutron star radius.  It is usually integrated radially outward for  $p$, $\rho$  and $m$   until it reaches $p\simeq 0$ at a star radius $R$.  A maximum value of the neutron star mass is obtained for a specific value of the central density. Typical results are plotted as in Figure \ref{NSs} with the star mass $M$ given in terms of the central density $\rho_c$,  or in terms of the radius $R$.  An EoS based on non-interacting neutron gas yields the maximum mass of a neutron star as $\sim 0.7M_{\odot}$, whereas a stiffer equation of state yields $\sim 3M_{\odot}$ \cite{bertulani2012neutron}. The horizontal band displays the  limits of the observed masses \cite{Rezzolla_2018}.  At very high pressures, Eq. \eqref{stars:tov} is quadratic in the pressure and if a star develops a too high central pressure, it will quickly develop an instability and will not be able to support itself against gravitational implosion. 

\subsection{The equation of state of neutron stars}

Neutron stars are almost exclusively made of neutrons with a small fraction, $\sim 1/100$, of electrons and protons. The neutron is transformed into a proton and an electron via the weak decay process ${\rm n} \rightarrow {\rm p} + e^- + \bar\nu_e$, liberating an energy of $\Delta E=m_n - m_p - m_e$ = 0.778 MeV which is carried away by the electron and the neutrino\footnote{In this and in the next equations we use $\hbar=c=1$.}.
In weak decay equilibrium,  as many neutrons decay as electrons are captured  in $p + e^- \rightarrow n + \nu_e$, which can be expressed in terms of the chemical potentials for each particle as $\mu_n = \mu_p + \mu_e $. The chemical potential is the energy required to add one particle of a given species to the system. The Pauli principle impedes decays when low-energy levels for the  proton, electron, or the neutron are already occupied.  Since the matter is neutral, the Fermi momenta of protons and electrons are the same,  $k_{F,p} = k_{F,e} $. 

A large number of experimental data on stable nuclei has obtained important results for symmetric nuclear matter ($Z=N$) such as an equilibrium number  density $\rho_0= 0.16$ nucleons/fm$^3$, and a binding energy per nucleon at saturation of $E/A = -16$ MeV, where $\epsilon(\rho)$ is the energy density. The saturation density $\rho_{0}$ corresponds to a Fermi momentum of $k_{F} = 263$ MeV/$c$,  small compared with $m_N = 939$ MeV/c$^2$ and thus justifying a non-relativistic treatment. A Taylor expansion of the energy per particle for asymmetric nuclear matter ($Z\ne N$) can be  done,
\begin{equation}
{E\over A}(\rho,\delta) = {E\over A}(\rho_{0},0)+{1\over 2} K_{0} x^{2}+{1\over 6} Q_{0} x^{3} + S(\rho)\delta^{2} + \cdots ,
\label{epsilon1}
\end{equation}
here $x=(\rho-\rho_{0})/3\rho_{0}$, $K_{0}$ is known as the incompressibility parameter, $Q_{0}$ the so-called skewness, and $S(\rho)$ is the symmetry energy which measures the contribution to the $E/A$ from the difference between $N$ and $Z$, parametrized as $\delta=(N-Z)/A$. 

For symmetric nuclear matter, the proton and neutron densities are equal, $\rho_n = \rho_p$, and the total nucleon density is $\rho = \rho_p + \rho_n = 2\rho_n$. The energy per nucleon is related to the energy density, $\epsilon(\rho)$, by means of $ \epsilon(\rho)= \rho E(\rho)/A $, which includes the nucleon rest mass, $m_N$. The density dependent function $E(\rho)/A - m_N$ has a minimum at $\rho = \rho_0$ with a value $E(\rho_{0})/A = -16$ MeV, obtained with
\begin{equation}
\frac{d}{d \rho}\left(\frac{E(\rho)}{A}\right) = \frac{d}{d \rho}\left(\frac{\epsilon(\rho)}{\rho}\right)= 0  {\ \rm\ at\ } \rho = \rho_0 \, . \label{eq:n0constraint}
\end{equation}
The EoS of homogeneous nuclear matter is the relation of the pressure and density, 
\begin{equation}
p(\rho,\delta)= \rho^{2}{d[\epsilon(\rho,\delta)/\rho] \over d\rho},
\end{equation}
with $\delta=(\rho_{n}-\rho_{p})/\rho$.

The value of incompressibility of nuclear matter, $K_{0}$, or the curvature of the energy per particle, or the derivative of the pressure  at the saturation density $\rho=\rho_0$,  is
\begin{eqnarray}
K_{0}= 9 \frac{d p(\rho)}{d \rho}=\left. 9\rho^{2}\frac{\partial^{2} (E/A)}{\partial\rho^{2}}\right|_{\rho_0}
=9 \left[ \rho^2 \frac{d^2}{d \rho^2} \left( \frac{\epsilon}{\rho} \right)\right]_{\rho_0} 
. \label{Kcomp}
\end{eqnarray} 
It has been extracted from the analysis of excitations of isoscalar giant monopole resonances in heavy ion collisions.

The symmetry energy $S(\rho)$ can also be expanded around $x=0$ 
\begin{equation}
S={1\over 2} \left. {\partial^2 E\over \partial \delta^2}\right|_{\delta=0}=J +Lx+{1\over 2} K_{sym}x^2 +\cdots \label{symenerg}
\end{equation} 
where $J=S(\rho_{0})$ is known as the bulk symmetry energy, $L$ determines the slope of the symmetry energy, and $K_{sym}$ is its curvature at the saturation density  $\rho=\rho_0$. The slope parameter is given by 
\begin{equation}
L=3\rho_{0}\left.{dS(\rho)\over d\rho}\right|_{\rho_{0}}. 
\end{equation}
Experimental values of masses, excitation energies and other nuclear properties have been compared to microscopic models yielding $J\approx 30$ MeV \cite{LI201829}.  The situation is not so clear for the slope parameter.  See Ref. \cite{RevModPhys.89.015007} for a recent review on the EoS of neutron stars.

\subsection{Microscopic theories of homogeneous nuclear matter}

\subsubsection{The EoS from Hartree-Fock mean field theory}

The analysis of experimental data is often done by comparison to microscopic calculations where, e.g.,  Skyrme interactions are used in traditional Hartree-Fock-Bogoliubiov calculations. Most of the Skyrme interactions are able to describe successfully a large number of nuclear properties, such as a  global description of nuclear masses and even-odd staggering of energies  \cite{Bertsch:PRC79.034306}.  Assuming that the nuclear interaction between nucleons $i$ and $j$  has a short range compared with the inter-nucleon spacing, the Skyrme interaction is an expansion which keeps at most terms quadratic in the nucleon momenta,
\begin{eqnarray}
v_{\rm Sk}&=&t_{0}\left( 1+x_{0}P_{\sigma} \right)\delta+{1\over 2}t_{1}(1+x_{1}P_{\sigma})(k^{2}\delta +\delta k'^{2}) \nonumber \\
&+& t_{2}\left( 1+x_{2}P_{\sigma} \right) {\bf k}\cdot \delta {\bf k}' + {1\over 6} t_{3}\left( 1+x_{3}P_{\sigma} \right) \rho^{\alpha} \delta \nonumber \\
&+& i W_{0} {\bf k}  ({\svec\sigma}_{i}+\svec\sigma_{j})\delta \cdot {\bf k}' \label{Skyrm}
\end{eqnarray}
where $\bf k$ and $\bf k'$ are the initial and final relative momenta of a colliding pair of nucleons, $P_{\sigma}$ is the spin-exchange operator between the two nucleons with spins $\svec\sigma$, $\delta = \delta({\bf r}_{i}-{\bf r}'_{j})$, ${\bf k}=(1/2i)(\svec\nabla_{I}-\svec\nabla_{j})$, acting on the wave function on the right,  ${\bf k}'$ is its adjoint, and $\rho({\bf r})$ is the local density at ${\bf r}={\bf r}_{1}+{\bf r}_{2}$. In this form, the Skyrme interaction has 10 parameters: $t_{i}$, $x_{i}$, $W_{0}$ and $\alpha$.  The Energy Density Functional, $E[\rho]$, can be easily obtained from the Skyrme functional yielding a precious dependence around the nuclear saturation density which can be used to infer properties of neutron stars \cite{BV:PRC.5.626}. 

In the Hartree-Fock  theory, the nuclear many-body wavefunction, $\Phi$, is described by a Slater-determinant of single-particle orbitals and obeys the Schr\"odinger equation  $H\Psi = E\Psi$. Each orbital wavefunction, $\phi_{k}$,  obeys an equation of the form
\begin{equation}
-{\hbar^{2} \over 2m}\nabla^{2}\phi_{k} + U({\bf r}_{1})\phi_{k}({\bf r}_{1})-\int d^{3}r_{2}W({\bf r}_{1},{\bf r}_{2})\phi_{k}({\bf r}_{2}) = \epsilon_{i}\phi_{k}({\bf r}_{1}), \label{HFeq}
\end{equation}
where $\epsilon_{k}$  are the single-particle energies. The potential $U({\bf r}_{1})$, is the direct contribution to the mean-field,
\begin{equation}
U({\bf r}_{1})=-\int d^{3}r_{2}v_{\rm Sk}({\bf r}_{1},{\bf r}_{2})\sum_{\epsilon_{k}<\epsilon_{F}}|\phi_{k}({\bf r}_{2})|^{2} ,
\end{equation}
where   $\epsilon_{F}$ is the Fermi energy. The third term in Eq. (\ref{HFeq}) arises due to the exchange potential
\begin{equation}
W({\bf r}_{1},{\bf r}_{2})=\sum_{\epsilon_{k}<\epsilon_{F}} v_{\rm Sk}({\bf r}_{1},{\bf r}_{2})\phi_{k}^{*}({\bf r}_{2})\phi_{k}({\bf r}_{1}).
\end{equation}
The set of N coupled equations for the N orbitals self-consistent field is nonlinear. One can solve them in an iterative way starting from some reasonable guess for a mean field where the particles are embedded, finding the single-particle eigenstates $\phi_{0k}$ and the eigenvalues $\epsilon_{0k}$, filling the lowest states and calculating the resulting fields $U$ and $W$ above. With the new Eq. (\ref{HFeq}) we repeat the whole cycle. However, the nonlinear HF equations have many solutions. Each of them provides a relative energy minimum as compared to the ``nearest neighbors" in the space of the Slater determinants. The iteration procedure does not determine which solution corresponds to an absolute minimum among Slater determinants. For example, spherical and deformed mean  field are possible as solutions with the same original interaction. An additional investigation should show which solution is energetically favorable. One can also look for the solutions with the distribution of the empty and  filled orbitals different from the normal Fermi gas in the ground state. For example, the solutions exist with some holes inside $\epsilon_F$. The self-consistent field determined by such a distribution of particles is different from the ground-state field. Therefore the single-particle states in these fields are not orthogonal which should be specially corrected.

Expressions for the energy density in infinite matter can be obtained by neglecting the Coulomb interaction and assuming plane waves for the orbitals. For uniform and spin-saturated nuclear matter, the EoS for arbitrary neutron and proton densities at zero temperature is given in terms of the Skyrme parameters by \cite{BV:PRC.5.626,Dutra:PRC.85.035201}
\begin{eqnarray}
\epsilon(\rho,\delta)&=&{3k_{F}^{2}\over 10 m}f_{5/3}+{1\over 8}t_{0}\rho\left[2(x_{0}+2)-(2x_{0}+1)f_{2}\right] \nonumber \\
&+&{1\over 48}t_{3}\rho^{\alpha+1}\left[2(x_{3}+2)-(2x_{3}+1)f_{2}\right] \nonumber \\
&+&{3\over 40}k_{F}^{2}\rho \Bigg\{ \left[t_{1}(x_{1}+2)+t_{2}(x_{2}+2)\right]f_{5/3}\nonumber \\
&+&{1\over 2}\left[t_{2}(x_{2}+2)-t_{1}(2x_{1}+1)\right]f_{8/3}\Bigg\}, \label{EoSSky}
\end{eqnarray}
where the first term is due to the kinetic energy density, $k_{F}=\left(3\pi^{2}\rho/2\right)^{2}$ and $f_{m}=[(1+\delta)^{m}+(1-\delta)^{m}]/2$.

Pairing is an important part of the total energy, in particular leading to phenomena such as the odd-even staggering effect of nuclear binding energies, and  the pairing energy needs to be added to Eq. (\ref{EoSSky}). It can be included in the HF method by using the Bardeen-Cooper-Schrieffer (BCS) theory \cite{BM:PR.110.936}, or its Hartree-Fock-Bogoliubov (HFB) extension \cite{blaizot1986quantum,Fetter2003Quantum}. The BCS theory is better described in the operator formalism. Then the ground state of the nucleus is defined as
\begin{equation}
\left|BCS\right> =\Pi_{k>0} \left( u_{k}+v_{k}a^{\dagger}_{k}a^{\dagger}_{\bar k}\right)\left|-\right>, \label{BCS}
\end{equation}
where $\left|-\right>$ is the vacuum with no particles, normalized as $\left<0|0\right> = 1$, $a^{\dagger}_{k}$ is the creation operator of the state $\left|k\right>$, $a^{\dagger}_{\bar k}$ is the creation
operator of the state 
\begin{equation}
\left|{\bar k}\right> =-1^{(j_{k}-m_{k})}\left| \epsilon_{k};n_{k}j_{k}l_{k},-m_{k}\right>,
\end{equation}
where $(n_{k}j_{k}l_{k},m_{k})$ are single-particle quantum numbers, $|v_{k}|^{2}$ is the probability that the state $\left|k\right>$ is occupied, and is related to $u_{k}$ by the condition $|u_{}|^2 + |v_{k}|^2 = 1$, and $u_{\bar k} = u_{k}$ and $v_{\bar k} = -v_{k}$.
The BCS state defined in Eq. (\ref{BCS}) is not an eigenstate of the particle-number operator and the BCS equations are obtained by applying the variational principle to the expectation value of the operator $H = H - \lambda N$ , where $H$ is the nuclear hamiltonian, $N$ is the particle number operator and $\lambda$ is a Lagrange multiplier. The parameters to be changed in the variational procedure now are the $v_{k}$ and $u_{k}$ coefficients. The application of the variational principle implies to get the  solutions of the set of BCS equations
\begin{equation}
\left(u_{k}^{2}+u_{k}^{2}\right)\Delta_{k}=2 u_{k}v_{k}\eta_{k} 
\end{equation}
where
\begin{equation}
\Delta_{k}=-{1\over \sqrt{2j_{k}+1}}\sum_{m}\sqrt{2j_{m}+1}u_{m}v_{m}\left<mm0|v_{{pair}}|kk0\right>,\label{uv2}
\end{equation}
and $\eta_{k}=\left<k|T|k\right>-\lambda$, where $T$ is the kinetic energy and a renormalization term was dropped. The matrix element with the pairing interaction, $v_{pair}$,  indicates that the single-particle wave functions are coupled to angular momentum zero.

The solution of the BCS equations provides the values of $v_{k}$ and $u_{k}$, which together with $\phi_{k}$ allows the evaluation of the expectation values of various ground state quantities related to the BCS ground state. The pairing added in the BCS model does not change the nuclear  field which means a lack of self-consistency.
The more general approach which is the most advanced version of the self-consistent HFB approximation \cite{blaizot1986quantum,Fetter2003Quantum}, widely used in condensed-matter physics in order to take into account simultaneously effects of Coulomb forces or impurities and superconducting pairing.  Pairing correlations are very important to describe single-particle motion and collective modes as rotation. Pairing modifies the distribution of particles over orbitals considerably.  As a result, the self-consistent  field is different than without pairing, influencing nuclear shapes, moments of inertia, mass parameters, transition probabilities and reaction cross sections. Since nuclear shape defines single-particle orbits and conditions for pairing, a complex interplay of various residual interactions is not accounted for in the HF approach which does not include pairing correlations on equal footing. The pairing added in the BCS model does not change the nuclear  field which means a lack of self-consistency. Therefore, the HFB approximation is required for accuracy in the description of nuclear properties. The HFB approximation a general canonical (Bogoliubov) transformation and introducing the concept of quasi-particles which does not conserve the particle number. In the second variant of HFB calculations (HFB+LN), one performs an approximate particle number projection using the Lipkin-Nogami (LN) method. For more details on the HFB approach, we refer to Refs. \cite{blaizot1986quantum,Fetter2003Quantum}.

A common parametrization of the pairing interaction is given by \cite{MSH:PRC.76.064316,BLS:PRC.85.014321}
\be
v_{pair}(1,2)= {v}_0 \,{g}_\tau[\rho,\beta\tau_z]\,\delta(\svec r_1-\svec r_2),
\label{eq:pairing_interaction} \ee
where $\rho=\rho_n+\rho_p$ is the
nuclear density and $\beta$ denotes here the asymmetry parameter $\beta=(\rho_n -\rho_p)/\rho$. One can introduced an isospin-dependence of the pairing interaction  through  the density-dependent term, $\mathrm{g}_\tau$,  determined by the  pairing gaps in nuclear matter. A convenient functional form, usually termed isoscalar + isovector pairing, is
\be
{g}_\tau[\rho,\beta\tau_z] =  1
-{f}_{s}(\beta\tau_z)\eta_{s} \left(\frac{\rho}{\rho_0}\right)^{\alpha_{s}} -{f}_{n}(\beta\tau_z)\eta_{n} \left(\frac{\rho}{\rho_0}\right)^{\alpha_{n}},
\label{eq:g1t}
\ee 
where $\rho_0$ is the saturation density of SNM and  ${f}_{s}(\beta\tau_z)=1-{f}_{n}(\beta\tau_z)$
with ${f}_{n}(\beta\tau_z)=\beta\tau_z=\left[\rho_{n}({\bf r})-\rho_{p}({\bf r})
\right]\tau_z/\rho({\bf r})$, with $\tau_{z}$ being the isospin operator. This form of the paring interaction introduces 5 additional parameters, $v_{0}$, $\eta_{s(n)}$ and $\alpha_{s(n)}$ in the HF calculation. Most often only two parameters are used, one for strength, and another for the radial dependence of the pairing interaction \cite{Bertsch:PRC79.034306}. 

The density dependence of the pairing interaction replicates the effect of pairing suppression at high density (momenta). The parameters $V_0, \eta$ are chosen so that one can describe different types of pairing called $volume$, $surface$ and $mixed$ pairing, reflecting pairing fields localized in the volume,  surface, or a mix of the two. The \textit{volume} interaction is not dependent on the density ($\eta=0$), and for this reason it is easier to handle.  The nuclear compressibility is strongly sensitive to the surface properties of the nucleus \cite{BLAIZOT1980171}.  Theoretical models using couplings with collective vibrations require pairing fields peaked at the surface of the nucleus \cite{PastorePRC2008} and, e.g., Ref. \cite{Bertsch:PRC79.034306} has shown that \textit{surface} pairing ($\eta=1$) reproduces nuclear masses with better accuracy as compared to other parametrizations. 

The density dependence of the pairing interaction gives rise to a rearrangement term in the single-particle Hamiltonian. This is because the energy functional in a Skyrme  approximation has the form: $ E= {E}_{kin}+ {E}_{Skyrme}+ {E}_{pair} + {E}_{Coul}$ with the respective kinetic, Skyrme, pairing, and  Coulomb terms. In the HFB method the single particle Hamiltonian is obtained from a functional derivative of the energy with respect to the density; and the contribution from ${E}_{pair}$ is usually called rearrangement term \cite{Avogadro:PRC.88.044319},
\begin{equation}
 h=  \frac{\delta {E}_{kin}}{\delta \rho} + \frac{\delta {E}_{skyrme}}{\delta \rho} + \frac{\delta {E}_{pair}}{\delta \rho}  + \frac{\delta  {E}_{coul}}{\delta \rho}.
\end{equation}
Similarly, the residual fields giving rise to the Quasi-particle Random Phase Approximation (QRPA) matrix (see below) are functions of the second derivatives of the densities and the rearrangement term is:
\begin{equation}
 \frac{\delta h_{rearr}}{\delta \rho} = \frac{ \delta } {\delta \rho } \left(  \frac{\delta  {E}_{pair} }{\delta \rho} \right). 
\end{equation}
Without an explicit density dependence of the pairing term, no rearrangement term appears either in the HFB or in the QRPA matrix, as is the case of the \textit{volume} pairing. But \textit{mixed} and \textit{surface} pairing parametrizations give rise to a non zero rearrangement term. 

 \begin{figure}[t]
\begin{center}
\includegraphics[height=2.6in]
{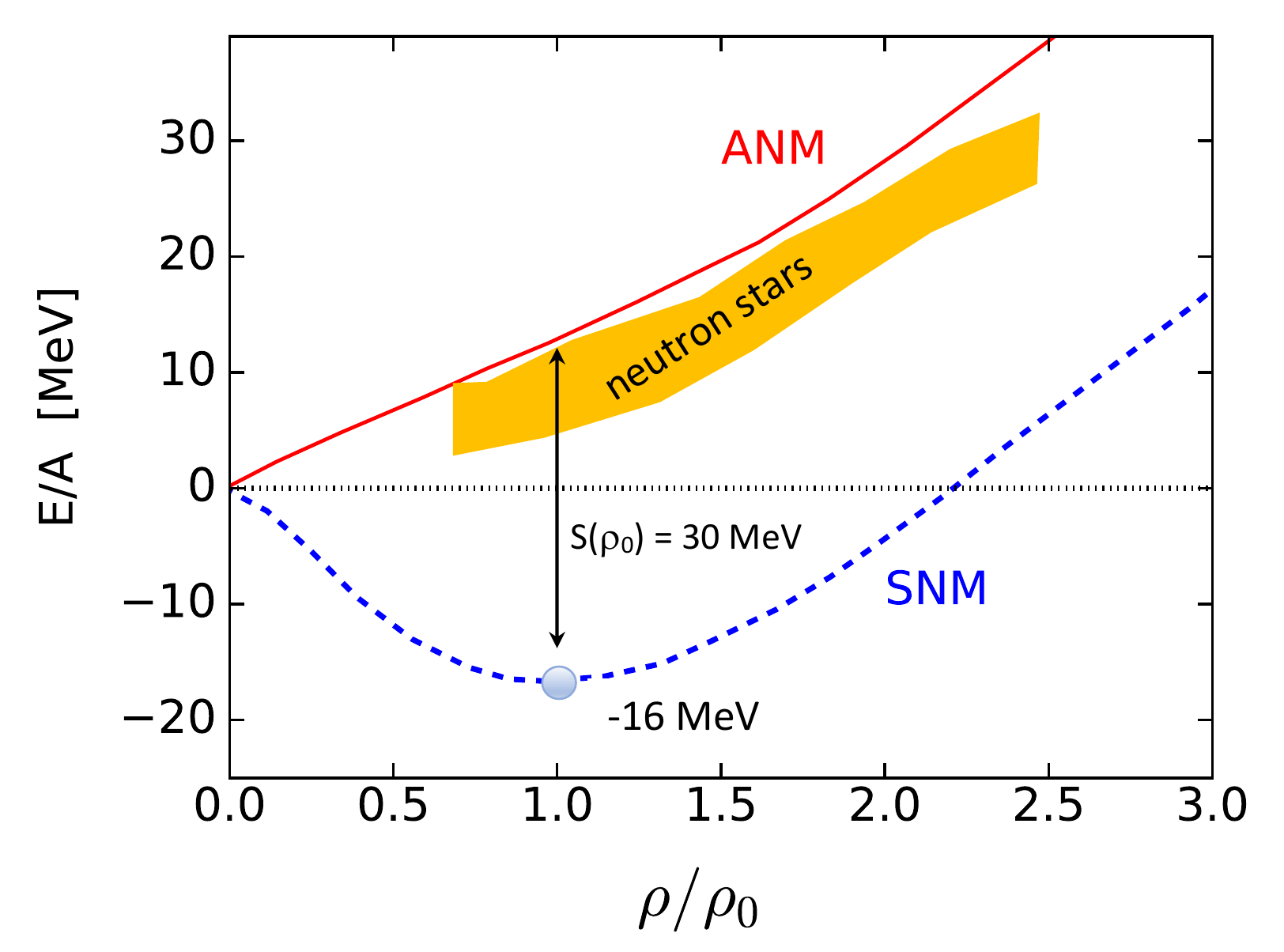} 
 \caption{Skyrme (SLy4) Hartree-Fock calculation of the nuclear matter (SNM) equation of state (EoS) compared to the result pure (ANM) neutron matter. The two curves are roughly separated by the bulk symmetry energy factor $S(\rho_0)$. The EoS for neutron stars is likely to be somewhat different than the ANM curve shown and lie somewhere within the hypothetical region shown in the figure.}
\label{EOSfig}
\end{center}
\end{figure}

In Fig. \ref{EOSfig} we show a Skyrme (SLy4) Hartree-Fock calculation of the nuclear matter (SNM) equation of state (EoS) compared to the result pure (ANM) neutron matter. The two curves are roughly separated by the bulk symmetry energy factor $S(\rho_0)$. The EoS for neutron stars is likely to be somewhat different than the ANM curve shown and lie somewhere within the hypothetical region shown in the figure.

\subsubsection{The EoS in relativistic mean field models}

Relativistic Mean Field (RMF) theories \cite{Serot:1986,Reinhard_1989,RING1996193,MengPRL.77.3963} are another way to calculate nuclear-matter properties. RMF is based on using the Euler-Lagrangian equations using a  Lagrangian for nucleon and mesons fields and their interactions. The Lagrangian density with nucleons described as Dirac particles interacting via the exchange of mesons and the photon $A^\mu$. The usual mesons considered are the scalar sigma ($\sigma$), which represents a large attractive field resulting from complex microscopic processes, such as uncorrelated and correlated two-pion exchange,  the
vector omega ($\bf \omega$) describing the short-range repulsion between the nucleons, and the iso-vector vector rho ($\ivec\rho$) carrying the isospin.  The ($\ivec\rho$) meson is responsible for  the  isospin asymmetry.  With the nucleon mass denoted by $m$ and the corresponding lessons masses  (coupling constants) denoted by $m_\sigma$ ($g_\sigma$), $m_\omega$ ($g_\omega$) and $m_\rho$ ($g_\rho$), one has (with $\hbar = c = 1$)
\begin{eqnarray}
{\cal L}&=&\bar{\psi}\ls i{\gamma^\mu}{\partial_\mu}-m-{g_\sigma}\sigma
 - g_\omega\gamma^\mu\omega_\mu - g_\rho \gamma^\mu {\ivec\tau}\cdot \ivec\rho_\mu \right. \nonumber \\
&-& \left. e\gamma^\mu\frac {1-\tau_3}{2}A_\mu \rs\psi \nonumber \\
&+&\frac{1}{2}\partial^\mu\sigma\partial_\mu\sigma-\frac{1}{2}m_\sigma^2\sigma^2 
-\frac{1}{4}\omega^{\mu\nu}\omega_{\mu\nu}
+\frac{1}{2}m_\omega^2\omega^\mu\omega_\mu \re
&-&\frac{1}{4}\ivec\rho^{\mu\nu}\cdot\ivec\rho_{\mu\nu}
+\frac{1}{2}m_\rho^2\ivec\rho^\mu\cdot\ivec\rho_\mu
-\frac{1}{4}A^{\mu\nu}A_{\mu\nu}, \label{rmfl}
\end{eqnarray}
where $\gamma^\mu$ are Dirac gamma matrices and
\begin{eqnarray}
\omega^{\mu\nu}&=&\partial^\mu\omega^\nu-\partial^\nu\omega^\mu\re
\ivec\rho^{\mu\nu}&=&\partial^\mu\ivec\rho^\nu-\partial^\nu\ivec\rho^\mu
\re
A^{\mu\nu}&=&\partial^\mu A^\nu -\partial^\nu A^\mu.
\end{eqnarray}
The arrows  are used to denote  isospin vectors. The basic RMF theory as stated above contains 4 parameters, namely, the $\sigma$ mass and the 3 nucleon-meson coupling constants ($g_\sigma, g_\omega, g_\rho$). The other masses are taken as their free values. Effective density dependences are often included through meson self-interaction terms. As in the HF-Skyrme case,  the remaining  parameters are determined by the fitting experimental observables.

The Dirac equation for a single nucleon is obtained by the variation of the Lagrangian density with respect to
$\bar\psi_{i}$, yielding
 \beq
 \ls \mbox{\boldmath$\alpha$}\cdot {\bf p} + \beta \left( m + \Sigma_{\mu}+\Sigma_S\right) \rs\psi_{i}=E_{i} \psi_{i}
 \eeq
where $\mbox{\boldmath$\alpha$} = \gamma_0 \mbox{\boldmath$\gamma$}$, and the nucleon self-energies $\Sigma_\mu$ and $\Sigma_S$ are
 \beqn
&&\Sigma_\mu = 
g_\omega\omega_\mu + g_\rho\ivec\tau\cdot\ivec\rho_\mu + e\frac{1-\tau_3}{2}A_\mu + \gamma_{\mu}j^{\mu} {1\over n}{\partial g_{\rho} \over  \partial n}\rho_{s}\sigma 
\nonumber \\
&&\Sigma_S   = g_\sigma\sigma .
 \eeqn
 The last term in $\Sigma_{\mu}$  is a rearrangement term due to the density dependence of the coupling between the sigma meson and the nucleon, and $\rho_{s}$ is the scalar density of nucleons, defined below.
The variation of the Lagrangian density with respect to the meson field operators yields the Klein-Gordon equations for mesons,
\beqn
\ls-\Delta + m_\sigma\rs \sigma &=& - g_\sigma\rho_s - g_2\sigma^2 - g_3\sigma^3\\
\ls-\Delta + m_\omega\rs \omega^{\mu} &=&   g_\omega j^{\mu} - c_3\omega^{\mu}(\omega^\nu \omega_{\nu})\\
\ls-\Delta + m_\rho  \rs \rho^{\mu}   &=&   g_\rho \bar{j}^{\mu}.
\eeqn
The nucleon spinors provide the relevant source terms
\beqn
\rho_s &=& \sum_i \bar{\psi}_i\psi_i n_i , \ \  j^{\mu}= \sum_i \bar{\psi}_i \gamma^{\mu} \psi_i n_i , \nonumber \\
\bar{j}^{\mu}&=& \sum_i \bar{\psi}_i \gamma^{\mu} \vec{\tau} \psi_i n_i , 
\eeqn
In infinite matter and at finite temperature, the Fermi-Dirac statistics imply that the occupation numbers $n_{i}$ of protons and neutrons are
\begin{equation}
n_{i}=\left( e^{(E_{i}-\mu)/k_{B}T} +1\right)^{-1},
\end{equation}
where $\mu$ is the chemical potential for a neutron (proton). For the EoS of neutron stars, one is mostly interested in $T\sim 0$. For a
spherical nucleus, there are no currents in the nucleus and the spatial vector components of $\omega_{\mu}$, $\rho_{\mu}$, and $A_{\mu}$ vanish. Thus, only the time-like components, $\omega_{0}$, $\rho_{0}$, and $A_{0}$ remain.

The introduction of pairing in RMF  is usually done similarly as in the  HF theory by using RMF + BCS, or an HFB-like procedure, as described previously. 
The energy density of uniform nuclear matter is, for $T=0$
\beqn
\epsilon &=& \sum_{i=n,p} \epsilon^{i}_{kin} + {1\over 2} \left[m_{\sigma}^2 \sigma^{2} + m_{\omega}^2 \omega_{0}^{2}+m_{\rho}^2 \rho^{0,3}\right]
\nonumber \\
&+&{1\over 3} g_{2}\sigma^{3}+{1\over 4} g_{3}\sigma^{4}+{1\over 3} c_{3}\omega_{0}^{2},
\eeqn
where
\beqn
\epsilon^{i}_{kin}={2\over (2\pi)^{3}}\int_{|k|<k_{F}} d^3 k \ E(k),
\eeqn
with effective mass $m^{*}=m+g_{\sigma}\sigma$, and $E(k)=\sqrt{k^{2}+m^{*2}}$. 

 \begin{figure}[t]
\begin{center}
\includegraphics[height=2.5in]
{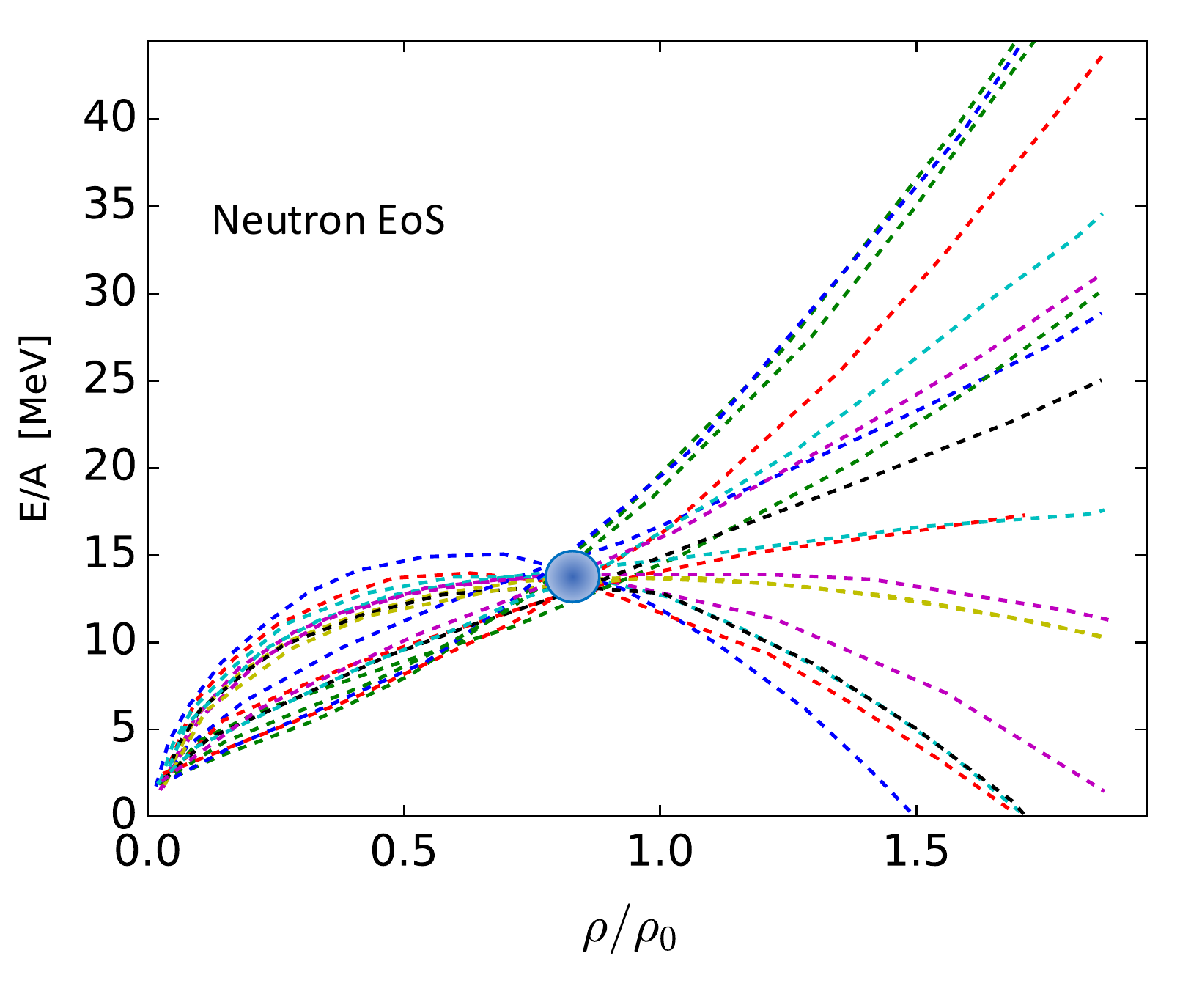} 
 \caption{Equation of state of pure (EoS) neutron matter as predicted by numerous Skyrme interactions. At the neutron saturation density they tend to agree, but diverge substantially as they depart from it. Such dispersive behavior of the numerous Skyrme parametrizations used in numerical calculations of the EoS was observed quite early \cite{PethickRavenhall.ARNPS1995,Brown.PRL.2000}. The relativistic mean field parameterizations also show a similar behavior for the several parameterizations used to describe nuclear properties.}
\label{ANMfig}
\end{center}
\end{figure}

In Fig. \ref{ANMfig} we show the equation of state of pure (EoS) neutron matter as predicted by numerous Skyrme interactions. At the neutron saturation density they tend to agree, but diverge substantially as they depart from it. Such dispersive behavior of the numerous Skyrme parametrizations used in numerical calculations of the EoS was observed quite early \cite{PethickRavenhall.ARNPS1995,Brown.PRL.2000}. The relativistic mean field parameterizations also show a similar behavior for the several parameterizations used to describe nuclear properties.

\subsubsection{Linear response theory and collective excitations}
Nuclear collective excitations are usually studied theoretically using the Random Phase Approximation (RPA), appropriate to describe  collective modes of small amplitude in a quantum many-body system. For  a small perturbation, the many-body ground state  is not a stationary state any longer. Instead it is a wave packet changing in time with different Fourier components of the packet. If the perturbation $V_{ext}$ is weak, the main component still corresponds to the ground state and excited states are admixed to it by the perturbation (linear approximation). Changing the frequency $\omega$ of the perturbation yields resonances in the response. They appear when real transitions with energy conservation, yielding the excitation spectrum of the system. In the next order, transitions between the excited states appear which in general show no resonance with the external  field. This is the (neglected in the RPA) ``noise" superimposed onto the coherent response, giving the origin of the term ``random phases".

The standard form of the RPA equations can be reached if one considers the ground state  filled as in the Fermi gas when the basis can be
subdivided into the particle ($p$) states, $n_{p} = 0$ if  $\epsilon_p > 0$, and hole ($h$) states, $n_{h} = 1$ if $\epsilon_h < 0$. If one considers the set $(ph) =k$ as a unified label of the $p$-$h$ excitation and introduces the notations for the matrix elements,
\be
(\rho_{\omega})_{ph}=X_{k}^{\omega},  \ \ \ \ \ \ \ (\rho_{\omega})_{hp}=Y_{k}^{\omega},
\label{densom}
\ee
where $\rho_{\omega}$ is a time independent matrix (Fourier transform of $\rho$, the density matrix) with matrix elements labeled by the pairs (ph) of particle-hole states. In terms of the HF ground state, $\left|0\right>$, the density matrix expressed in an arbitrary basis is $\rho_{ph}=\left<0|a^{\dagger}_{h}a_{p}|0\right>$.

The canonical RPA equations are
\be
 \begin{pmatrix}
A      & B   \\
 B^{*}     & A^{*} 
\end{pmatrix} 
 \begin{pmatrix}
 X^{\omega}   \\
Y^{\omega}
\end{pmatrix} 
= \hbar \omega 
 \begin{pmatrix}
1      & 0   \\
0     & -1
\end{pmatrix} 
 \begin{pmatrix}
 X^{\omega}   \\
Y^{\omega}
\end{pmatrix}
\label{rpaeq}
\ee
with the transition amplitudes,
\begin{eqnarray}
A_{kk'} &\equiv& A_{ph,p'h'}=\epsilon_{k}\delta_{kk'}+ \left<ph'|V_{ext}|p'h\right> \nonumber \\
B_{kk'} &\equiv& B_{ph,p'h'}=\left<pp'|V_{ext}|h'h\right>,
\label{rpaeq2}
\end{eqnarray}
Where $\epsilon_{k} = \epsilon_{p} -\epsilon_{h}$ is the energy of the bare p-h excitation. This eigenvalue problem does not correspond to the diagonalization of a hermitian operator (because of the matrix $\sigma_{z}$ in the right hand side). Therefore, one cannot guarantee that the eigenvalues $\hbar \omega$ are real. An imaginary part of frequency would signal the instability of the mean field placed in the base of the RPA.
The backward amplitude $Y_{k}$ incorporates correlations in the ground state and reveal the presence of the holes below $\epsilon_{F}$ and of the particles above $\epsilon_{F}$ in the actual ground state which does not coincide with the HF vacuum.

As in the HBF theory, the Quasi-particle Random Phase Approximation (RPA) is based on Bogoliubov transformed operators ${b,b^{\dagger }}$ that are introduced to annihilate and create a ``quasiparticle''  with a well-defined energy, momentum and spin but as a quantum superposition of particle  and hole state. They carry coefficients $u$ and   $v$ given by the eigenvectors of a Bogoliubov matrix. The framework of the time dependent superfluid local density approximation (TDSLDA) \cite{Bulgac1288,StetcuPRC.2011,BulgacARNPS2013,StetcuPRL2015,BulgacPRL2016,BulgacPSS2019}  is more powerful  because it allows one to study large amplitude collective motion.  This is an extension of the Density Functional Theory (DFT) to superfluid nuclei that can assess the response to external time-dependent fields.  The time evolution of the nucleus  is determined by the time-dependent mean field
\begin{eqnarray}
\lefteqn{i\hbar\frac{\partial}{\partial t} 
\left  ( \begin{array} {c}
  U({\bf r},t)\\  
  V({\bf r},t)\\ 
\end{array} \right ) \nonumber }\\
&& =
\left ( \begin{array}{cc}
h({\bf r},t)-\mu&\Delta({\bf r},t)\\
\Delta^*({\bf r},t)&-h^* ({\bf r},t)+\mu
\end{array} \right )  
\left  ( \begin{array} {c}
  U({\bf r},t)\\
  V({\bf r},t)\\ 
\end{array} \right ),
\end{eqnarray}
where $h(\mathbf{r},t)=\partial E(\mathbf{r},t)/\partial \rho$ is the single-particle Hamiltonian, $\Delta({\bf r},t)=\partial E(\mathbf{r},t)/\partial \rho_{{pair}}$ is the pairing field, and $\mu$ is the chemical potential. Both are calculated self-consistently using an
energy functional such as a Skyrme interaction.   A Fourier transform of the time evolution of the nuclear density yields the energy spectrum.

\subsection{The incompressibility modulus and skewness parameter of nuclear matter}
Nuclei display collective excitation modes which have been studied for more than 6 decades.  Baldwin and Klaiber  \cite{BaldwinPR71.3,BaldwinPR73.1156} reported the existence of highly collective modes in photoabsorption experiments. A decade before that, Bothe and Gentner \cite{Bothe1937} had already  obtained very large cross sections for the photoproduction of radioactive elements on numerous targets, by two orders of magnitude larger than predicted by theory, what was a first indication that such resonances might exist. They used high-energy (15 MeV) photons at the time. We now understand that the origin of these absorption resonances are due to the collective response of the nuclei to the  photon electric dipole (E1). The dipole  collective states were also earlier predicted by Migdal \cite{migdal1945quadrupole}. 
Later, giant resonances were further studied with photoabsorption processes with increased accuracy Refs.  \cite{BermanRMP47.713}. The centroid of giant resonances  are located above the particle emission threshold, mostly decaying by neutron emission. The large Coulomb barrier in nuclei usually prevents charged particle decay. Therefore, photoabsorption cross sections are usually observed from neutron yields  \cite{Aumann2005,AumannPRC.51.416}.  

 \begin{figure}[t]
\begin{center}
\includegraphics[height=2.5in]
{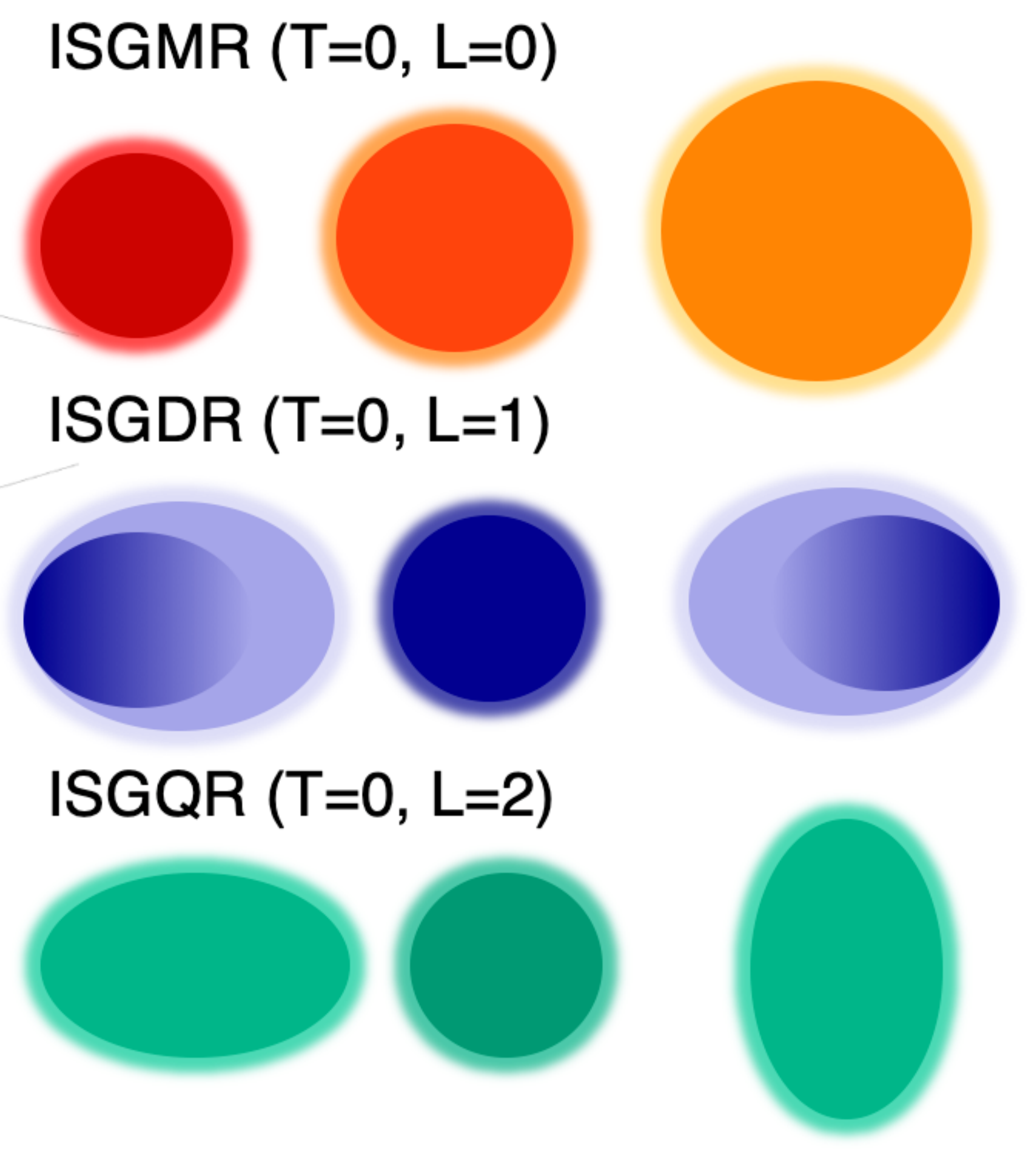} 
 \caption{Schematic view of giant resonance vibrations in nuclei. The isoscalar giant monopole resonance (ISGMR) in a spherical nucleus consists of a breathing mode-like vibration, the Isovector Giant Dipole Resonance (IVGDR), here shown for a deformed nucleus, consists of vibrations of protons against neutrons, and the Isoscalar Giant Quadrupole Resonance is like a prolate-oblate vibration mode in the nucleus.}
\label{giantR}
\end{center}
\end{figure}

 The incompressibility of nuclear matter, Eq.(\ref{Kcomp}), has been extracted from several experiments. The giant resonances, very collective nuclear vibrations, of several multi polarities (see Figure \ref{giantR}) has been very valuable for such studies. The isoscalar giant monopole resonance (ISGMR) in a spherical nucleus consists of a breathing mode-like vibration, while the Isovector Giant Dipole Resonance (IVGDR), here shown for a deformed nucleus, consists of vibrations of protons against neutrons, and the Isoscalar Giant Quadrupole Resonance is like a prolate-oblate vibration mode in the nucleus. In particular, the study of the isoscalar giant monopole resonance (ISGMR) has been one of the best tools \cite{YoungbloodPRL77} to extract the magnitude of $K_0$ \cite{ringschuck80}. Results in the range $K_{0} = 210-240$ MeV have emerged from such analysis 
\cite{Colo:PRC.70.024307,Piekarewicz:PRC.69.041301,BLAIZOT1980171,Avogadro:PRC.88.044319,Khan:PRL.109.092501}.  Not only Skyrme-type calculations, but also those based on a relativistic mean field approach, e.g., in Ref. \cite{NikisicPRC.2008}  show  good estimates of the ISGMR centroids with the same value of the incompressibility modulus.

The energy centroid of the ISGMR is related with the nuclear incompressibility modulus $K_A$ which is the equivalent for a nucleus of the elastic constant of a spring. Nuclei with a low value of $K_A$ are called  ``soft nuclei''  and those with high values of $K_A$ as called ``stiff nuclei''. The relation between $K_A$ and the energy centroid of the ISGMR is  \cite{BLAIZOT1980171}
\begin{equation}
  E_{ISGMR} =\hbar \sqrt{\frac{ K_A}{m_{N}\langle r^2 \rangle}},
\end{equation}
where $m_{N}$ is the nucleon mass, and $\langle  r^2 \rangle $  is the mean square radius of the nucleus.

Extracting the value of the incompressibility of nuclear matter $K_{0}$  from $K_A$ is strongly model dependent \cite{BLAIZOT1980171, ShlomoPRC1993}. Therefore, the most reliable way to obtain $K_0$ is from a microscopic calculation which reproduces accurately the experimental data for  a large number of nuclei. To extract the centroid of the giant resonance ($E_{ISGMR}$)
one uses the ratios $m_1/m_0$, $\sqrt{m_1/m_{-1} }$ and $\sqrt{m_3/m_1}$, where the moments are defined as
\begin{equation}\label{eq:moment}
 m_k= \int_0^{\infty} E^{k} S(E)dE, 
\end{equation}
where the strength is 
\begin{equation}
S(E)= \sum_j |\langle 0|F_0|j \rangle|^{2}\delta(E-E_j), \label{sfunc}
\end{equation}
 where $|0\rangle$ is the ground state and $|j\rangle$ is an eigenstate with energy $E_j$ of the QRPA. 
The monopole operator is of the form:
\begin{equation}
F_0 = \sum_{i=1}^{A} r_i^2 .
\end{equation}
The integral of Eq. (\ref{eq:moment}) should run from zero to infinity, but, in practice, the QRPA calculations yield discrete values for the eigenstates and the integral of Eq. (\ref{eq:moment}) reduces to a finite sum.

The results of HFB calculations reported on Refs. \cite{Avogadro:PRC.88.044319,Bertulani.PRC.100.015802} are shown in Table \ref{NMp}.
20\% of the nuclei investigated in Refs.  \cite{Avogadro:PRC.88.044319,Bertulani.PRC.100.015802} are well explained with the SLy5 interaction, and 10\% with the SkM* interaction. For the majority of the nuclei under investigation  the centroid energy of the ISGMR is better reproduced using the soft interaction Skxs20 ($K_{0}\approx$ 202 MeV) in contrast to the generally accepted value for $K_{NM}\approx 230$ MeV.  Therefore, there is still some uncertainty in the generally accepted value of the incompressibility of nuclear matter.

The  skewness parameter $Q_{0}$ is more difficult to ascertain and estimated values in the range $-500  \leq Q_{0} \leq  100$ MeV \cite{Cai2017}. The lack of accurate knowledge of the high powers in the Taylor expansion means that densities extrapolated to values well below and well above the saturation density tend to yield conflicting results \cite{Brown:PRL.85.5296}. Therefore, there is a strong interest in the literature to pinpoint those Skyrme interactions that better describe neutron matter properties.

\subsection{The slope parameter of the EoS}
Both the HFB as well as the RMF method are the abundantly used in the literature to describe nuclear properties and to study, by extrapolation, the properties of neutron stars. As an example, we show in Table \ref{NMp} the predictions of a few Skyrme models (there are hundreds of them)   \cite{BEINER197529,DOBACZEWSKI1984103,REINHARD1995467,Brown:PRC.58.220,CHABANAT1998231,Reinhard:PRC.60.014316,GORIELY2005425,Dutra:PRC.85.035201} for some basic quantities needed for the EoS of neutron stars. Whereas $K_{0}$ and $J$ tend to consistently agree among the models, the slope parameter $L$ can differ widely. 

\begin{table}[ht]
\begin{center}
\begin{tabular}{|c|c|c|c||c|c|c|c|c|}
\hline\hline
 & $K_{0}$ & $J$   &  $L$ &  & $K_{0}$ & $J$   &  $L$   \\ 
\hline 
SIII & 355. & 28.2 & 9.91& SLY5 & 230. & 32.0 & 48.2 \\
\hline 
SKP & 201. & 30.0 & 19.7& SKXS20 & 202. & 35.5 & 67.1\\
\hline
SKX & 271. & 31.1 & 33.2 & SKO & 223. & 31.9 & 79.1 \\
\hline
HFB9 & 231. &  30.0 & 39.9   & SKI5 & 255. & 36.6 & 129.      \\
\hline
\hline
\end{tabular}
\caption{Predictions by some Skyrme models for the properties of nuclear matter at the saturation density  in MeV units. 
The parameters for the Skyrme forces were taken from \cite{BEINER197529,DOBACZEWSKI1984103,REINHARD1995467,Brown:PRC.58.220,CHABANAT1998231,Reinhard:PRC.60.014316,GORIELY2005425,Dutra:PRC.85.035201}}. \label{NMp}
\end{center}
\end{table}

Because the  theoretical models from Refs. \cite{BEINER197529,DOBACZEWSKI1984103,REINHARD1995467,Brown:PRC.58.220,CHABANAT1998231,Reinhard:PRC.60.014316,GORIELY2005425,Dutra:PRC.85.035201}, as well as from numerous other Skyrme and RMF models,  have been fitted to reproduce nuclei at laboratory conditions, it is far from clear which value of $L$ should be adopted  in the EOS of  neutron stars. Some constraints based on observations and new laboratory experimental data have been used in a few references, see, e.g., Ref. \cite{Dutra:PRC.85.035201} to eliminate some ``outdated'' Skyrme or RMF functionals. But there is still a long way to go to reach a consensus on what are the best models to be extrapolated to neutron stars.

In fact, it is easy to understand the strong dependence of the neutron EoS on the symmetry energy  $S$. For pure neutron matter,  $\delta =1$,  and at the saturation density $\rho \sim \rho_0$ one has $p = L\rho_0/3$. Hence, the neutron matter and its response to gravitational pressure is directly connected to the slope parameter $L$. It is also worthwhile to mention that the explosion of a core-collapse supernova is strongly depends on the symmetry energy of the nuclear EOS. For more discussion of these subjects see, e.g., Refs. \cite{Lattimer536,glendenning2000compact,WEBER2005193,Haensel:2007yy,Lattimer:2012,bertulani2012neutron,Hebeler_2013,Bender:RMP.75.121,Bertsch:PRC79.034306,Goriely:PRC89.054318,NIKSIC2011519,Avogadro:PRC.88.044319,Matsuo:PRC.91.034604,Zhao:PRL.115.022501,Liang:PRL.101.122502,Meng:PRC.73.037303,MENG2006470,Zhou:PRL.91.262501,Meng:PRL.77.3963,Meng:PRL.80.460,Zhao:PRL.107.122501}.  

\subsubsection{Nuclear radii}
\begin{figure}[t]
\begin{center}
\includegraphics[height=1.in]
{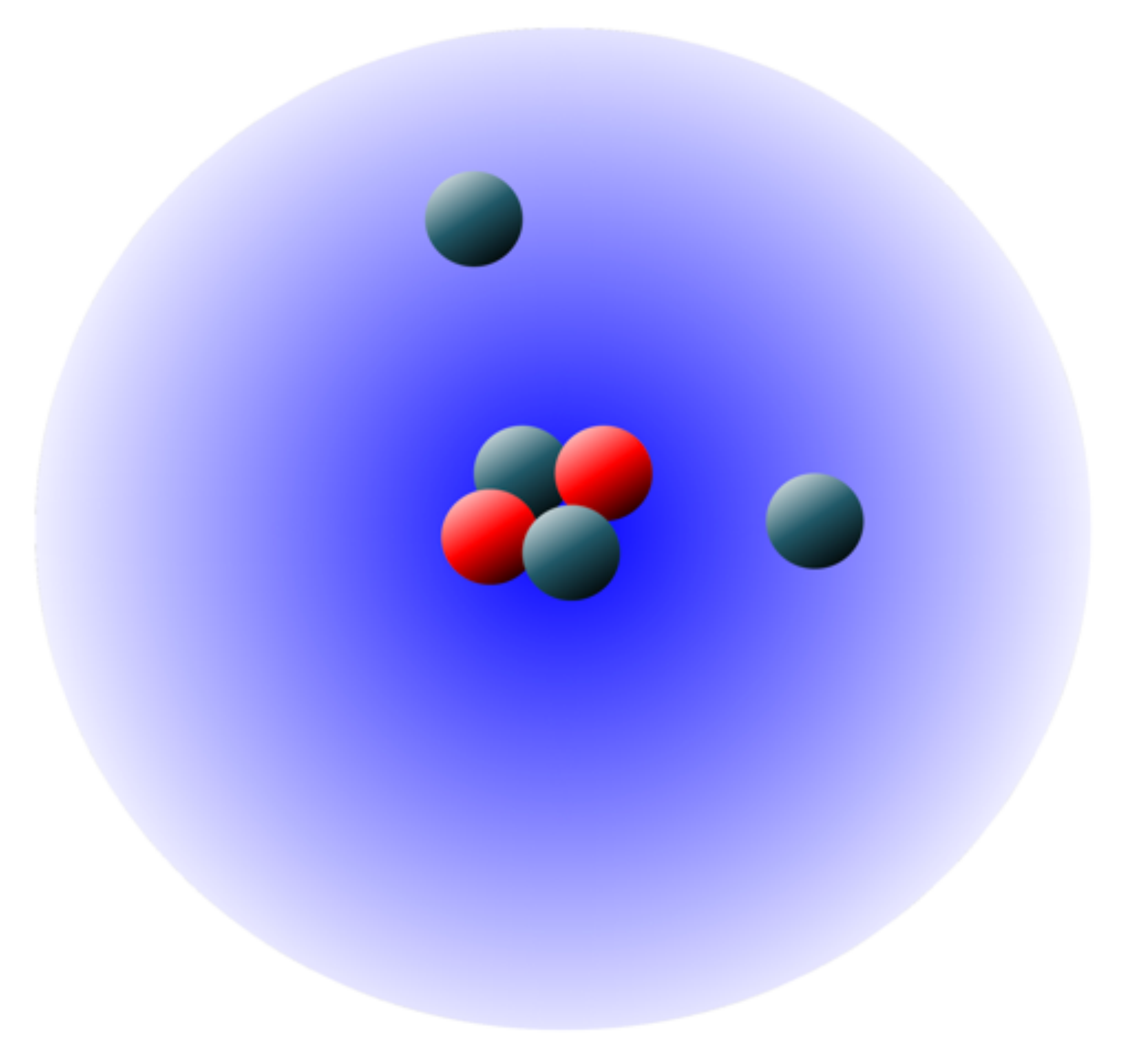} 
\includegraphics[height=2.5in]
{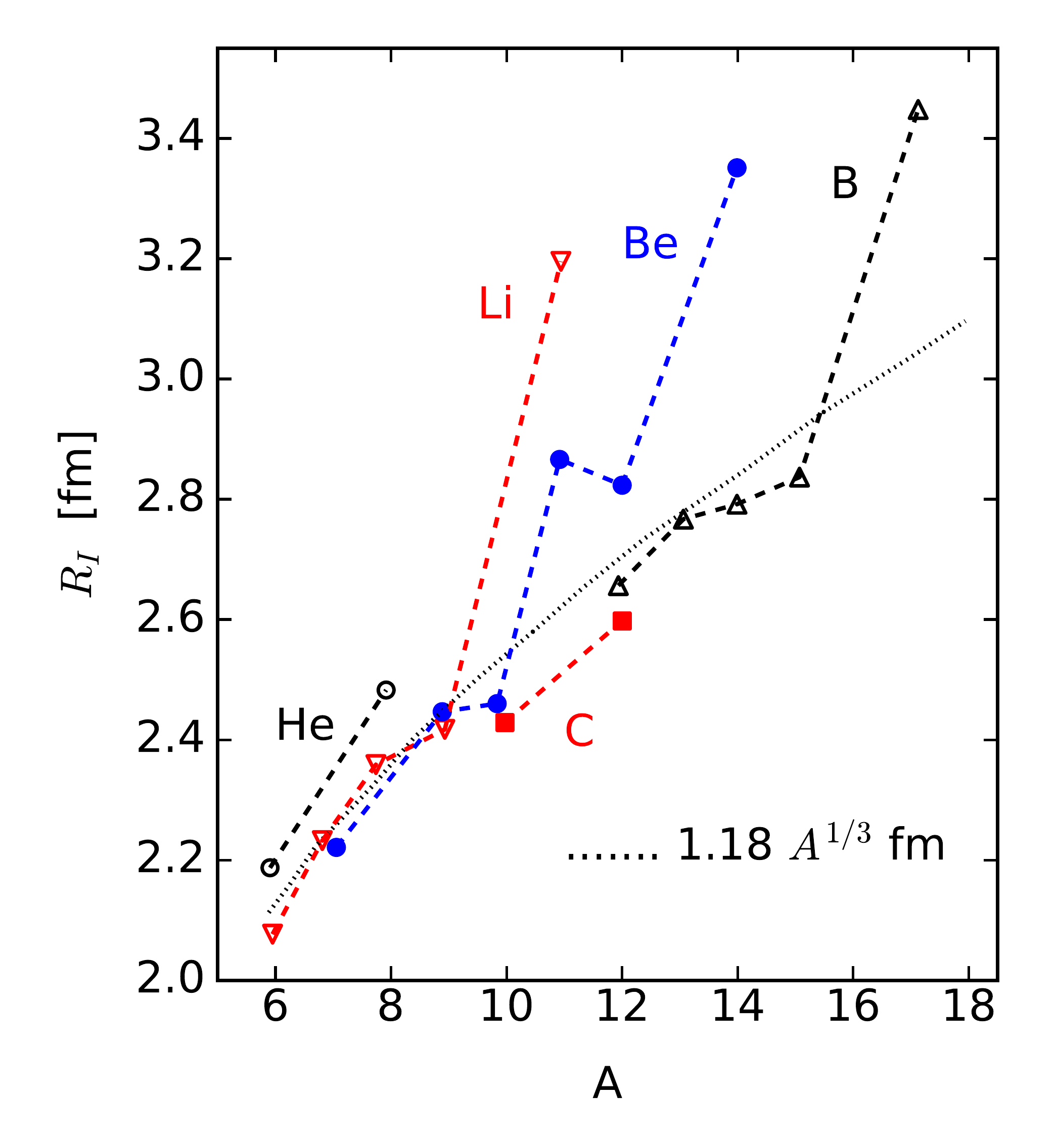} 
 \caption{{\it Left:} An artistic drawing of a $^{11}$Li nucleus, showing the two neutrons building a halo around a $^{4}$He nucleus core. {\it Right:} Interaction radii of light nuclei (errors bars not shown). The large deviation from the expected radii of a nucleus (dotted line) is evident for the so-called ``halo'' nuclei.}
\label{halofig}
\end{center}
\end{figure}

Mean field theories such as HFB and RMF can obtain physical observables of nuclear interest such as the root mean square radius,  the charge radius, and the neutron skin thickness. The nuclear charge radius  can be measured experimentally by the scattering of electrons or muons by the nucleus. As for the neutron distribution, hadronic probes are often used. It was in this way that abnormally large radii of light nuclei were found.  In Figure \ref{halofig} (left) an artistic drawing of a $^{11}$Li nucleus is shown, with the two neutrons building a halo around a $^{9}$Li nucleus core. The interaction radius, $R_{I}$ is defined as the radius in the interaction cross section, $\sigma = \pi R_{I}^{2}$, for reactions leading to any number of removed nucleons. Interaction radii of light nuclei are plotted  (errors bars not shown) in Figure \ref{halofig} (right). The large deviation from the expected radii of a nucleus (dotted line) is evident for the so-called ``halo'' nuclei \cite{TANIHATA1985380,TanihataPRL.55.2676,TANIHATA1988592}. 

The unexpected large size of the $^{11}$Li nucleus was also made clear by measuring the momentum distributions of $^{9}$Li fragments in nucleon-removal reactions \cite{KobayashiPRL.60.2599}. A superposition of two approximate gaussian-shaped distributions was used to reproduce the data. The wider peak was connected to the knockout of neutrons from a tightly bound $^{9}$Li core, while the narrower peak was thought to arise from the knockout of the loosely bound valence neutrons in $^{11}$Li. The two-neutron separation energy in $^{11}$Li is small, of the order of 300 keV, explaining why their removal only slightly perturbs the $^{9}$Li core, leading to a natural explanation of the narrow component of the momentum distribution. In fact, a narrow momentum distribution is related to a large spatial extent of the neutrons, which is a due to their small separation energy. These na\"\i ve conclusions were later confirmed with Coulomb breakup experiments \cite{KOBAYASHI198951} and with theory [17,18]. The influence of nuclear haloes in astrophysics is observed in numerous reactions. Perhaps, the most celebrated one is the $^7$Be(p,$\gamma$)$^8$B reaction, of relevance for the production of energetic solar neutrinos via the $^8$B decay. Theoretically, it is found that an accurate value of the S-factor (at $E_{cm}\sim 20$ keV) for continuum to bound state radiative capture is only possible if one integrates the nuclear matrix elements up to distances of $r=200$ fm  \cite{Bertulani1987}.

 \begin{figure}[t]
\begin{center}
\includegraphics[height=2.2in]
{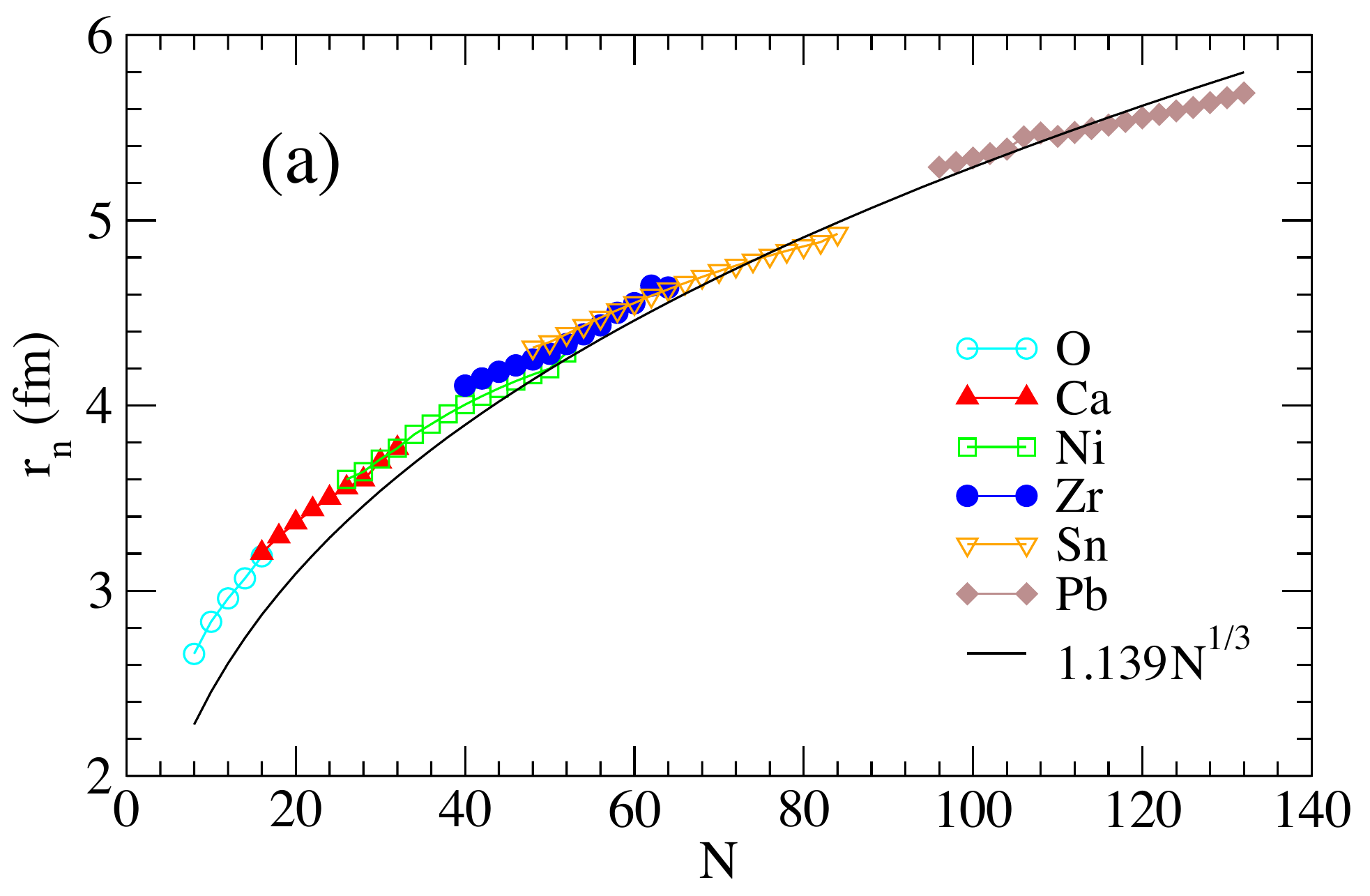} 
\includegraphics[height=2.2in]
{Rn.pdf} 
 \caption{Neutron (a) and proton (b)  radii of various isotopes and isotones calculated by means of the HF + BCS model with isoscalar + isovector pairing interaction \cite{BLS:PRC.85.014321}. The solid lines are the empirical fits used in Ref. \cite{MENG2006470}.}
\label{rnrp}
\end{center}
\end{figure}

Recently, isotope-shift measurements have been performed for He, Li and Be isotopes to determine their charge radii \cite{LuRMP85.1383,MassPRL122.182501}. The isotope shift, $\delta \nu_{IS}$, has contributions from the mass shift , $\delta \nu_{MS}$, and the field shift , $\delta \nu_{FS}$, i.e., , $\delta \nu_{IS}=\delta \nu_{MS}+\delta \nu_{FS}$. The latter contribution is proportional to the charge radius of the nucleus, $\delta \nu_{FS}  = C \left<r\right>_c^2$. To obtain the radius with accuracy a variational calculation is done for the electronic wavefunction, and QED corrections are included. The measurements show good agreement with ab-initio nuclear structure calculations \cite{MassPRL122.182501}.  Such experiments open the window to explore a larger number of radii of light nuclei providing valuable constraints to refine current nuclear models.  The technique has been extended to heavy nuclei and in particular,  the mean-square nuclear charge radii has been measured along the even-A tin isotopic chain $^{108-134}Sn$  by means of collinear laser spectroscopy at ISOLDE/CERN using several atomic transitions \cite{GorgesPRL122.192502}.

For medium-heavy and heavy nuclei, matter radii are not so well known. In particular,  for rare nuclear isotopes not only matter radii but also information on charge radii is scarce, despite the new data obtained with isotope-shift experiments mentioned above.  Most of our knowledge on these nuclear properties come from microscopic mean field theories, such as those we described previously. For example in Figure \ref{rnrp}, from Ref.  \cite{BLS:PRC.85.014321}, we show the neutron (a) and proton (b)  radii of various isotopes and isotones calculated by means of the HF + BCS model with isoscalar + isovector pairing interaction. Solid lines are empirical fits used in Ref. \cite{MENG2006470}. One notices in Figure  \ref{rnrp}(a) that the HF + BCS model yields larger neutron radii for nuclei with $N < 40$ than the simple phenomenological formula $R_{n}=1.139N^{1/3}$ fm but yields a smaller radii for nuclei with $N > 120$. In Figure  \ref{rnrp}(b) the proton radii for $N = 20$, 28, 40, 50, 82, and 126 isotones are shown as a function of proton number $Z$. The simple $R_{p}=1.263Z^{1/3}$ fm dependence is also shown. The simple formula reproduces rather well the HF + BCS calculations and could be used as a good starting point to describe the isospin dependence of nuclear charge radii in neutron rich nuclei. However, as expected from phenomenological models, one sees a deviation with the HF + BCS model, especially for heavy $N=50$ and $N=82$ isotones.

 \begin{figure}[htb]
\begin{center}
\includegraphics[height=2.5in]{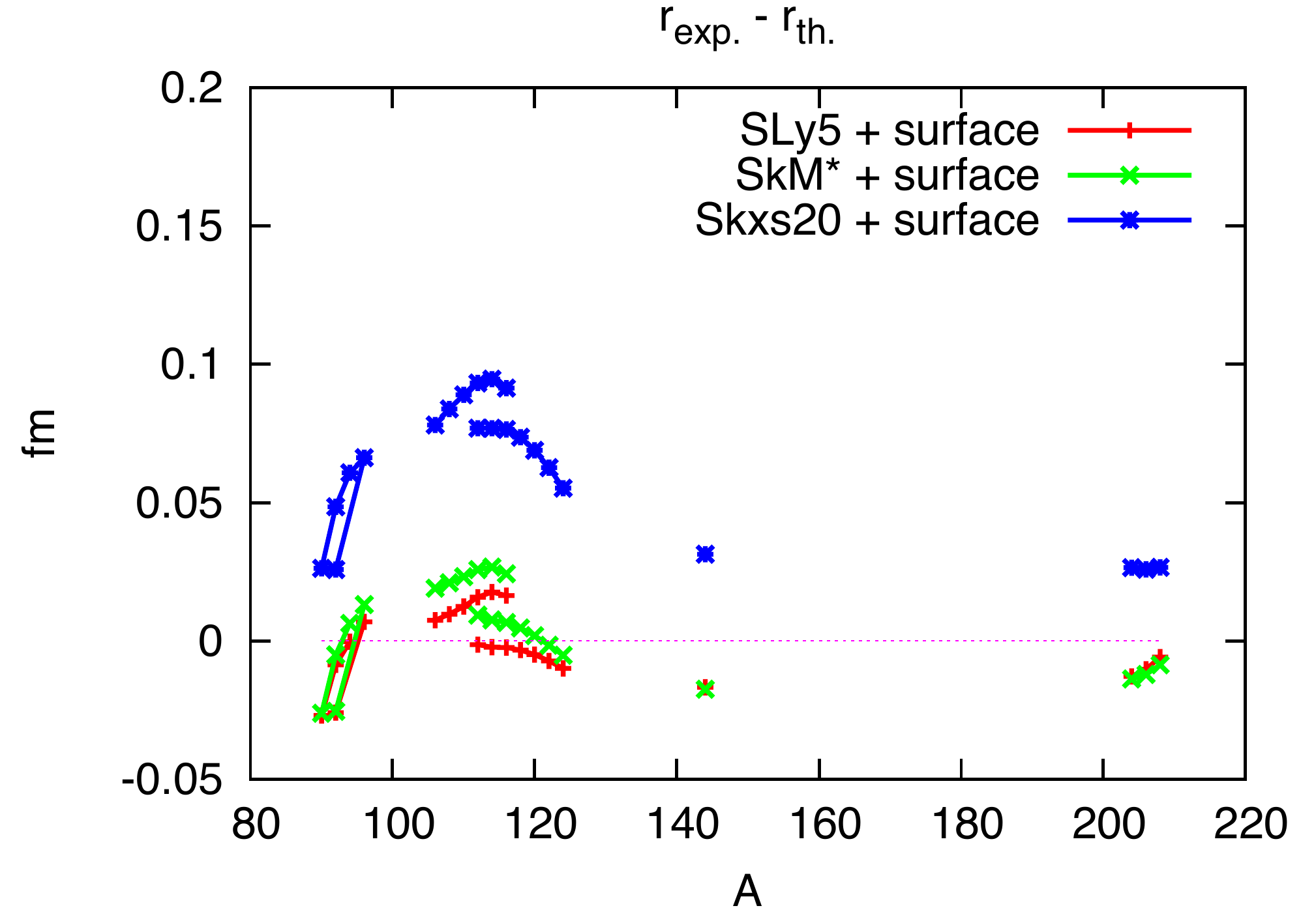} 
 \caption{Difference between the experimental \cite{ANGELI2004185} and the theoretical charge radii of Zr, Mo, Cd, Sn, Sm and Pb isotopes(isotopes of the same element are connected by a curve) \cite{Avogadro:PRC.88.044319}.}
\label{raddifig}
\end{center}
\end{figure}

The differences between known cases of charge radii of Zr, Mo, Cd, Sn, Sm and Pb isotopes and calculations based on the three Skyrme interactions with surface pairing are shown in Figure \ref{raddifig} \cite{Avogadro:PRC.88.044319}.  The determination of nuclear charge radii in very neutron-rich nuclei is one of the most challenging experimental goals. It would be an excellent test of microscopic theories for the nucleus.  An electron-rare-isotope ion collider would be the best possibility, as we discuss later.

\subsubsection{Neutron skins}

A neutron excess in a nucleus leads to a larger neutron pressure than that due to the protons. Such a neutron pressure contributes to the energy per nucleon which then becomes a function of the nuclear density and its nucleon asymmetry. Neutron skins in nuclei are therefore expected to be directly correlated to the symmetry energy and the slope parameter of the nuclear matter EoS. This is expected even for a large stable nucleus such as $^{208}$Pb. The neutron skin in nuclei is defined as 
\begin{equation}
\Delta r_{np}\equiv r_{n}-r_{p} = \left<r_{n}^{2}\right>^{1/2}- \left<r_{p}^{2}\right>^{1/2}.
\end{equation}

Very little is known about neutron skin in nuclei. There are few experimental data using, e.g., antiprotonic atoms \cite{TrzcPRL.87.082501}. The antiproton annihilation method consists of the study of the residual nuclei with one unit mass number smaller than that of the target $A_{T}$. The assumption is that such residues originate from events in which all produced pions miss the target, and the target remains with a very low excitation energy so that compound-nucleus evaporation or fission does not occur. When both the $A_{T}-1$ products are $\gamma$-emission unstable, their  yields  are determined by standard nuclear-spectroscopy techniques. These yields are directly related to the proton and neutron densities at the annihilation region in the nucleus. The radial distance of this position in the nucleus, is assumed to be almost independent of the target atomic number $Z$ and is compared to calculations \cite{WycechPRC.54.1832} based on antiproton-nucleus optical potentials, generated by the so-called t-$\rho$ approximation, and the antiproton atomic orbits together with the proton orbitals in the nucleus involved in the annihilation process. 

 \begin{figure}[t]
\begin{center}
\includegraphics[height=2.7in]
{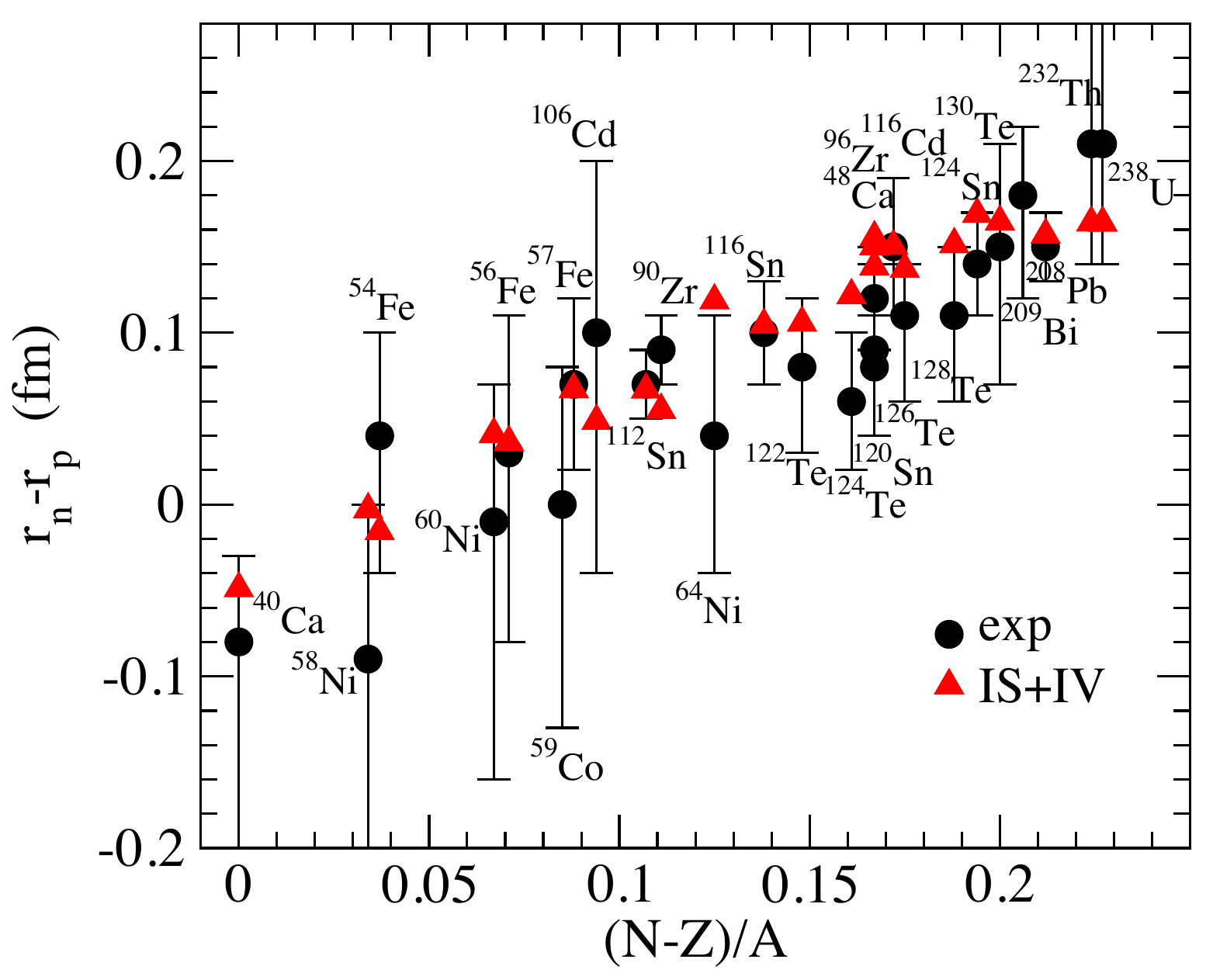} 
 \caption{Neutron skin as a function of isospin parameter $\delta = (N-Z)/A$ calculated by means of the HF + BCS model with (isoscalar + Isovector) IS + IV interaction \cite{BLS:PRC.85.014321}. Experimental data are taken from Ref. \cite{TrzcPRL.87.082501} based on the annihilation of antiprotons at the surface of the nuclei.}
\label{trcz}
\end{center}
\end{figure}

The antiprotonic atoms annihilation data is not very accurate, as shown in Figure \ref{trcz} by the filled circles for the neutron skin in several nuclei as a function of isospin parameter 
\be
\delta = {{N-Z}\over A}
\ee
and with empirical values obtained by antiprotonic atom experiments in a wide range of nuclei from $^{40}$Ca to $^{238}$U \cite{TrzcPRL.87.082501}. It is clear that the statistics are not very good. Also shown  in the figure (triangles) are results of calculations based on the HF + BCS model with Isoscalar + Isovector (IS + IV) pairing interaction \cite{BLS:PRC.85.014321}. It is evident that the data can accommodate a wide range of Skyrme parameters and is not very constraining. 

 \begin{figure}[t]
\begin{center}
\includegraphics[height=2.6in]
{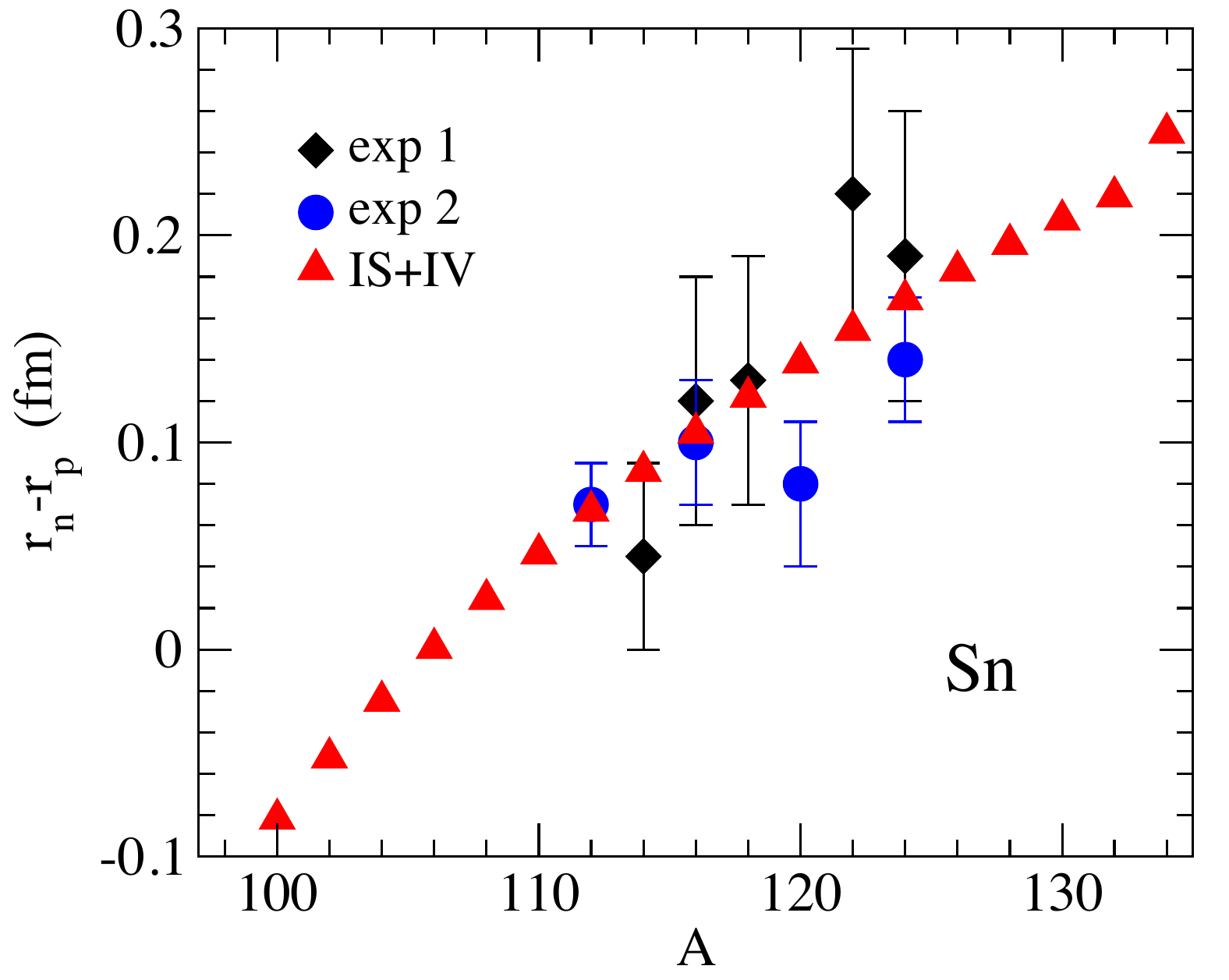} 
 \caption{ Combined experimental data for the neutron skin for Sn isotopes. are taken from Refs. \cite{KrasznahorkayPRL.82.3216,TrzcPRL.87.082501}. The triangles represent numerical results  obtained with the HF + BCS model with IS + IV interaction \cite{BLS:PRC.85.014321}. }
\label{Snrnrp}
\end{center}
\end{figure}

Another tool to investigate the neutron-skin thickness is the excitation of the spin-dipole resonance (SDR) \cite{KrasznahorkayPRL.82.3216}. It is based on the idea that the total $L=1$ strength of the SDR is sensitive to the neutron-skin thickness \cite{GAARDE199079, Alford2002}. The SDR is be strongly excited in the (p,n) reaction in inverse kinematics using radioactive nuclear beams. A few radioactive isotopes have been studied using this technique. In Figure \ref{Snrnrp} we show the combined experimental data of neutron skin for Sn isotopes \cite{KrasznahorkayPRL.82.3216}.  The triangles represent numerical results  obtained with the HF + BCS model with IS + IV interaction \cite{BLS:PRC.85.014321}. It is again evident that the data are not good enough to distinguish the most adequate Skyrme interactions.

The liquid drop model \cite{MYERS1969395,MYERS1974186} is perhaps the best way to understand the correlation between the neutron skin and the slope parameter of the symmetry energy \cite{ROCAMAZA201896}. In the droplet model the neutron skin is given by
\begin{equation}
\Delta r_{np}= \sqrt{3 \over 5} \left[ t-{e^2 Z \over 70 J}+{5\over 2R}(a_{n}^{2}-a_{p}^{2})\right], \label{ld1}
\end{equation}
where $J = S(\rho_{0})$ is the volume term of the symmetry energy, Eq. (\ref{symenerg}), $R$ is the mean nuclear radius and $a_{n(p)}$ is the surface diffuseness for neutrons (protons). The  second term in the equation above is due to the Coulomb repulsion of protons.  One usually neglects the term contains the surface diffuseness (i.e., setting $a_{n}=a_{p}$). The first term in Eq. (\ref{ld1}) depends in leading order on the asymmetry parameter $\delta=(N-Z)/A$,
\begin{equation}
t= {3 r_{0}\over 2} {J\over Q} {{\delta - \delta_{c}}\over {1+(9J/4Q)A^{-1/3}}}, \label{tld1}
\end{equation}
where $r_{0}$ pertains to the relation $R=r_{0}A^{1/3}$ fm, 
\begin{equation}
\delta_{c}= {e^{2}Z \over 20Jr_{0}A^{1/3}},
\end{equation}
and $Q$ is the so-called surface stiffness coefficient, being a parameter of the droplet model involving the dependence of the surface tension energy on the isospin and neutron skin thickness. Therefore, it is not obvious from the transcendental relation, Eq. (\ref{tld1}) that the neutron skin thickness is proportional to the slope parameter. 

A bit of phenomenology can help to illuminate the $\Delta r_{np}$ vs. $L$ correlation, as shown in Ref. \cite{WardaPRC.80.024316}. The key to the proof is to identify \cite{ROCAMAZA201896}
$${J\over Q}{1\over ({1+(9J/4Q)A^{-1/3}})} \sim \left({S(A)\over J} -1 \right)A^{1/3},$$ 
where 
$$S(A) = {J\over {1+(9J/4Q)A^{-1/3}}}$$ 
is the droplet model symmetry energy for a finite nucleus.
For heavy nuclei $S(A) \sim S(\rho)$ \cite{WardaPRC.80.024316}. Therefore, one gets,
\begin{equation}
t= {2 r_{0}\over 3 J} L\left[ 1 -x{K_{sym}\over 2L} \right] x A^{1/3}({\delta - \delta_{c}})\label{tld2}
\end{equation}
with the parameters defined as in Eqs. (\ref{epsilon1}) and (\ref{symenerg}). Therefore, to lowest order, $t = (2r_{0}/3J)L\delta$, showing a linear correlation between the neutron skin and the slope parameter, assuming the validity of the assumptions taken above.

\begin{figure}[t]
\begin{center}
{\includegraphics[width=9cm]{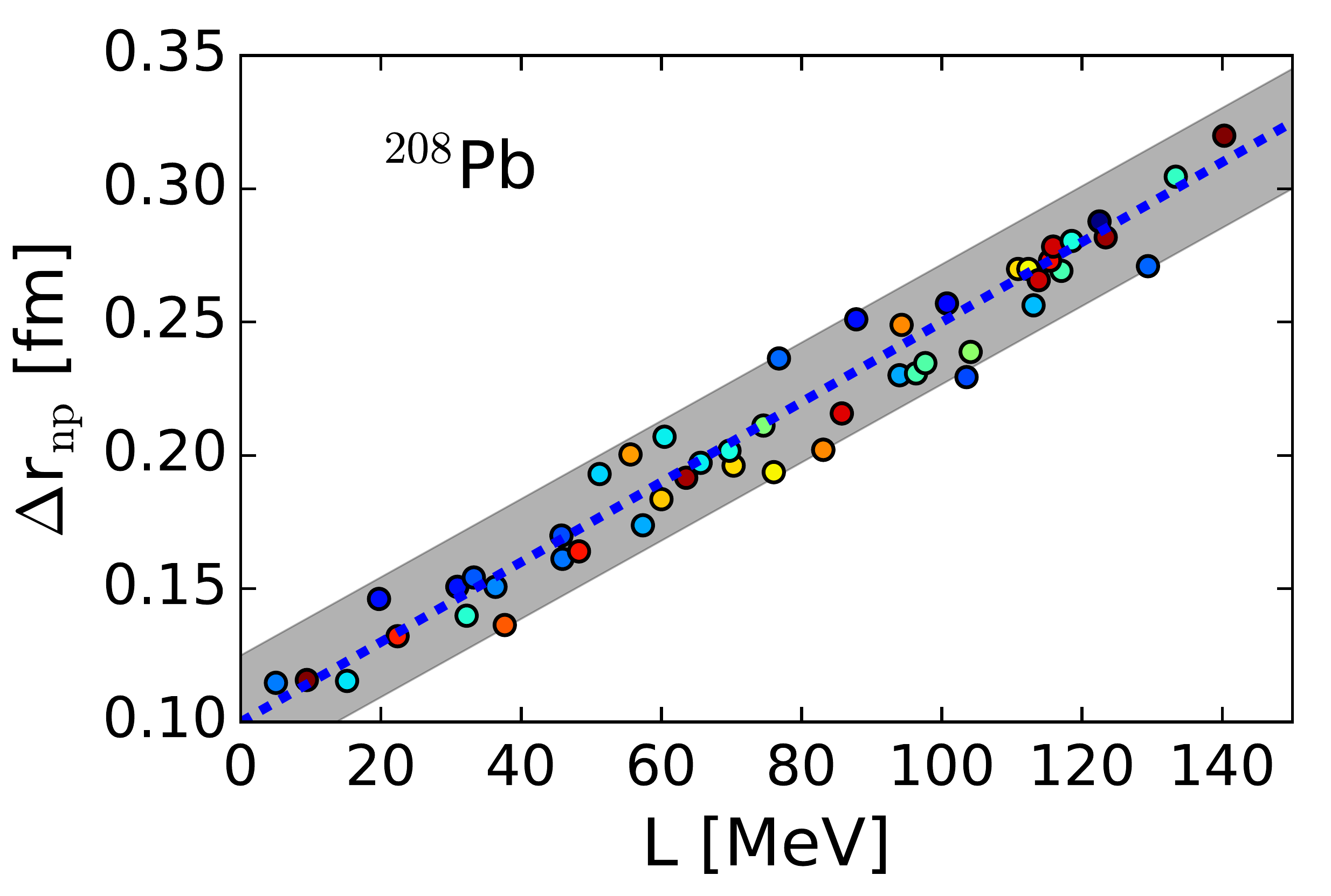}}
\end{center}
\vspace{-0.5cm}
\caption{Neutron skin thickness of $^{208}$Pb as a function of the slope parameter of the nuclear symmetry energy at saturation as predicted by numerous nuclear mean-field interactions. The different colored circles represent calculations with a particular interaction, either Skyrme or Gogny-type, or RMF interactions. The shaded area reveals the uncertainty range.}
\label{skinL}
\end{figure}

In Fig. \ref{skinL} we show the neutron skin thickness of $^{208}$Pb as a function of the slope parameter of the nuclear symmetry energy at saturation, as predicted by numerous nuclear mean-field interactions. The mean field calculations shown in the plot range from Skyrme zero-range forces, as in Eq. (\ref{Skyrm}) and Gogny finite range forces \cite{Decharge1980,BERGER1991365,CHAPPERT2008420,Than2011,Loan.PRC2011,GorielyPRL102.242501,Sellahewa2014,ChenPRC.90.044305,Boquera2017}. In the Gogny models, some of the delta functions in Eq. (\ref{Skyrm}) are replaced by gaussians with widths representing the range of nuclear forces. Also shown in the figure are results for relativistic forces based on effective field theory Lagrangians with meson self-interactions, density-dependent meson-nucleon couplings, and point couplings, as in Eq. (\ref{rmfl}) \cite{LalazissisPRC.55.540,SUGAHARA1994557,HorowitzPRL.86.5647,Lalazissis2005,TypelRopke2010,BaoShen2014,Fortin2016,PaisPRC.94.015808,tolos.centelles.ramos2017,Negreiros2018}. 

\begin{figure}[t]
\centerline{
\includegraphics[width=1.05\columnwidth]{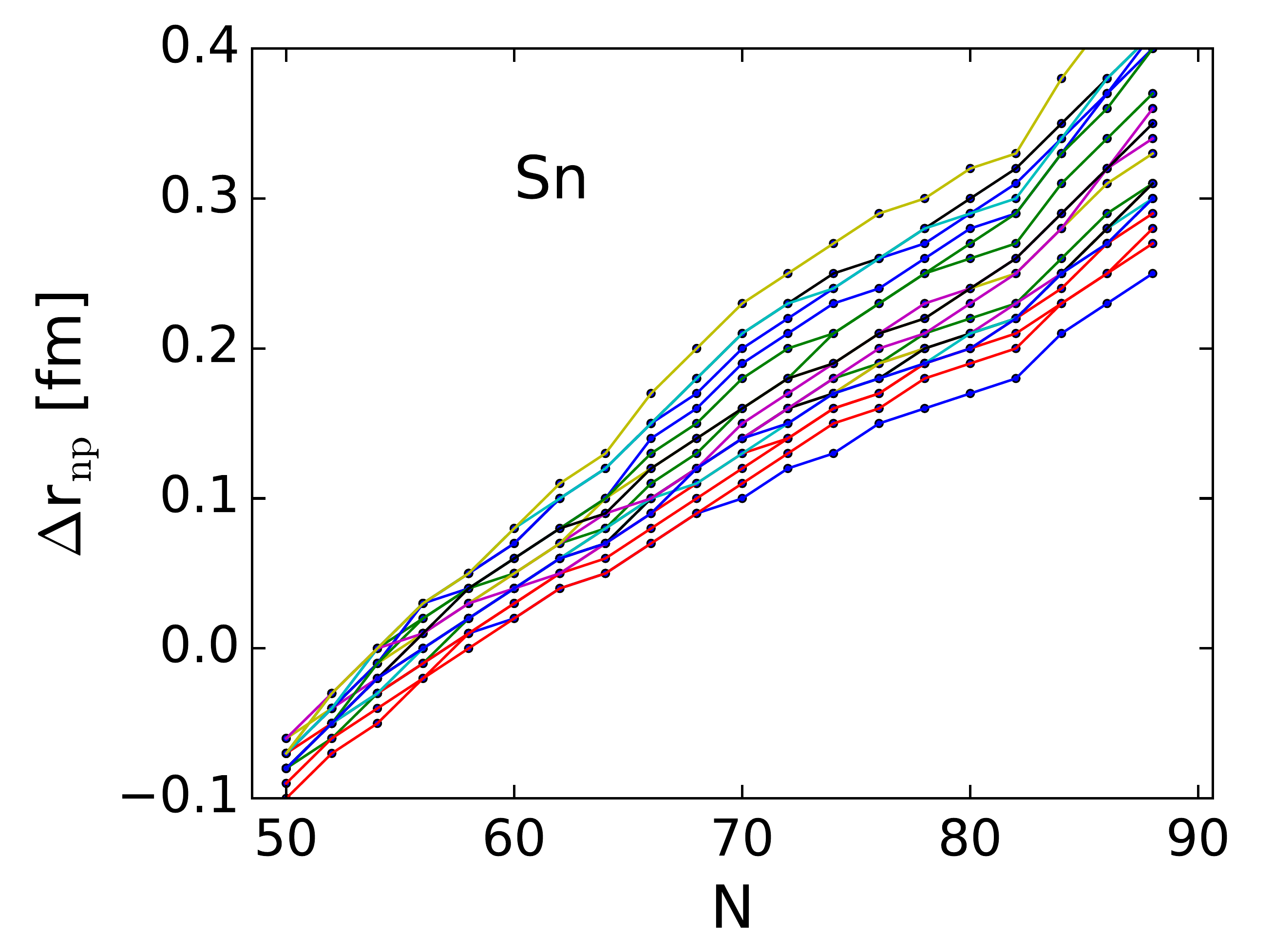}}
\caption{The points represent neutrons skin, $\Delta r_{np}$ calculated for tin isotopes with the 23 Skyrme interactions \cite{Bertulani.PRC.100.015802}. Each one of the lines correspond to one of the interactions and are also guide to the eyes.}
\label{nskinsn2}
\end{figure}

In Figure \ref{nskinsn2} we show results for the neutron skin, $\Delta r_{np}$ calculated with more than 20 Skyrme interactions \cite{Bertulani.PRC.100.015802}. It is visible that neutron skins calculated with different Skyrme interactions tend to spread out as the neutron number increases. The same is observed for Ni and Pb isotopes \cite{Bertulani.PRC.100.015802}. Calculations for the stable tin isotopes with masses $A=116, \ 118,\ 120$, yield neutron skin in the range $0.1-0.3$ fm, depending on the Skyrme force adopted. 

\begin{figure}[t]
\begin{center}
{\includegraphics[width=9cm]{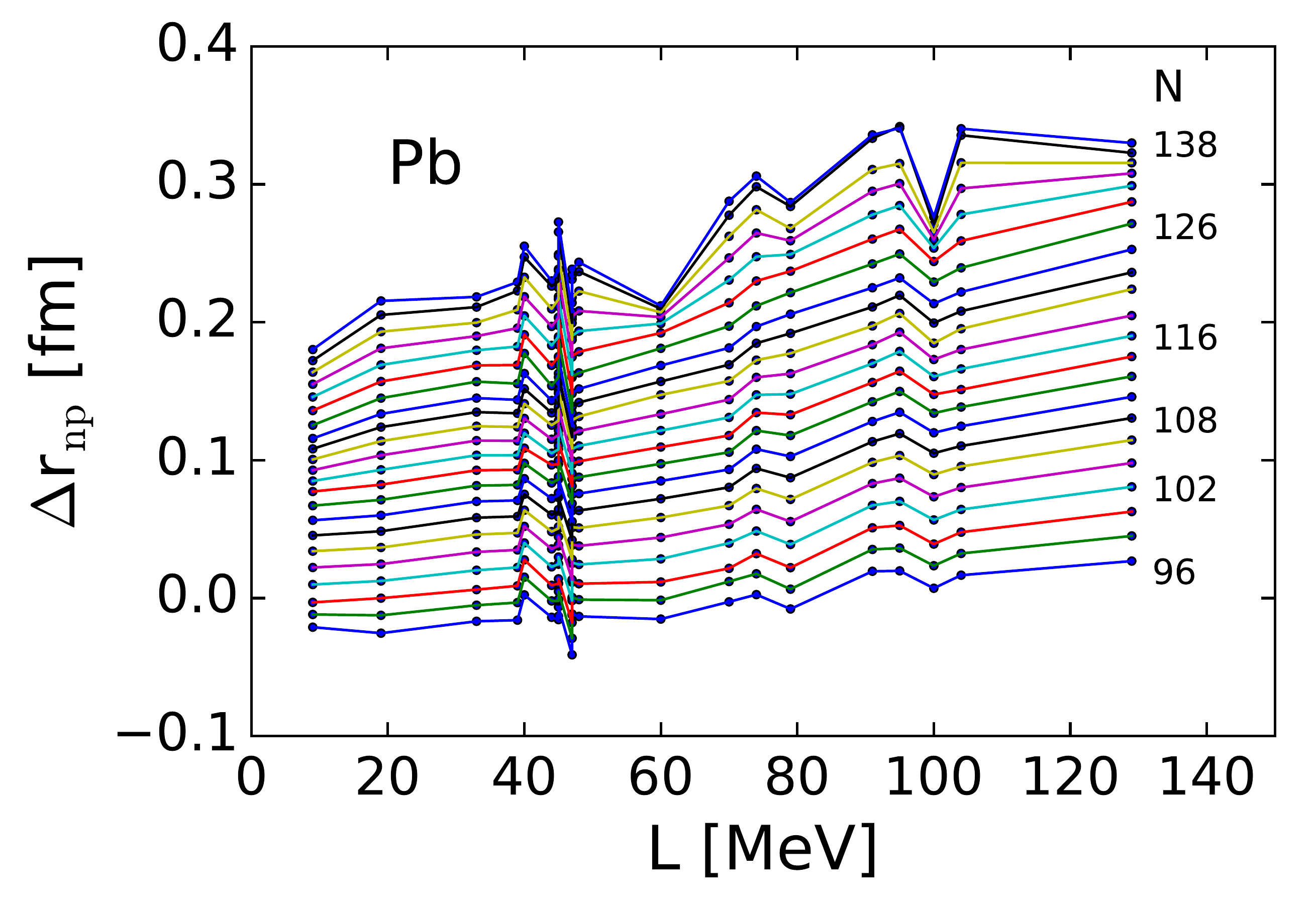}}
\end{center}
\vspace{-0.5cm}
\caption{Neutrons skin, $\Delta r_{np}$, calculated with numerous Skyrme interactions \cite{Dutra:PRC.85.035201,BEINER197529,DOBACZEWSKI1984103,REINHARD1995467,Brown:PRC.58.220,CHABANAT1998231,Reinhard:PRC.60.014316,GORIELY2005425,Dutra:PRC.85.035201,STOITSOV200543,STOITSOV20131592,KOHLER1976301,BARTEL198279,Bennour1989,CHABANAT1997710,Brown:1998,Agrawal2003} as function of the predicted slope parameter $L$. The lines are guides to the eyes. Each of the connected curves corresponds to a different lead isotope with neutron number $N$ \cite{Bertulani.PRC.100.015802}.}
\label{leadiso}
\end{figure}

As observed in Fig. \ref{skinL},  the predictions for the slope parameter are all over the place within the range $L\sim 5$-140 MeV. It is also evident that a near linear correlation exists between the neutron skin and the slope parameter within a 5\% uncertainty in $\Delta r_{np}$ for a given value of $L$ and a 15\% uncertainty in $L$ for a given value of $\Delta r_{np}$. The question is: what are the best microscopic models? As it stands the Figure \ref{skinL}  implies an overall theoretical uncertainty of  the neutron skin of $^{208}$Pb within the range $\Delta r_{np}=0.1$-35 fm and for the slope parameter the uncertainty is about $L=5-140$ MeV. 

\begin{figure}[t]
\begin{center}
{\includegraphics[width=9cm]{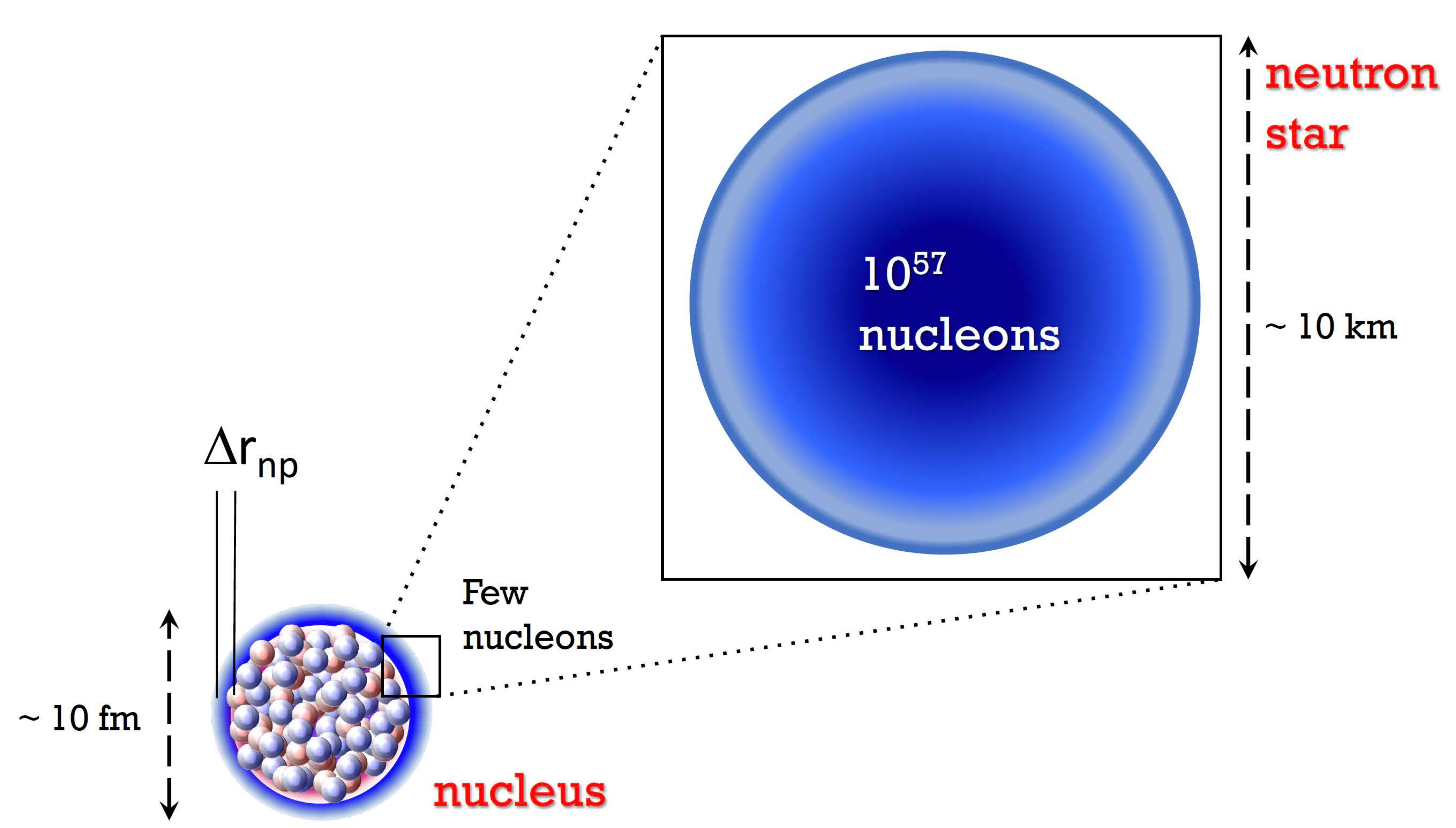}}
\end{center}
\vspace{-0.5cm}
\caption{The neutrons form a neutron skin in a heavy nucleus, e.g., Pb,  within a  have a range of $0.1-0.2$ fm. Studying the experimental value of neutron skins and their predictions from microscopic models, one hopes to clarify nuclear interactions in many-nucleon systems, the EoS of asymmetric nuclear matter,  and possibly the properties of neutron stars, with $\sim 10^{57}$ neutrons.}
\label{neutronskin}
\end{figure}

In Fig. \ref{leadiso} we show the neutrons skins, $\Delta r_{np}$, calculated with numerous Skyrme interactions \cite{Dutra:PRC.85.035201,BEINER197529,DOBACZEWSKI1984103,REINHARD1995467,Brown:1998,CHABANAT1998231,Reinhard:PRC.60.014316,GORIELY2005425,STOITSOV200543,STOITSOV20131592,KOHLER1976301,BARTEL198279,Bennour1989,CHABANAT1997710,Agrawal2003} as function of the predicted slope parameter $L$. The lines are guides to the eyes. Each of the connected curves corresponds to a different lead isotope with neutron number $N$ \cite{Bertulani.PRC.100.015802}. One observes that the nearly linear correlation shown in Fig. \ref{skinL} also varies as the number of neutrons is changed. As expected, greater variations are observed for nuclei with larger neutron excess. Based on the results, it is apparent that we are faced with the quest to extrapolate the physics of neutron skins due to a few extra neutrons to that of a neutron stars with $10^{57}$ nucleons, as schematically shown in Figure \ref{neutronskin}. It is a quest that requires accurate experiments in symbiosis with well developed microscopic nuclear theories.

\subsubsection{Dipole polarizability and electromagnetic response}
The electric polarizability of a nucleus is a measure of the tendency of the nuclear charge distribution to be distorted by an external electric field, e.g. 
\begin{equation}
\alpha_{D}\sim  {{\rm electric \  dipole \  moment} \over {\rm external \  electric \  field}}.
\end{equation}
The action of an external field $F e^{i\omega t}+F^{\dagger}e^{-I\omega t}$, of dipolar form,
\begin{equation}
F_{{JM}}=\sum_{k}^{A}r^{J}Y_{JM}({\bf r})\tau_{z}(k), \ \ \  (L=1, \ \ {\rm for \ dipole}),
\end{equation} 
yields a response proportional to the static dipole polarizability
\begin{equation}
\alpha_{D}= {8\pi e^{2}\over 9}\sum_{i}{1\over E}\left| \left< i|F_{1}|0\right>\right|^{2}={8\pi e^{2}\over 9}m_{-1},\label{responf}
\end{equation}
where $m_{-1}$ is the inverse-energy-weighted moment (see Eq. (\ref{eq:moment})) of the strength function defined as in Eq. (\ref{sfunc}). Isovector energy-weighted sum rules (EWSR) for dipole and quadrupole excitations are, respectively,
\begin{equation}
m_{1}^{(D)}= {\hbar^{2}\over 2m_{N}}{NZ\over A} (1+\kappa_{D}),
\end{equation}
and
\begin{equation}
m_{1}^{(Q)}= {\hbar^{2}\over 2m_{N}}{50\over 4\pi} A\left<r^{2}\right>(1+\kappa_{Q}),
\end{equation}
where $m_{N}$ is the nucleon mass, A, N, and Z are the charge, neutron and charge number, and $\kappa_{j}$ is a correction due to the momentum dependence of the nucleon-nucleon interaction.

Using a different notation, the electric polarizability is proportional to the inverse energy weighted sum rule of the electric dipole response in the nucleus, i.e.,
\begin{equation}
\alpha_{D}= {8\pi e^{2}\over 9} \int {dE\over E} {dB(E1;E)\over dE} = {\hbar c\over 2\pi^{2}} \int {dE\over E^{2}} \sigma_{abs}(E), \label{alphadeq}
\end{equation}
where $E$ is the photon energy, 
\begin{equation}
\int dE {dB(E1;E)\over dE} \equiv \sum_{j<{j}_{max}}B(E1;E_{0}\rightarrow E_{j}) \label{dbdedef}
\end{equation} 
is known as the response function to the external field, and $\sigma_{abs}(E)$ is the photo-absorption cross section. $B(E1;E_{0}\rightarrow E_{j})$ is the reduced transition strength from an initial state $0$ to a final state $j$ in the nucleus induced by the photo-absorption.  The notation $dB/dE$ for the response function is often used when many excited states within an energy interval $dE$ are accounted for, such as in transitions to the continuum ($E_{j}> E_{threshold}$).

The dipole polarizability $\alpha_{D}$ can be extracted from isovector excitations of the nuclei, such as Coulomb excitation experiments. For stable nuclei the dipole response is mostly concentrated in the isovector giant dipole resonance (IVGDR), exhausting almost 100\% of the energy-weighted sum rule. One understands this excitation mode as an oscillation of neutrons against protons where the symmetry energy, $S$ in Eq. (\ref{epsilon1}) acts as a restoring force. Studies of symmetry energy using dipole polarizability and its relation to the symmetry energy were already known in the literature \cite{KRIVINE1982281,AHRENS1975479}. A renewed interest in  isovector excitations arose due to the possibility  that in neutron-rich nuclei a softer symmetry energy, changing slowly with density, predicts larger values for the dipole polarizability at the lower densities. This means that the quantity $m_{-1}$, proportional to $\alpha_{D}$ is highly sensitive to the density dependence of the symmetry energy and suggests a correlation so that the larger skin $\Delta r_{np}$  the larger $\alpha_{D}$  \cite{ReinhardPRC.81.051303}. 

It is worthwhile noting that Migdal \cite{migdal1945quadrupole} obtained the dipole polarizability assuming a sharp-edged liquid drop
energy density and found that
\begin{equation}
m_{-1}={A\left<r^{2}\right>\over 48J }.
\end{equation} 
This approximation implies a change in the proton density which varies linearly with the position in the dipole ($z$) direction. It induces then a ``tilting'' collective mode rather different in nature from the ``sliding'' Goldhaber-Teller mode \cite{MEYER1982269}. Using the liquid drop model with surface corrections, Ref. \cite{MEYER1982269} was able to show that the dipole polarizability can be related to the parameters of the droplet model as
\begin{equation}
\alpha_{D}\equiv m_{-1}={A\left<r^{2}\right>\over 48J }\left[ 1- x{L\over J}-\left(3M-2{L^{2}\over K_{0}}\right){\delta^{2} \over J}+{15 \over 4A^{1/3}}{J \over Q}\right],
\end{equation}
where $M$ is a coefficient specifying the deviation from a quadratic dependence on $\delta$. Using the same arguments leading from Eq. (\ref{tld1}) to Eq. (\ref{tld2}) one is able to show that \cite{CentellesPRC.82.054314}
\begin{equation}
\alpha_{D}\simeq {\pi e^{2}A\left<r^{2}\right>\over 54J }\left[ 1+ {5\over 2} {\Delta r_{np}+\sqrt{3\over 5}{e^{2}Z\over 70J} - \Delta r_{np}^{surf}\over \left<r^{2}\right>^{{1/2}}(\delta - \delta_{c})}\right],
\end{equation}
where $\Delta r_{np}^{surf}\sim  0.09$ fm for $^{208}$Pb, is almost constant for the EDFs. This shows that, in the LDM, $\Delta r_{np}$ (and $L$) are better correlated with $\alpha_{D}J$ than directly with $\alpha_{D}$ for a heavy nucleus \cite{WardaPRC.80.024316,CentellesPRC.82.054314}.

\begin{figure}[t]
\begin{center}
{\includegraphics[width=8cm]{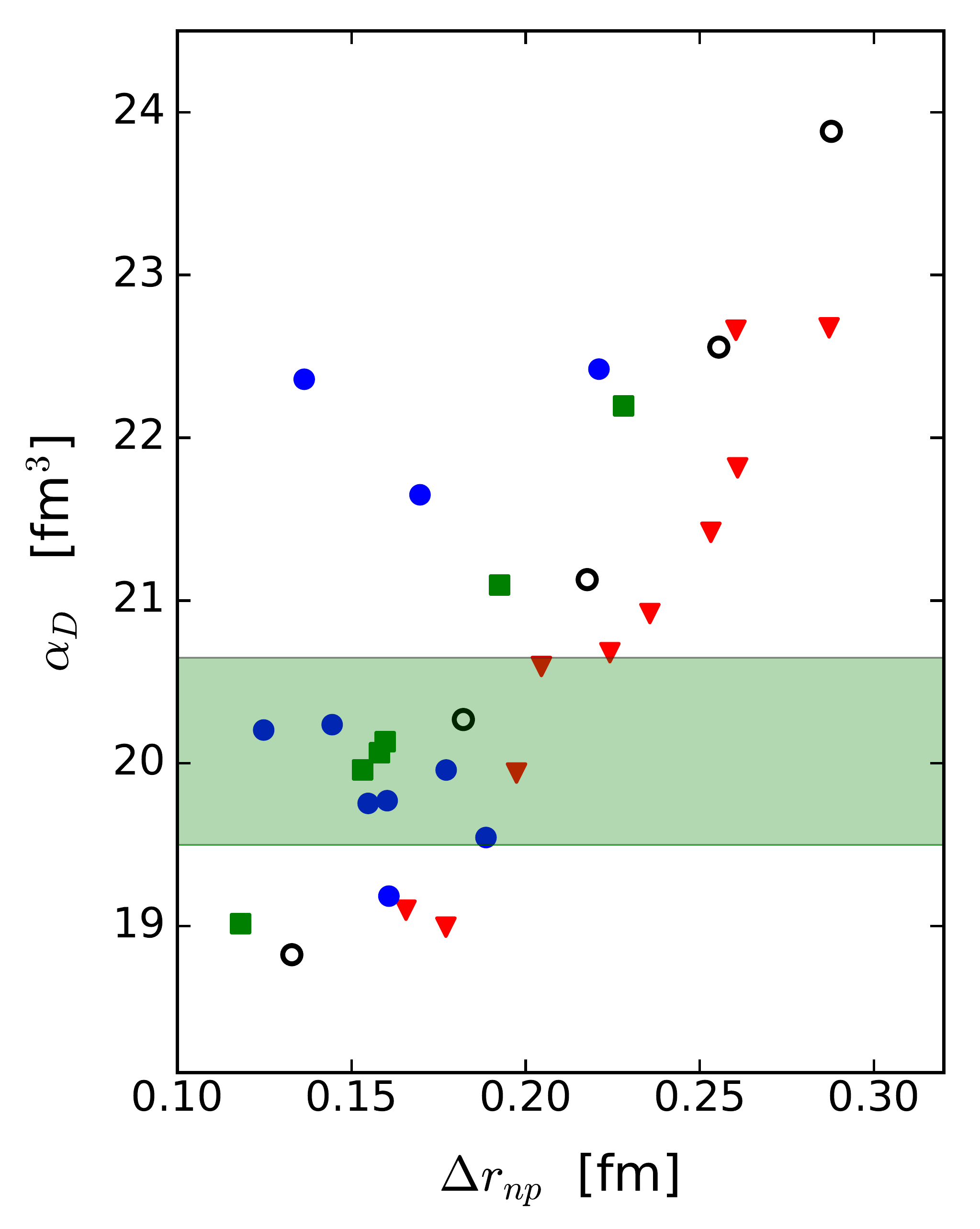}}
\end{center}
\vspace{-0.5cm}
\caption{Predictions for the dipole polorizabitly and for the neutron skin of $^{238}$Pb \cite{PiekarewiczPRC.85.041302}. The filled circles are Skyrme predictions with the interactions SIII \cite{BEINER197529}, SkI3 and SkI4 \cite{REINHARD1995467}, SkM \cite{BARTEL198279}, SkO \cite{Reinhard:PRC.60.014316}, SkP \cite{DOBACZEWSKI1984103}, SkX \cite{Brown:PRC.58.220}, SLy4 and SLy6 \cite{CHABANAT1997710,CHABANAT1998231}, Sk255 \cite{Agrawal2003}, BSk17 \cite{GorielyPRL.102.152503}, LNS \cite{CaoPRC.73.014313}, and UNEDF0 and UNEDF1 \cite{STOITSOV20131592}.The filled triangles are results from the NL3/FSU interaction \cite{PiekarewiczPRC.83.034319,AgrawalPRC.81.034323}, the open circles are DD-ME interaction \cite{VretenarPRC.68.024310,Lalazissis2005}. The filled squares are predictions from the Skyrme-SV interaction \cite{KlupfelPRC.79.034310}. The experimental constraint for the dipole polarizability from a Coulomb excitation experiment \cite{TamiPRL.107.062502} is shown as a horizontal band.}
\label{alphad}
\end{figure}

In Fig. \ref{alphad} we plot the predictions for the dipole polorizabitly and for the neutron skin of $^{238}$Pb \cite{PiekarewiczPRC.85.041302}. The filled circles are Skyrme predictions with the interactions SIII \cite{BEINER197529}, SkI3 and SkI4 \cite{REINHARD1995467}, SkM \cite{BARTEL198279}, SkO \cite{Reinhard:PRC.60.014316}, SkP \cite{DOBACZEWSKI1984103}, SkX \cite{Brown:PRC.58.220}, SLy4 and SLy6 \cite{CHABANAT1997710,CHABANAT1998231}, Sk255 \cite{Agrawal2003}, BSk17 \cite{GorielyPRL.102.152503}, LNS \cite{CaoPRC.73.014313}, and UNEDF0 and UNEDF1 \cite{STOITSOV20131592}.The filled triangles are results from the NL3/FSU interaction \cite{PiekarewiczPRC.83.034319,AgrawalPRC.81.034323}, the open circles are DD-ME interaction \cite{VretenarPRC.68.024310,Lalazissis2005}. The filled squares are predictions from the Skyrme-SV interaction \cite{KlupfelPRC.79.034310}. The experimental constraint for the dipole polarizability from a Coulomb excitation experiment \cite{TamiPRL.107.062502} is shown as a horizontal band. The first clear observation is that the linear dependence of the dipole polarizability on the basis of the interactions used is not as clear as expected. The second is that the experiment of Ref. \cite{TamiPRL.107.062502} extracted a neutron skin of $\Delta r_{np}=1.56^{+0.025}_{0.021}$ fm, whereas the average $\Delta r_{np}=(0.168\pm 0.022)$ fm, is extracted from mean field calculations \cite{PiekarewiczPRC.85.041302} based on all interactions mentioned previously. This value that is close to the one of the experiment   of Ref. \cite{TamiPRL.107.062502}.
 
\begin{figure}[t]
\begin{center}
{\includegraphics[width=9cm]{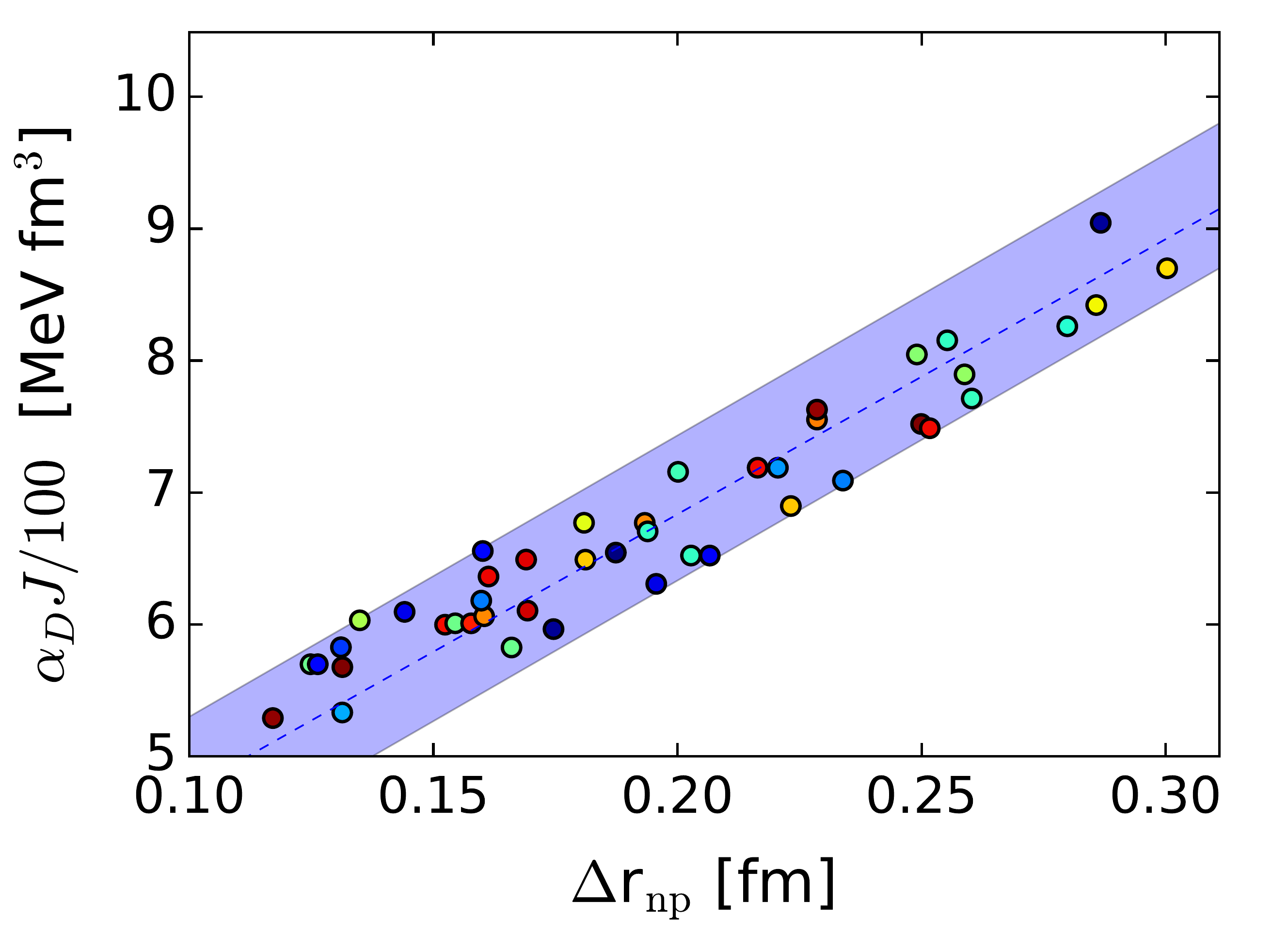}}
\end{center}
\vspace{-0.5cm}
\caption{Same as in Fig. \ref{alphadJ}, but with the dipole polarizability multiplied by the volume term of the symmetry energy, $J$.}
\label{alphadJ}
\end{figure}

Figure \ref{alphad} shows  a significant amount of scatter between the results for different functionals. However, as described in details in Ref. \cite{ROCAMAZA201896}, a much better correlation can be obtained if one multiplies the polarizability, $\alpha_{D}$ by the volume term of the symmetry energy, $J$. This is clearly seen in Figure \ref{alphadJ} where we plot all data displayed Fig. \ref{alphad} but now in the form of the product $\alpha_{D} J$. The evident correlation between $\alpha_{D} J$ and $\Delta r_{np}$ has indeed been confirmed by microscopic calculations \cite{ROCAMAZA201896}.

\begin{figure}[t]
\begin{center}
{\includegraphics[width=9cm]{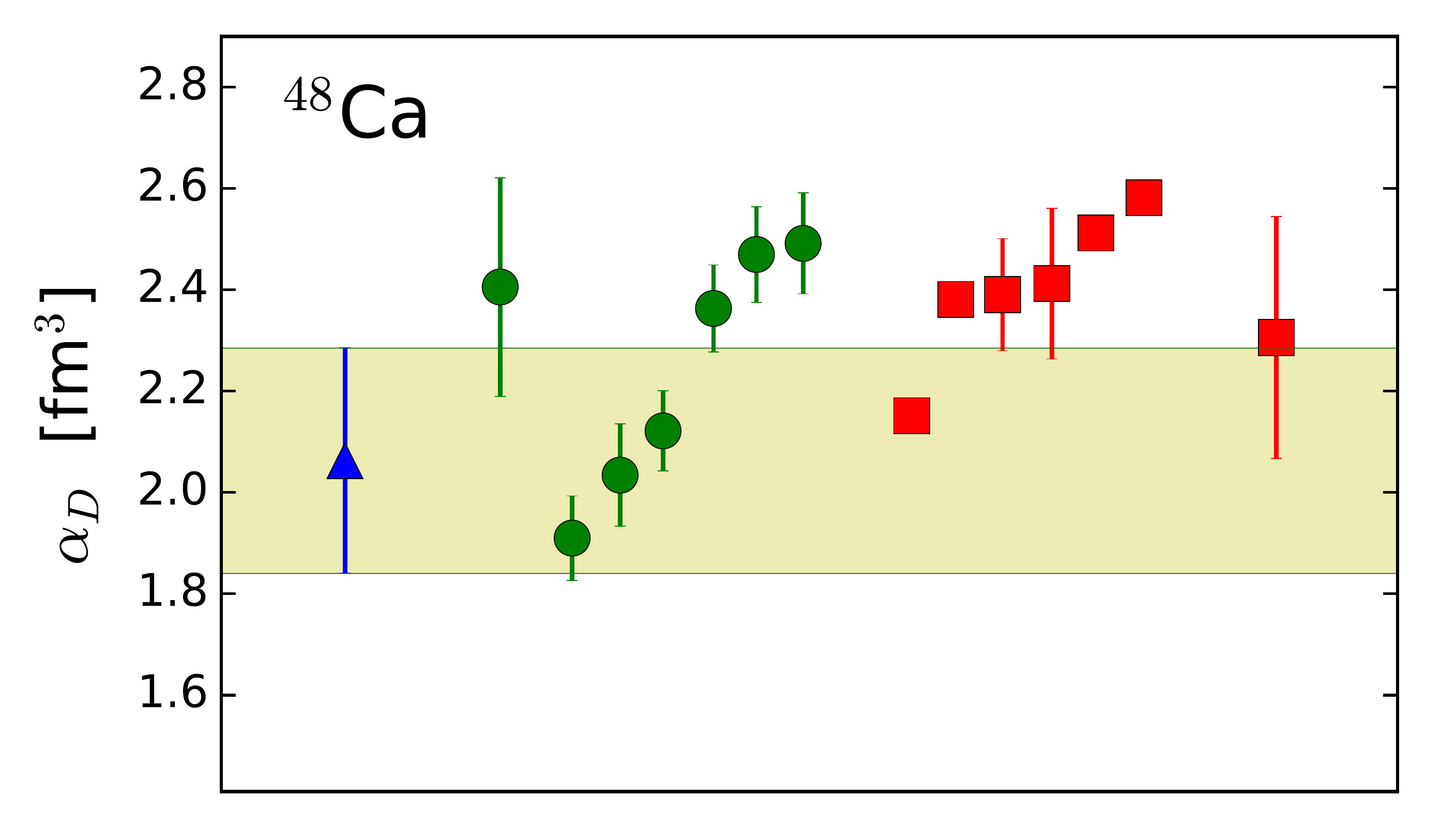}}
\end{center}
\vspace{-0.5cm}
\caption{Electric dipole polarizability in $^{48}$Ca as predicted by chiral-EFT calculations (circles) and mean-field models (squares).  More details on the $\chi$-EFT calculations, see Ref. \cite{Hagen:2016}. The experimental data \cite{BirkhanPRL.118.252501} is represented by the solid triangle and the experimental error by the horizontal band.}
\label{alphad2}
\end{figure}

In Fig. \ref{alphad2} we show the electric dipole polarizability in $^{48}$Ca as predicted by chiral-EFT  \cite{Hagen:2016} calculations (circles) and mean-field models (squares).  The experimental data \cite{BirkhanPRL.118.252501} is represented by the solid triangle and the experimental error by the horizontal band. Based only on these interactions and functionals, the authors of Ref. \cite{BirkhanPRL.118.252501} extract a neutron skin in $^{48}$Ca of 0.14-0.20 fm, which, together with the ab-initio $\chi$-EFT results imply that the volume symmetry energy ranges within $J=28.5$-33.3 MeV and a slope parameter of $L = 43.8$-48.6 MeV, which sets a very accurate constraint for the slope parameter. This surprisingly tight constraint still needs to be confirmed by other experimental and theoretical developments. 

\section{Indirect methods with relativistic radioactive beams}

\subsection{Electron scattering}

\subsubsection{Elastic electron scattering}
Since at the energies considered here, the electron can be treated as a point-like particle, its interaction with the nucleus is extremely well described by quantum electrodynamics (QED), and its momentum transfer can be used as a variable quantity for a given energy transfer, they are ideal probes of nuclear  charge distributions, transition densities, and nuclear response functions. 

Under the conditions of validity of the plane-wave Born approximation (PWBA), the elastic electron scattering cross section in the laboratory  is
\begin{equation}
\left(  \frac{d\sigma}{d\Omega}\right)  _{\mathrm{PWBA}}=\frac{\sigma_{M}}{1+\left(  2E/M_{A}\right)  \sin^{2}\left(  \theta/2\right)  }\ |F_{ch}%
\left(  q\right)  |^{2}, \label{PWBA}%
\end{equation}
where $\sigma_{M}=(e^{4}/4E^{2})\cos^{2}\left(  \theta/2\right) \sin^{-4}\left(  \theta/2\right)  $ is known as the Mott cross section. The  denominator accounts for a recoil correction, $E$ is the electron total energy, $M_{A}$ is the nucleus mass and $\theta$\ is the electron scattering angle. $q=2k\sin\left(  \theta/2\right)  $ is the momentum transfer, $\hbar k$ is the electron momentum, and $E=\sqrt{\hbar^{2}k^{2}c^{2}+m_{e}^{2}c^{4}}$. The Mott cross section has a dependence on the momentum transfer as
\begin{equation}
\sigma_{M} \sim {E\over q^{4}} . \label{mott}
\end{equation}

The charge form factor $F_{ch}\left(  q\right)  $ is, for a spherical charge distribution,
\begin{equation}
F_{ch}\left(  q\right)  =\int_{0}^{\infty}dr\ r^{2}j_{0}\left(  qr\right) \rho_{ch}\left(  r\right)  . \label{form}
\end{equation}

\begin{figure}[t]
\begin{center}
{\includegraphics[width=8cm]{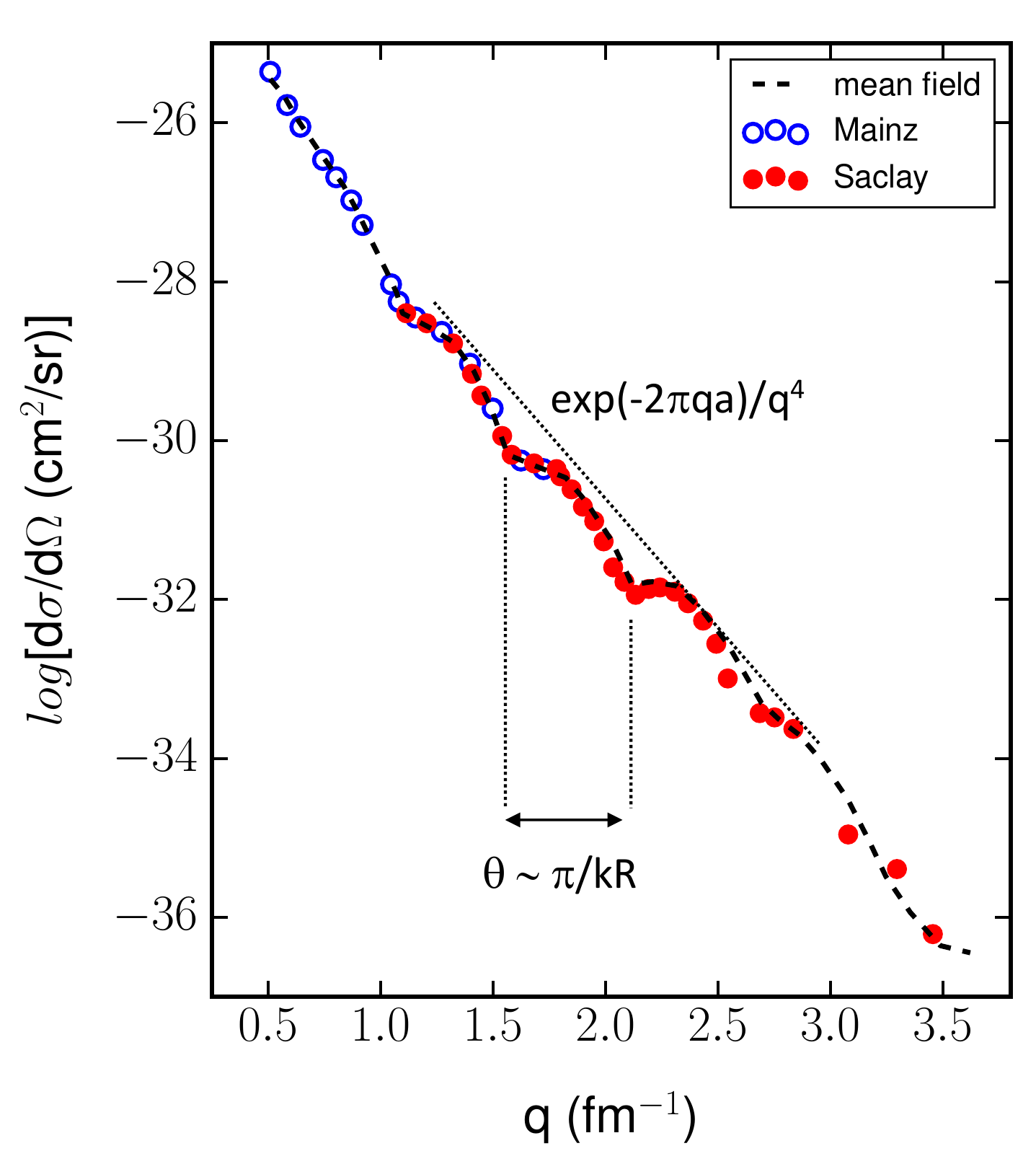}}
\end{center}
\vspace{-0.5cm}
\caption{
Data on elastic electron scattering off Pb from the Saclay and Mainz experimental groups. The dashed lines are form factors calculated with Skyrme-Hartree-Fock theory \cite{Cavendon1987}. The dotted lines show that the exponential decay of the cross sections by many orders of magnitude are due to the nuclear diffuseness while the dips are reflect the nuclear radius. 
}
\label{elastice}
\end{figure}

Elastic electron scattering essentially measures the Fourier transform of the charge distribution through the form factor $F_{ch}$. The physics displayed in the cross section is better understood if we take for simplicity the one-dimensional case. For light nuclei, well described by a Gaussian distribution, we have
\begin{equation}
F_{ch}\left(  q\right) \sim \int e^{iqx} \rho(x) dx \sim \int dx {e^{iqx} \over a^2+x^2} =
{\pi\over a} \ e^{-qa}, \end{equation}
While, for a heavy nucleus,  the density $\rho$ is better described by a Fermi function or Woods-Saxon charge distribution, yielding
\begin{equation}
F_{ch}\left(  q\right) \sim\int dx {e^{iqx} \over 1+e^{(x-R)/a}} \sim (4\pi)  \sin (qR) \ e^{-\pi qa},
\end{equation}
valid for $R\gg a$, and $ qa \gg 1$. Upgrading the plane waves to eikonal wavefunction (see text ahead), one has 
\begin{eqnarray}
F_{ch}\left(  q\right) &\sim& \int db b J_0(qb) [1-e^{i\chi(b)}] \nonumber \\ &\sim& \int db b {J_0(qb)  \over 1+\exp[({b-R\over a})]} \nonumber \\
&\sim& {R\over q} J_1(qR) \exp(-\pi q a),
\end{eqnarray} 
where $J_n$ are Bessel functions of order $n$. Therefore, in the differential cross sections for elastic electron scattering the distance between minima  is a direct measure of the nuclear size, whereas their exponential decay reflects the surface diffuseness of the charge distribution. This is shown in Figure \ref{elastice} for the electron scattering off lead. The dashed lines are form factors calculated with Skyrme-Hartree-Fock theory \cite{Cavendon1987}. The dotted lines show that the exponential decay of the cross sections by many orders of magnitude are due to the nuclear diffuseness while the dips are reflect the nuclear radius.

One can also deduce that, at low-momentum transfers,  Eq. \eqref{form} yields to leading order
\begin{equation}
F_{ch}\left(  q\right)  /Z=1-\frac{q^{2}}{3!}\left\langle r_{ch}^{2}\right\rangle + \frac{q^{4}}{5!}\left\langle r_{ch}^{4}
\right\rangle +{\cal O}(q^{6}), \label{formexp}
\end{equation}
where $\left< r^{n}\right>$ is the $n$-th moment of the charge density distributions,
\begin{equation}
\left< r^{n}\right>=\int r^{n }\rho(r) d^{3}r \label{rnmom}.
\end{equation}
At the lowest order, elastic electron scattering  at low-momentum transfers obtains the root mean squared radius of the charge distribution, $\left\langle r_{ch}^{2}\right\rangle ^{1/2}$. With increasing data at higher momentum transfers, more details of the charge distribution are probed.

\subsubsection{Inelastic electron scattering}
Inelastic electron scattering is another powerful tool for infer the properties of nuclear excited states, such as their spins, parities, and transition strengths between ground and excited states (e.g. Refs. \cite{Barber:1962,eisenberg1988excitation}).  In the plane-wave Born approximation (PWBA), one obtains for the cross section for inelastic electron scattering  \cite{Barber:1962,eisenberg1988excitation}
\begin{align}
\frac{d\sigma}{d\Omega}  &  =\frac{8\pi e^{2}}{\left(  \hbar c\right)  ^{4}
}\left(  \frac{p^{\prime}}{p}\right)  \sum_{L}\left\{  \frac{EE^{\prime}
+c^{2}\mathbf{p\cdot p}^{\prime}+m^{2}c^{4}}{q^{4}}\left\vert {\mathcal M}\left(
{\bf q};CL\right)  \right\vert ^{2}\right. \nonumber\\
&  \left.  +\frac{EE^{\prime}-c^{2}\left(  \mathbf{p\cdot q}\right)  \left(
\mathbf{p}^{\prime}\cdot\mathbf{q}\right)  -m^{2}c^{4}}{c^{2}\left(
q^{2}-q_{0}^{2}\right)  ^{2}}\right.\nonumber \\
&\times \left. \left[  \left\vert {\mathcal M}\left(  {\bf q};ML\right)
\right\vert ^{2}+\left\vert {\mathcal M}\left(  {\bf q};EL\right)  \right\vert
^{2}\right]  \right\}  \label{PWBA2}
\end{align}
where $J_{i}$ $\left(  J_{f}\right)  $ $\ $is the initial (final) angular momentum, $\left(  E,\mathbf{p}\right)  $ and
($E^\prime ,\mathbf{p}^{\prime}$) are the initial and final four-momenta of the electron, and $\left(  q_{0},\mathbf{q}\right)
=\left(\Delta E/{\hbar c},\Delta \mathbf{p}/{\hbar}\right)  $ is the four-momentum transfer in the reaction. ${\cal M}_{ij}\left(  q;\Pi L\right)  $
are matrix elements for  Coulomb ($C$), electric ($E$) and magnetic ($M$) multipolarities, $\Pi=C,E,M$, respectively. For small momentum transfers and scattering at forward angles, electron scattering involves the same matrix elements as scattering by real photons and by
Coulomb excitation \cite{BERTULANI1988299}. Both in Coulomb excitation as in the scattering by real photons  the energy and momentum transfers are related by $|\Delta {\bf p}| =\Delta E/c$, whereas in electron scattering  the  momentum and energy transfer are varied independently, allowing for an additional information on the nuclear response. 

At very forward angles and small energy transfers, one can use the Siegert theorem \cite{Siegert1937,Sachs1951}  to prove that the  electric and Coulomb form factors appearing in Eq. (\ref{PWBA2}) are proportional to each other. Then, for electric multipole excitations, one can write \cite{BertulaniPRC.75.024606}
\begin{equation}
\frac{d\sigma}{d\Omega dE_{\gamma}}=\sum_{L}\frac{dN_{e}^{(EL)}\left(
E,E_{\gamma},\theta\right)  }{d\Omega dE_{\gamma}}\ \sigma_{\gamma}
^{(EL)}\left(  E_{\gamma}\right)  , \label{EPA}
\end{equation}
valid  for excitation energies $ E_\gamma \ll \hbar c/R$, with the ``equivalent photon number'' given by \cite{BertulaniPRC.75.024606}
\begin{eqnarray}
&&\frac{dN_{e}^{(EL)}\left(  E,E_{\gamma},\theta\right)  }{d\Omega dE_{\gamma}}
  =\frac{4L}{L+1}\frac{\alpha}{E}\left[  \frac{2E}{E_{\gamma}}\sin\left(
\frac{\theta}{2}\right)  \right]  ^{2L-1}\nonumber\\
&  \times&\frac{\cos^{2}\left(  \theta/2\right)  \sin^{-3}\left(
\theta/2\right)  }{1+\left(  2E/M_{A}c^{2}\right)  \sin^{2}\left(
\theta/2\right)  }\nonumber \\
&\times& \left[  \frac{1}{2}+\left(  \frac{2E}{E_{\gamma}}\right)
^{2}\frac{L}{L+1}\sin^{2}\left(  \frac{\theta}{2}\right)  +\tan^{2}\left(
\frac{\theta}{2}\right)  \right]  . \label{EPAE}%
\end{eqnarray}
Therefore, under these conditions (very forward angles and small energy transfers) the cross section for electron scattering are proportional to the cross sections for real photons inducing electric multipolarity $EL$ excitations, $\sigma_{\gamma}^{(EL)}\left(  E_{\gamma}\right)$. 
Because it is nearly impossible to study photo-nuclear cross sections with real photons and radioactive nuclei as targets, this might be a way to access this information in electron-ion facilities, if they are ever realized in practice  \cite{etde-21504514,Koop2012,Suda2012,Ohnishi_2015,SUDA20171}.  

\subsubsection{Electron radioactive-ion collider}
Electron radioactive-beam colliders have been proposed and already exist in some nuclear physics facilities (for a recent review, see \cite{SUDA20171}). 

To investigate the asymmetry properties of bulk  nuclear matter, a  systematic study of the charge-density distributions, of nuclei with large proton-neutron asymmetry is necessary. Together with information about the hadronic matter distribution in nuclei, one can get information on the nuclear matter incompressibility \cite{Gambhir1986,Wang.PRC.60.034310},  the bulk of the nuclear symmetry energy \cite{Vretenar.EPJA2004}, and its slope parameter to extrapolate to the pure neutron matter equation of state and its dependence on density \cite{Brown.PRL.2000}. Despite heroic efforts, this era of nuclear physics is still in its infancy as an electron-radioactive-beam collider with the properties needed to investigate unstable nuclei is still unavailable \cite{SUDA20171}.  

Since the electron scattering theory is  very well understood,  electron scattering is potentially a great tool  for inferring the charge distribution in unstable nuclei in a rare isotope ion-electron collider. In conjunction with hadronic probes, such as elastic proton scattering, a better study of the evolution of neutron skins in nuclei can be achieved.   The implications of elastic and inelastic electron scattering in determining the properties of light, neutron-rich nuclei have been studied in Ref. \cite{Bertulani_2006,BertulaniPRC.75.024606,BERTULANI2007366}. But while the electron-ion collider is not available, other sources of information on the difference between neutron and proton distribution in neutron-rich nuclei need to be used.

\subsubsection{The electron parity violation scattering}

In order to include weak interaction scattering in electron scattering, the electromagnetic potential $Ze^{2}/r$ has to be generalized to
\begin{equation}
V(r)= ke^{2}g_{e}g_{T}{\exp{(-Mr)}\over 4\pi r} . \label{pvio1}
\end{equation}
where here we use the notation $\hbar=c=1$, $g_{e}$ is the electron charge and $g_{T}$ is the charge of the target particle in units of the electron charge $e$. The constant $k$ denotes the strength of the coupling, with $M$ being the mass of the exchanged particle, where for $M=0$ one has the usual  electromagnetic interaction, while, for  $M = M_Z$ (mass of $Z_{0}$-boson) the interaction is due to the neutral weak current. One also has  $k = 1$ for electromagnetism and $k = (\sin \theta_{W} \cos \theta_{W})^{2}$ for weak interaction scattering. 
The exchanged particle in the neutral weak interaction, i.e. the $Z_{0}$ boson, has both vector and axial vector couplings. For spinless nuclei, the net axial coupling to the nucleus is absent. Moreover, the $Z_{0}$ has a much larger coupling to the neutron than the proton and it also contains
a large axial coupling to the electron. That is the reason for a parity-violating amplitude in electron scattering.

In Table  \ref{parviot} we list the electromagnetic and weak charges, or coupling constants, of the electron and light quarks.

\begin{table}[ht]
\begin{center}
\label{parviot}
\begin{tabular}{|c|c|c|c|}
\hline\hline
Particle & $g_{em}$ & $g_{V}$   &  $g_{A}$    \\ 
\hline 
$e^{-}$& -1 &  $-{1\over 4} +\sin^{2} \theta_{W}$ &${1\over 4}$\\
\hline 
$u$ & ${2\over 3}$ & ${1\over 4} -{2\over 3}\sin^{2} \theta_{W}$ &$-{1\over 4}$ \\
\hline
$d,s$& $ - {1\over 3}$ &$ -{1\over 4} +{1\over 3}\sin^{2} \theta_{W} $&${1\over 4}$ \\
\hline
\hline
\end{tabular}
\caption{The electromagnetic and weak charges, or coupling constants, of the electron and light quarks.}. 
\end{center}
\end{table}

In longitudinally polarized electron scattering, the weak charge of a relativistic electron depends on its helicity, so that $g_{R} \neq g_{L}$, where $g_{R}$ and $g_{L}$ now denote the the charge of an electron with right-handed and left-handed helicity, respectively. Therefore, it is more convenient to write them in terms of the vector and axial-vector weak charges, $ g_{R} = g_{V} + g_{A}$ and $g_{L} = g_{V}-g_{A}$. The weak and electromagnetic charges of the electrons and relevant quarks are given in Table 1. The electroweak mixing angle is known to high precision, as $\theta_{W}=0.23116 \pm 0.00013$ \cite{Nakamura2010}.

With the potential of Eq. (\ref{pvio1}), the scattering cross section of Eq. (\ref{PWBA}) becomes (neglecting recoil corrections)
\begin{equation}
\left(  \frac{d\sigma}{d\Omega}\right)  _{\mathrm{PWBA}}=\left(\frac{2kg_{e}g_{T}}{q^{2} + M^{2} }\right)^{2}\ |F_{ch}
\left(  q\right)  |^{2} \cos^{2}\left({\theta \over 2}\right), \label{piov2}
\end{equation}
When the momentum transfer is small so that $q\ll M_{Z}$, the weak interaction is negligible compared to the electromagnetic charge,  so that the charge form factor is an sum of form factors for light-quarks distributions in the nucleus:
\begin{equation}
F_{ch}^{(e.m.)}/e={2\over 3} F_{u}- {1\over 3} \left(F_{d}+F_{s}\right).\label{piov3}
\end{equation}

In experiments aiming at studying the effects of  parity violating electron scattering one measures the asymmetry
\begin{equation}
A_{PV}={d\sigma_{R}-d\sigma_{L}\over d\sigma_{R}-d\sigma_{L}}.\label{piov4}
\end{equation}
for polarized electrons with right and left helicity. If this quantity is different than zero, it measures the strength of parity violation. In this case, it is dominated by the interference between the weak and electromagnetic amplitudes, so that
\begin{eqnarray}
A_{PV}&=&-{q^{2}\over 2 (M_{Z}\cos \theta_{W}\sin\theta_{W})^{2} }\nonumber 
\\ 
&\times& {({1\over 4} -{2\over 3}\sin^{2} \theta_{W})F_{u}+ (-{1\over 4} +{1\over 3}\sin^{2} \theta_{W})(F_{d}+F_{s})\over {2\over 3} F_{u}- {1\over 3} \left(F_{d}+F_{s}\right)}\nonumber \\
\label{piov5}
\end{eqnarray}
This provides a different combination of quark distributions in the nucleon than Eq. (\ref{piov3}) and is one of the reasons why parity-violating electron scattering is a useful tool in nuclear physics.

A formulation following the same path as described above can be used to directly connect the scattering observables to the neutron and proton densities in the nucleus \cite{HorowitzPRC.63.025501}. To a good approximation, the total interaction of the electron and a nucleus, including the electromagnetic, $V_{C}$, and the axial potential, $A$, arising from the parity violating term,  can be written as 
\begin{equation}
{\cal V}(r) = V_{C}(r) + \gamma_{5} A(r),\label{piov6}
\end{equation}
where 
 \begin{equation}
A(r)={G_{F}\over 2^{3/2}} \left[(1-4\sin^2\theta_{W})\rho_{p}(r) - \rho_{n}(r)\right]\label{piov7}
\end{equation}
The electron-nucleus interaction via the axial potential is of order one eV while the electromagnetic interaction with the nucleus is of order MeV. Therefore, it only becomes important in parity violating scattering. Since $\sin^{2} \theta_{W}\sim 0.23$,  the factor $(1-4\sin^2\theta_{W}$) is small and $A(r)$ depends mainly on the neutron distribution $\rho_{n}(r)$.

As shown previously, the presence of the  parity-violating asymmetry quantity in Eq. (\ref{piov4}) will involve the interference between $V_{C}(r)$
and $A(r)$ and one gets for it a simple expression
 \begin{equation}
A_{PV}(r)={G_{F}Q^{2}\over 4\pi e^{2}\sqrt{2}} \left[4\sin^2\theta_{W}-1+ {F_{n}(Q)\over F_{p}(Q)}\right],\label{piov8}
\end{equation}
where $Q$ is the four-momentum transfer, $Q^{2}=q_{\mu}q^{\mu}>0$, and $F_{n}(F_{n})$ is the neutron(proton) form factor. Therefore, the parity-violating asymmetry is proportional to $Q^{2}/M^2_Z$ because $G_{F}\propto 1/M^2_Z$. Besides, since $1-4\sin^2\theta_W \ll 1$ and $F_p(q)$ is known in a nucleus such as $^{208}$Pb the asymmetry  yields a direct measure of $F_n(q)$ and is a good way to study the neutron skin of the nucleus \cite{HorowitzPRC.63.025501}.

Based on these ideas, an experiment named the Lead Radius Experiment (PREX)  for electron scattering experiments on $^{208}$Pb was carried out at the Jefferson Lab (JLab) for the doubly-magic nucleus whose first excited state was discriminated by high resolution spectrometers. The neutron skin in $^{208}$Pb was obtained by an analysis of the parity violating asymmetry, yielding $\Delta r_{np}(^{208}{\rm Pb})=0.33^{+0.16}_{-0.18}$ fm \cite{Horowitz2014,michaels2015electroweak}. If we compare this experimental result to the plot in Fig. \ref{skinL}, we see that it practically agrees with any of the mean-field model, and a wide range of the slope parameter $L$. The PREX experiment was a benchmark run and a proposal exists for the Calcium Radius Experiment (CREX) electron scattering on $^{48}$Ca. For this nucleus, microscopic nuclear theory calculations have been developed and shown to be sensitive to poorly constrained three-nucleon forces \cite{Hagen:2016}. The realization of this experiment can set a tighter constraint on the slope parameter of the ANM EoS. 

\subsection{Direct reactions at relativistic energies}

\subsubsection{Nucleus-nucleus scattering at high energies}

Direct reactions in nucleus-nucleus collisions at relativistic energies have a fundamental problem: No-one really knows how to treat accurately relativistic many-body collisions, when retardation, simultaneity,  and the cumulative dynamics of local (strong) and long-range (Coulomb) pair-wise interactions are treated consistently. Evidently, non-relativistic DWBA methods are inappropriate to study nuclear excitation and other inelastic process at laboratory energies of 100 MeV/nucleon and above, where Lorentz contraction and other relativistic effects at the level of 10\% or more factor in the kinematics and in the dynamics of the system. In particular, it is highly doubtful if the concept of an optical potential, a non-relativistic concept, can be adapted to a covariant theory involving four-dimensional potentials.   

The relativistic treatment of direct reactions at high-energy collisions is amenable to a reasonable theoretical treatment only if a sudden approximation is justified. In this case, the theoretical description of the relativistic scattering waves can be separated from the internal structure of the nuclei which then can be treated in a non-relativistic fashion.

The best way to treat ``soft'' collisions between composite particles at high energies is the formalism of the Glauber theory \cite{glauber1959lectures}. The wavefunction of a high energy projectile after it interacts with a target is given by
 \begin{equation}
\left| \Psi (t) \right> = e^{iQ_\mu x^{\mu}}S(b) \left|\psi({\bf r}_{1}, {\bf r}_{2},\cdots, {\bf r}_{A_{P}}\right>\label{eik1}
\end{equation}
where $S(b)$ is the eikonal ``survival'' amplitude, also known as ``eikonal scattering matrix'', and $\left|\psi\right>$ is the internal wave function of the projectile. In this way, its wavefunction is factorized in a scattering times an internal part. Anti-symmetrization of projectile and target nucleons is neglected. Moreover, $Q_{\mu}=(E_{f}-E_{I},{\bf p}_{f}-{\bf p}_{I})$ is the four-momentum transfer and $x_{\mu}=(t,{\bf r})$ are the corresponding coordinates. Assuming that the total energy of the projectile is much larger than the energy transfer in the collision, i.e., $E_{I}=E_{f}$, one has the stationary wavefunction
 \begin{equation}
\left| \Psi \right> = e^{I{\bf q}\cdot {\bf r}}S(b) \left|\psi({\bf r}_{1}, {\bf r}_{2},\cdots, {\bf r}_{A_{P}},\right>\label{eik2}
\end{equation}
where ${\bf q}$ is the momentum transfer. $S(b)$ can also be cast in the form $S(b) = \exp(i \chi (b) )$, where $\chi(b)$ is known as the eikonal phase $\chi_{R} + i \chi_{i}$, with a real imaginary part $\chi_{i}$ accounting for inelasticity. In the eikonal theory (for more details, see Ref. \cite{Bertulani:2004}, the projectile survival probability is given by
 \begin{equation}
{\cal P}_{survival}(b)=|S(b)|^{2}.\label{eik3}
\end{equation}

The eikonal scattering wavefunction for a potential $U({\bf r})$ at high energies is given by \cite{glauber1959lectures,Bertulani:2004}
 \begin{equation}
\psi_{eik}^{(+)}({\bf r})=\exp\Big[i\big({\bf k}\cdot{\bf r}+\chi({\bf b},z)\big)\Big],\label{eik3b}
\end{equation}
where
 \begin{equation}
\chi(\mathbf{b},z)=-{\frac{1}{\hbar\mathrm{v}}}\ \int_{-\infty}^{z}U(\mathbf{b}, z')\ dz'\  .
\end{equation}
But here, again, we run into the problem we mentioned previously  that if $U$ is taken as an ordinary optical potential, the formalism is not in compliance with the laws of transformation in special relativity. 

The good news is that the scattering part of the wavefunction in Eq. (\ref{eik1}) is Lorentz invariant because $S(b)$ only depends on the transverse coordinate $b$, usually associated with an impact parameter. But further simplifications need to be done when dealing with a many-body system. First, the survival amplitude,  $S(b)$, is cast as an incoherent sum of scattering matrices for each individual nucleon nucleon collision, i.e.,
\begin{equation}
S(b)=\prod_{j}e^{i\chi_{j}(b_{j})}=e^{i\sum_{j}\chi_{j}(b_{j})}, \label{eik4}
\end{equation}
where  $\chi_{j}(b_{j})$ is now the eikonal phase acquired in a binary collision $j$. Notice that the assumption of incoherent collisions do not play well with an exact relativistic treatment of the collision, as simultaneity is invoked, and retardation is neglected. But the advantage of using only the transverse directions to the beam is that the expression is Lorentz invariant.

The above formulation is difficult to implement as one needs to account for the transverse distances $b_{j}$ between all nucleons participating in the collision. An often used approximation is called the Optical Limit (OL) of the Glauber theory amounting to an average over all the collisions what means that a continuous ``frozen'' nucleon density can be used to calculate $S(b)$. In other words
 \begin{equation}
S_{OL}(b)=e^{i\chi_{OL}(b)}= \left< e^{i\sum_{j}\chi_{j}(b_{j})}\right>. \label{eik5}
\end{equation}
As explained in Ref. \cite{Bertulani:2004}, one can obtain the OL eikonal phase by relating it to the observables in nucleon-nucleon  collisions so that
\begin{equation}
\chi_{OL}(b)= \int \rho_{P}({\bf r}')\Gamma({\bf s}-{\bf r}'+{\bf r}'')\rho_{T}({\bf r}'') d^{3}r'd^{3}r'' ,\label{eik6}
\end{equation} 
where $\rho_{P}$($\rho_{T}$) is the projectile(target) density and the ``profile function'' $\Gamma({\bf b})$. For most cases, involving spherically symmetric densities, one can use the Fourier transform of the nuclear densities, yielding the simple expression
\begin{equation}
\chi_{OL}(b)= \int \tilde{\rho}_{P}(q')\,\tilde{\Gamma}(q)\,\tilde{\rho}_{T}(q) \,J_0(qb)\,q \,dq ,\label{eik7}
\end{equation} 
where $J_{0}$ is the ordinary Bessel function of zeroth-order, and the profile function in momentum space
\begin{equation}
\tilde{\Gamma}(q)=\frac{i +\alpha_{NN}}{4\pi }\sigma_{NN}e^{-\beta_{NN}q^2} .\label{eik8}
\end{equation}
In the equation above, $\sigma_{NN}$ is the total nucleon-nucleon cross section,  $\alpha_{NN}$ is the ratio of the real to the imaginary part of the $NN$-scattering amplitude, and $\beta_{NN}$ is the momentum dependence parameter.

The parameters of the nucleon-nucleon cross scattering amplitudes for $E_{lab} \geq100$ MeV/nucleon are shown in Table \ref{nnpar}, extracted from Refs. \cite{Ray:1979,HUSSEIN1991279}. An isospin average of these quantities can be done in most practical cases, while in other situations they can be used to separate cross sections sensitiveness to the proton and neutron densities separately.

\begin{table}[ht]
\begin{center}
\label{nnpar}
\begin{tabular}{|c|c|c|c||c|c|c|c|c|}
\hline\hline
$E_{lab}$ & $\sigma_{pp}$ & $\alpha_{pp}$ & $\beta_{pp}$ & $\sigma_{pn}$ & $\alpha_{pn}$
& $\beta_{pn}$\\\hline
\lbrack MeV] & [fm$^{2}$] &  & [fm$^{2}$] & [fm$^{2}$] &  & [fm$^{2}$]\\\hline
40 & 7.0 & 1.328&  0.385&  21.8&  0.493&  0.539\\\hline
60&  4.7 & 1.626 & 0.341 & 13.6 & 0.719 & 0.410\\\hline
80&  3.69&  1.783&  0.307 & 9.89&  0.864&  0.344\\\hline
100 & 3.16&  1.808&  0.268&  7.87 & 0.933&  0.293\\\hline
120&  2.85&  1.754&  0.231&  6.63&  0.94&  0.248\\\hline
140&  2.65 & 1.644&  0.195&  5.82 & 0.902 & 0.210\\\hline
160 & 2.52&  1.509&  0.164&  5.26&  0.856&  0.181\\\hline
180 & 2.43&  1.365 & 0.138&  4.85 & 0.77&  0.154\\\hline
200 & 2.36 & 1.221 & 0.117&  4.54&  0.701&  0.135\\\hline
240 & 2.28&  0.944&  0.086&  4.13&  0.541&  0.106\\\hline
300 & 2.42&  0.626&  0.067&  3.7 & 0.326&  0.081\\\hline
425 & 2.7&  0.47&  0.078&  3.32&  0.25&  0.0702\\\hline
550 & 3.44&  0.32&  0.11&  3.5 & -0.24&  0.0859\\\hline
650 & 4.13&  0.16&  0.148&  3.74&  -0.35&  0.112\\\hline
700&  4.43&  0.1&  0.16&  3.77&  -0.38 & 0.12\\\hline
800 & 4.59 & 0.06&  0.185&  3.88&  -0.2&  0.12\\\hline
1000 & 4.63&  -0.09&  0.193&  3.88&  -0.46 & 0.151\\\hline
2000 & 4.67 &  0. & 0.12 & 3.88 & -0.50 & 0.151\\\hline
\end{tabular}
\caption{Parameters \cite{Ray:1979,HUSSEIN1991279,IbrahimPRC.77.034607} for the nucleon-nucleon amplitude, as given by
Eq. (\ref{eik8}). The compilation of the interpolated values to several energies and adapted to collisions at lower energies were taken from Ref. \cite{IbrahimPRC.77.034607}.}\label{annan}
\end{center}
\end{table}

The Coulomb interaction also adds to the eikonal phase. In the OL, the total eikonal phase is $\chi=\chi_{OL}+\chi_{C}$, where  the Coulomb
eikonal phase, $\chi_{C}$ is to first-order
\begin{equation}
\chi_{C}(b)=2\eta\ln(kb)\ ,\label{intro4}
\end{equation}
where $\eta=Z_{1}Z_{2}e^{2}/\hbar\mathrm{v}$, $Z_{1}$ and $Z_{2}$ are the charges of projectile and target, respectively, $v$ is their relative
velocity, $k$ their wavenumber in the center of mass system. Eq. \ref{intro4} reproduces the exact Coulomb scattering amplitude when used in the calculation of the elastic scattering with the eikonal approximation \cite{Bertulani:2004}:
\begin{equation}
f_{C}(\theta)={\frac{Z_{1}Z_{2}e^{2}}{2\mu v^{2}\ \sin^{2}(\theta/2)}}
\ \exp\Big\{-i\eta\ \ln\Big[\sin^{2}(\theta/2)\Big]+i\pi+2i\phi_{0}
\Big\}\label{fctheta}
\end{equation}
where 
\begin{equation}
\phi_{0}=\arg\Gamma(1+i\eta/2)=-\eta C+\sum_{j=0}^{\infty}\left(  {\frac{\eta}{j+1}}-\arctan
{\frac{\eta}{j+1}}\right)  \ , \label{elast6}
\end{equation}

The Coulomb phase in Eq. (\ref{intro4}) diverges at $b=0$, but since the strong absorption suppresses scattering at small impact parameters this fact does not matter numerically.  One can also assume a uniform charge distribution with radius $R$ and the Coulomb phase is finite for $b=0$:
\begin{eqnarray}
\chi_{C}(b)  & =&2\eta\ \Bigg\{\Theta(b-R)\ln(kb)+\Theta(R-b) \nonumber \\
&\times& \Bigg[\ln(kR)+\ln(1+\sqrt{1-b^{2}/R^{2}})\nonumber\\
& -&\left. \left. \sqrt{1-b^{2}/R^{2}}-{\frac{1}{3}}(1-b^{2}/R^{2})^{3/2}%
\right]\right\}\ ,\label{chico_0}%
\end{eqnarray}
where $\Theta$ is the step function. Assuming a Gaussian distribution of charge radius $R$ for light nuclei, the Coulomb phase becomes
\begin{equation}
\chi_{C}(b)=2\eta\ \Big\{\ln(kb)+{\frac{1}{2}}E_{1}(b^{2}/R^{2}%
)\Big\}\ ,\label{chico2}%
\end{equation}
where the error function $E_{1}$ is defined as
\begin{equation}
E_{1}(x)=\int_{x}^{\infty}{\frac{e^{-t}}{t}}\ dt\ ,\label{chico3}%
\end{equation}
which also also converges, as $b\rightarrow0$. 

At intermediate energy collisions ($E_{lab} \simeq50$ MeV/nucleon),  a correction due to the Coulomb deflection can be done by replacing the  impact parameter $b$ within Eq. (\ref{eik7}) by the distance of closest approach in Coulomb scattering, 
\begin{equation}
b^{\prime}={a_{0}\over \gamma}+\sqrt{\left({a_{0}\over \gamma}\right)^{2}+b^{2}} \ ,\label{intro10}%
\end{equation}
where $a_{0}=Z_{1}Z_{2}e^{2}/m\mathrm{v}^{2}$ is half the distance of closest approach in a head-on collision of point charged particles. This correction leads to a considerable improvement of the eikonal amplitudes for the scattering of heavy systems in collisions at intermediate energies.  The Lorentz factor $\gamma=(1-v^2/c^2)^{-1/2}$ is remnant of the dynamic correction of the elastic Coulomb scattering, as shown in Ref. (\cite{KumarPRC.96.034605}). 

It is also useful for consistency purposes to extract optical potentials by using an inversion method of the eikonal phases.  In this approach, one uses the Abel transform \cite{glauber1959lectures}
\begin{equation}
U_{opt}(r)={\frac{\hbar\mathrm{v}}{i\pi r}}{\frac{d}{dr}}\int_{r}^{\infty}{\frac{\chi(b)}{(b^{2}-r^{2})^{1/2}}}\ r\ dr\ . \label{opt9}
\end{equation}
This procedure has been tested in Ref. \cite{VitturiPRC.36.1404} leading to effective potentials with tails very close to those obtained with phenomenological potentials of Ref. \cite{ALAMANOS198437}. Under certain approximations, and for Gaussian density
distributions, the potential obtained through Eq. (\ref{opt9}) coincides with that obtained with the double folding procedure \cite{VitturiPRC.36.1404}.

\subsubsection{Elastic scattering}\label{sec:elastic_th}

Using the scattering waves in the eikonal approximation, one can show that the elastic scattering amplitudes  are given by \cite{glauber1959lectures,Bertulani:2004}
\begin{equation}
f_{el}(\theta)=ik\ \int_{0}^{\infty}db\ b\ J_{0}(qb)\ \Big\{1-\exp
\Big[i\chi(b)\Big]\Big\}\ , \label{elast3}
\end{equation}
where $q=2k\sin(\theta/2)$,  $\theta$ is the scattering angle and $\chi$ includes contribution of both nuclear and Coulomb scattering. The elastic
scattering cross section is
\begin{equation}
{\frac{d\sigma_{el}}{d\Omega}}=\Big|f_{el}(\theta)\Big|^{2}\ . \label{elast3b}%
\end{equation}
Adding and subtracting the Coulomb amplitude, $f_{C}(\theta)$ in eq. \ref{elast3}, one gets
\begin{equation}
f_{el}(\theta)=f_{C}(\theta)+ik \int_{0}^{\infty}db b J_{0}(qb) \exp \Big[i\chi_{C}(b)\Big]\Big\{1-\exp\Big[i\chi(b^{\prime}%
)\Big]\Big\}\ ,\label{simplif}
\end{equation}
where we replaced $b$ in $\chi_{OL}(b)$ by $b^{\prime}$ as given by Eq. (\ref{intro10}) to account for the nuclear recoil. In contrast to Eq. (\ref{elast3}), Eq. (\ref{simplif}) converges quickly because the argument in the integral drops to zero at large distances.

\begin{figure}[t]
\begin{center}
{\includegraphics[width=8cm]{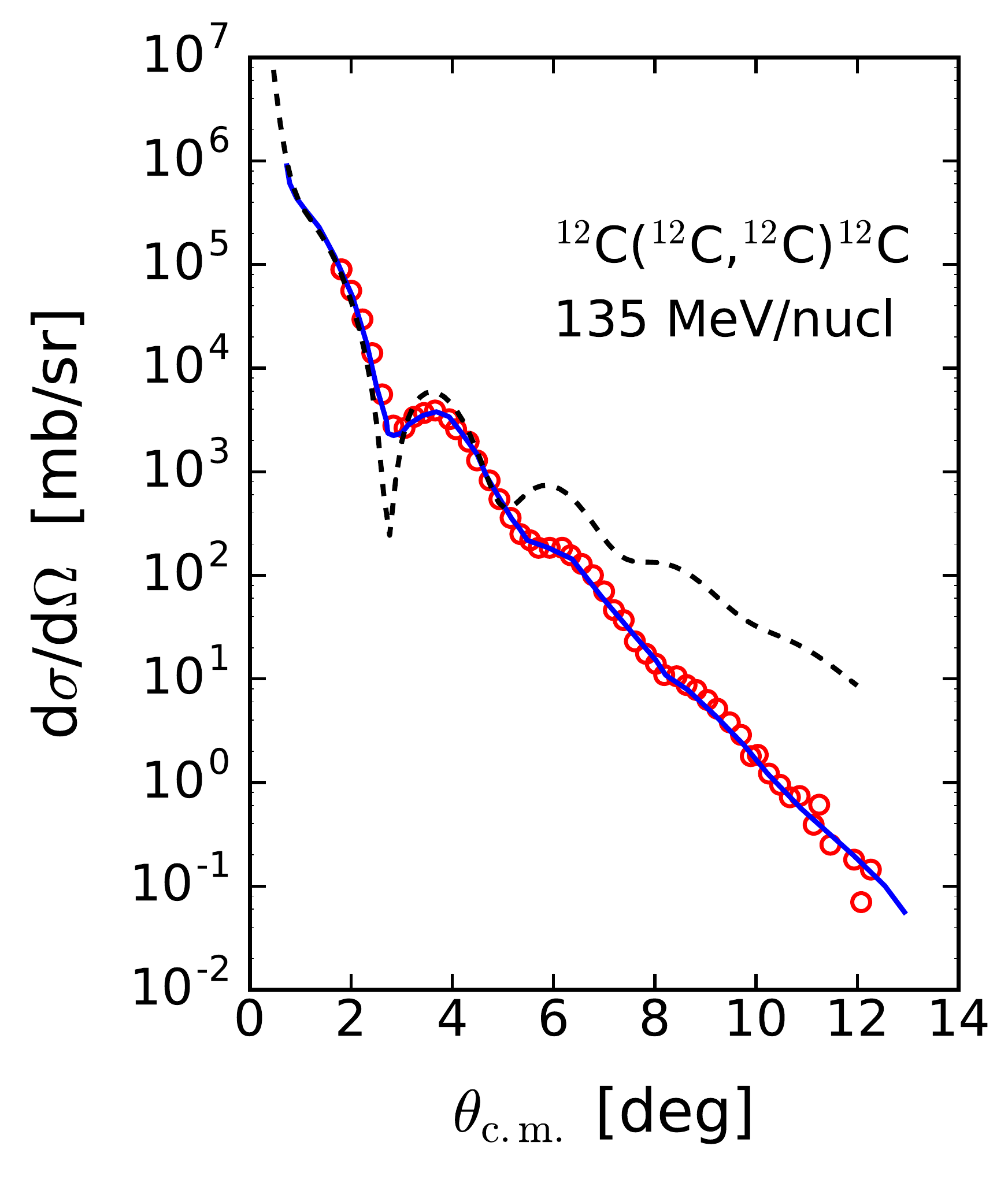}}
\end{center}
\vspace{-0.5cm}
\caption{
Experimental data on the differential cross section for the elastic scattering of $^{12}$C from $^{12}$C at  135 MeV/nucleon. The solid curve shows the result of the non-relativistic optical model calculation, whereas the dashed curve uses the Glauber formalism described in the text \cite{ICHIHARA1994278}.
}
\label{12c12c}
\end{figure}

For proton-nucleus elastic scattering, the cross sections acquire two components for spin-up and spin-down protons. The eikonal elastic scattering cross section becomes \cite{glauber1959lectures,Bertulani:2004}
\begin{equation}
{\frac{d\sigma_{el}}{d\Omega}}=\Big|F(\theta)\Big|^{2}+\Big|G(\theta
)\Big|^{2}\ ,
\end{equation}
where the spin-up scattering amplitude is
\begin{eqnarray}
F(\theta)&=&f_{C}(\theta)+ik\ \int_{0}^{\infty}db\ b\ J_{0}(qb)\ \exp
\Big[i\chi_{C}(b)\Big]\nonumber \\
&\times& \bigg\{1-\exp\Big[i\chi(b)\Big]\ \cos\Big[kb\ \chi
_{S}(b)\Big]\bigg\}
\end{eqnarray}
and the spin-down scattering amplitude is
\begin{equation}
G(\theta)=ik\ \int_{0}^{\infty}db\ b\ J_{1}(qb)\ \exp\Big[i\chi_{C}(b)+i\chi(b)\Big]\ \sin\Big[kb\ \chi_{S}(b)\Big]\ .
\end{equation}
In the equation above $q=2k\sin(\theta/2)$, where $\theta$ is the scattering angle,  $J_{0}$ ($J_{1}$) is the zero (first) order Bessel function. The eikonal phase $\chi_{S}$ is given by \cite{glauber1959lectures,Bertulani:2004}
\begin{equation}
\chi_{S}(\mathbf{b})=-{\frac{1}{\hbar\mathrm{v}}}\ \int_{-\infty}^{\infty}U_{SO}(\mathbf{b},\ z)\ dz\ , \label{elast20}
\end{equation}
where the spin-orbit potential is given by $({\bf s}\cdot {\bf L})U_{SO}(r)$, with ${\bf r}=({\bf b},z)$. However, as we discussed previously, this approach is not compatible with the relativistic transformation laws, which require four-potentials to describe interactions.

In Fig. \ref{12c12c} we show the experimental data on the differential cross section for the elastic scattering of $^{12}$C from $^{12}$C at  135 MeV/nucleon. The solid curve shows the result of the non-relativistic optical model calculation, whereas the dashed curve uses the Glauber formalism described in the text \cite{ICHIHARA1994278}. Despite the good agreement of the experimental data with a non-relativistic DWBA model, the Glauber calculation is superior in physics ingredients, as no parameters were introduced to fit the data, except for the observables in nucleon-nucleon scattering entering the profile function, Eq. (\ref{eik8}).  One sees that at large scattering angles the reproduction of the experimental data is not so good. Modifications of the theory to do a better job at higher energies has been introduced by several authors and, in general, the Glauber theory and its improvements have done a very good job in reproducing elastic scattering in high-energy heavy ion collisions (see, e.g., Refs. \cite{Hebborn2018,Book:Ber04}).

\begin{figure}[t]
\begin{center}
{\includegraphics[width=8.5cm]{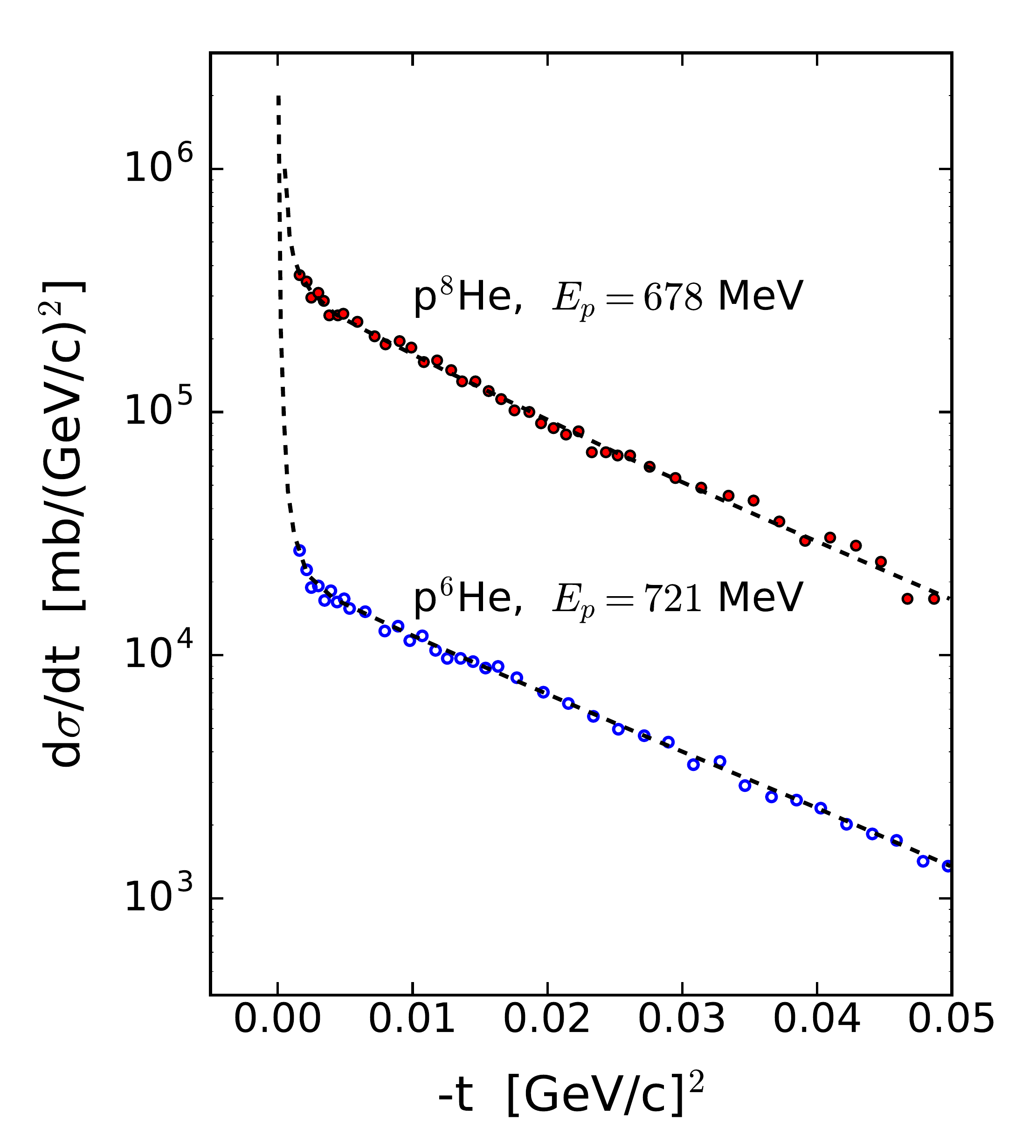}}
\end{center}
\vspace{-0.5cm}
\caption{\label{6he8he}
Differential cross sections, $d\sigma/dt$ versus the four-momentum transfer squared ($-t=Q^{2}$), for p$^4$He, p$^6$He and p$^8$He elastic scattering at energies $E_p = 628$ and 721  MeV, respectively.  The data are the experimental values from Refs. \cite{AlkhazovPRL78.2313,NEUMAIER2002247}.  The calculated cross sections are based on the Glauber theory for elastic scattering using nucleon-nucleon scattering data as input \cite{ALKHAZOV2002269}.
}
\end{figure}

Fig. \ref{6he8he}  shows the differential cross sections, $d\sigma/dt$ versus the four-momentum transfer squared ($-t=Q^{2}$), for p$^4$He, p$^6$He and p$^8$He elastic scattering at energies $E_p = 628$ and 721  MeV, in inverse kinematics, respectively.  The data are the experimental values from Refs. \cite{AlkhazovPRL78.2313,NEUMAIER2002247}.  The calculated cross sections are based on the Glauber theory for elastic scattering using nucleon-nucleon scattering data as input \cite{ALKHAZOV2002269}. The analysis of these data were useful to determine the nuclear matter distribution in $^{6}$He and $^{8}$He.

\subsubsection{Dirac phenomenology}\label{sec:dirac_phenomenology}

A very successful phenomenological approach was developed in Ref. \cite{ArnoldPRC.23.1949} based on an optical potential consisting of two parts:  $U_{0}(r)$, transforming like the time-like component of a Lorentz four-vector; and a second potential, $U_{S}(r)$, is a Lorentz scalar. 
$U_{0}$ and $U_{S}$ are regarded as effective interactions  arising from nucleons interacting via the meson exchange, folded with proton and neutron densities.  They depend  on the masses and coupling constants can be  of the neutral vector $\omega$ and scalar $\sigma$ bosons of the one boson exchange (OBE)  model of the two- nucleon interaction.  The Dirac equation for the scattering wave for the proton-nucleus system is
\begin{equation}
\Big\{ \mbox{\boldmath$\alpha$}\cdot {\bf p} + \beta \Big[ m+U_{S}(r) \Big] + \Big[ U_{0}(r) + V_{C}(r) \Big] \Big\} \Psi({\bf r}) = E \Psi({\bf r}) , \label{direcph}
\end{equation}
where, $m$ is the proton mass, $E$ the nucleon total energy in the c.m. frame, and $\alpha_{k} \ (k=1,2,3)$ and $\beta$ are Dirac matrices. One problem with this method is the difficulty to disentangle the contributions of $\rho_{0}(r)$ from $\rho_{S}(r)$, including medium effects on the meson couplings and masses.

An effective Schr\"odinger equation can be obtained from this procedure, which still carries the essence of relativistic corrections and has been shown to be very successful to describe proton-nucleus scattering at high energies \cite{ArnoldPRC.23.1949,KurthPRC.49.2086,CooperPRC.47.297,ShimPRC.59.317}.  The Dirac equation (\ref{direcph}) can be rewritten as two coupled equations for the upper ($\Psi_u$) and lower ($\Psi_l$) components of the Dirac wave function, $\Psi ({\bf r})$. Keeping only the two upper components of the wavefunction, using the definition $\Psi_u = \sqrt{B}\phi$, where
$$B(r)={m+U_S +E-U_0 -V_{C} \over m+E}, $$
one obtains the coupled equations
\begin{equation}
\left[p^2 + 2E\left(U_{cent} +U_{SO} \mathbf{\mbox{\boldmath$\sigma$}} \cdot {\bf L} \right) \right]\phi ({\bf r}) =\left[(E-V_C )^2- m^2\right] \phi ({\bf r})
\label{effd}
\end{equation}
with
\begin{equation}
U_{cent}  =  \frac{1}{2E} \Big[ 2EU_0 +2mU_S -U_0^{2}+U_S^2-2V_c U_0 +U_{D} \Big] \label{cent}
\end{equation}
where the Darwin potential  is given by
\begin{equation}
U_D(r) =-  \frac{1}{2r^2 B} \frac{\partial }{\partial r}\left(r^2 \frac{\partial B}{\partial r}\right)+\frac{3}{4B^2}\left(\frac{\partial B}{\partial r}\right)^2 ,
\end{equation}
and the spin-orbit potential by
$$ U_{SO}(r)= -{1\over 2EBr} {\partial B \over \partial r}.$$

The potentials $U_{0}$ and $U_{S}$ are treated exactly in the same way as the non-relativistic optical potentials, usually parametrized in terms of a sum of real and imaginary Woods-Saxon potentials. However, as an inheritance of their relativistic nature, their depths are quite different than those in low-energy scattering. Typically, e.g., for  p+$^{40}$Ca at $E_{p} = 200$ MeV one has  ${\rm Re} U_{0} \sim -350$ MeV, ${\rm Im} U_{0} \sim -100$ MeV, ${\rm Re} U_{S} \sim -550$ MeV, and ${\rm Im} U_{S} \sim -100$ MeV \cite{ArnoldPRC.23.1949}. There are therefore big cancelations in Eq. (\ref{cent}), leading a central depth of $U_{cent}$ compatible with the non-relativistic models. But relativity also introduces modifications in $U_{cent}$ rendering them different than a simple Woods-Saxon form. Moreover, the approach is more consistent than in the non-relativistic case, as there is a prescription on how to get the spin-orbit potential out of $U_{0}$ and $U_{S}$.  Hence in the Dirac approach, the spin-orbit potential appears naturally when one reduce the Dirac equation to a Schr\"{o}dinger-like second-order differential equation, while in the non-relativistic Schr\"{o}dinger approach, one has to insert the spin-orbit potential by hand.

\begin{figure}[t]
\begin{center}
{\includegraphics[width=8cm]{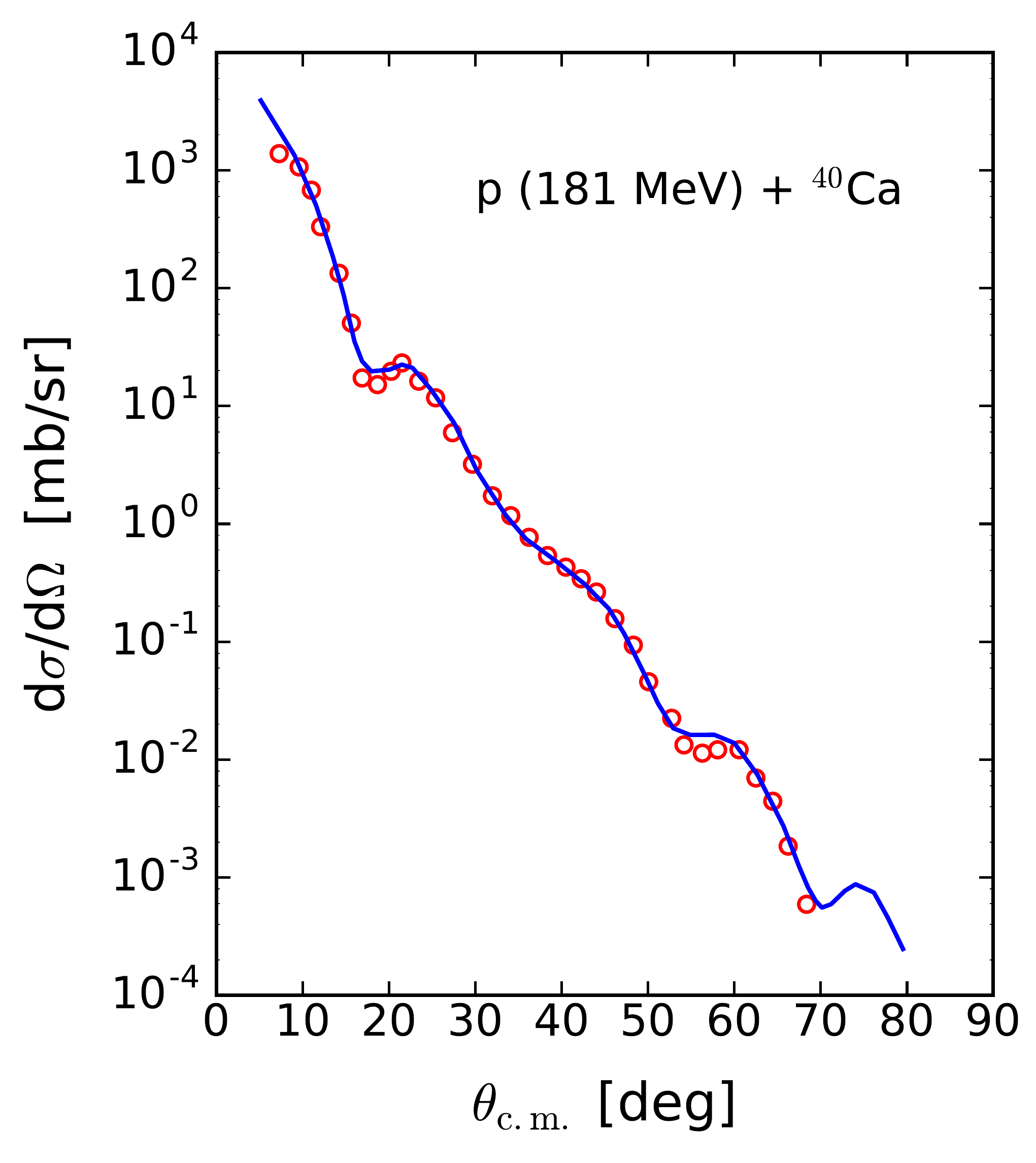}}
\end{center}
\vspace{-0.5cm}
\caption{\label{diracfig}
Elastic p-$^{40}$Ca cross sections at 181 MeV. The  curves are the results of the relativistic Dirac phenomenology  optical model analysis \cite{ArnoldPRC.23.1949}.
}
\end{figure}

Fig. \ref{diracfig} shows the elastic p-$^{40}$Ca cross sections at 181 MeV. The  curves are the results of the relativistic optical model analysis \cite{ArnoldPRC.23.1949}. The agreement with the data is excellent. Many other examples can be found in the literature. For relativistic unstable isotopes, data can be inferred from inverse kinematics using proton targets. So far, most data has been obtained at low energies. At high energies a few reactions have been carried out with light projectiles, such as $^8$Be+p at 700 MeV \cite{KOROLEV2018200}. In this case, the data displays an exponential smooth decrease with angle due to the rather transparent charge distribution in the light nucleus. 

Elastic scattering data is a simple way to access sizes, density profiles, and other geometric features of nuclei.  For example, the beautiful exponential decrease of the cross sections with the nuclear diffuseness is clearly seen in elastic scattering data at large energies.  In contrast, inelastic scattering requires many other pieces of information about the intrinsic nuclear properties and are sensitive to the models used to describe nuclear excitation. Often, the coupling to many excitation channels has to be considered. This  contrasts to the nice features of elastic scattering as a probe of the nuclear geometry and density profiles.

\begin{figure}[t]
\begin{center}
{\includegraphics[width=9.cm]{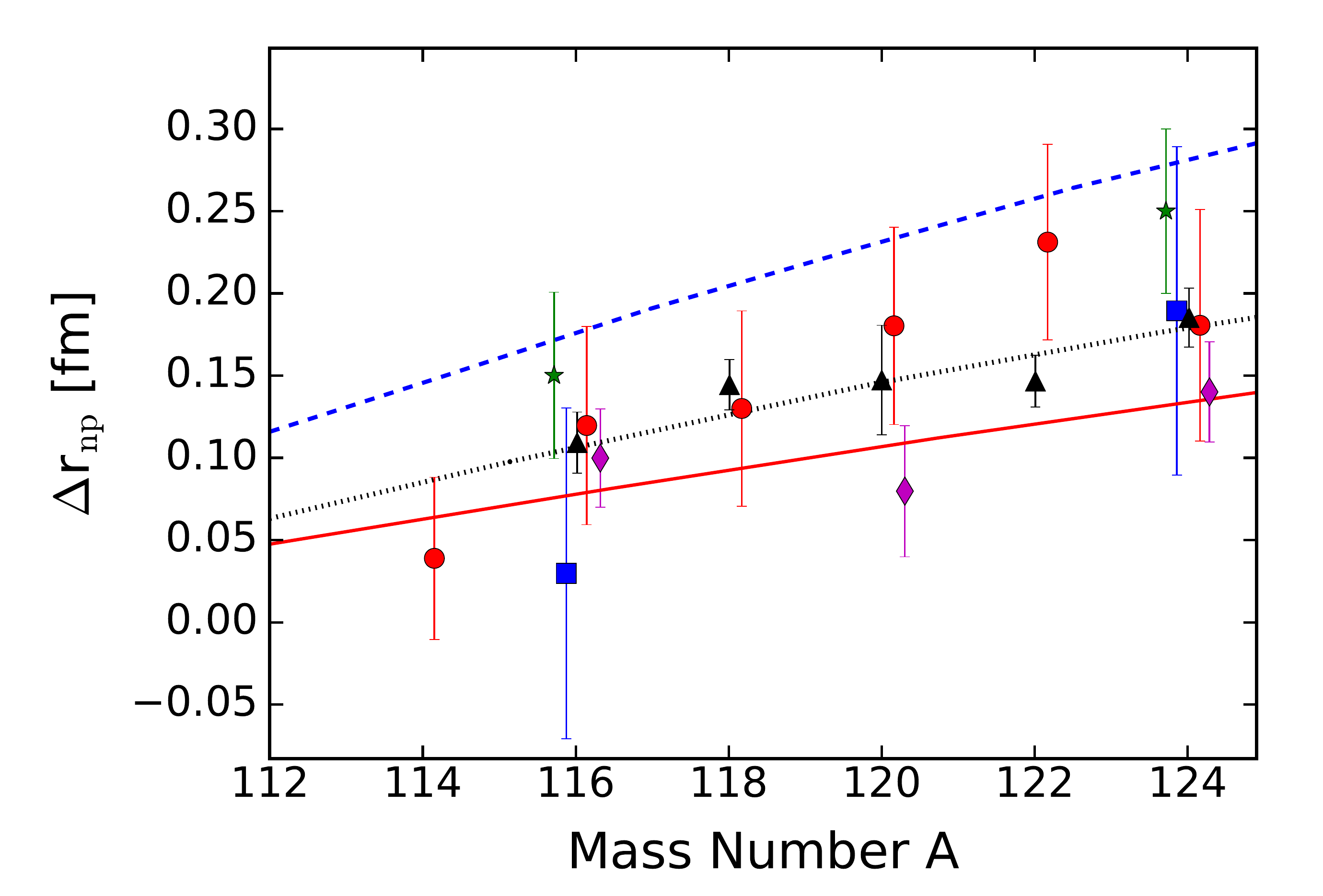}}
\end{center}
\vspace{-0.5cm}
\caption{\label{terafig}
Neutron skin thickness of tin isotopes obtained by various experimental methods, namely, elastic proton scattering at 295 MeV \cite{TerashimaPRC.77.024317} (black triangles),  elastic proton elastic scattering at 800 MeV \cite{RayPRC.19.1855} (green stars), giant dipole resonance \cite{KRASZNAHORKAY1994521}, spin dipole resonance \cite{KrasznahorkayPRL.82.3216} (red circles), and antiprotonic x-ray atoms \cite{TrzcPRL.87.082501}. The dashed blue line represents the RMF predictions of Ref. \cite{LALAZISSIS19991}, whereas the solid red line are HF calculations using the SkIII parametrization \cite{TAJIMA199623} and the dotted black line use the SkM* parametrization \cite{FriedrichPRC.33.335}.}
\end{figure}

In Ref. \cite{TerashimaPRC.77.024317},  the neutron density distributions of tin isotopes have been studied by measuring the cross sections and analyzing powers for proton elastic scattering at 295 MeV. The relativistic Love-Franey interaction was tuned to explain the proton elastic scattering off
 $^{58}$Ni whose density distribution is well known. The compiled results for this experiments added to other data is shown in Figure \ref{terafig}. The RMS radii of the point proton and neutron density distributions were extracted and compared with theoretical mean-field calculations. Skyrme-Hartree-Fock calculations using the SkM* parametrization were in good agreement with the RMS radii of both point proton and neutron density distribution, as shown in Fig. \ref{terafig}. Neutron skin thickness of tin isotopes obtained by various experimental methods, namely, elastic proton scattering at 295 MeV \cite{TerashimaPRC.77.024317} (black triangles),  elastic proton elastic scattering at 800 MeV \cite{RayPRC.19.1855} (green stars), giant dipole resonance \cite{KRASZNAHORKAY1994521}, spin dipole resonance \cite{KrasznahorkayPRL.82.3216} (red circles), and antiprotonic x-ray atoms \cite{TrzcPRL.87.082501}. The dashed blue line represents the RMF predictions of Ref. \cite{LALAZISSIS19991}, whereas the solid red line are HF calculations using the SkIII parametrization \cite{TAJIMA199623} and the dotted black line use the SkM* parametrization \cite{FriedrichPRC.33.335}. The results show a clear increase in neutron skin thickness with mass number, although the values obtained were not as large as what some RMF models predict.
 
\subsubsection{Total nuclear reaction cross sections}

Since the survival probability is defined as in Eq. (\ref{eik3}), it is evident that the reactions cross sections is given by 
\begin{eqnarray}
\sigma_{R}&=&2\pi \int db b \Big[1-\big|S(b)\big|^{2}\Big] \nonumber \\
&=&2\pi \int db b \left[1-\left|\left<\Psi_{0}\Big|\prod_{j}\exp\Big(i\chi_{j}({\bf b}_{j}\Big)\Big|\Psi_{0}\right>\right|^{2}\right],\nonumber \\
\label{sigreac}
\end{eqnarray}
where $\Psi_{0}$ is the projectile+target intrinsic (translation-invariant)  ground state wave functions accounting for the position of the nucleons, and $b_{i}$ is the projection onto the collisional transverse plane of the relative coordinates between the nucleons in a binary collision $j$. The product accounts for all possible binary collisions.

Only in a few cases  the equation (\ref{sigreac}) is solved exactly, as, e.g., in Ref. \cite{VargaPRC.66.034611}. Usually the reaction cross section is calculated with help of the optical limit of the Glauber model \cite{glauber1959lectures}. It is worth noticing that the reaction cross section includes any channel out of the elastic one and therefore includes the excitation of the colliding nuclei. We will discuss these cases later, but first we concentrate on the the case where Eq. (\ref{sigreac}) applies, which some authors prefer to call interaction cross section. As we discussed previously in connection with the data presented in Fig. \ref{halofig}. Often, experimental results are compared to the phenomenological formula $\sigma_{R}=\pi r_{0}^{2}(A_{P}^{1/3}+A_{T}^{1/3})^{2}$. And this was used to interpret the data shown in Fig. \ref{halofig}, leading to the discovery of halo nuclei.

\begin{figure}[t]
\begin{center}
{\includegraphics[width=8cm]{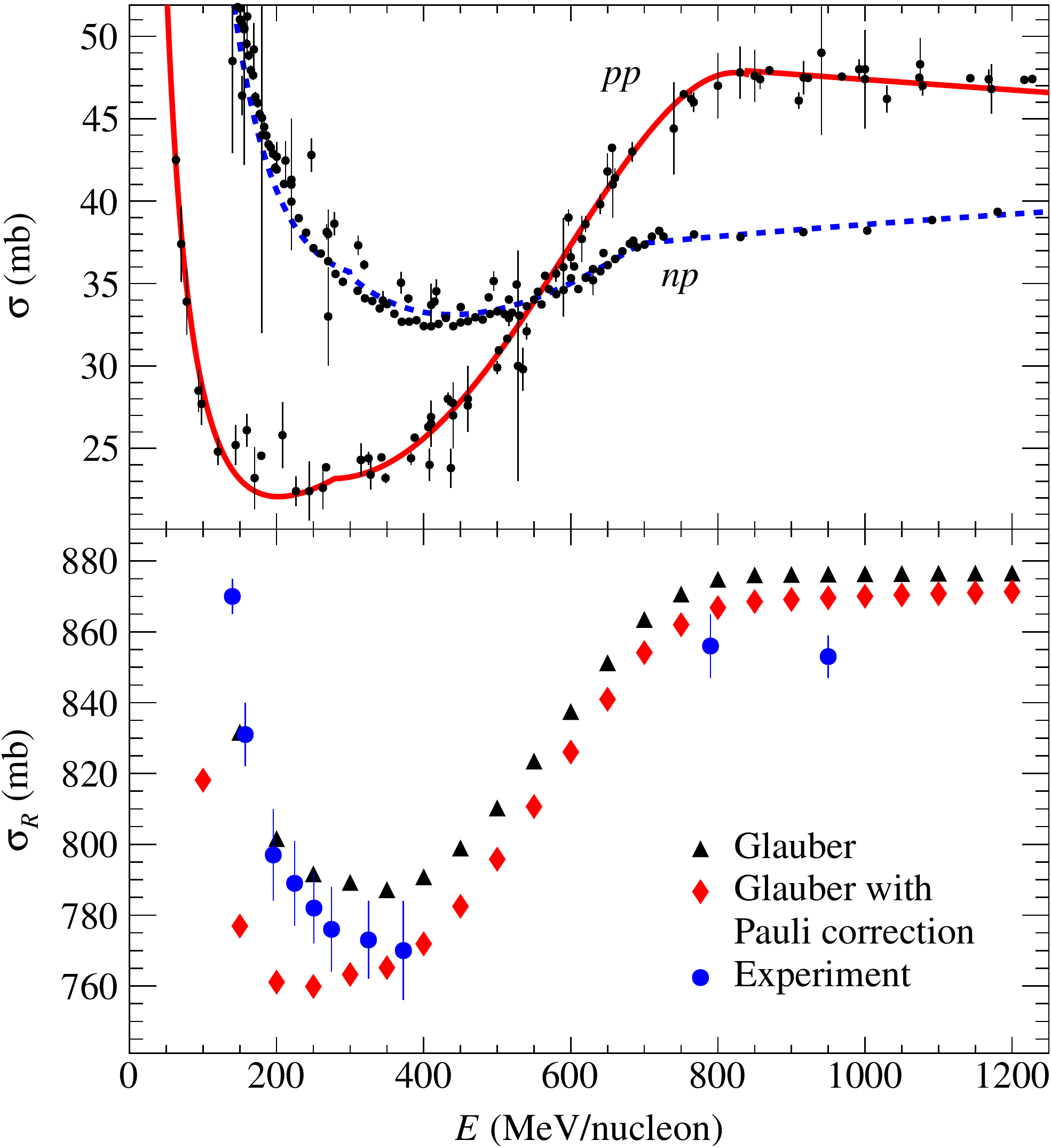}}
\end{center}
\vspace{-0.5cm}
\caption{\label{xsvse}
Nucleon-nucleon (top panel) and the total reaction cross sections for $^{12}$C on $^{12}$C (bottom panel) as a function of projectile energy. The blue points are data from Refs. \cite{TakechiPRC.79.061601} ($100$ to $400$ MeV/nucleon), 790 MeV/nucleon) \cite{Tani1990}, and  \cite{OZAWA2001599} (950 MeV/nucleon). Black triangles represent the result of a parameter-free eikonal calculation using the Glauber optical limit. The red diamonds include the effects of Pauli blocking \cite{Schindler2017}.
}
\end{figure}

Experimental data on reaction cross sections for nucleus-nucleus collisions at high energies are not very abundant, specially for radioactive nuclear species. In Fig. \ref{xsvse} we plot the nucleon-nucleon (top panel) and the total reaction cross sections for $^{12}$C on $^{12}$C (bottom panel) as a function of projectile energy. The blue points are data from Refs. \cite{TakechiPRC.79.061601} ($100$ to $400$ MeV/nucleon), 790 MeV/nucleon) \cite{Tani1990}, and  \cite{OZAWA2001599} (950 MeV/nucleon). Black triangles represent the result of a parameter-free eikonal calculation using the Glauber optical limit. The red diamonds include the effects of Pauli blocking \cite{Schindler2017}.

Fig. \ref{xsvse} shows the free total nucleon-nucleon cross sections   \cite{YaoJPG2006}. A very useful chi-square fit of the experimental data, yields the expressions obtained in Ref. \cite{BertulaniConti10}, 
\begin{equation}
\sigma_{pp}=
\left\{
\begin{array}
[c]{c}%
19.6+{4253/ E} -{ 375/ \sqrt{E}}+3.86\times 10^{-2}E \\
({\rm for }\ E < 280\  {\rm MeV}) \\ \; \\
32.7-5.52\times 10^{-2}E+3.53\times 10^{-7}E^3  \\
-  2.97\times 10^{-10}E^4  \\
({\rm for }\   280\ {\rm MeV}\le E < 840\  {\rm MeV}) \\ \; \\
50.9-3.8\times 10^{-3}E+2.78\times 10^{-7}E^2 \\
 +1.92\times 10^{-15} E^4  \\
({\rm for}\  840 \ {\rm MeV} \le E \le 5 \ {\rm GeV})\end{array}
\right.
\label{signn1}
\end{equation}
for proton-proton collisions, and
\begin{equation}
\sigma_{np}=
\left\{
\begin{array}
[c]{c}%
89.4-{2025/ \sqrt{E}}+{19108/ E}-{43535/ E^2}
\\
 ({\rm for }\ E < 300\  {\rm MeV}) \\ \; \\
14.2+{5436/ E}+3.72\times 10^{-5}E^2-7.55\times 10^{-9}E^3
 \\
 ({\rm for }\   300\ {\rm MeV}\le E < 700\  {\rm MeV}) \\ \; \\
33.9+6.1\times 10^{-3}E-1.55\times 10^{-6}E^2 \\
 +1.3\cdot 10^{-10}E^3\\
 ({\rm for}\  700 \ {\rm MeV} \le E \le 5 \ {\rm GeV}) \end{array}
\right.
\label{signn2}
\end{equation}
for proton-neutron collisions, where $E$ is the laboratory energy. The fits are represented by dotted and solid curves in the top panel of Fig. (\ref{xsvse}).

The experimental data for the total reaction cross section for $^{12}$C on $^{12}$C as a function of beam energy is shown  as filled circles in the bottom  panel of Fig. \ref{xsvse}. Its is worth noticing that few data exist in the energy region of most relevance for the study of radioactive nuclei in the present and planned rare isotope facilities. The optical limit the Glauber theory was used to obtain the black triangles using the above parameterization for the NN cross sections. It is interesting  to notice that medium corrections such as the inclusion of the effects of the Pauli principle \cite{BertulaniConti10} (shown in the figure as filled diamonds) are relevant even at high energies because one of the nucleons in the binary collision can end up in a low energy state, blocked by other nucleons occupying that state. The incorporation of in medium effects in nucleon-nucleon collisions is thus a relevant (see also, Ref. \cite{ChenPRC.87.054616}). 

\begin{figure}[t]
\begin{center}
{\includegraphics[width=8cm]{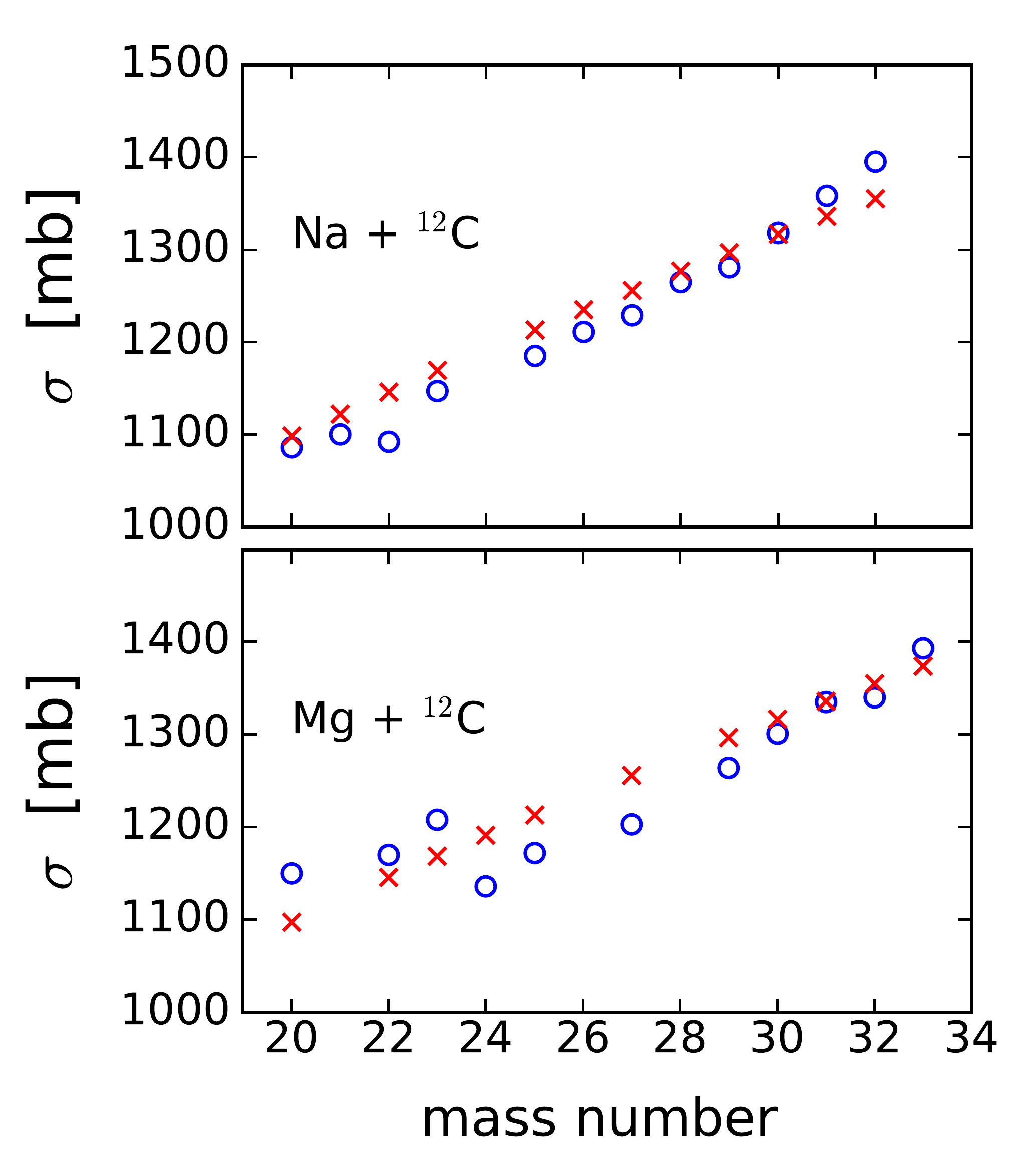}}
\end{center}
\vspace{-0.5cm}
\caption{\label{reacxs}
{\it Top}: The measured interaction cross sections for $^{20-32}$Na (open circles) projectiles at 950 MeV/nucleon on a C target \cite{SUZUKI1998661}. The crosses are the Glauber calculations using HFB densities. {\it Bottom:} Same as top panel, but for $^{20-33}$Mg projectiles at the same bombarding energy.
}
\end{figure}

In Fig. \ref{reacxs} we show measured interaction cross sections for $^{20-32}$Na and $^{20-33}$Mg (open circles) projectiles at 950 MeV/nucleon on a C target \cite{SUZUKI1998661}. The crosses are the Glauber optical limit calculations using HFB densities as inputs in Eq. (\ref{eik6}). The agreement with the experimental data is quite good. In fact, one can turn this argument around and use the experimental data and adjust the matter radius in the densities to extract densities which can be used as a constraint to theoretical models \cite{SUZUKI1998661}. 

\subsection{Coulomb excitation}

\begin{figure}[tb]
\begin{center}
{\includegraphics[width=8cm]{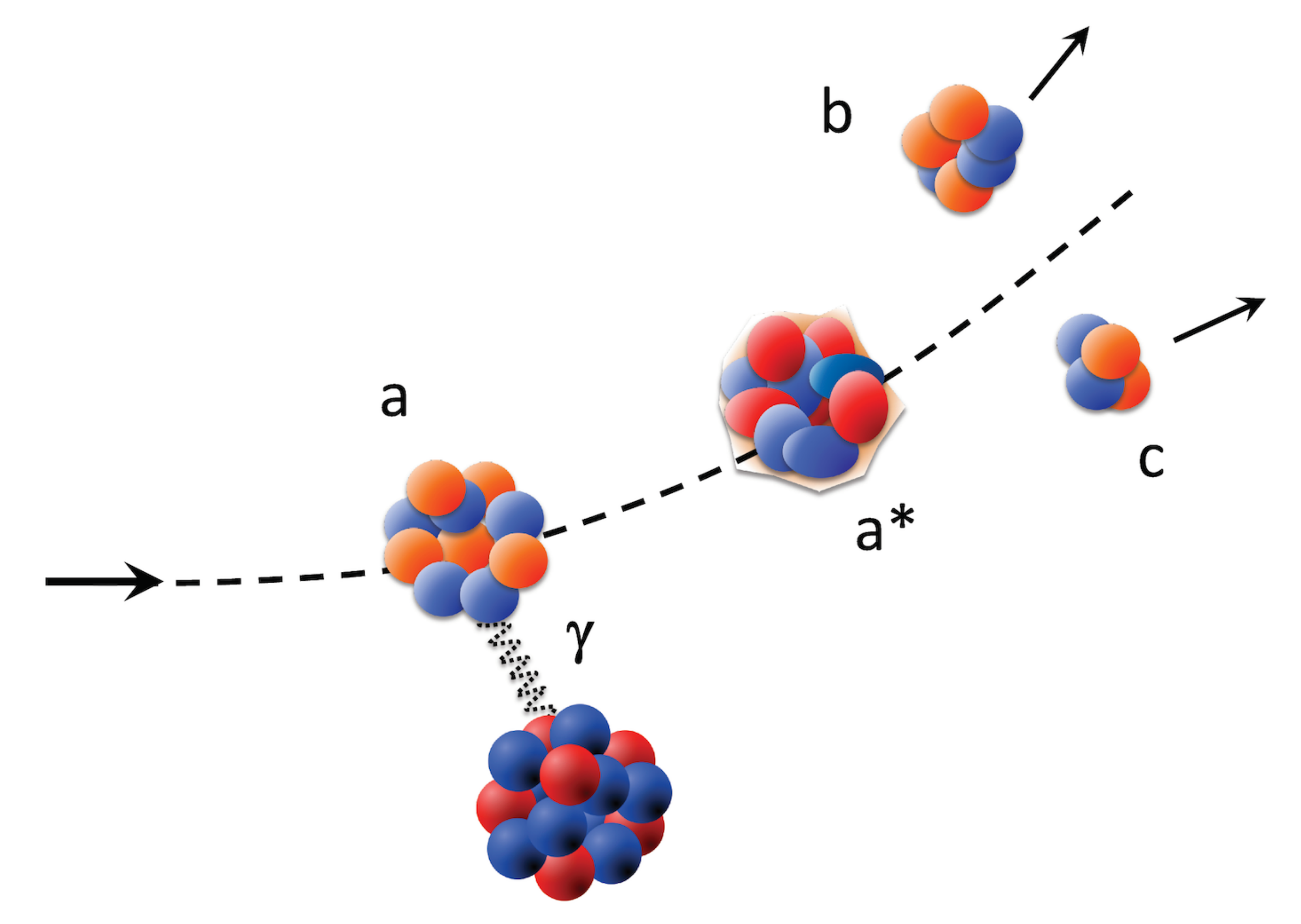}}
\end{center}
\vspace{-0.5cm}
\caption{\label{reacxsc}
Schematic description of the Coulomb excitation of a projectile $a$ leading to an excited nucleus, $a^*$, or a direct breakup of the projectile.}
\end{figure}

As with the case of electron scattering, Coulomb excitation (Fig. \ref{reacxsc}) is one of the most useful probes of nuclear structure because the interaction is well-known. However, unless the collision occurs ate very low energies, the excitation or breakup caused by the strong interaction can lead to large contributions, and also to nuclear-Coulomb interference. Experimentalists investigate the best cases for the bombarding energies, angular distributions, and excitation energies that can be safely described as Coulomb excitation.

Coulomb excitation cross sections are directly related to the photonuclear cross sections by means of  \cite{BERTULANI1988299} 
\begin{equation}
\frac{d\sigma _{C}\left( E_{x}\right) }{dE_{x}}=\sum_{E\lambda }\frac{%
n_{E\lambda }\left( E_{x}\right) }{E_{x}}\sigma _{E\lambda }^{\gamma }\left(
E_{x}\right) +\sum_{M\lambda }\frac{n_{M\lambda }\left( E_{x}\right) }{E_{x}}%
\sigma _{M\lambda }^{\gamma }\left( E_{x}\right) \;,  \label{sigmac}
\end{equation}
where $\sigma {_{\pi \lambda }^{\gamma }}$ $\left( E_{x}\right) \;$ are the
photonuclear cross sections for the multipolarity $\pi \lambda $  and $E_{x}$ is the excitation energy.

The photonuclear cross sections are related to the reduced matrix elements, for the excitation energy $E_{x}$, through the relation \cite{BERTULANI1988299} 
\begin{equation}
\sigma _{\gamma }^{\pi \lambda }(E_{x})=\frac{(2\pi )^{3}(\lambda +1)
}{\lambda \left[ (2\lambda +1)!!\right] ^{2}}\left( \frac{E_{x}}{\hbar c}
\right) ^{2\lambda -1}\frac{dB\left( \pi \lambda, E_{x}\right)}{dE_{x}}  \label{(1.2)}
\end{equation}
where $dB/dE_{x}$ is defined in Eq. (\ref{dbdedef}). 
For differential cross sections one obtains 
\begin{equation}
\frac{d\sigma_C (E_{x})}{d\Omega }=\frac{1}{E_{x}}\sum\limits_{\pi \lambda }
\frac{dn_{\pi \lambda }}{d\Omega }(E_{x},\theta )\sigma _{\gamma }^{\pi
\lambda }(E_{x}),  \label{(1.6)}
\end{equation}
where $\Omega$ denotes to the solid scattering angle.

At high energies, above $E_{lab}=100$ MeV/nucleon, the eikonal formalism developed in Ref. \cite{BERTULANI1993158} yields for the virtual photon numbers in Eq. \eqref{sigmac} 
\begin{equation}
n_{\pi \lambda }(E_x )=Z_{1}^{2}\alpha \ {\frac{\lambda \bigl[(2\lambda +1)!!\bigr]^{2}}{(2\pi )^{3}\ (\lambda +1)}}\ \sum_{m}\ |G_{\pi \lambda
m}|^{2}\ g_{m}(E_x)\;,  \label{n}
\end{equation}
and 
\begin{equation}
g_{m}(E_x )=2\pi  \biggl({\frac{E_x }{\gamma \hbar v}}\biggr)^{2} \int db b K_{m}^{2}\biggl({\frac{E_x b}{\gamma \hbar v}}\biggr) \exp \bigl\{-2\ \chi_{I}(b)\bigr\}\;,  \label{g}
\end{equation}
where $\chi _{I}(b)$ is the imaginary part of the eikonal phase and the functions $G_{\pi \lambda m}(c/v)$ were obtained in Ref. \cite{WINTHER1979518} and, for the lowest multipolarities are given by
\bea
G_{E11}\left(  x\right)  &=&\frac{1}{3}\sqrt{8\pi}x=-G_{E1,-1}\left(  x\right)
\nonumber \\ G_{E10}\left(  x\right)  &=&-i\frac{4}{3}\sqrt{\pi}\left(  x^{2}
-1\right)  ^{1/2},
\eea
\be
G_{E22}\left(  x\right)  =-\frac{2}{5}\sqrt{\frac{\pi}{6}}x\left( x^{2}-1\right)  ^{1/2}=G_{E2,-2}\left(  x\right)  ,
\ee
\be
G_{E21}\left(  x\right)  =i\frac{2}{5}\sqrt{\frac{\pi}{6}}\left( 2x^{2}-1\right)  =iG_{E2,-1}\left(  x\right)  ,
\ee
\be
G_{E20}\left(  x\right)  =\frac{2}{5}\sqrt{\pi}x\left(  x^{2}-1\right)^{1/2},
\ee
for the electric  $E1$ and $E2$ excitations, and
\be
G_{M11}\left(  x\right)  =-i\frac{1}{3}\sqrt{8\pi}=G_{M1,-1}\left(  x\right),\quad G_{M10}\left(  x\right)  =0,
\ee
for the $M1$ excitations.

\begin{figure}[tb]
\begin{center}
{\includegraphics[width=9cm]{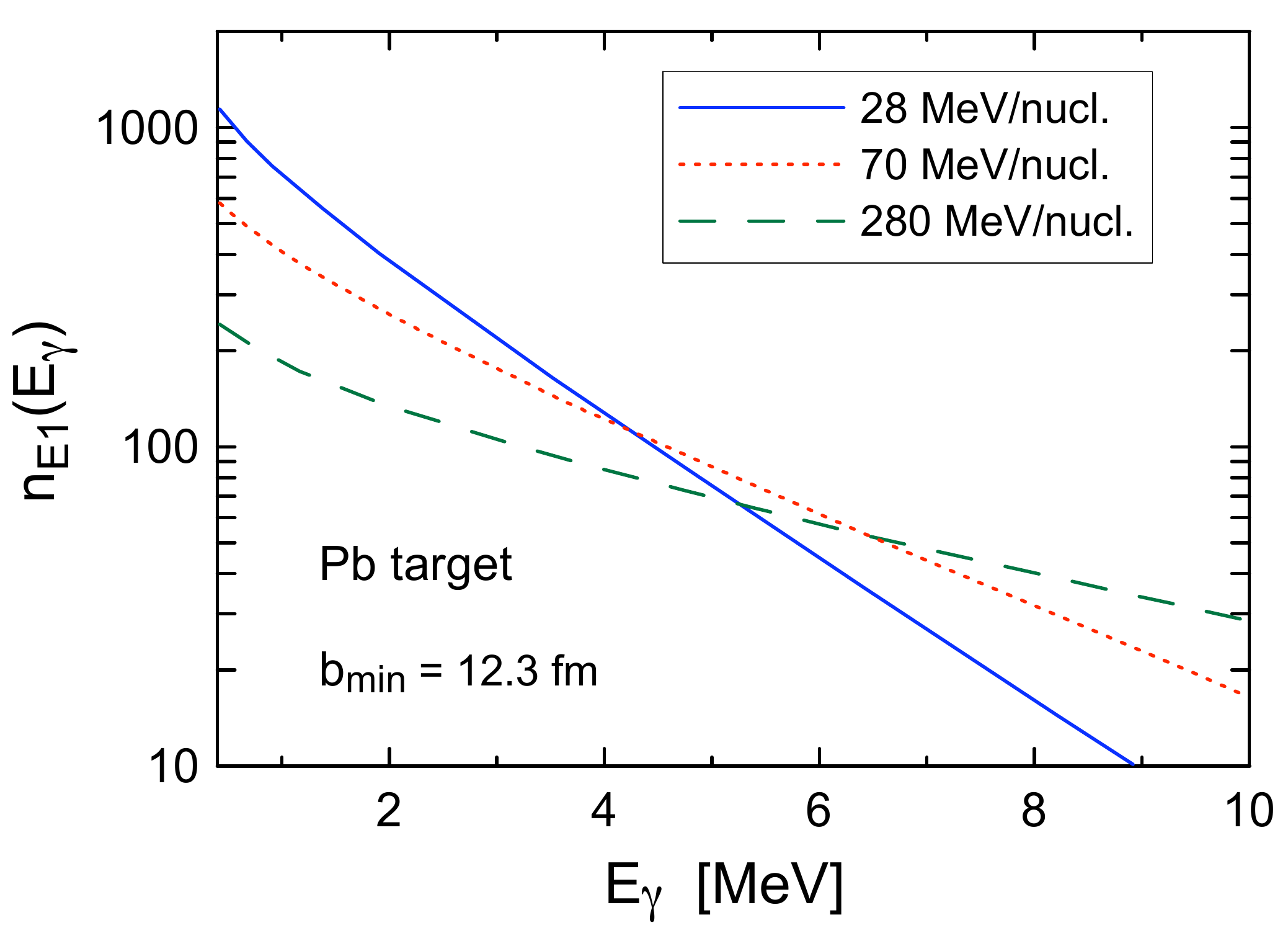}}
\end{center}
\vspace{-0.5cm}
\caption{\label{vpnfig}
Number of virtual photons for the E1 multipolarity, as ``seen'' by a projectile flying by a lead target at impact parameters $b>  12.3$ fm, at three  projectile energies.}
\end{figure}

In Figure \ref{vpnfig} we show a calculation (with $E_\gamma \equiv E_x$) of the virtual photons for the $E1$ multipolarity, ``as seen'' by a  
projectile passing by a lead target at impact parameters  $b>12.3$ fm, at three projectile energies. As the bombarding energy  increases, virtual photons with larger energies become available for the reaction. The number of states accessed in the excitation process is concomitantly increased. 

The cross sections for Coulomb excitation are usually larger for electric dipole (E1) excitations and for the isovector giant dipole resonance (GDR). It leads overwhelmingly to neutron decay and can be calculated as \cite{BERTULANI1988299,AumannPRC.51.416}
\begin{equation}
\sigma_{C}^{-n}=\int {dE\over E} n_{E1}(E) \sigma_{\gamma}^{GDR}(E),
\end{equation}
where the equivalent photon number is obtained from
\begin{eqnarray}
n_{E1}(E) &=&{2Z_{T}^2\alpha \over \pi}\left( {\omega c\over \gamma v^2}\right)^{2}\int db\ b \nonumber \\
&\times& \left[K_1^2(x) + {1\over \gamma^{2}}K_{0}^{2}(x) \right] \Lambda(b), \label{ne1}
\end{eqnarray}
with  $v$ equals to the projectile velocity, the Lorentz contraction factor given by $\gamma = (1-v^2/c^2)^{-1/2}$, $\alpha$ being the fine-structure constant and $K_{n}$ the modified Bessel function of nth-kind as a function of $x=\omega b/ \gamma v$. We denote the excitation energy by $E=\hbar \omega$. The photo-nuclear cross sections $\sigma_{\gamma}^{GDR}$ can be parametrized by a Lorentzian shape 
\begin{equation}
\sigma_{\gamma}^{GDR}(E)=\sigma_{0}{E^{2}\Gamma^{2}\over (E^{2}-E_{GDR}^{2})^{2}+E^{2}\Gamma^{2}},
\end{equation}
where $E_{GDR} = 31.2A_{P}^{-1/3} + 20.6A_{P}^{-1/6}$ is a fit to the mass dependence of the centroid of the experimentally observed GDR. It is a mixture of the mass dependence predicted by the hydrodynamical Goldhaber-Teller and Steinwedel-Jensen models  \cite{goldhaber:1948:PREV,jensen:1950:ZFN}. The parameter $\sigma_0$ can be chosen to yield the Thomas-Reiche-Kuhn (TRK) sum rule 
\begin{equation}
\int dE \sigma_{\gamma}^{GDR}(E)=60{N_{P}Z_{P}\over A_{P}}\ {\rm MeV \ mb}, \label{gdr}
\end{equation}
which is a nearly model independent results for the nuclear response to a dipole operator \cite{eisenberg1988excitation}. $N_P$($Z_P$) are the neutron(charge) number of the excited projectile.

The width $\Gamma$ of the GDR is more complicated to explain. It has a strong dependence on the nuclear shell structure. Experimental systematics provides widths ranging within $4-5$ MeV for a closed shell nucleus and can grow to 8 MeV for a nucleus between closed shells. Rare nuclear isotopes are not easy to investigate experimentally using photo nuclear reactions. One often adopts a microscopic theoretical model such as the random phase approximation (RPA)  \cite{BERTULANI1999139}. One can also use a simple phenomenological parameterization of the GDR width in the form, $\Gamma_{GDR} = 2.51 \times 10^{-2} E_{GDR}^{1.91}$ MeV, with $E_{GDR}$ in units of MeV.    The centroid of the ISGQR can be taken as $E_{ISGQR}=62/A_{P}^{1/3}$ MeV. 

\begin{figure}[t]
\begin{center}
{\includegraphics[width=9cm]{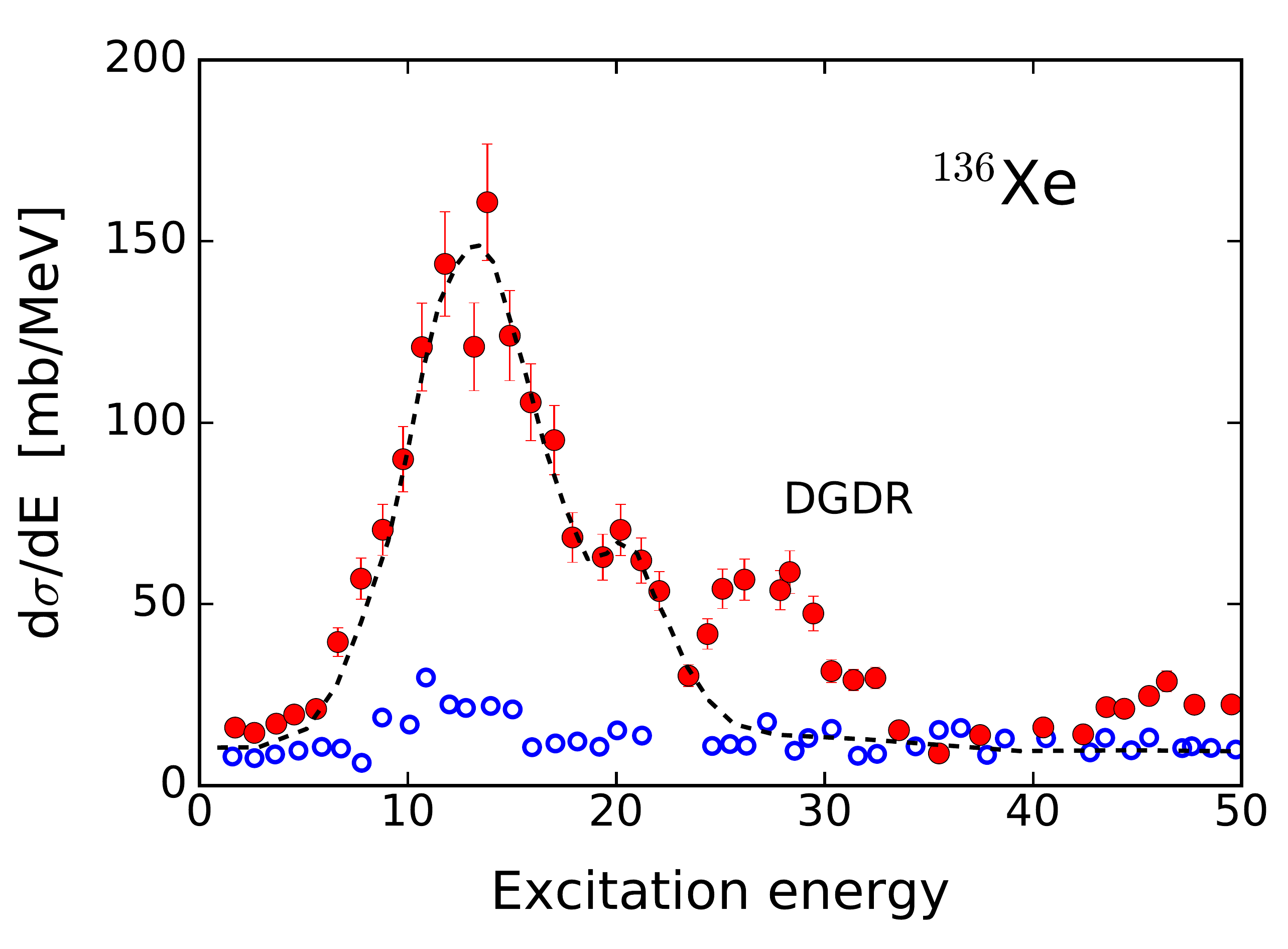}}
\end{center}
\vspace{-0.5cm}
\caption{\label{dgdr}
Experimental cross sections for the excitation of $^{136}$Xe (700 MeV/nucleon) projectiles incident on lead (solid circles) and carbon targets (open circles). The dashed curve is a calculation including the excitation of isoscalar and isovector giant quadrupole resonances and the isovector giant dipole resonance (IVGDR). Altogether, these resonances compose the large bump in the spectrum. The double giant dipole resonance (IVGDR) is identified as the bump at double the energy of the IVGDR. Data are from Ref. \cite{SchmidtPRL.70.1767}.}
\end{figure}

One of the most dramatic findings in the application of relativistic Coulomb excitation was the discovery of the Double Giant Dipole Resonance (DGDR).  It was in fact first found in pion scattering at the Los Alamos Pion Facility \cite{MordechaiPRL.61.531}. The excitation of the DGDR can be described as a two-step mechanism induced by the pion-nucleus interaction. Using the Axel-Brink hypotheses, the cross sections for the excitation of the DGDR with pions tend to agree within the experimental possibilities. Five years after the DGDR excitation experiments with pions were revealed, the first DGDR Coulomb excitation experiments  were carried out at the GSI facility in Darmstadt/Germany \cite{SchmidtPRL.70.1767,RitmanPRL.70.533}. The excitation of multiple giant resonances had been predicted a few years before in Ref. \cite{BAUR1988313,BaurPRC.34.1654}. In Fig. \ref{dgdr} we show the result of one of these experiments \cite{SchmidtPRL.70.1767}, which detected neutron decay channels of giant resonances excited with relativistic projectiles. The excitation spectrum of relativistic $^{136}$Xe projectiles incident on Pb are compared with the spectrum using C targets. The comparison proves that nuclear contribution to the excitation is small for large-Z targets. An additional experiment \cite{RitmanPRL.70.533} used the photon decay of the DGDR as a probe. A bump in the spectra of coincident photon pairs was observed around an energy twice as large as the GDR centroid energy in $^{208}$Pb targets excited with relativistic $^{209}$Bi projectiles. The advantages of relativistic Coulomb excitation of heavy ions over other probes (pions, nuclear excitation, etc) was clearly demonstrated in several GSI experiments, and reviewed in Refs. \cite{Aumann1998,BERTULANI1999139}. 

\begin{figure}[t]
\begin{center}
{\includegraphics[width=9cm]{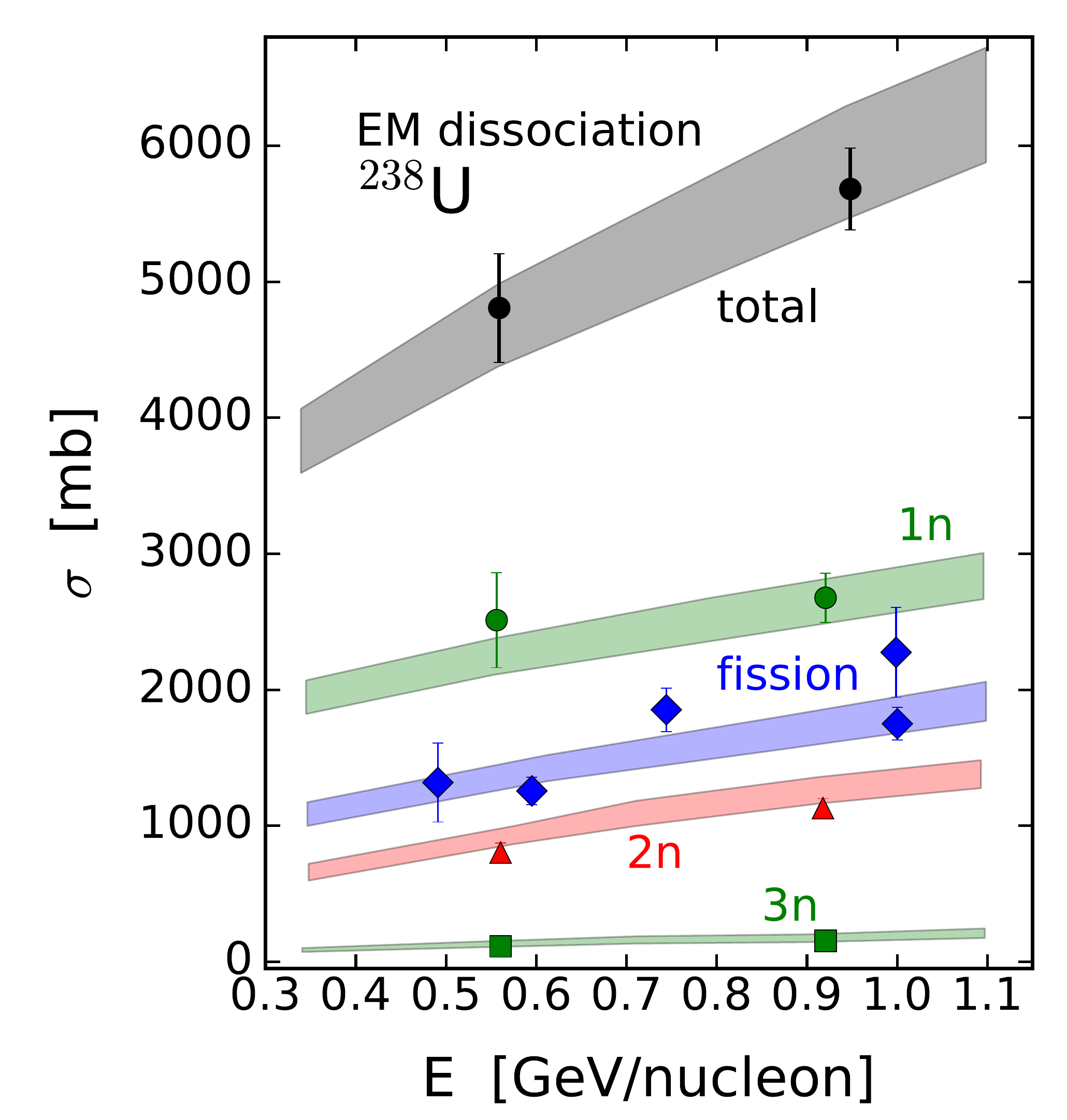}}
\end{center}
\vspace{-0.5cm}
\caption{\label{emdis}
Cross section for electromag­netic dissociation of $^{238}$U. The Coulomb fis­sion cross sections (diamonds), mea­sured with $^{208}$Pb beams on $^{238}$U-targets are from Ref. \cite{Polikanov1994}, whereas the fission cross sections, as well as the xn cross sections  are obtained in inverse kinematics with $^{238}$U ­beams \cite{Armbruster1996}. The data obtained in Ref. \cite{Rubehn1995} with Au-targets were scaled to the data obtained with Pb targets. The curves show theoretical calculations using two sets of experimental GDR parameters as an input. }
\end{figure}

It was earlier on recognized \cite{Bertulani1884} that the electromagnetic excitation of nuclei in relativistic heavy ion collisions lead to large cross sections because of the excitation of giant resonances. The decay of these resonances, including the DGDR, lead to several decay channels, such as xn evaporation and fission. This was demonstrated by comparison off experiment to theory  in Ref. \cite{AumannPRC.51.416,AUMANN1996321}.  Therefore, EM dissociation leads to large interaction cross sections, which can easily measure in inverse kinematics. In Fig. \ref{emdis} we show the cross section for electromag­netic dissociation of $^{238}$U \cite{AUMANN1996321}. The Coulomb fis­sion cross sections (diamonds), mea­sured with $^{208}$Pb beams on $^{238}$U-targets are from Ref. \cite{Polikanov1994}, whereas the fission cross sections, as well as the xn cross sections  are obtained in inverse kinematics with $^{238}$U ­beams \cite{Armbruster1996}. The data obtained in Ref. \cite{Rubehn1995} with Au-targets were scaled to the data obtained with Pb targets. The curves show theoretical calculations using two sets of experimental GDR parameters as an input. Hence, it is clear that,  except for light targets, one should expect that EM dissociation will always be relevant in any nucleus-nucleus collision at relativistic energies and might be either used as a spectroscopic tool, or become a background for processes where the effects of the strong interaction are studied.

\begin{figure}[t]
\begin{center}
{\includegraphics[width=9cm]{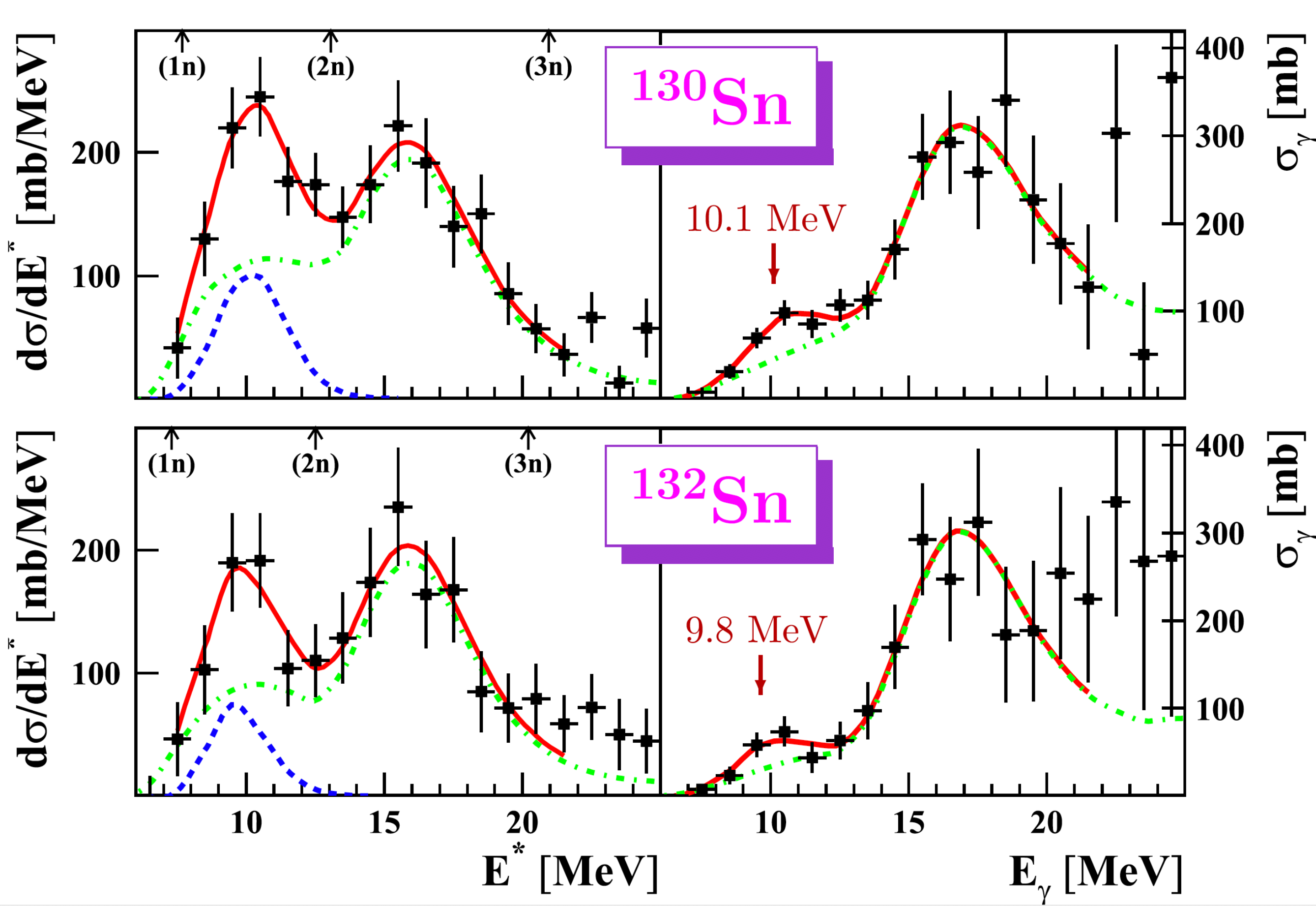}}
\end{center}
\vspace{-0.5cm}
\caption{\label{coulexsn}
{\it Left panels:} Differential cross sections, with respect to excitation energy $E^{*}$, obtained with electromagnetic dissociation of $^{130}$Sn and $^{132}$Sn. The arrows indicate the neutron-separation thresholds. {\it Corresponding right panels:} Deduced photo-neutron cross sections. The curves represent fitted Gaussian (blue dashed line) and Lorentzian (green dash-dotted line) distributions, assigned to the Pygmy Dipole Resonance (PDR) (centroid indicated by an arrow) and Giant Dipole Resonance (GDR), respectively, and their sum (red solid line), after folding with the detector response. Data are from Ref. \cite{adrich:2005:PRL}.}
\end{figure}

In Fig. \ref{coulexsn} in the left panels, we show the differential cross sections, with respect to excitation energy $E^{*}$, obtained with electromagnetic dissociation of $^{130}$Sn and $^{132}$Sn. Data are from Ref. \cite{adrich:2005:PRL}. The arrows indicate the neutron-separation thresholds. In the right panels we show the deduced photo-neutron cross sections. The curves represent fitted Gaussian (blue dashed line) and Lorentzian (green dash dotted line) distributions, assigned to the Pygmy Dipole Resonance (PDR) (centroid indicated by an arrow) and Giant Dipole Resonance (GDR), respectively, and their sum (red solid line), after folding with the detector response.

\subsubsection{The Coulomb dissociation method}

The idea behind the Coulomb dissociation method is relatively simple \cite{BAUR1986188}. The (differential, or angle-integrated) Coulomb breakup cross  section for $a+A\rightarrow b+c+A$ follows from Eq. \eqref{sigmac}. It can be rewritten as
\begin{equation}
{d\sigma_{C}^{\pi\lambda }(E_\gamma)\over
d\Omega}={dn^{\pi\lambda}(E_\gamma;\theta;\phi)\over dEd\Omega}\  \sigma_{\gamma+a\
\rightarrow\ b+c}^{\pi\lambda}(E_\gamma),\label{CDmeth}
\end{equation}
where $E_\gamma$ is the energy transferred from the relative motion to the breakup,  $\sigma_{\gamma+a\ \rightarrow\
b+c}^{\pi\lambda}(E_\gamma)$ is the photo-dissociation cross section for
the multipolarity ${\pi\lambda}$ and photon energy $E_\gamma$, and $j_{k}$ are the spins of the particles involved.  Using time reversal invariance,
the radiative capture cross section $b+c\rightarrow a+\gamma$ from $\sigma_{\gamma+a\ \rightarrow\ b+c}
^{\pi\lambda}(E_\gamma)$, becomes
\begin{equation}
\sigma^{\pi\lambda}_{b+c\rightarrow a+\gamma}(E_\gamma)=
{2(2j_a+1)\over (2j_b+1)(2j_c+1)}{k^2\over k_\gamma^2}\sigma_{\gamma+a\ \rightarrow\ b+c}%
^{\pi\lambda}(E_\gamma),
\label{CDmeth5}
\end{equation}
where $k^2=2m_{bc}(E_\gamma-S)$ with $S$ equal to the separation energy, and $k_\gamma=E_\gamma/\hbar c$.

\begin{figure}[t]
\begin{center}
{\includegraphics[width=9cm]{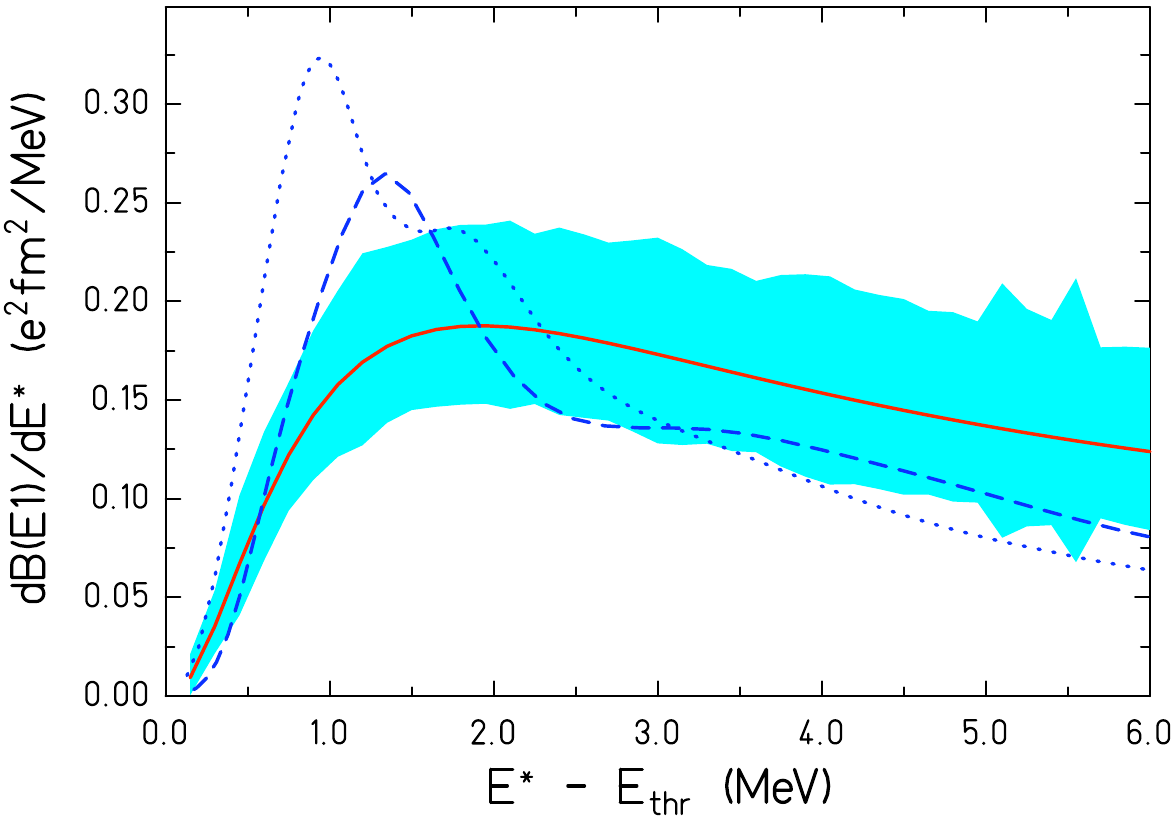}}
\end{center}
\vspace{-0.5cm}
\caption{\label{6he2n}
 Electric dipole response function for $^6$He. The shaded area represents the experimental results from the Coulomb dissociation experiment  reported in Ref. \cite{AumannPRC59.1252}. Dashed and dotted lines correspond to calculations using three-body models Refs. \cite{DANILIN1998383,CobisPRL79.2411}. }
\end{figure}

 The Coulomb Dissociation (CD) method was originally proposed in Ref. \cite{BAUR1986188}. It has been tested  successfully in a number of reactions of interest to astrophysics. The best known case is the reaction $^{7}$Be$(p,\gamma)^{8}$B, first studied in Ref. \cite{MotobayashiPRL73.2680}, followed by  numerous other similar experiments.    As an example of the application of the CD method, we quote the two-neutron capture on $^4$He that may play a role in the post-collapse phase in type-II supernovae. A nucleosynthesis bottleneck prevents formation of nuclei with $A \ge 9$ from nucleons and $\alpha$-particles and the reaction $^4$He$(2n,\gamma)^6$He could  be a solution to bridge the instability gap at $A=5$. It is worth mentioning that one believes that  the most probable candidate is the ($\alpha n,\gamma$) process in a type-II supernova. A Coulomb dissociation experiment to study this reaction is shown in Figure \ref{6he2n} with the  electric dipole response function for $^6$He. The shaded areas represent the experimental results from the Coulomb dissociation experiment reported in Ref. \cite{AumannPRC59.1252}. The dashed and dotted lines correspond to calculations with  three-body models of Refs. \cite{DANILIN1998383,CobisPRL79.2411}. The experimental analysis of this experiment concluded that 10\% of the CD cross section proceeds via the formation of $^5$He, with an  estimate of 1.6 mb MeV for the photoabsorption cross section of $^6$He$(\gamma,n)^5$He, which agrees with theoretical calculations \cite{Efros1996}. One has concluded that the cross sections for formation of $^5$He and $^6$He via one (two) neutron capture by $^4$He are not large enough to compete with the  ($\alpha$n, $\gamma$) capture process  \cite{Aumann2005}.  This  attests the usefulness of the CD method to help us understanding basic questions of relevance for nuclear astrophysics.

\begin{figure}[t]
\begin{center}
{\includegraphics[width=9cm]{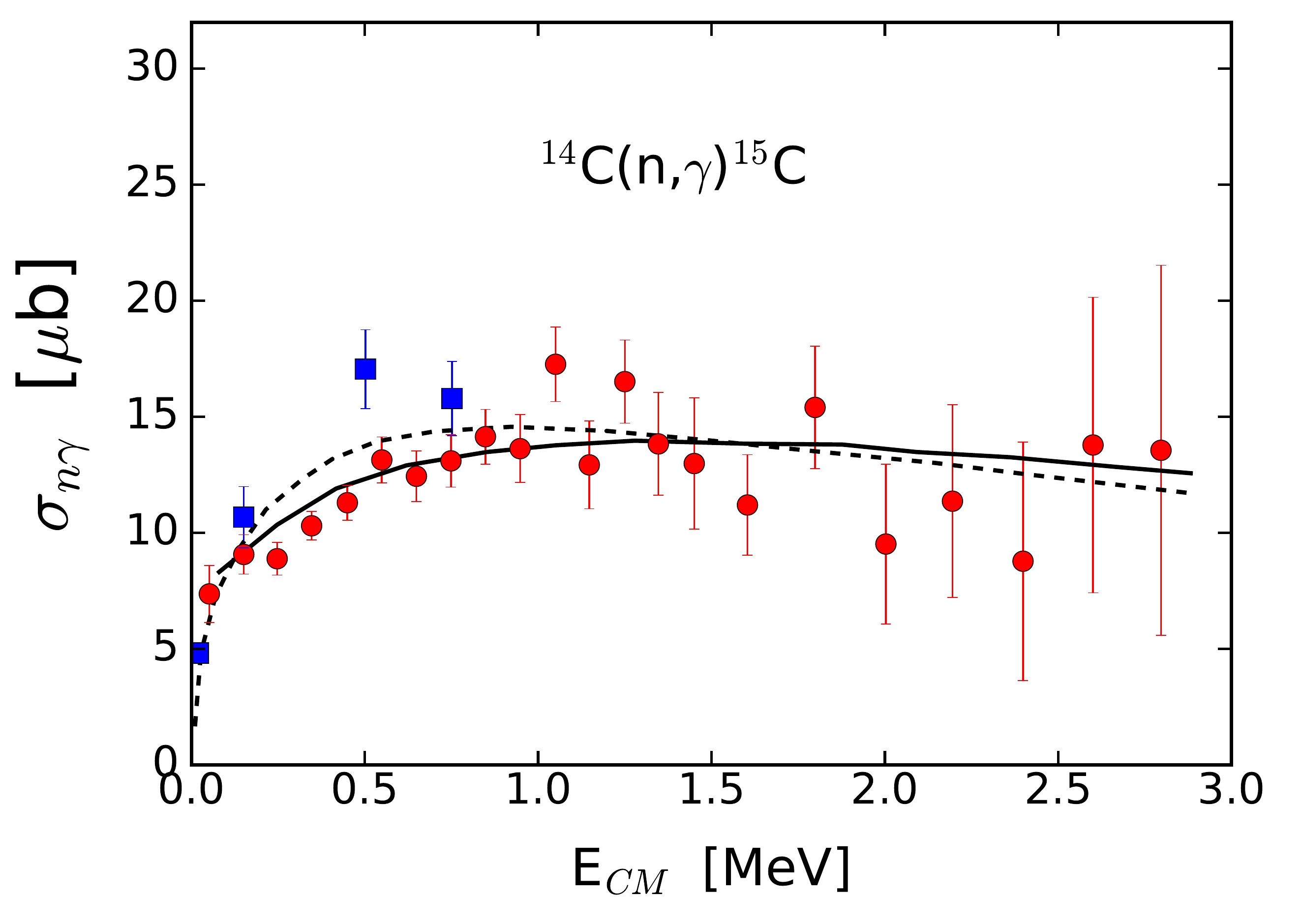}}
\end{center}
\vspace{-0.5cm}
\caption{\label{14Cng}
Neutron capture cross section of $^{14}$C leading to the $^{15}$C ground state. Solid red circles are the results of the experiment of Ref. \cite{Aumann_2013}, and the blue squares represent those from a direct capture measurement \cite{ReifarthPRC77.015804}. The dot-dashed curve is a calculation based on the direct radiative capture model, whereas the solid curve is the same calculation but includes experimental
resolution.  }
\end{figure}

In Fig. \ref{14Cng} we show the neutron capture cross section of $^{14}$C leading to the $^{15}$C ground state. The solid red circles are the results of the experiment of Ref. \cite{Aumann_2013}, and the blue squares represent those from a direct capture measurement \cite{ReifarthPRC77.015804}. The dot-dashed curve is a calculation based on the direct radiative capture model, whereas the solid curve is the same calculation including the experimental
resolution. The experimental result is consistent with a final state ground state capture to $^{15}$C being a halo state with a dominant s-wave
component. The agreement of the Coulomb breakup result with the direct capture measurement suggests that Coulomb breakup is a good
method to obtain neutron capture cross sections involving radioactive nuclei. A word of caution is that in the inverse reaction of Coulomb breakup the reaction is restricted to the ground state. For $^{15}$C, there is only one excited state at 0.74MeV, and the neutron capture to this
state is estimated to be very small \cite{Wiescher1990ApJ,DESCOUVEMONT2000559}. Such fortunate situations may be less probable for heavier neutron-rich nuclei, but we may expect lower level densities near closed shells, such as N = 50 and 82, where Coulomb breakup can be used.  Nuclei in this region of the chart may be relevant to the r-process, and thus of importance for studies using Coulomb breakup.

\begin{figure}[t]
\begin{center}
{\includegraphics[width=8.5cm]{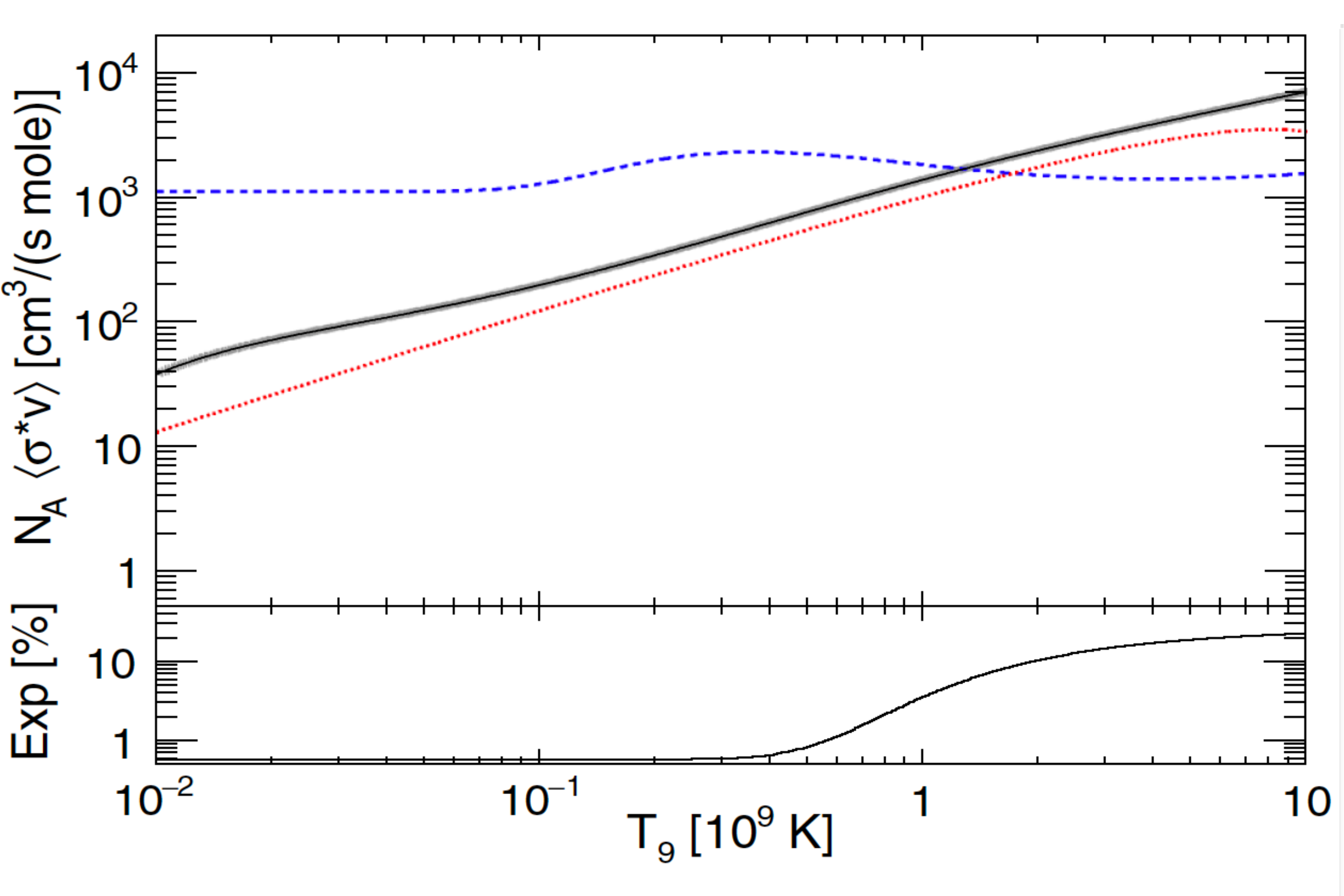}}
\end{center}
\vspace{-0.5cm}
\caption{\label{17cng}
Reaction rate for neutron capture on $^{17}$C with respect to the stellar temperature in $T_9$ units (billion K). The data from Ref. \cite{HeinePRC95.014613} (grey band) are compared to rate calculations using the Hauser-Feshbach theory \cite{Sasaqui2005} (dashed blue line) and a direct capture model \cite{HerndlPRC60.064614} (dotted red line). The lower panel shows the percentage contribution of experimental data for transitions to the ground state in $^{18}$C.}
\end{figure}

 Another example concerns the $^{17}$C$(n,\gamma)^{18}$C reaction at astrophysical energies. Elements heavier than iron are thought to be  created in reactions in the slow (s-) and rapid (r-)neutron capture processes \cite{KaeppelerRMP83.157,ARNOULD200797} since they are not suppressed by the Coulomb barrier at low energies \cite{B2FH1957}. An evidence is that the abundance pattern observed in ultra metal-poor stars \cite{Burris_2000,Sneden1996ApJ,Westin_2000,Hill2002AA} that are attributed to the r-process is remarkably close to solar abundance within $56 \le Z \le 76$, suggesting the existence of a generic production mechanism. Several possible scenarios with nucleosynthesis flows  are sensitive to reaction rates of light neutron-rich nuclei existent in core-collapse Type II supernova (SN) explosions or neutron star mergers \cite{Sasaqui2005}.  The final heavy element abundances were found to change up to an order of magnitude as compared to calculations without light nuclei and the neutron capture on $^{17}$C was considered critical as the rate was solely based on Hauser-Feshbach calculation. Up to recently, no experimental information on the neutron capture cross sections of $^{17}$C was available. The reaction was studied in Ref. \cite{HeinePRC95.014613} using the CD method and the experimental analysis yielded the results shown in Fig. \ref{17cng} for the reaction rates in a r-process site. The figure plots the reaction rate for neutron capture on $^{17}$C with respect to the stellar temperature in $T_9$ units (billion K). The data from Ref. \cite{HeinePRC95.014613} (grey band) are compared to rate calculations using the Hauser-Feshbach theory \cite{Sasaqui2005} (dashed blue line) and a direct capture model \cite{HerndlPRC60.064614} (dotted red line). The lower panel shows the percentage contribution of experimental data for transitions to the ground state in $^{18}$C.
  
 \subsection{Relativistic coupled-channels method}
 
 \subsubsection{Continuum states and relativistic corrections}
 
Eq. \eqref{CDmeth} is obtained with first-order perturbation theory. One also assumes that the nuclear contribution is small,
or that it can be extracted in other types of experiments. For example, $^8$B has a small proton separation energy ($\approx 140$ keV).  Multiple-step, or higher-order effects, are  important for such loosely-bound systems \cite{BERTULANI1992163,BERTSCH1993136}. A calculation of reorientation effects in the break up of halo nuclei was first presented in Ref. \cite{BERTULANI1992163}.  In the (non-relativistic) projectile frame of reference, the time-dependent Schr\"{o}dinger equation due to a (relativistic) electromagnetic interaction for its internal wave function expanded  in terms of bound states and continuum states, $\Psi (t)=\sum_{j\ell m}a_{j\ell m}\,\phi _{j\ell m}$, with energies $E_{j}$, yields to the coupled-channel equations 
\begin{equation}
i\hbar {\frac{da_{j\ell m}}{dt}}=\sum_{j^{\prime }\ell ^{\prime }m^{\prime}}\left<\phi _{j\ell m}\left|V_{ex}\right|\phi _{j^{\prime }\ell
^{\prime }m^{\prime }}\right> \;a_{j^{\prime }\ell^{\prime }m^{\prime }}\;e^{-i(E_{j}^{\prime }-E_{j})t/\hbar }.
\label{BC9235}
\end{equation}
where we reserve the index $j=0$ for the ground state and $V_{ex}$ is a combination of time dependent scalar $\phi({\bf r},t)$ and vector potentials, ${\bf A}({\bf r},t)$, so that $\left< |V_{ext}|\right> \equiv \delta \rho  \phi+\delta {\bf j}\cdot {\bf A}$, with $\delta \rho$ and $\delta {\bf j}$ being transition densities and currents.

\begin{figure}[t]
\begin{center}
{\includegraphics[width=9cm]{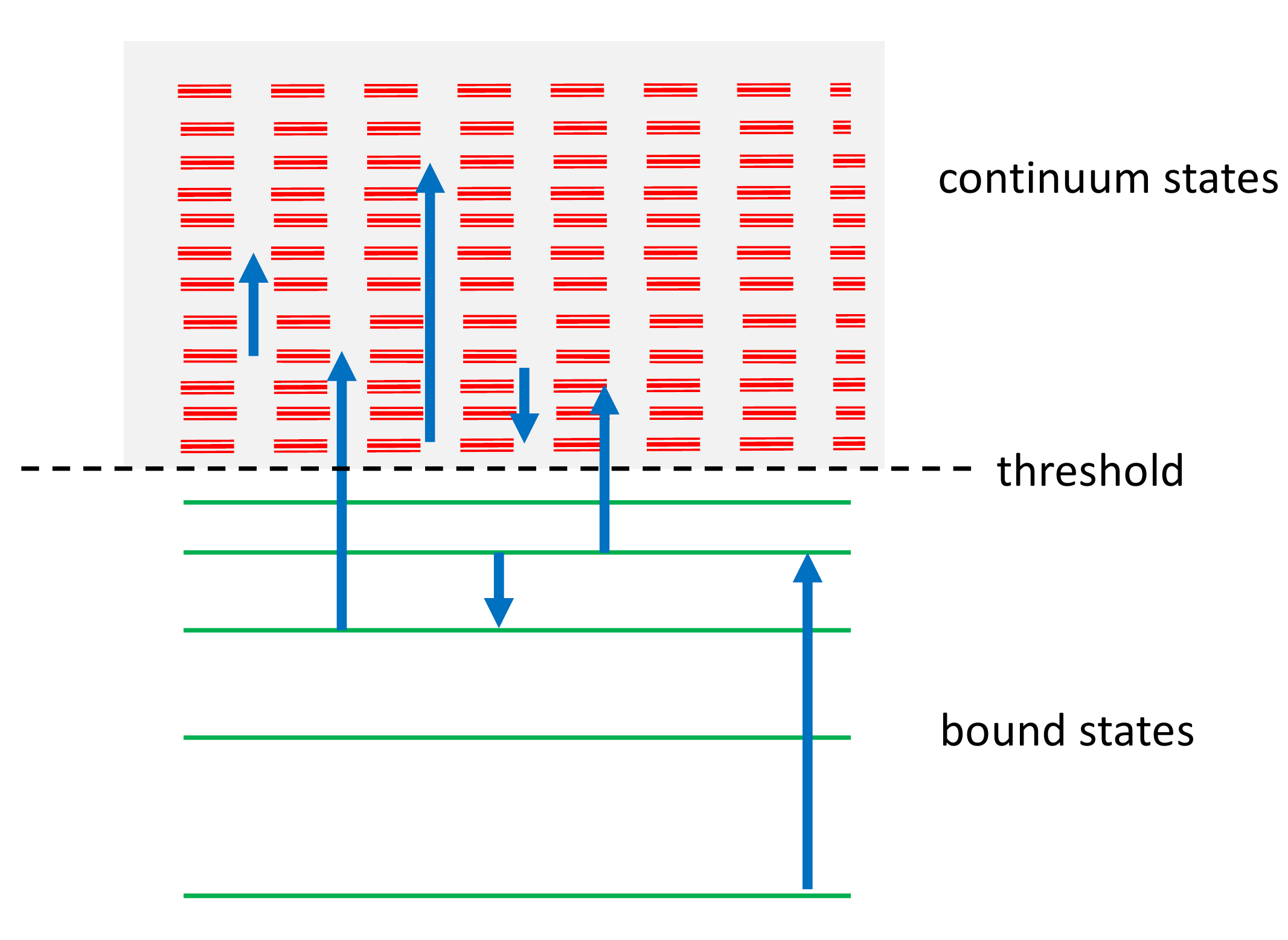}}
\end{center}
\vspace{-0.5cm}
\caption{\label{cdcc}
Schematic description of a coupled-channels calculation including the discretized continuum states. The arrows represent transitions between states.}
\end{figure}

The continuum discretization can be done in several ways. An obvious way is to use pseudo-states, e.g., selected states generated by the same  potential as the bound states, or by the use basis of time-dependent discrete states are defined as  
\begin{eqnarray}
\left| \phi _{k\ell m}\right\rangle &=&e^{-iE_{k \ell m}t/\hbar }\left| 0\right\rangle
,\;\;\;\mathrm{if}\;\;E_{k\ell m}<0\\ \nonumber
&=&e^{-iE_{k}t/\hbar }\int \Gamma _{k}(E)\;\left| E\ell
m\right\rangle,\;\;\;\mathrm{if} \;\;E_{k\ell m}>0  \label{BC9231}
\end{eqnarray}
where $\left|E\ell m \right>$ being continuum wavefunctions of the projectile fragments with good energy and angular momentum quantum numbers $E\ell \,m$. The function $\Gamma _{k}(E)$ is assumed to peak at an energy $E$ in the continuum. This leads to nearly discrete states $\left|\phi _{k\ell m}\right>$ allowing for an easy implementation of the coupled-states calculations.  At relativistic energies, the projectile moves at forward angles with nearly constant velocity. Thus one can replace $z=vt$ in the above equations to obtain a series of coupled-equations in coordinate space. As shown below, the amplitudes $a_{j\ell m}$ are become the scattering matrices $S_{j\ell m}$. The coupled-channels approach with the inclusion of the continuum is commonly know as Continuum Discretized Coupled-Channels (CDCC) method and is schematically shown in Fig. (\ref{cdcc}) where the arrows represent transitions between the states due to the perturbative field $V_{ex}$.

The relativistic Coulomb potential for the multipolarity $\Pi L$ ($\Pi = E$ or $M$; $L=1,2, \cdots$)  as seen by the projectile P passing by a target T are given in Ref. \cite{BertulaniPRL94.072701}:
\begin{equation}
V_{{\rm E1}\mu}=\sqrt{\frac{2\pi}{3}}\xi Y_{1\mu}\left(  \mathbf{\hat{\mbox{\boldmath$\xi$}}}\right)  \frac{\gamma Z_{\rm T}e}{\left(
b^{2}+\gamma ^{2}z^{2}\right)^{3/2}}\left\{
\begin{array}
[c]{c}
\mp b\ \ (\mathrm{if}\ \ \ \mu=\pm1)\\ \sqrt{2}z\ \ (\mathrm{if}\ \ \ \mu=0)\
\end{array}
\right.  \label{eq6}
\end{equation}
for the E1 (electric dipole) field, and
\begin{align}
V_{{\rm E2}\mu}= &  \sqrt{\frac{3\pi}{10}}\xi^{2}Y_{2\mu}\left(\mathbf{\hat {\mbox{\boldmath$\xi$}}}\right)  \frac{\gamma
Z_{\rm T}e}{\left(  b^{2}+\gamma^{2}z^{2}\right)  ^{5/2}}\nonumber\\
&  \times\left\{
\begin{array}
[c]{c}
b^{2}\ \ \ \ (\mathrm{if}\ \ \ \mu=\pm2)\\
\mp(\gamma^2+1)bz\ \ \ \ (\mathrm{if}\ \ \ \mu=\pm1)\\
\sqrt{2/3}\left(  2\gamma^{2}z^{2}-b^{2}\right)  \ \ \ \ (\mathrm{if}%
\ \ \ \mu=0)\
\end{array}
\right.  \label{eq7}
\end{align}
for the E2 (electric quadrupole) field. The intrinsic coordinate between the nucleons are denoted by $\mathbf{\mbox{\boldmath$\xi$}}$ and $b$ is the impact parameter (or transverse coordinate) in the collision of P and T, which is defined by $b=\sqrt{x^2+y^2}$ with ${\bf R} = (x,y,z)$, the relative coordinate of P from T in the Cartesian representation. The Lorentz contraction factor is denoted by $\gamma=\left(
1-v^{2}/c^{2}\right)^{-1/2}$, where $v$ is the velocity of P. These relations are obtained with so-called far-field approximation \cite{EsbensenPRC65.024605} with $R$ assumed to be always larger than $\xi$. The coupling potentials in E-CDCC are obtained with Eqs.~(\ref{eq6})--(\ref{eq7}) as shown below. The derivation of these equations is described in Ref.~\cite{BERTULANI2003317}.

In addition, for magnetic dipole excitations,
\begin{equation}
V_{{\rm M1}\mu}(b,z,\mathbf{{\mbox{\boldmath$\xi$}}})=i\sqrt{\frac{2\pi}{3}}{\cal M}_{1\mu}\left(  \mathbf{{\mbox{\boldmath$\xi$}}}\right)
{v \over c}\frac{\gamma Z_{\rm T}e}{\left(b^{2}+\gamma ^{2}z^{2}\right)^{3/2}}\left\{
\begin{array}
[c]{c}%
\pm b\  (\mathrm{if}\ \mu=\pm1)\\
0 (\mathrm{if} \ \mu=0)\
\end{array}
\right.  \label{relM1}%
\end{equation}
where
${\cal M}_{1\mu}(  \mathbf{{\mbox{\boldmath$\xi$}}})$
is the intrinsic M1 operator.

A second method used to handle higher-order effects in high energy collisions is to solve the time-dependent Schr\"odinger equation in the frame of reference of the excited nucleus by a lattice discretization of the four dimensional space.  This method was developed in Ref.  \cite{BERTSCH1993136} and amounts to expand the  wavefunction is a spherical basis, $$\Psi (\mathbf{r_{k}},t)={1\over r_{k}}\sum_{\ell m}u_{\ell m}(r_{k},t)Y_{\ell m}(\hat{\mathbf{r}}_{k}),$$ use the orthogonal properties of the spherical harmonics and obtain the time-dependent equation in the lattice
\begin{eqnarray}
& &\left[ \Delta_{k}-\frac{\ell (\ell +1)}{r_{k}^{2}}-\frac{2\mu _{cx}}{ \hbar }V_{ex}(r_{k})\right] \;u_{\ell m}(r_{k},t)\nonumber \\ &+&\sum\limits_{\ell ^{\prime
}m^{\prime }}S_{\ell ^{\prime }m^{\prime }}^{(\ell m)}\Big(\Pi L; r_{k},t\Big)\ u_{\ell ^{\prime}m^{\prime }}(r_{k},t)=-\frac{2\mu _{cx}}{\hbar }{\partial u_{\ell m}(r_{k},t) \over
\partial t}, \nonumber \\
\label{(3.10b)}
\end{eqnarray}
where $S_{\ell ^{\prime }m^{\prime }}^{(\ell m)}\Big(\Pi L; r_{k},t\Big)$ is a source function of multipolarity $\Pi L$ which couples the different partial waves and depends on the perturbative potential $V_{ex}$. For more details, see Refs. \cite{BERTSCH1993136,BertulaniPRC49.2839,ESBENSEN1995107}.

In Refs. \cite{BertulaniPRL94.072701,OgataPTP.121.1399,OgataPTP.123.701} a proposal has been made to include relativistic dynamical effects also using non-relativistic nuclear potentials and extending them to include relativistic dynamics in an approximate way. The starting point are the non-relativistic eikonal-CDCC (E-CDCC) equations \cite{OgataPRC68.064609,OgataPRC73.024605} for a three-body reaction between the  and the target
\begin{equation}
\dfrac{i\hbar^2}{E}k\dfrac{d}{d z}\psi_{c}^{(b)}(z) = \sum_{c'}{\mathfrak{F}}^{(b)}_{cc'}(z) \;
\psi_{c'}^{(b)}(z) \ e^{i\left(k_{c'}-k_c \right) z}, \label{cceq4}
\end{equation}
where $c$ are the channel indices \{$i$, $\ell$, $m$\}; $i>0$
($i=0$) denotes the $i$th discretized-continuum (ground) state, and $\ell$ and $m$\ are, respectively, the orbital angular momentum
and its projection on the $z$-axis taken to be parallel to the incident beam. The internal spin indices are suppressed  for simplicity.
Note that in Eq.~(\ref{cceq4}) $b$ is relegated to a superscript since it is not a dynamical variable. The reduced coupling potential ${\mathfrak{F}}^{(b)}_{cc'}(z)$ is given by
\begin{equation}
{\mathfrak{F}}^{(b)}_{cc'}(z)
=
{\cal F}^{(b)}_{cc'}(z)
-\dfrac{Z_{\rm P}Z_{\rm T}e^2}{R}\delta_{cc'},
\label{FF1}
\end{equation}
where
\begin{equation}
{\cal F}^{(b)}_{cc'}(z)
=
\left\langle
\Phi_{c}
|
U_{\rm CT}+U_{\rm vT}
|
\Phi_{c'}
\right\rangle_{\bm \xi}
=
{\cal F}^{{\rm nucl}(b)}_{cc'}(z)+{\cal F}^{{\rm Coul}(b)}_{cc'}(z),
\label{FF2}
\end{equation}
\begin{eqnarray}
{\cal F}^{{\rm nucl}(b)}_{cc'}(z)
&=&
\left\langle
\Phi_{c}
|
U^{\rm nucl}_{\rm CT}+U^{\rm nucl}_{\rm vT}
|
\Phi_{c'}
\right\rangle_{\bm \xi},
\\
{\cal F}^{{\rm Coul}(b)}_{cc'}(z)
&=&
\left\langle
\Phi_{c}
|
U^{\rm Coul}_{\rm CT}+U^{\rm Coul}_{\rm vT}
|
\Phi_{c'}
\right\rangle_{\bm \xi},
\label{NRC}
\end{eqnarray}
where v and C denote two clusters composing the projectile.

$\Phi_c({\bm \xi})$ are the internal wave functions of P,  ${\bm \xi}$ are the coordinate of v relative to C, and $U_{\rm CT}$ ($U_{\rm vT}$) is the potential between C (v) and T including nuclear and Coulomb parts. The multipole expansion for each term on the right-hand-side of Eq.~(\ref{FF2}) is
\begin{eqnarray}
{\cal F}^{{\rm nucl}(b)}_{cc'}(z)&=&
\sum_\lambda {\cal F}^{{\rm nucl}(b)}_{cc',\lambda}(z),
\\
{\cal F}^{{\rm Coul}(b)}_{cc'}(z)&=&
\sum_\lambda {\cal F}^{{\rm Coul}(b)}_{cc',\lambda}(z).
\end{eqnarray}
The explicit form of the nuclear multipoles is given in Ref.~\cite{OgataPRC73.024605}.

In order to include the dynamical relativistic effects  the conjecture similar to that in Ref. \cite{FESHBACH1977110},  one can make the replacements \cite{OgataPTP.121.1399,OgataPTP.123.701}
\begin{equation}
{\cal F}^{\lambda (b)}_{cc'}(z) \rightarrow \gamma f_{\lambda,m-m'}
{\cal F}^{\lambda (b)}_{cc'}(\gamma z) .\label{FF3}
\end{equation}
The factor $f_{\lambda,\mu}$ is equal to unity for nuclear couplings, while for Coulomb couplings one sets
\begin{equation}
f_{\lambda,\mu}
=
\left\{
\begin{array}{cl}
1/\gamma & \quad (\lambda=1, \mu=0) \\
(\gamma^2+1)/(2\gamma)   & \quad (\lambda=2, \mu=\pm1) \\
1        & \quad ({\rm otherwise}) \\
\end{array}
\right.
\label{FF4}
\end{equation}
following Eqs.~(\ref{eq6}) and (\ref{eq7}). Correspondingly, 
\begin{equation}
\dfrac{Z_{\rm P}Z_{\rm T}e^2}{R}\delta_{cc'}
\rightarrow
\gamma\dfrac{Z_{\rm P}Z_{\rm T}e^2}{\sqrt{b^2+(\gamma z)^2}}\delta_{cc'}
\label{FF5}
\end{equation}
in  (\ref{FF1}). 
With these changes we note that Eq.~(\ref{cceq4}) is Lorentz covariant, as desired.

Solving Eq.~(\ref{cceq4}) under the boundary condition
\begin{equation}
\lim_{z \to -\infty}\psi_{c}^{(b)}(z)=\delta_{c0},
\end{equation}
where $0$ denotes the incident channel, one gets the following the excitation amplitude of the channel state $c$ by solving the coupled-channels equations (\ref{BC9235}) with
\begin{equation}
{\cal A}_{c0}(b) \equiv
\displaystyle
{\lim_{z \to \infty}}\psi_{c}^{(b)}(z)
\end{equation}
Incorporating nuclear absorption at small impact parameters implies that the eikonal S-matrix in the coupled-channels calculations yield
\be
S_{c}(b) = {\cal A}_{c0}(b) \exp\Big\{-Im\left[\chi_{OL}(b)\right]\Big\} \label{scba}
\ee
 with $\chi_{OL}(b)$ given by Eq. (\ref{eik7}).
 
 The angular distribution of the inelastically scattered particles can be obtained from the S-matrix above, yielding
\begin{equation}
f_{c}(\theta)= ik\int_{0}^{\infty}db b  J_{m}(qb) S_{c}(b) . \label{angp1}
\end{equation}
The inelastic scattering cross section is obtained by an average over the initial spin and a sum over the final spin:
\begin{equation}
{\frac{d\sigma_{inel}}{d\Omega}} = \frac1{2\ell_{0}+1} \sum_{m_{0},m_f}|f_{c}(\theta)|^{2} \ . \label{angp91}
\end{equation}

 \subsubsection{Nuclear excitation of collective modes \label{necm}}
 
Using the collective particle-vibrator coupling model the matrix element for the transition $j\longrightarrow k$ becomes \cite{SATCHLER1987215}
\begin{eqnarray}
&<\ell_{k}m_{k}|U_{N(\lambda\mu)} |\ell_{j}m_{j}>=-\displaystyle{{\delta_{\lambda}}\over{\sqrt{2\lambda+1}}}\ <\ell_{k}
m_{k}|Y_{\lambda\mu}|\ell_{j}m_{j}>\nonumber \\
&\times \ \ Y_{\lambda\mu}(\hat{\mathbf{r}})\ U_{\lambda}(r) \label{VfiN}
\end{eqnarray}
where $\delta_{\lambda}$ is the vibrational amplitude (deformation parameter), $U_{\lambda}(r)$ is the transition potential, and
\begin{eqnarray}
<\ell_{k}m_{k}|Y_{\lambda\mu}|\ell_{j}k_{j}>&=&(-1)^{\ell_{k}-m_{k}}\ \left[  {\frac{
(2I_{k}+1)(2\lambda+1)}{4\pi(2\ell_{j}+1)}}\right]  ^{1/2}\nonumber \\ &\times& \left(  {{{{{ {
\genfrac{}{}{0pt}{}{\ell_{k} }{-m_{k}}
} }}}}}{{{{{ {
\genfrac{}{}{0pt}{}{\lambda}{\mu}
} }}}}}{}{{{{{ {
\genfrac{}{}{0pt}{}{\ell_{j} }{m_{j}}
} }}}}}\right)  \left(  {{{{{ {
\genfrac{}{}{0pt}{}{\ell_{k} }{0}
} }}}}}{{{{{ {
\genfrac{}{}{0pt}{}{\lambda}{0}
} }}}}}{{{{{ {
\genfrac{}{}{0pt}{}{\ell_{j} }{0}} }}}}}\right)  \ . \label{WE}
\end{eqnarray}

The transition potentials can be related to the real part of the optical potentials, $U=Re[U_{opt}]$, originated from low energy collisions. Notice that one does not include the imaginary part of the optical potentials because the inelastic cross section already includes the absorption part in the factor $\exp[i\chi_{OL}(b)]$ in Eq. (\ref{scba}).  The transition densities are given by \cite{SATCHLER1987215}
\be
\delta \rho_{0}(r)=-\alpha_{0}\left(3U+r{\frac{d}{dr}}\right)\rho_{0} (r),
\ee
for isoscalar monopole excitations,
\be
\delta \rho_{1}^{{IV}}(r)=\beta_{1}^{IV} \left({\frac{d}{dr}}+\frac{1}{3} R \frac{d^{2}}{dr^{2}}\right)\rho_{0} (r)
\ee
for isovector dipole excitations, and 
\be
\delta \rho_{l}(r)=\beta_{l} {\frac{d}{dr}}\rho_{0} (r)
\ee
for isoscalar giant quadrupole and higher multipoles. $\alpha_0$ and $\beta_i$ are the  vibrational amplitudes. $R$ is the nuclear radius at ${1}/{2}$ the central nuclear density and $\rho_{0} (r)$ are the ground state densities.The same equation above can be used for isovector giant resonances of $l\ge 2$ multi polarities. For a thorough discussion on the subject, see Ref. \cite{SATCHLER1987215}. 

The corresponding transition potentials are \cite{SATCHLER1987215}
\begin{equation}
U_{0}(r)=-\alpha_{0}\left(3+r{\frac{d}{dr}}\right)U(r)\ , \label{U0}
\end{equation}
for isoscalar monopole,
\begin{equation}
U_{1}^{{IV}}(r)=\delta_{1}\left({\frac{d}{dr}}+\frac{1}{3} R \frac{d^{2}}{dr^{2}}\right)U(r)\ , \label{U1}
\end{equation}
for isovector dipole, and
\begin{equation}
U_{l}(r)=\delta_{l}{\frac{dU(r)}{dr}}\ , \label{U2}%
\end{equation}
for $l\ge 2$. In these equations, $\delta_{i}=\beta_{i}R$, are known as deformation lengths. 

The matrix elements  can be calculated using well-known sum-rules leading to relations between the deformation length, and the nuclear sizes and the excitation energies. For isoscalar excitations one has the sum rules \cite{SATCHLER1987215}
\begin{equation}
\alpha_{0}^{2}= 2 \pi\ {\frac{\hbar^{2} }{m_{N}}} \ {\frac{1}{<r^{2}> A E_{x}}} \ , \ \ \ \ \  \delta_{\lambda\geq2}^{2} = {\frac{2 \pi}{3}}
\ {\frac{\hbar^{2} }{m_{N}}} \ \lambda\ (2\lambda+1) \ {\frac{1}{A E_{x}}}
\label{deform1}
\end{equation}
where $A$ is the atomic number, $\left<r^{2}\right>$ is the r.m.s. radius of the nucleus, and $E_{x}$ is the excitation energy.

For dipole isovector excitations \cite{SATCHLER1987215}
\begin{equation}
\delta_{1}^{2}= {\frac{\pi}{2}} \ {\frac{\hbar^{2} }{m_{N}}} \ {\frac{A }{NZ}}
\ {\frac{1}{E_{x}}}\ ,
\end{equation}
where $Z$ ($N$) the charge (neutron) number. The transition potential in this case is modified from Eq. (\ref{U1}) to account for the isospin dependence \cite{SATCHLER1987215}. It becomes
\begin{equation}
U_{1}^{{IV}}(r)=-\Lambda\ \Big( {\frac{N-Z }{A}} \Big) \ \Big( {\frac{dU }{dr}}
+ {\frac{1}{3}} \ R \ {\frac{d^{2} U }{dr^{2}}} \Big) \ ,
\end{equation}
where the factor $\Lambda$ depends on the difference between the proton and
the neutron matter radii as
\begin{equation}
\Lambda{\frac{2(N-Z)}{3A}} = {\frac{R_{n}-R_{p} }{{\frac{1}{2}} \ (R_{n}%
+R_{p})}} = {\frac{\Delta r_{np} }{R}} \ . \label{deform11}%
\end{equation}
The strength of isovector excitations increases with the neutron skin $\Delta r_{np}$ which is larger for neutron-rich nuclei.

As shown in Ref. \cite{HarakehPRC23.2329}, for the isoscalar giant dipole resonance,  the transition density in the deformed model is expressed as
\bea
\delta \rho^{(IS)} (r) &=& -{\beta_{1}^{(IS)} \over R\sqrt{3}}\Bigg[ 3r^{2}{d\over dr} +10r - {5\over 3} \left<r^{2}\right>{d\over dr}  \nonumber \\
&-&\epsilon \left( r{d^{2}\over dr^{2}} +4 {d\over dr} \right) \Bigg]\rho_{0} (r), \label{harakeh}
\eea
where 
\be
\epsilon = \left({4\over E_{2}}+{5\over E_{0}}\right){\hbar^{2}\over 3mA},
\ee
with $A$ being the nuclear mass, $E_{2}$ and $E_{0}$ are the excitation energies of the giant isoscalar quadrupole and monopole resonances, respectively. In Eq. (\ref{harakeh}) $\left< \right>$ means average value.

The collective coupling parameter for the isoscalar dipole resonance is
\be
\beta_{1}^{(IS)}={6\pi \hbar^{2}R^{2}\over m_{N}AE_{x}}\left( 11\left<r^{4}\right> - {25\over 3} \left<r^{2}\right>^{2} -10 \epsilon \left<r^{2}\right> \right).
\ee
Analogously, the transition potential can be written, for isoscalar dipole excitations
\bea
U_{1}^{(IS)} (r) &=& -{\delta_{1}^{(IS)} \over R\sqrt{3}}\Bigg[ 3r^{2}{d\over dr} +10r - {5\over 3} \left<r^{2}\right>{d\over dr}  \nonumber \\
&-&\epsilon \left( r{d^{2}\over dr^{2}} +4 {d\over dr} \right) \Bigg]U (r), \label{harakeh2}
\eea

The reduced transition probability for electromagnetic transitions is defined by \cite{eisenberg1988excitation}
\begin{equation}
B\left(  \pi\lambda; i\rightarrow j\right)  = {\frac{1}{2I_{i}+1}} \ \left|
<I_{j}||\mathcal{M}(\pi\lambda)||I_{i}> \right|  ^{2} \ , \label{Bvalue}%
\end{equation}
and neglecting the spin of the initial state $i$, these matrix elements can be related to the deformation parameters  $\alpha_0$ and $\delta_\lambda$ by means of the sum-rules \cite{SATCHLER1987215}
\begin{equation}
B(E0) = \left[  {\frac{3 ZeR^{2}}{10\pi}} \right]  ^{2} \alpha_{0}^{2}\ ,
\ \ \ \ \ \ B(E1) = {\frac{9}{4\pi}}\left(  {\frac{NZe}{A}}\right)  ^{2}
\delta_{1}^{2} \ ,
\end{equation}
and
\begin{equation}
B(E\lambda)_{\lambda\geq2} = \left[  {\frac{3 }{4\pi}} ZeR^{\lambda
-1}\right]  ^{2} \delta_{\lambda}^{2} \ ,
\end{equation}

\subsubsection{Angular distribution of $\gamma$-rays}

After the excitation with energy $\hbar \omega$, the state $\left|  I_{f}\right\rangle $ can decay by gamma emission to  $\left|  I_{g}\right\rangle $. There are complications due to the fact that the nuclear levels are not only populated through Coulomb excitation, but also by conversion and $\gamma
$-transitions from higher states. To compute the angular distributions one must know the parameters $\Delta_{l}\left(  i\longrightarrow j\right)  $ and $\epsilon_{l}\left( i\longrightarrow j\right)  $\, for $l\geq1$ \cite{AlderWinther1965},
\begin{equation}
\epsilon_{l}^{2}\left(  i\longrightarrow j\right)  =\alpha_{l}\left(
i\longrightarrow j\right)  \Delta_{l}^{2}\left(  i\longrightarrow j\right)  ,
\label{casc1}
\end{equation}
where $\alpha_{l}$ is the total $l$-pole conversion coefficient, and
\begin{eqnarray}
\Delta_{\pi l}&=&\left[  \frac{8\pi\left(  l+1\right)  }{l\left[  \left(2l+1\right)  !!\right]  ^{2}}\frac{1}{\hbar}\left(  \frac{\omega}{c}\right)
^{2l+1}\right]  ^{1/2}\left(  2I_{j}+1\right)  ^{-1/2} \nonumber \\
&\times& \left\langle I_{j}\left\|  i^{s(l)}\mathcal{M}(\pi l)\right\|  I_{i}\right\rangle \ ,
\label{gad0_1}
\end{eqnarray}
with $s(l)=l$ for electric $\left(  \pi=E\right)  $ and $s(l)=l+1$ for magnetic $\left(  \pi=M\right)  $ transitions. The square of $\Delta_{\pi l}$
is the $l$-pole $\gamma$-transition rate (in $s^{-1}$). $\mathcal{M}(\pi l)$ is the electromagnetic operator.

In the non-relativistic case \cite{AlderWinther1965}, the gamma ray angular distributions after the excitation depend on the frame of reference used. Here the z-axis corresponds to the beam axis and considering a simple situation in which the $\gamma$-ray is emitted directly from the final excited state $f$ to a lower state $g$ observed experimentally it was demonstrated in Ref. \cite{BERTULANI2003317} 
that the angular dependence of the $\gamma$-rays is given explicitly by the spherical coordinates $\theta$ and $\phi$ of the photon momentum vector \textbf{k} by
\begin{equation}
W\left(  \theta\right)  =\sum_{ \overset{ k=\mathrm{even}}{M_{i},M_{f},
l,l^{\prime}} } \left(  -1\right)  ^{M_{f}}\left|  a_{i\longrightarrow
f}\right|  ^{2} F_{k}\left(  l,l^{\prime},I_{g},I_{f}\right)    \nonumber
\end{equation}
\begin{equation}
\times 
\left(
\begin{array}
[c]{ccc}%
I_{f} & I_{f} & k\\
M_{f} & -M_{f} & 0
\end{array}
\right)
\sqrt{2k+1}P_{k}\left(  \cos\theta\right)  \Delta_{l}\Delta^{*}
_{l^{\prime}}\;,
\end{equation}
where $P_{k}$ are Legendre polynomials, and
\begin{align}
F_{k}\left(  l,l^{\prime},I_{g},I_{f}\right)   &  =\left(  -1\right)
^{I_{f}-I_{g}-1}\sqrt{\left(  2l+1\right)  \left(  2l^{\prime}+1\right)
\left(  2I_{f}+1\right)  \left(  2k+1\right)  }\nonumber\\
&  \times\left(
\begin{array}
[c]{ccc}%
l & l^{\prime} & k\\
1 & -1 & 0
\end{array}
\right)  \left\{
\begin{array}
[c]{ccc}
l & l^{\prime} & k\\
I_{f} & I_{f} & I_{g}
\end{array}
\right\}  \ .
\end{align}

The angular distribution of $\gamma$-rays described above is in the reference frame of the excited nucleus. To obtain the distribution in the laboratory one has to perform the transformation
\begin{equation}
\theta_{L}=\arctan\left\{  {\frac{\sin\theta}{\gamma\left[  \cos\theta+
\beta\right]  }}\right\}  \ ,
\end{equation}
and
\begin{equation}
W (\theta_{L}) = \gamma^{2} \left(  1+ \beta\cos\theta\right)  ^{2} W
(\theta)\ ,
\end{equation}
where $\gamma$ is the Lorentz contraction factor, and $\beta=\sqrt{1-1/\gamma^{2}}$. The photon energy in the laboratory is $E^{ph}_{L}=\gamma E^{ph}_{cm}\left(  1 +\beta\cos\theta\right)  $.

\subsection{Excitation of pygmy resonances}

Giant resonances have been observed all along the nuclear chart. Pygmy resonances, on the other way, have been observed mainly in neutron-rich nuclei. Its first observation was reported in 1961 with the observation of a large number of bunched unbound states  in $\gamma$-rays emitted after neutron capture \cite{BartholomewARNPS11.120161.001355}. The name pygmy resonance, or pygmy dipole resonance (PDR), was first used in 1969 in connection with calculations of neutron capture cross sections \cite{Brzosko1969}\footnote{We are grateful to Riccardo Raabe (KU Leuven) for sharing this information with us.}.  The theoretical treatment of the PDR as a collective excitation mode in the nucleus was first published in Ref. \cite{MohanPRC3.1740} using a three fluid model with a proton and another neutron fluid in the same orbitals, and an additional fluid of neutron excess interacting weakly with the rest. The model predicted that the neutron excess fluid oscillates against a core with $N = Z$. Finally, the first experimental proposal to measure the effect of Coulomb excitation of  pygmy resonances was submitted in 1987 in Japan at the JPARC facility \cite{Kubono1987}. The relevance for the existence of pygmy resonances for nucleosynthesis processes was reported in Refs. \cite{GORIELY199810,PhysRevC.91.044318,LARSEN201969}.

The interest on excitation of pygmy resonances in exotic nuclei surge in the 1990s when a narrow peak associated with the small separation energies was reported in Coulomb dissociation experiments \cite{IekiPRL70.730,SackettPRC48.118}. A few years earlier Refs. \cite{BERTULANI1988615,BERTULANI1991751,BertulaniPRC46.2340}, a simple expression for the response function emerges for electric multipoles was obtained using simple Yukawa or Hulthen functions for bound states and plane waves for the continuum,  
\begin{eqnarray}
{dB_{E L}(E) \over dE}&=& C_{L}{2^{L-1} \over \pi^{2}} (2L+1)(L!)^{2}\left( { \hbar^{2} \over \mu} \right)^{L} \nonumber \\
&\times& Z^{2}e_{L}^{2}{\sqrt{S} (E-S)^{L+1/2}\over E^{2L+2}},\label{dbde1}
\end{eqnarray}
where $C_{L}$ is a constant depending on how the wave functions are normalized, $e_{L}$ is the effective charge for the multipolarity $L$, $\mu$ is the reduced mass and $S$ is the separation energy of a two-cluster  nucleus. In the case of the  electric dipole (E1) excitation,
\begin{eqnarray}
{dB_{E 1}(E) \over dE}&=& C_{1}{3 Z^{2}e_{L}^{2}\hbar^{2}\over \mu \pi^{2}}  {\sqrt{S} (E-S)^{3/2}\over E^4}.\label{dbde2}
\end{eqnarray}
Eqs. (\ref{dbde1}) and (\ref{dbde1}) predict a maximum of the response function at  $E_{x}=8S/5$ and a width $\Gamma \sim E_{x}$. They have been used in numerous experimental analyses \cite{NAKAMURA1994296,SHIMOURA199529} and compared with predictions of other theoretical models \cite{TeruyaPRC43.R2049,BertulaniPRC46.2340,OtsukaPRC49.R2289,Manju:EPJA2019}.  

Adding final state interactions complicates the theoretical predictions based on Eq. (\ref{dbde1}). Considering the transition from a bound single-particle s-state to a continuum p-wave, as in the Coulomb breakup of $^{11}$Be, and using the notation $d\sigma /d\Omega \sim |\mathcal{I}_{s\rightarrow p} I^{2}$, Ref. \cite{BertulaniPRC.75.024606} finds
\begin{align}
\mathcal{I}_{s\rightarrow p}    \simeq\frac{(E-S_{n})^{3/4}}{E^{2}}\left[  1+\left(  \frac{\mu}{2\hbar^{2}}\right)
^{3/2}\frac{\sqrt{S_{n}}\left(  3E_{r}-2S_{n}\right)  }{-1/a_{1}+\mu r_{1}%
E_{r}/\hbar^{2}}\right]  , \label{isp}%
\end{align}
where $S_{n}$ is the neutron separation energy, and the effective range expansion of the phase shift, $\delta$, $k^{2l+1}\cot
\delta\simeq-1/a_{l}+r_{l}k^{2}/2,$ was used to obtain the correction included by the second term within brackets. For $l=1$, $a_{1}$ is the \textquotedblleft scattering volume\textquotedblright\ (units of length$^{3}$) and $r_{1}$ is the \textquotedblleft effective momentum\textquotedblright\ (units of 1/length). Their interpretation is not as simple as the $l=0$ effective range parameters.

For a three-cluster nucleus, such as $^{11}$Li or $^{6}$He, three-body models are evidently necessary,   yielding  for the E1 multipolarity \cite{Pushkin_1996,ZHUKOV1993151,DANILIN1998383,BertulaniPRC.75.024606}
\be 
{dB_{E 1}(E) \over dE} \propto {(E-S_{2n})^{3}\over E^{11/2}} (1 + FSI),
\ee
where the final state interaction (FSI) term is given by an integral over hyper angles \cite{BertulaniPRC.75.024606}. Without the FSI term, the E1 response in the three-body model would imply a peak at around $E \simeq1.8S_{2n}$.

\begin{figure}[ptb]
\begin{center}
{\includegraphics[width=8.5cm]{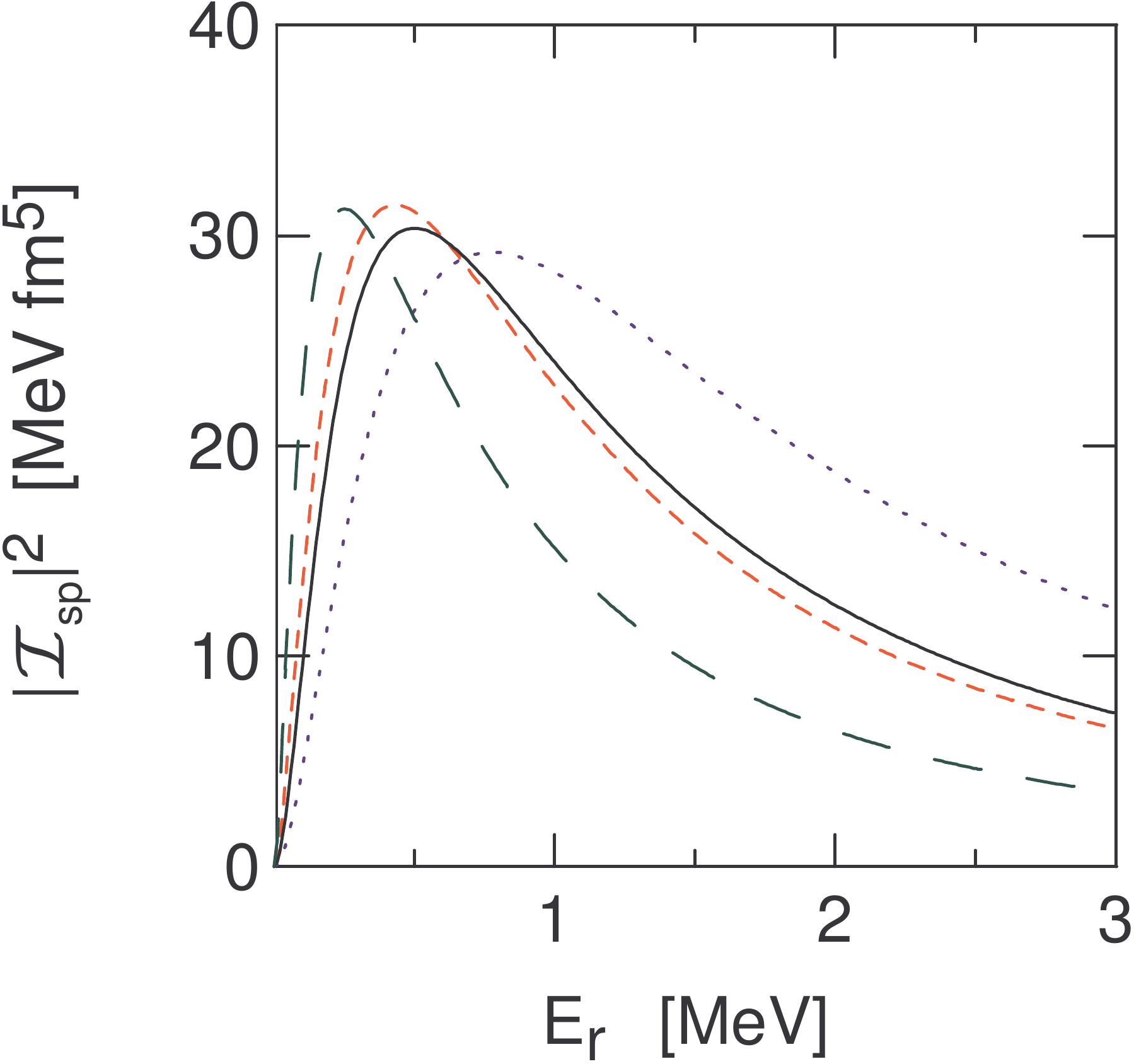}}
\end{center}
\caption{$\left\vert \mathcal{I}_{s\rightarrow
p}\right\vert ^{2}$ \ calculated using eq. \ref{isp}, assuming different values of the scattering length and effective range for the n-$^{10}$Be scattering in the final state \cite{BertulaniPRC.75.024606}.}
\label{isp2}
\end{figure}

\begin{figure}[ptb]
\begin{center}
{\includegraphics[width=8.5cm]{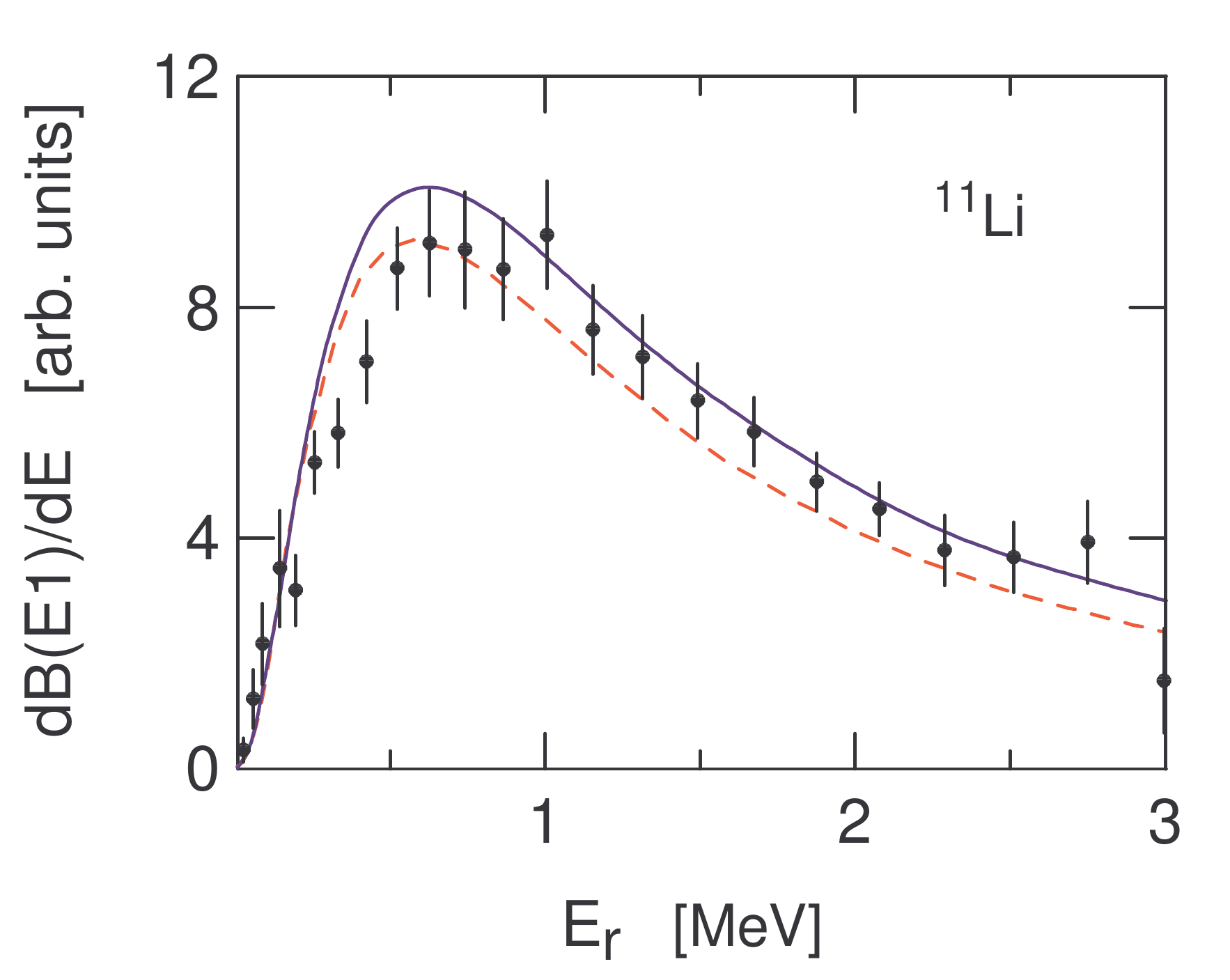}}
\end{center}
\caption{Comparison between the calculation of the response function (in arbitrary units) in $^{11}$Li  using zero phase-shifts, $\delta_{nn}=0$ and $\delta_{nc}=0$, (dashed line), or including the effects of final state interactions (continuous line) \cite{BertulaniPRC.75.024606}. The experimental data are from Ref. \cite{SHIMOURA199529}.}
\label{dbdeli11}
\end{figure}

In both two-body and three-body halos, the FSI model shifts appreciably the peak in the EM response.Therefore, the separation energy still roughly determines the peak location of the E1 response but at a different energy and with a different width. It was also shown that final state interactions can substantially change the location of the low energy peak \cite{BertulaniPRC.75.024606,TYPEL2005247}. This is clear from Fig. \ref{isp2}, where the function $\mathcal{I}_{s\rightarrow p}$ is plotted for different bu reasonable adopted values for the scattering length and effective range \cite{BertulaniPRC.75.024606}. In Fig. \ref{dbdeli11} we show a comparison between the calculation of the response function (in arbitrary units) in $^{11}$Li  using zero phase-shifts, $\delta_{nn}=0$ and $\delta_{nc}=0$, (dashed line), or including the effects of final state interactions (continuous line) \cite{BertulaniPRC.75.024606}. The experimental data are from ref. \cite{SHIMOURA199529}.

Following the first calculations on pygmy resonances based on two- and three-body models, it was soon realized that the the role of the continuum and of the nuclear contribution was crucial to explain the experimental data on the EM response obtained in breakup experiments \cite{BERTULANI1992163,BertulaniPRC49.2839,CHATTERJEE2000477,KidoPRC53.2296,BanerjeePRC65.064602,MARGUERON2002105,IBRAHEEM2005414,BertulaniPRL94.072701,MoroPRC73.044612,OgataPTP.123.701,OgataPRC68.064609,RavinderPRC86.061601,Hagino:ENFSL2013,Broglia_2016,Alamanos2017,Broglia_2019}. Calculations based on effective field theories started to appear recently \cite{CapelPRC98.034610}. 

\begin{figure}[ptb]
\begin{center}
{\includegraphics[width=4.7cm]{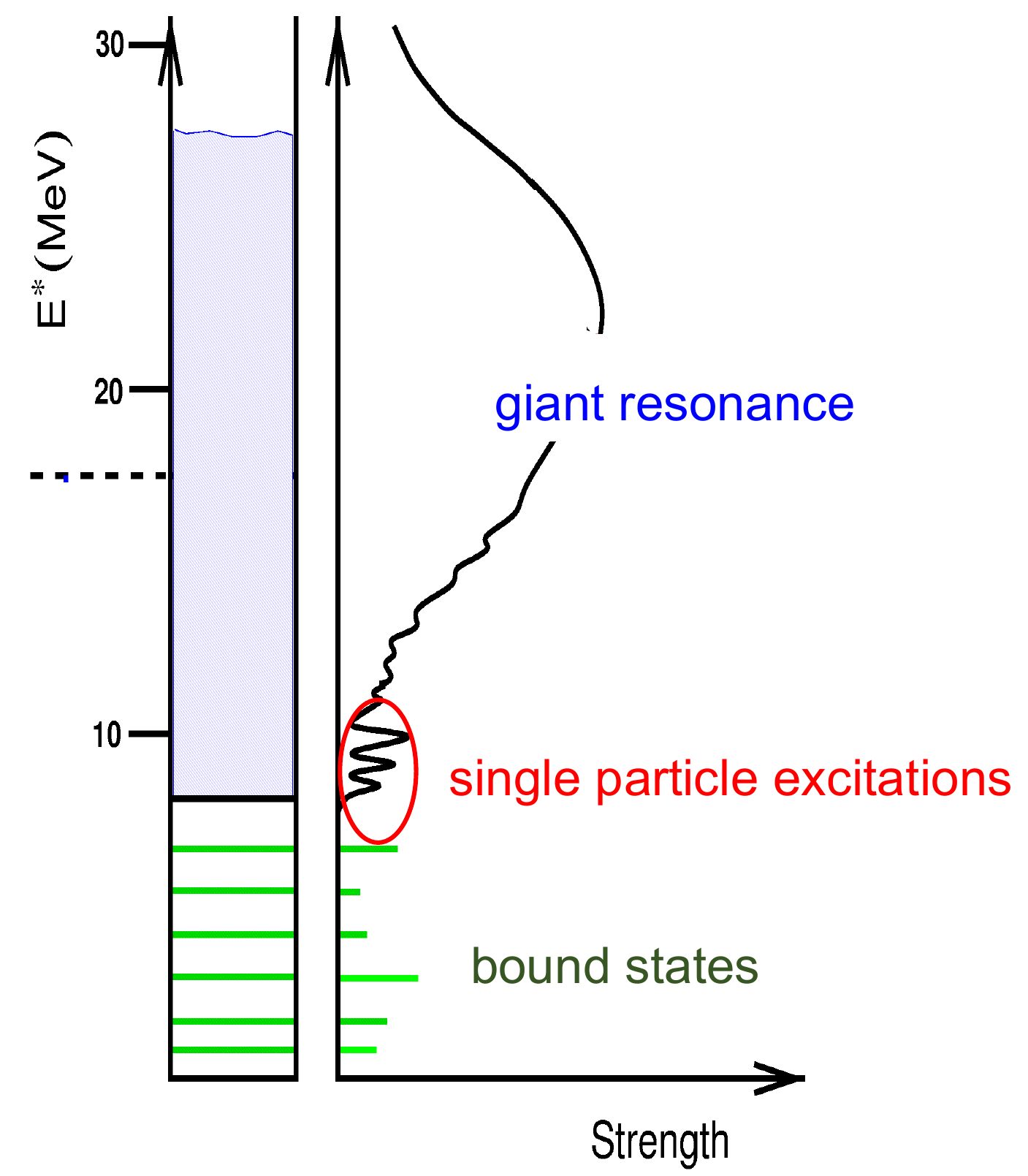}\includegraphics[width=4cm]{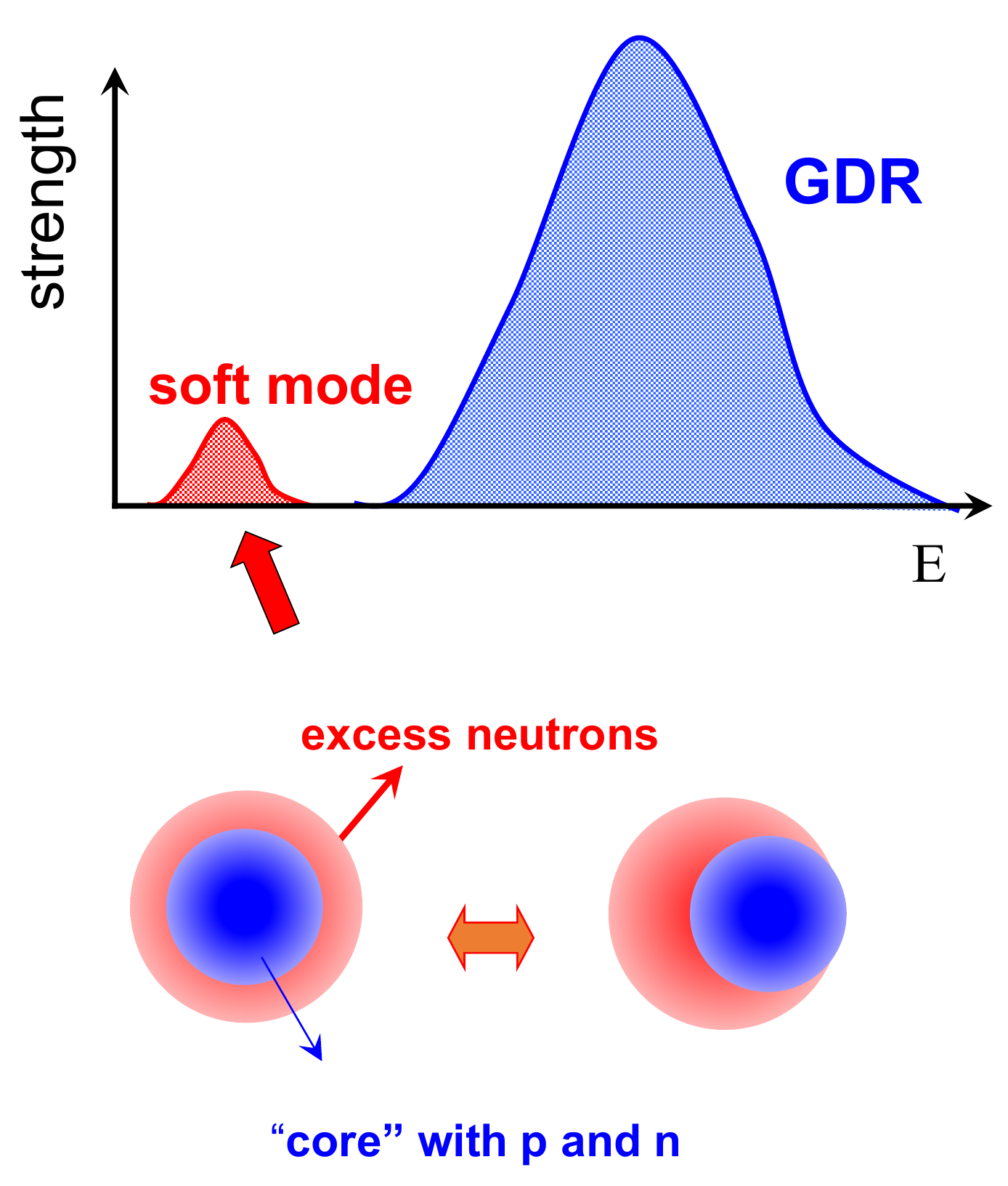}}
\end{center}
\caption{Schematic view of the excitation of giant resonances (left) as a broad and large peak in the excitation spectrum of nuclei around 10-20 MeV. The pygmy resonances are viewed as a collective excitation of a neutron ``mantle'' vibrating against a tightly bound nuclear core (left). Besides the giant resonance, here a Giant Dipole Resonance (GDR) at higher energies (right), one should see a concentrated bump at lower energies due to the pygmy states.}
\label{pygmyfig}
\end{figure}

Because the dipole response at low energies were also observed for heavier nuclei, the old idea that pygmy resonances could have a collective character emerged again and revived old theoretical predictions \cite{MohanPRC3.1740,Suzuki101143/PTP.83.180,VanIsackerPRC.45.R13}. Extending  the Goldhaber-Teller (GT) \cite{GoldhaberPR74.1046} and  Steinwedel-Jensen (SJ) \cite{jensen:1950:ZFN} hydrodynamical models to pygmy resonances in ``soft'' neutron-rich nuclei, and using the admixture of the two pictures as in Ref. \cite{MeyersPRC15.2032}, the radial transition density of pygmy resonances can be described by \cite{BertulaniPRC.75.024606}
\begin{eqnarray}
\delta \rho(r) = \sqrt{4\pi \over 3} R \left[Z_{\scriptscriptstyle GT}\alpha_{\scriptscriptstyle GT} {d\over dr} + Z_{\scriptscriptstyle SJ}\alpha_{\scriptscriptstyle SJ}{K\over R} j_{1}(kr)\right]\rho_{0}(r) , \label{GTSJ1}
\end{eqnarray}
where $Z_{i}$ are the GT and SJ effective charges  \cite{BertulaniPRC.75.024606}, $\alpha_{i}$ are amount of GT and SJ admixture, with  $\alpha_{\scriptscriptstyle GT}+\alpha_{\scriptscriptstyle SJ}=1$, $K=9.93$ and $R$ is the (mean (sharp density) nuclear radius. $j_{1}(kr)$ is the spherical Bessel function of order 1, with $k=2.081/R$. The transition density in Eq. \ref{GTSJ1} na\"ively describes a pygmy resonance as a collective dipole vibration of protons and neutrons ({\it soft dipole mode}), as shown schematically in Figure \ref{pygmyfig}. 

For the Goldhaber and Teller model \cite{GoldhaberPR74.1046} one has $\alpha_{SJ}=0$, and the centroid energy of the IVGDR is obtained as,
 \begin{eqnarray}
E_{IVGDR}= \left({3 {\cal S} \hbar^{2}\over 2 aRm_{N}}\right)^{{1/2}}, \label{GTSJ}
\end{eqnarray}
where $a$ is the nuclear distribution diffuseness, $m_{N}$  is the nucleon mass, and ${\cal S}$ is not the nucleon separation energy but the energy needed to extract one neutron from the proton environment. This is roughly is the part of the potential energy due to the neutron-proton interaction within the nuclear environment, also roughly proportional to the {\it symmetry energy} $\propto (N-Z)/A$.  Goldhaber and Teller \cite{GoldhaberPR74.1046}  used ${\cal S} = 40$ MeV, $a=1-2$ fm and to obtain $E\simeq 10-20$ MeV for a medium heavy and heavy nuclei, in rough agreement with the experimentally found centroid energies of the IVGDR.  

It is not straightforward to extend the arguments of Goldhaber and Teller discussed in the preceding paragraph to the case of neutron rich nuclei. But the quantity  $\cal S$ should somehow relate to the symmetry energy. This could be obtained within a microscopic model because the pygmy resonances display fine-structures that seem to be much more dependent on the coupling of phonon states with complex configurations than in giant resonances. In fact, the collectivity of the pygmy resonance is not well established. Hydrodynamical models are surely unfit for halo nuclei.  We could  approximate  ${\cal S}$ to the separation energy, ${\cal S} = S$ and expect that the product $aR$ is proportional to $S^{-1}$. Then, naively, we would get $E_{PDR}\sim \beta S$, with $\beta\sim 1$. For $^{11}$Li, we take $a=1-2$ fm, $R=3$ fm, and $S = 0.3$ MeV, one gets $E_{PDR} = 1-2$ MeV, which is also compatible with experimental data.  But, evidently, hydrodynamical models  lack accuracy and can only be used to give physics insight of the problem. Microscopic models, often relying on the linear response theory, are a more accurate method to describe giant resonances and pygmy resonances on the same foot. 

\begin{figure}[ptb]
\begin{center}
{\includegraphics[width=8.5cm]{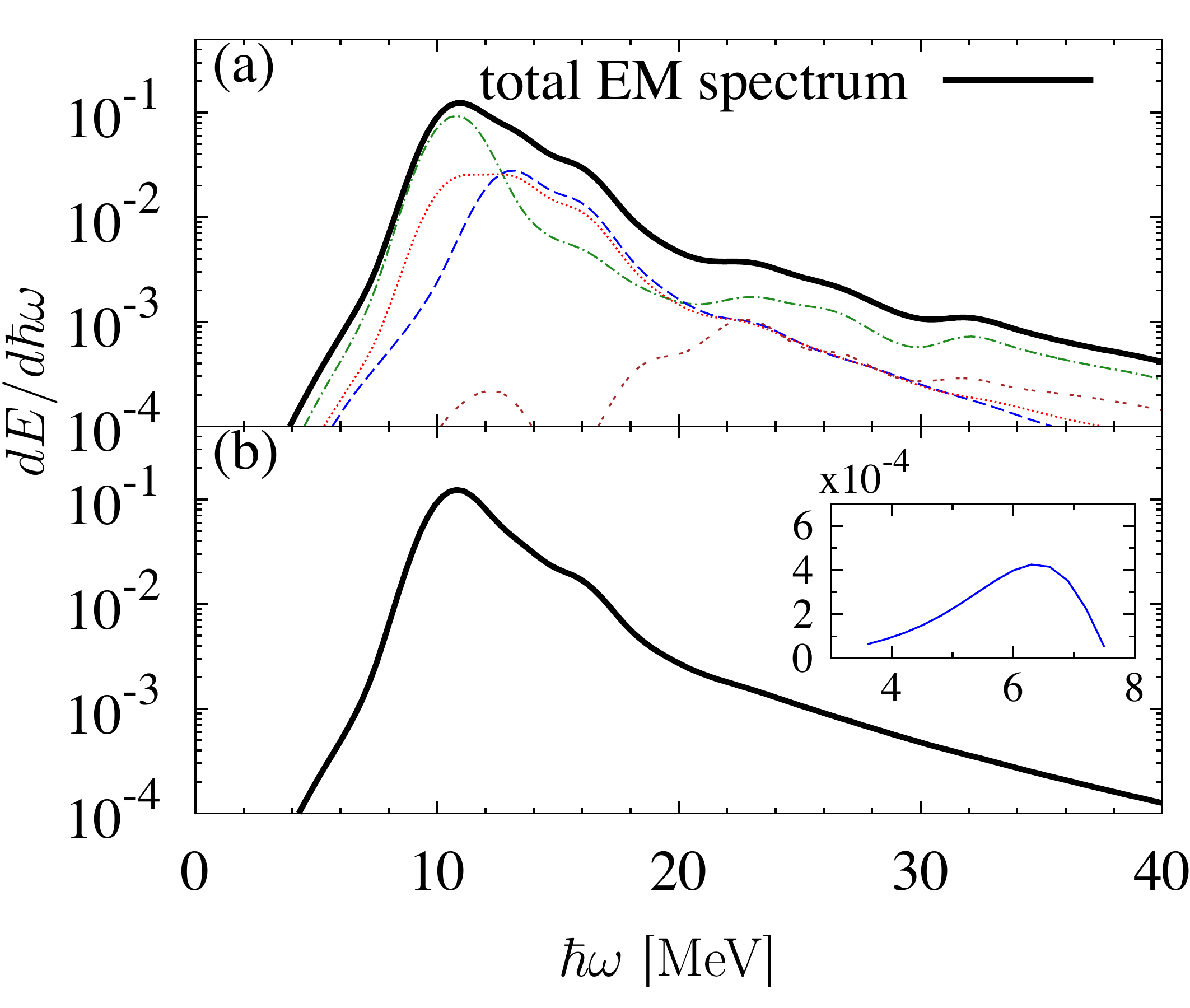}}
\end{center}
\caption{{\it Top:} Total energy spectrum (solid line) of emitted EM radiation, $dE/d\hbar \omega$, for $^{238}$U + $^{238}$U collision at the impact parameter $b = 12.2$ fm. The total quadrupole contribution is shown by a double-dotted line. The other curves  are contributions from the three target-projectile orientations. {\it Bottom:} The electric dipole radiation only emitted from the target nucleus. The inset shows the pygmy resonance contribution to the emitted spectrum, only visible in the main figure as a slope change at low energies \cite{StetcuPRL2015}.}
\label{tdslafig}
\end{figure}

The electromagnetic response in weakly-bound nuclei was first studied using the RPA equations done in Ref. \cite{BertschPRC41.1300}, based on the continuum RPA model developed in Ref. \cite{SHLOMO1975507}. Later it was also used in Refs. \cite{BertulaniPRC46.2340,TeruyaPRC43.R2049} to study the effects of clustering and higher multipole EM  response. The RPA model results for halo nuclei again showed a pronounced peak at small excitation energies, again interpreted as a pygmy resonance, but in fact it was just the effect of a small separation energy in the nuclei. For medium heavy and heavy nuclei different RPA models  have shown that basically all neutron-rich nuclei display a visible bunching in the response at low energies which has again been attributed to the pygmy resonance. Relativistic mean field models also yield similar results, e.g. in Refs. \cite{CATARA199786,CATARA1997449,VRETENAR2001496,SARCHI200427,PiekarewiczPRC73.044325,LiangPRC75.054320,PaarRPP:2007,LITVINOVA2007111,TsonevaPRC77.024321,BarbieriPRC77.024304,PaarPRL103.032502,InakuraPRC80.044301,LanzaPRC79.054615,CarbonePRC81.041301,RocaMazaPRC85.024601,VretenarPRC85.044317}.  More recently the powerful TDSLA method, described previously has been used to study pygmy resonances and also nuclear large amplitude collective motion \cite{Bulgac:JPCS2008,Bulgac1288,StetcuPRC.2011,BulgacARNPS2013,StetcuPRL2015,BulgacPRL2016}.  First TDSLDA calculations for relativistic Coulomb excitation in a collision of $^{238}$U + $^{238}$U were reported in Ref. \cite{StetcuPRL2015}.  One has reported considerable amount of electromagnetic strength occurs at low energies, around $E_x \sim 7$ MeV. This additional structure was  attributed to  the excitation of the pygmy dipole resonance (PDR). This is shown in Fig. \ref{tdslafig} where in the upper panel one sees the total energy spectrum (solid line) of emitted EM radiation, $dE/d\hbar \omega$, for $^{238}$U + $^{238}$U collision at the impact parameter $b = 12.2$ fm. The total quadrupole contribution is shown by a double-dotted line. The other curves  are contributions from the three target-projectile orientations. In the bottom panel one sees the electric dipole radiation only emitted from the target nucleus. The inset shows the pygmy resonance contribution to the emitted spectrum, only visible in the main figure as a slope change at low energies \cite{StetcuPRL2015}.

As we have discussed previously the excitation of exotic nuclei by electrons is not possible within the next years. Theoretically,   the electric pygmy dipole resonance in electron scattering has been studied at large angles at existing facilities, e.g. at the S-DALINAC in Darmstadt, Germany \cite{2019arXiv190408772P}. They demonstrate that the excitation of pygmy resonance states in (e,e') reactions is predominantly of transversal character for large scattering angles. They were also able to extract the fine structure of the pygmy states at low excitation energies. Electron scattering with light and loosely bound nuclei could be the first sort of inelastic scattering studied experimentally if an electron-radioactive-ion collider might become available \cite{etde-21504514,SUDA20171,Karataglidis2017}. In Ref. \cite{BERTULANI2005203} that,  for an electron energy $E$, the total cross section for the dissociation of a two-body cluster halo nuclei is
\begin{equation}
\sigma_{e}(E_{e}) =64\sqrt{2} \pi  {e_{eff}^{2} \over \mu c^{2}S} \ln \left({E_{e} \over S}\right),
\end{equation}
where $\mu$ is the reduced mass and $e_{eff}$ is the effective charge. This equation predicts an inverse separation energy  dependence which helps inelastic electron scattering  for very loosely bound nuclei. For example, if we assume  $S =100$ keV, $E_{e}=10$ MeV,  $e_{eff}= e$, and $\mu  = m_{N}\sim 10^3$ MeV, we get  $\sigma_{e} =25$ mb. The cross section also increases, despite very slowly, with the electron energy which brings no real advantage for high energy electrons.  

\subsection{Extracting dipole polarizabilities}

\begin{figure}[t]
\begin{center}
{\includegraphics[width=7.5cm]{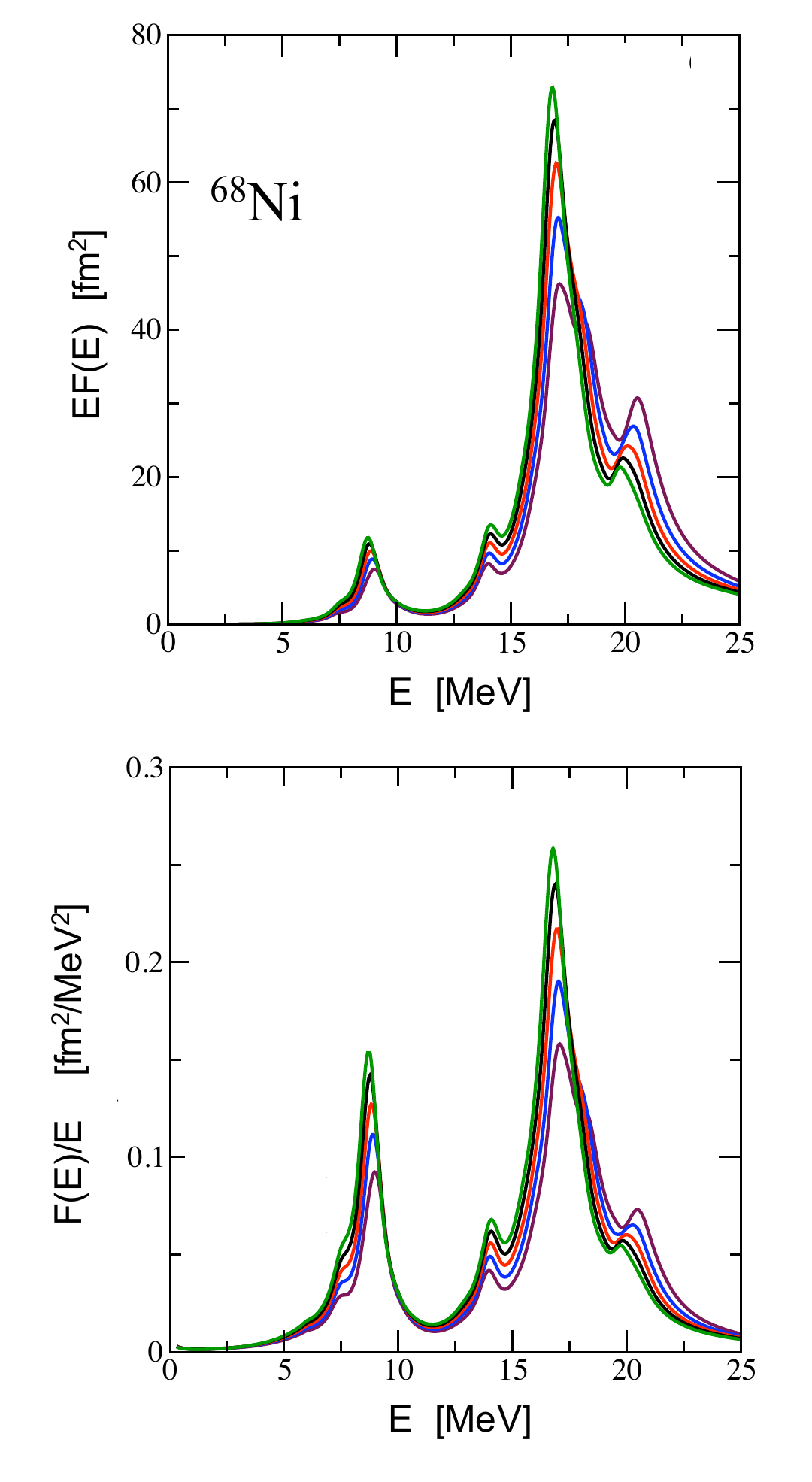}}
\end{center}
\caption{ {\it Top:} The energy weighted dipole response, as defined in Eq. (\ref{responf}) $^{68}$Ni computed with the RMF and a family of FSU interactions \cite{PiekarewiczPRC.83.034319}. {\it Bottom:} The inverse energy weighted dipole response using the same interactions.}
\label{68nialphad}
\end{figure}

As we discussed previously, the nuclear dipole polarizability $\alpha_{D}$ defined as in Eq. (\ref{alphadeq}) is a useful quantity to constrain the symmetry energy \cite{ReinhardPRC.81.051303,TamiPRL.107.062502}.  The dipole polarizability are usually extracted from Coulomb excitation experiments. The advantage of Coulomb excitation is that, for the E1 multipolarity, the virtual photon numbers entering Eq. (\ref{sigmac}) have a $n_{E1}\sim \ln(1/E)$ dependence with the excitation energy. Together with the $1/E$ factor this favors the low energy part of the spectrum thus enhancing the features of the pygmy resonances. Therefore, Coulomb dissociation is nearly proportional to the nuclear dipole polarizability at low energies.  This is becomes evident from Fig. (\ref{68nialphad}), where on the left we show the energy weighted dipole response, as defined in Eq. (\ref{responf}) $^{68}$Ni computed with the RMF and a family of FSU interactions \cite{PiekarewiczPRC.83.034319}. On the right we see that the inverse energy weighted dipole response using the same interactions, entering in the calculation of the dipole polarizability, Eq. (\ref{68nialphad}), is largely enhanced at low energies.

The extract of the dipole polarizability and the exploration of their connection to the slope parameter is reported in, e.g.,  Refs. \cite{TamiPRL.107.062502,adrich:2005:PRL,wieland:2009:PRL,wieland:2018:PRC,SchwengnerPRC78.064314,IwamotoPRL108.262501,RyezayevaPRL89.272502,PoltoratskaPRC85.041304,zilges:2005:PPNP,VOLZ20061,SavranPRL97.172502,savran:2018:PLB,EndresPRC80.034302,EndresPRL105.212503,TonchevPRL104.072501,Hagen:2016,BirkhanPRL.118.252501,rossi:2013:PRL}. As mentioned in relation to Fig. (\ref{alphad2}), a large range of values, of the order of 20-30\%, has been found in experiment studying $\alpha_{D}$. This uncertainty is not small enough to  large to constrain most of the energy functionals stemming from Skyrme and relativistic models. Developments on nuclear reaction theory are also necessary to obtain the desired accuracy in the experimental analyses \cite{BRADY2016553}.

In the Coulomb excitation of pygmy resonances there is a large excitation probability at small impact parameters, with a strong coupling between the  pygmy and giant resonances. This leads to dynamical effects changing the transition probabilities and cross sections for the excitation of the PDR. Such effects have been observed in the case of the excitation of double giant dipole resonances (DGDR) \cite{BERTULANI1988299,Aumann1998,BERTULANI1999139}. The DGDR is a consequence of higher-order effects in relativistic Coulomb excitation due to the large excitation probabilities of giant resonances in heavy ion collisions at small impact parameters and a strong dynamical coupling between the usual giant resonances and the DGDR arises \cite{BertulaniPRC53.334}.   A study of this effect in the excitation of the PDR using the relativistic coupled channels (RCC) equations introduced in Eq. (\ref{BC9235}) was done in Ref. \cite{BRADY2016553}.  The resonances were described by  Lorentzian functions  centered at the energies  $E_{PDR}$ for pygmy dipole resonances and $E_{GDR}$ ($E_{GQR}$) for the isovector (isoscalar) giant dipole (quadrupole)  resonances. 

The excitation of  $^{68}$Ni on $^{197}$Au and $^{208}$Pb targets at  600 and 513 MeV/nucleon were investigated. These reactions have been experimentally investigated in Refs. \cite{wieland:2009:PRL,rossi:2013:PRL}. In the first experiment a pygmy dipole resonance in $^{68}$Ni was identified at $E_{PDR} \simeq 11$ MeV with a width of $\Gamma_{PDR} \simeq 1$ MeV, exhausting about 5\% of the Thomas-Reiche-Kuhn (TRK) energy-weighted sum rule. The identification was done by observing the excitation and decay via gamma emission. Another experiment  found that the PDR centroid energy was located at 9.55 MeV, with a 2.8\% fraction of the TRK sum rule, and a width of 0.5 MeV. The PDR identification was done by observing the neutron decay channel of the PDR. Ref. \cite{BRADY2016553} studied the effects of the coupling between giant resonances and the PDR, on the excitation function d$\sigma$/dE. The centroid energy $E_{PDR}=11$ MeV consistent with Refs. \cite{wieland:2009:PRL,rossi:2013:PRL} was used with a width at half maximum of 2 MeV, adopted from theoretical calculations \cite{VRETENAR2001496,BertulaniPRC.75.024606,BERTULANI2007366,PaarRPP:2007,KREWALD-IJMPE2009,Ponomarev_2014,PapakonstantinouPRC89.034306}  and not from the experimental data \cite{wieland:2009:PRL,rossi:2013:PRL}.  For the (isovector) 1$^-$ giant dipole resonance (GDR) it was assumed $E_{GDR}= 17.2$ MeV and $\Gamma_{GDR}=4.5$ MeV and for the (isoscalar) 2$^+$ giant quadrupole resonance (GQR) the values $E_{GQR}=15.2$ MeV and $\Gamma_{GQR}=4.5$ MeV were used. 

\begin{figure}[t]
\begin{center}
\includegraphics[scale=0.37]{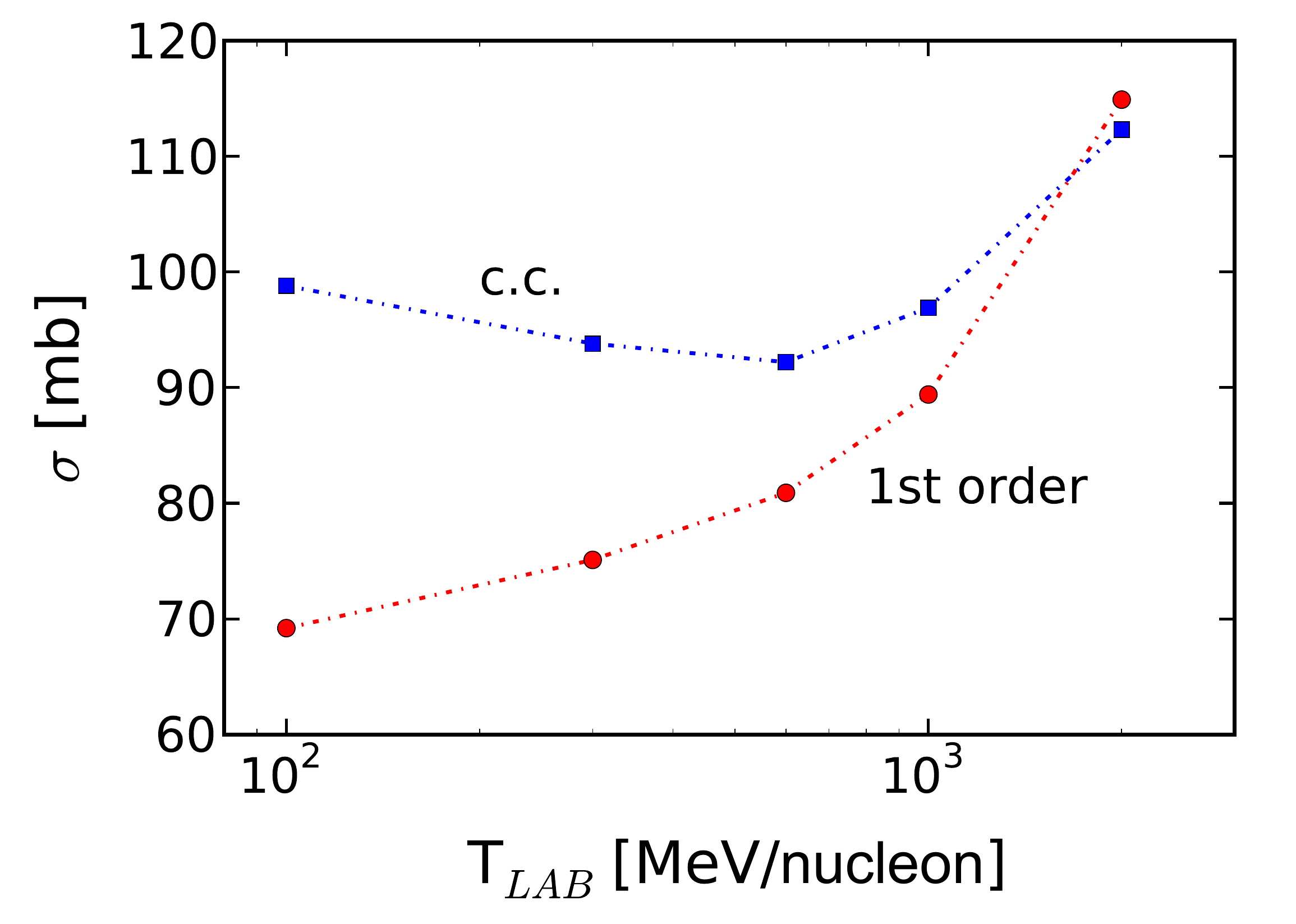}
\caption{Coulomb excitation cross sections of the PDR in  $^{68}$Ni projectiles incident on $^{197}$Au targets as a function of the bombarding energy. The filled circles are calculations using first-order perturbation theory and the filled squares represent coupled-channel calculations.}
\label{CS}
\end{center}
\end{figure} 

In Figure \ref{CS} one sees the Coulomb excitation cross sections of the PDR in  $^{68}$Ni projectiles incident on $^{197}$Au targets as a function of the bombarding energy. The filled circles are calculations using first-order perturbation theory and the filled squares represent coupled-channel calculations.  The deviation is clearly more pronounced at lower energies. Around 600 MeV/nucleon the PDR excitation cross section changes from 80.9 mb obtained with the virtual photon method to 92.2 mb obtained with the coupled-channels method. This implies that the extracted PDR strength from the experimental data would have an appreciable change of 14\% with a a reduction by nearly the same amount in the excitation  strength.  

Coupled-channels calculations were also performed for the reaction  $^{68}$Ni+$^{208}$Pb at 503 MeV/nucleon, corresponding to the experiment of Ref. \cite{rossi:2013:PRL}. The Coulomb excitation cross section of the PDR in $^{58}$Ni were  found to be 57.1 mb to first order, but including couplings to the giant resonances increase the cross section  to 61.4 mb,  a small but still relevant 7.5\% correction. The value of the dipole polarizability $\alpha_D$ extracted from the experiment in Ref. \cite{rossi:2013:PRL} is $3.40$ fm$^3$ while to reproduce the experimental cross section with our dynamical calculations we have $\alpha_D = 3.27$ fm$^3$, a small but non-negligible correction. If a linear relationship between the dipole polarizability and the neutron skin is assumed \cite{PiekarewiczPRC.83.034319},  a reduction of the neutron skin from 0.17 fm, as reported in Ref. \cite{rossi:2013:PRL}, to 0.16 fm is  expected. This correction still lies within the experimental error \cite{rossi:2013:PRL}. 

Therefore, we conclude that due to the large Coulomb excitation probabilities of giant resonances in heavy ion collisions at energies around and above 100 MeV/nucleon,  the excitation of the PDR is also appreciably modified due to the coupling between the $1^-$ and $2^+$ states.  In the future it might be possible to carry out nearly ``ab-initio" calculations based on a microscopic theory, coupled with a proper reaction mechanism. A known alternative, already used in previous studies of multiphonon  resonances \cite{BERTULANI1988299}, is to  use individual states calculated with the RPA or other microscopic together with higher order perturbation theory. Advanced mean-field time-dependent method such as that developed in Ref. \cite{StetcuPRL2015} will also help clarify the reaction dynamics. 

\subsection{Extracting neutron skins from nucleus-nucleus collisions}

\subsubsection{Removing nucleons from nuclei at high energies}

Previously, we have discussed several ways to extract the neutron skin of nuclei and correlate it to the EoS in neutron stars. Here we will describe a method to extract it from measurements of fragmentation reactions. The method is based on the theory described in Ref. \cite{Book:Ber04,MillerARNPS2007}. We will treat all nucleons in the same way irrespective if they are protons or neutrons. Latter we show how this can be easily extended to discern protons from neutrons.

\begin{figure}[t]
\begin{center}
\includegraphics[height=1.8in]{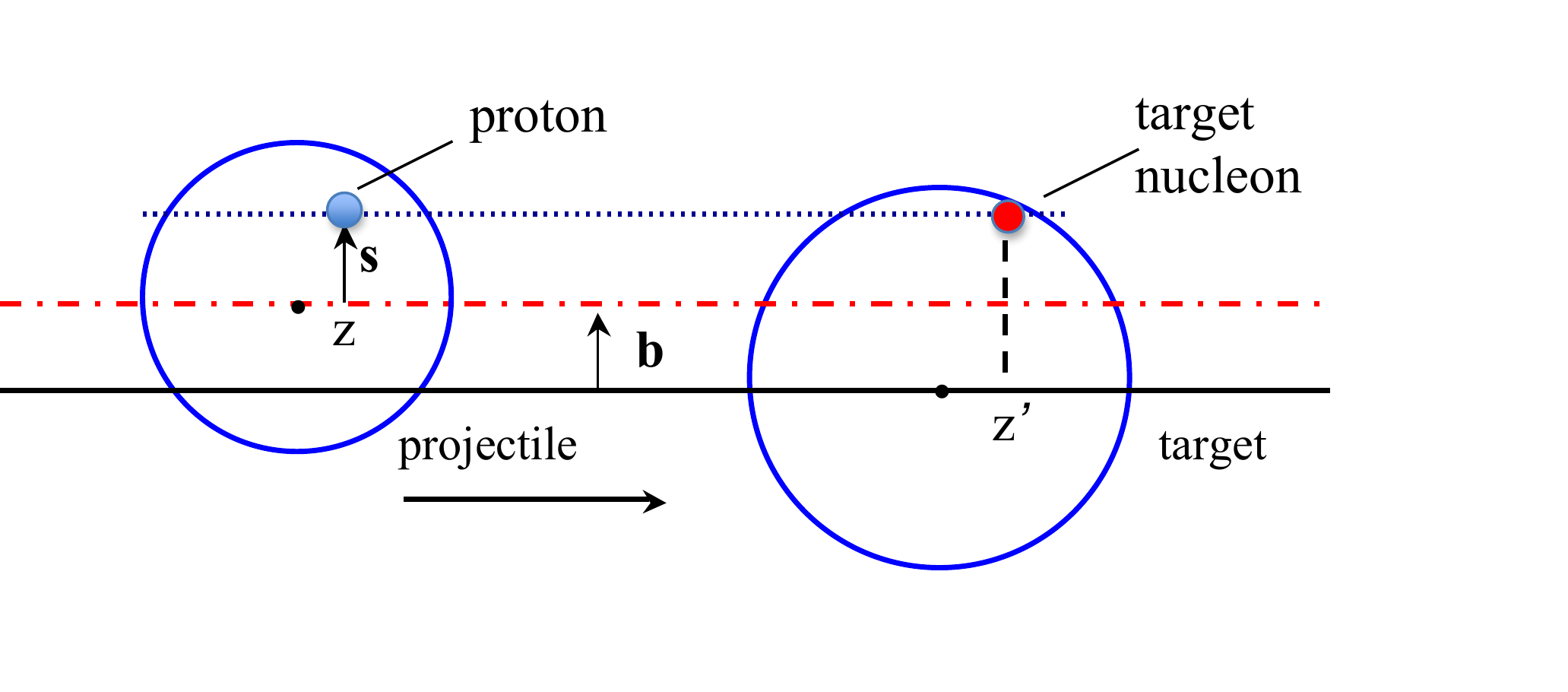}
\caption{Collision geometry for a projectile proton hitting a target nucleon.}
\label{nncolls}
\end{center}
\end{figure}

Let us assume, as shown in Fig. \ref{nncolls}, that a nucleus-nucleus collision occurs at an impact parameter b.  The probability of having a nucleon-nucleon collision within the transverse element area $d\mathbf{b}$  is defined as $t\left(  \mathbf{b}\right)  d\mathbf{b}$, where $t\left(  \mathbf{b}\right) $ is known as the \textit{thickness function}. It is normalized so that
\begin{equation}
\int t\left(  \mathbf{b}\right)  d\mathbf{b}=1. \label{apa1}
\end{equation}
For unpolarized projectiles $t\left(  \mathbf{b}\right)  =t\left( b\right)  $. Frequently, one uses $t\left( \mathbf{b}\right)  \simeq\delta\left(  \mathbf{b}\right) $, which simplifies the calculations considerably.

Since the total transverse area for nucleon-nucleon collisions is given by the nucleon-nucleon cross section $\sigma_{NN}$, the probability of an inelastic nucleon-nucleon collision occurring within this area is  $t\left(  \mathbf{b}\right) \sigma_{NN}$. The probability of finding a nucleon in $d\mathbf{b}_{B}dz_{B}$ is given by $\rho\left(  \mathbf{b}_{B},z_{B}\right) d\mathbf{b}_{B}dz_{B}$, where the nuclear density is normalized to
unity:
\begin{equation}
\int\rho\left(  \mathbf{b}_{B},z_{B}\right)  d\mathbf{b}_{B}dz_{B}\;=1\;.
\label{apa2}
\end{equation}

Using these definitions, it is easy to derive \cite{Book:Ber04} the probability of occurrence of $n$ nucleon-nucleon collisions in a nucleus-nucleus collision at impact parameter \textbf{b}. Denoting by $A(B)$ the number of nucleons in nucleus $A(B)$, this probability becomes
\begin{equation}
P\left(  n,\mathbf{b}\right)  =\left(
\begin{array}
[c]{c}
AB\\
n
\end{array}
\right)  \left[  T\left(  \mathbf{b}\right)  \sigma_{NN}\right]  ^{n}\left[
1-T\left(  \mathbf{b}\right)  \sigma_{NN}\right]  ^{AB-n}\;. \label{apa9}
\end{equation}
The first term is the number of combinations for the  occurrence of $n$ collisions out of $AB$ possible nucleon-nucleon encounters. The second term is the probability of  $n$ collisions, while the last term is the probability of  $AB-n$ misses. 

For nucleus-nucleus collisions, the thickness function $T\left( \mathbf{b}\right)  $, can be related to the corresponding thickness function for nucleon-nucleon collisions by means of
\begin{equation}
T\left(  \mathbf{b}\right)  =\int\rho\left(  \mathbf{b}_{B},z_{B}\right)
d\mathbf{b}_{B}dz_{B}\;\rho\left(  \mathbf{b}_{A},z_{A}\right)  d\mathbf{b}
_{A}dz_{A}\;t\left(  \mathbf{b}-\mathbf{b}_{A}-\mathbf{b}_{B}\right)  \;,
\label{apa5}%
\end{equation}
with $\int T\left(  \mathbf{b}\right)  d\mathbf{b}=1$. 

The total probability, or differential cross section, is given by
\begin{equation}
\frac{d\sigma}{d^{2}b}=\sum_{n=1}^{AB}P\left(  n,\mathbf{b}\right)
=1-\left[  1-T\left(  \mathbf{b}\right)  \sigma_{NN}\right]  ^{AB},
\label{apa10}%
\end{equation}
with the total nucleus-nucleus cross section given by
\begin{equation}
\sigma=\int d^{2}b\;\left\{  1-\left[  1-T\left(  \mathbf{b}
\right)  \sigma_{NN}\right]  ^{AB}\right\} \label{apa11}.
\end{equation}

The quantity
\begin{equation}
\left|S({\bf b})\right|^2 = \left[  1-T\left(  \mathbf{b}
\right)  \sigma_{NN}\right]  ^{AB} \label{apas11}
\end{equation}
is the square of the scattering matrix, for reasons that become clear when one derives this equation using the eikonal approximation.
In the optical limit of the eikonal approximation, described previously, where a nucleon of projectile undergoes only one collision in the target nucleus,
\begin{equation}
\left|S({\bf b})\right|^2 \simeq \exp\left[-ABT\left(  \mathbf{b}
\right)  \sigma_{NN}\right]  \label{apas11b}
\end{equation}
Note that if the densities are normalized to the number of nucleons, instead of Eq. \eqref{apa2}, then the equation above is exactly the same used to obtain the  ``survival probability'' in the ``optical limit of the Glauber theory''. Evidently, Eq. \eqref{apa11} is superior than  using the approximation \eqref{apas11b}. Eq.  \eqref{apas11b} is derived in terms of pure statistical theory, while Eq. \eqref{apas11b} is an approximation when $T\left(  \mathbf{b}\right)  \sigma_{NN}\ll 1$.

The individual thickness functions for the nucleus $A$ can be defined as
\begin{equation}
T_{A}\left(  \mathbf{b}_{A}\right)  =\int\;\rho\left(  \mathbf{b}_{A}
,z_{A}\right)  dz_{A}\;, \label{apa7}
\end{equation}
and similarly for the nucleus $B$. Then we can rewrite
\begin{equation}
T\left(  \mathbf{b}\right)  =\int d\mathbf{b}_{A}d\mathbf{b}_{B}\;T_{A}\left(
\mathbf{b}_{A}\right)  T_{B}\left(  \mathbf{b}_{B}\right)  \;t\left(
\mathbf{b}-\mathbf{b}_{A}-\mathbf{b}_{B}\right)  \;. \label{apa8}
\end{equation}

The probability of $n$ nucleons in $A$ colliding with $m$ nucleons in $B$, using $t\left(  \mathbf{b}\right)  \simeq\delta\left(  \mathbf{b} \right)  $ becomes  \cite{Book:Ber04}
\begin{align}
P\left(  n,m,\mathbf{b}_{A}\mathbf{,b}_{B}\right)   &  =\left(
\begin{array}
[c]{c}%
A\\
m
\end{array}
\right)  \left(
\begin{array}
[c]{c}%
B\\
n
\end{array}
\right)  \left[  T_{B}\left(  \mathbf{b}_{B}\right)  \sigma_{NN}\right]
^{n}\left[  1-T_{B}\left(  \mathbf{b}_{B}\right)  \sigma_{NN}\right]
^{B-n}\nonumber\\
&  \times\left[  T_{A}\left(  \left|  \mathbf{b}-\mathbf{b}_{B}\right|
\right)  \sigma_{NN}\right]  ^{m}\left[  1-T_{A}\left(  \left|  \mathbf{b}%
-\mathbf{b}_{B}\right|  \right)  \sigma_{NN}\right]  ^{A-m}\;.\nonumber\\
&  \label{apa18}%
\end{align}

The ``abrasion-ablation model'' for nuclear fragmentation \cite{Bowman1973}  is based on this equation. This model can also be extended to account for the isospin dependence of the nucleon-nucleon collisions, as we show below. One can interpret $m$ as the number of holes created in the nucleon orbitals in the target.  {\it Ablation} refers to the decay phase of the nuclei after the ``holes" are created in the nucleon orbitals. The equations above can also be deduced quantum-mechanically using the eikonal approximation  \cite{Book:Ber04}. The derivation presented here is much simpler and only uses classical probability concepts.  This model has been shown to describe rather well the fragment yields in relativistic heavy ion collisions (see, for example, \cite{HufnerPRC12.1888,CarlsonPRC46.R30}).

\subsubsection{Isospin dependence}

We assume again that $\rho_p^{P}$ and $\rho_n^{T}$ are the projectile and target proton and neutron densities, normalized so that
\begin{equation}
\int d^3r \rho_p^T(r) = 1, \ \ \ \ \ \ \ {\rm and} \ \ \ \ \int d^3r \rho_n^T(r) = 1 ,
\end{equation}
while $\sigma_{pp}$ and $\sigma_{pn}$ are the total (minus Coulomb) proton-proton and proton-neutron scattering cross sections, respectively.

The cross section for the production of a fragment $(Z,N)$ from a projectile $(Z_P,N_P)$  due to nucleon-nucleon collisions is given by 
\begin{eqnarray}
&\sigma(N_{P},Z_{P};N,Z)=
\left(
\begin{array}{c}Z_P \\ Z
\end{array}
\right)
\left(
\begin{array}{c}N_P \\ N
\end{array}
\right)
\int d^2 b \left[ 1-P_p(b)\right]^{Z_P-Z} \nonumber \\
&\times P_p^Z(b) \left[ 1-P_n(b)\right]^{N_P-N}P_n^N(b), \label{sigma}
\end{eqnarray} 
where $b$ is the collision impact parameter. The binomial coefficients in front of the integral account for all  combinations selecting $Z$ protons out of the $Z_P$ projectile protons, and similarly for the neutrons \cite{Book:Ber04,MillerARNPS2007}. The probabilities for single nucleon survival are denoted by $P_p$ for protons and $P_n$ for neutrons. The probability that a projectile proton does not collide with target nucleons is given by  \cite{Book:Ber04,MillerARNPS2007}
\begin{eqnarray}
&P_p(b)=\int dzd^2s \rho_p^P({\bf s},z) \exp\left[ -\sigma_{pp} Z_T\int d^2s \rho_p^T({\bf b-s},z) \right. \nonumber \\
&-\left. \sigma_{pn} N_T\int d^2s \rho_n^T({\bf b-s},z) \right],  \label{ppb}
\end{eqnarray} 
where $\sigma_{pp}$ and $\sigma_{np}$ are the proton-proton (Coulomb removed) and proton-neutron total cross sections. A fit of experimental data in the energy range of $10$ to $5000$\,MeV is often used, as in Eqs. (\ref{signn1}) and (\ref{signn2}) (see top panel of Fig. \ref{6he8he}).

The primary yields in this model depend on the incident energy only through the nucleon-nucleon cross sections in the absorption factors. At high energies $E/A > 100$ MeV/nucleon), these cross sections have nearly the same value. One expects that differences in proton and neutron scattering are smallest at such energies. Such differences will be more evident at lower energies where the proton-neutron cross section is about three times the  proton-proton and neutron-neutron cross sections.  

As shown in Refs. \cite{Bertulani1884,BertulaniJPG2001,ChenPRC87.054616}, the  Pauli blocking projection yields an average nucleon-nucleon cross section for two Fermi gases with relative momenta ${\bf k}_0$ given by  
\begin{align}
\sigma_{NN}(k,\rho_1,\rho_2)   =\int{\frac{d^{3}k_{1}d^{3}k_{2}}{(4\pi
k_{F1}^{3}/3)(4\pi k_{F2}^{3}/3)}}
{\frac{2q}{k_0}}\ \sigma_{NN}^{free}(q)\
{\frac{\Omega_{Pauli}}{4\pi
}}\ , \label{ave}%
\end{align}
where  $2\mathbf{q}%
=\mathbf{k}_{1}-\mathbf{k}_{2}-\mathbf{k}_0$. The momentum of a single nucleon is denoted by ${\bf k}_i$.  
Pauli-blocking enters through the restriction that the magnitude of the final nucleon momenta,
$|\mathbf{k^{\prime}}_{1}|$ and $|\mathbf{k^{\prime}}_{2}|$, lie outside the
Fermi spheres, with radii, $k_{F1}$ and $k_{F2}$. This leads to a limited fraction of the solid angle into which the nucleons can scatter, $\Omega_{Pauli}$. 

The numerical calculations can be simplified if we assume that  the free nucleon-nucleon cross section entering Eq. \eqref{ave} is isotropic \cite{Bertulani1884,BertulaniJPG2001,ChenPRC87.054616}.
 This is a rough approximation because the anisotropy of the free NN cross section is markedly manifest at large energies.
 In the isotropic case, a formula which fits the numerical integration in Eq. \eqref{ave} is \cite{Bertulani1884,BertulaniJPG2001,ChenPRC87.054616}
\begin{eqnarray}
\sigma_{NN}(E,\rho_1,\rho_2) &=&\sigma_{NN}^{free}(E){1 \over 1+1.892\left({\displaystyle{|\rho_1-\rho_2|\over \tilde{\rho}\rho_0}}\right)^{2.75}}\nonumber \\
&\times& 
\left\{
\begin{array}
[c]{c}%
\displaystyle{1-{37.02 \tilde{\rho}^{2/3}\over E}}, \ \ \   {\rm if} \ \ E>46.27 \tilde{\rho}^{2/3}\\ \, \\
\displaystyle{{E\over 231.38\tilde{\rho}^{2/3}}},\ \ \ \ \  {\rm if} \ \ E\le 46.27 \tilde{\rho}^{2/3}\end{array}
\right.
\label{VM1}
\end{eqnarray}
where $E$ is the laboratory energy in MeV, $\tilde{\rho}=(\rho_1+\rho_2)/\rho_0$,  with $\rho_0=0.17$ fm$^{-3}$, and $\rho_i(r)$ is the local density at position $r$ within nucleus $i$.

\subsubsection{Ablation model for neutron removal}
The differential yield of fragments in which $N$ neutrons are removed  from the projectile is,
\begin{eqnarray}
\sigma_{-N}
= \sum_{N^\prime>N} \omega(N_P,Z_P,N^\prime,E^*; N)\sigma (N_{P},Z_{P};N',Z_{P})
&  \label{abrasion}%
\end{eqnarray}
where $\omega(N_P,Z_P,N^\prime,E^*; N)$ is the probability that the nucleus $(N_P,Z_P)$ with $N^\prime$ neutrons removed and  excitation energy $E^*$, ends up with $N$ neutrons after evaporation.

The removal of $N^\prime$ neutrons in the first stage leads to  ``primary yields''. After evaporation the nucleus is left with $N_x$ neutrons, called by ``secondary yields''. It is the secondary yields which are usually measured in experiments. 

If evaporation is neglected (primary yields), the total cross section for neutron removal with all protons remaining is given by
\begin{equation}
\sigma_{\Delta N}=\sigma_{\rm all \; n\;  decay\; channels}^{{Z_P \ {\rm survives}}}
= \int d^{2}b   [P_p({\bf b})]^{Z_P}\left\{1- \left[P_n({\bf b})\right]^{N_{P}}\right\}\;.  \label{abrasionc}
\end{equation}
This means that the probability that $Z_P$ protons survive while all possible neutron removal occurs is equal to the probability that all protons survive (irrespective to what happens to any neutron)  minus the probability that all protons and neutrons survive, simultaneously. 
 
Following the same procedure as above,  we can obtain a general formula for the  the total reaction cross section \cite{Book:Ber04}
\begin{equation}
\sigma_{\rm R}
= \int d^{2}b \  \left\{1- \left[P_p({\bf b})\right]^{Z_{P}}\left[P_n({\bf b})\right]^{N_{P}}\right\}
\;.
 \label{abrasionct}
\end{equation}
The reaction cross section is then due to the probability that anything happens, i.e., the unity, minus the probability that all protons and neutrons survive.

For spherical nuclei, the probabilities $P_n({\bf b})$ and $P_p({\bf b})$ do not depend on the direction of the impact parameter. For deformed nuclei, the calculation is much more complicated, as the orientation of the nuclei have to be taken into account and properly averaged.

It is clear from Eq. \eqref{abrasion} that the calculation is not as simple as shown in the previous sub-section. Also note that $\sum_{N^\prime=1}^{N} \omega(N_P,Z_P,N^\prime,E^*; N) \neq 1$, because the nucleus can end up with $N^{\prime}>N$ and also proton evaporation can occur although with a much smaller probability due to the Coulomb barrier.

However, if after one, or multiple, neutron-knockout all decay fragments of the same element are detected, then evaporation is not relevant in Eq. \eqref{abrasion}, except for a small uncertainty due to proton, $\gamma$, $\alpha$, emission probabilities. Then one can use Eq. \eqref{abrasionc}, which is purely geometric and only depends on the densities.

\begin{figure}[t]
\begin{center}
\includegraphics[height=1.in]{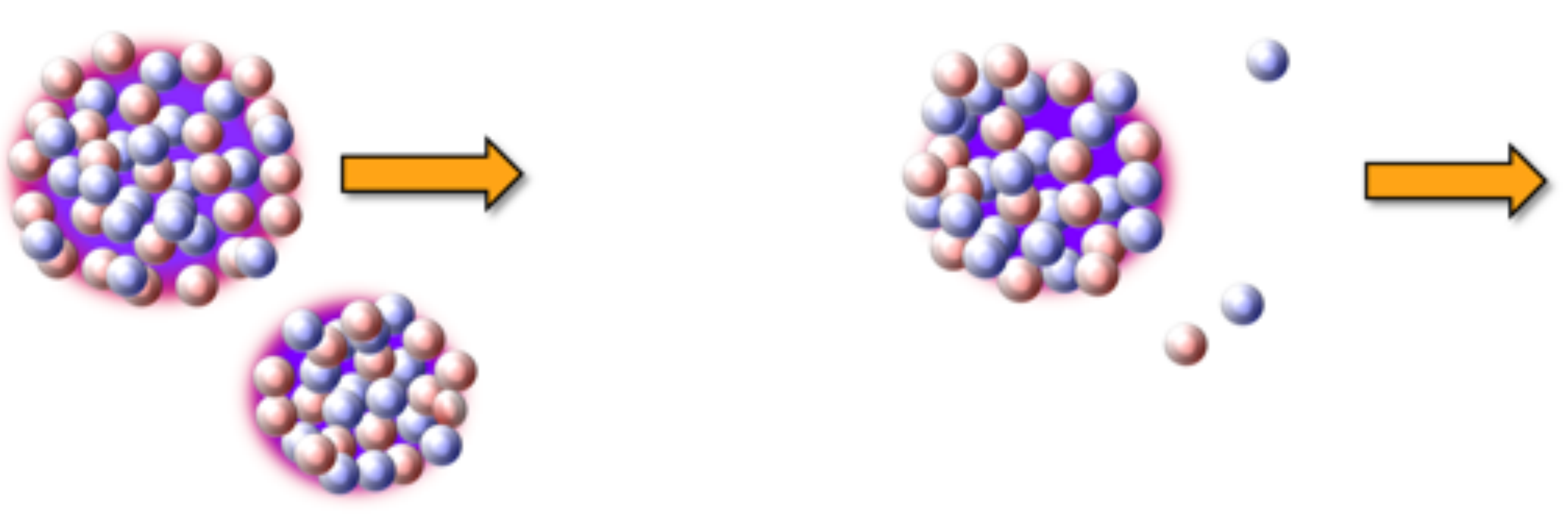}
\caption{Schematic view of fragment production in nucleus-nucleus collisions at high energies.}
\label{fragprod}
\end{center}
\end{figure}

\subsubsection{Total neutron removal cross sections as a probe of the neutron skin}
In Ref. \cite{AumannPRL119.262501} the total neutron removal cross sections has been proposed as a probe of the neutron skin in nuclei, as shown schematically in Fig. \ref{fragprod}. The exploratory study chose the neutron-rich part of the tin isotopic chain and specifically the reactions Sn+$^{12}$C. The density of $^{12}$C was obtained from a model-independent elastic scattering analysis up to large momentum transfers, $q^2$,  using the Fourier Bessel expansion \cite{OffermannPRC44.1096} and extrapolated by using a Whittaker function at very large radii. The $rms$ radius  of $^{12}$C  was taken as the published best value of 2.478(9)\ fm \cite{OffermannPRC44.1096}. We  proton and neutron densities was assumed to be the same.

\begin{figure}[t]
\includegraphics[width=0.49\columnwidth]{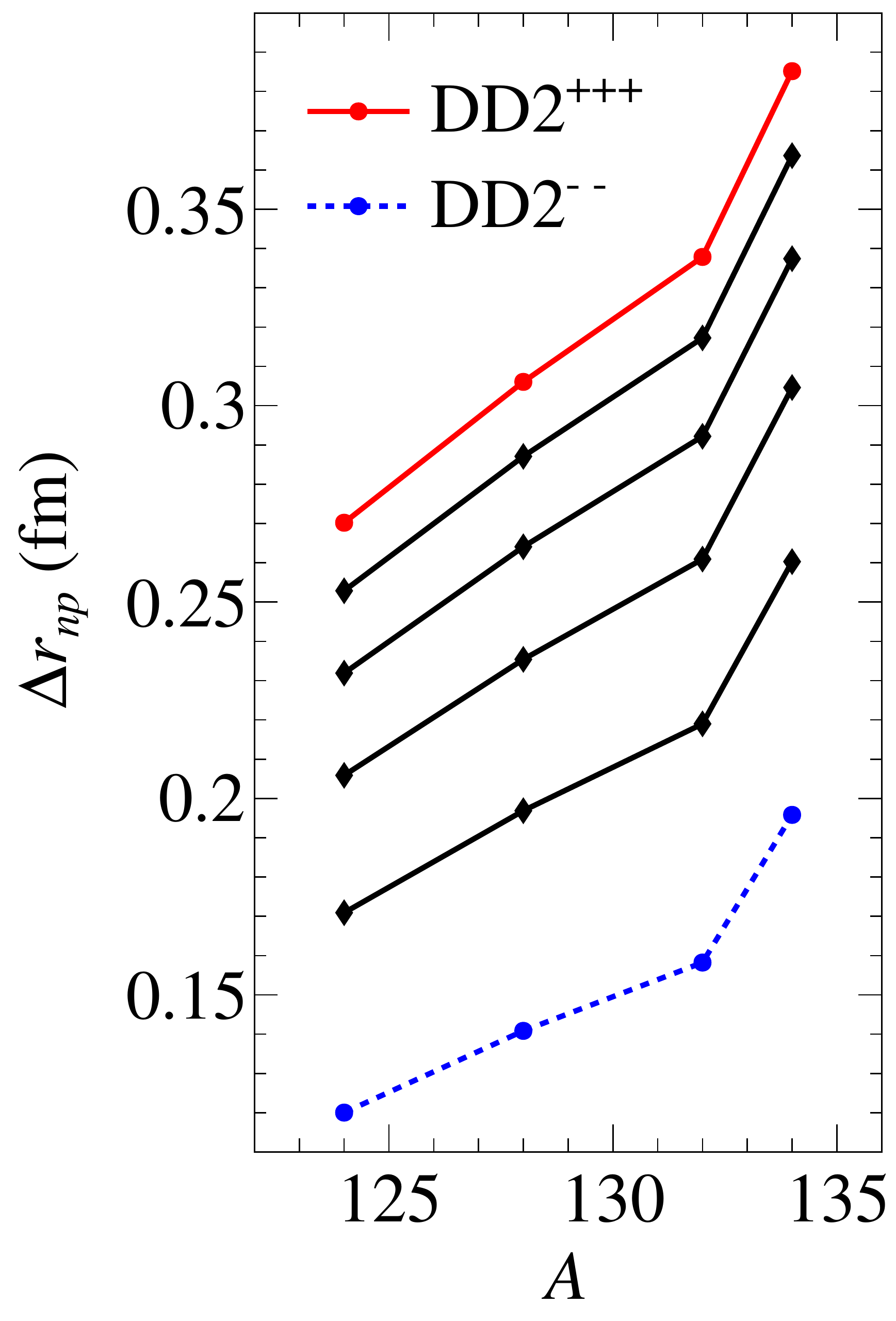}
\includegraphics[width=0.49\columnwidth]{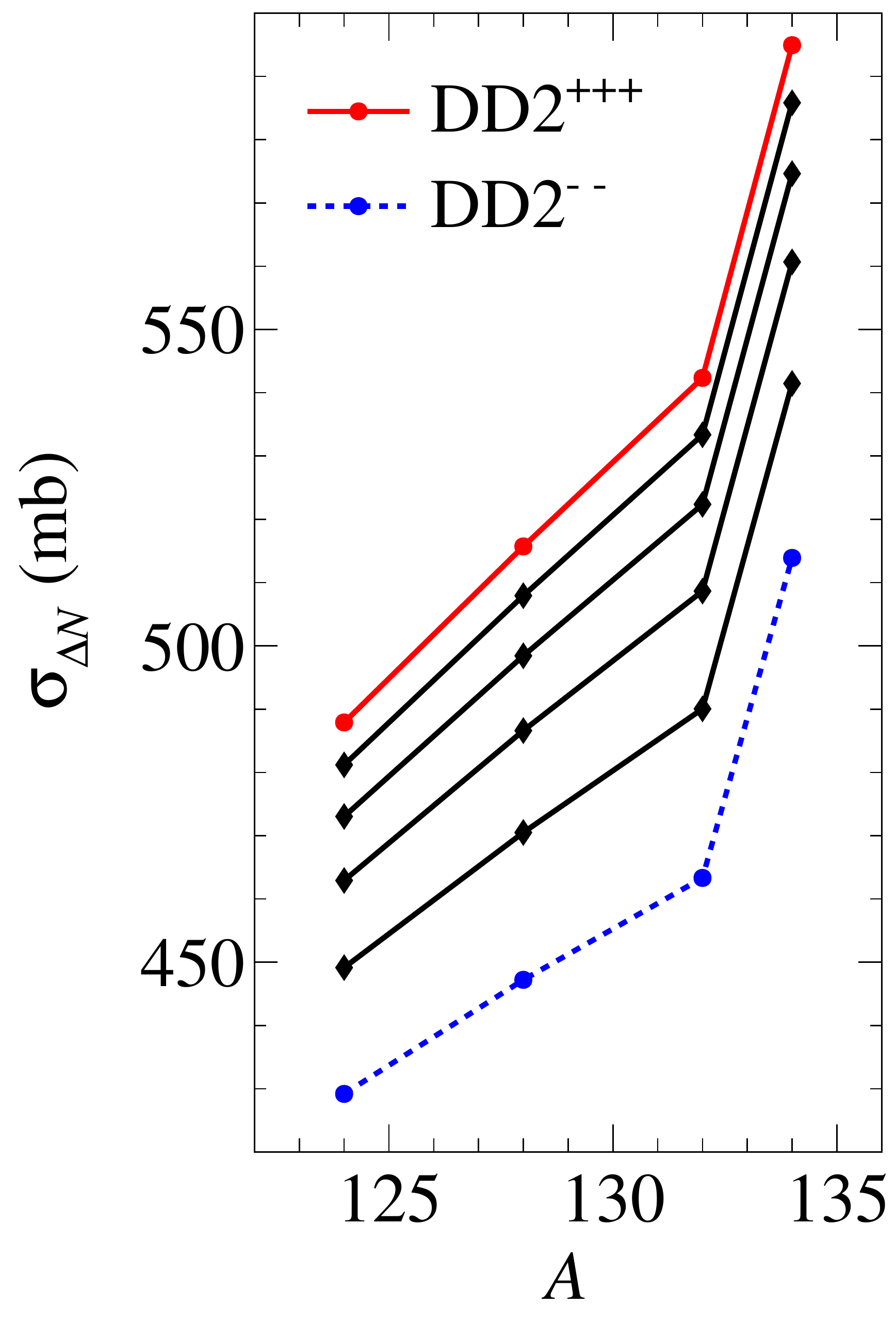}
\caption{Neutron-skin thickness $\Delta r_{np}$ (left) and the related neutron-removal cross sections $\sigma_{\Delta N}$ (right) for Sn isotopes  predicted by RMF models based on the use of the DD2 interaction \cite{TypelPRC81.015803,TypelPRC89.064321}. The slope parameter $L$  varies between 25 MeV (for the DD2$^{--}$in interaction ) and 100 MeV (for DD2$^{+++}$) \cite{TypelPRC89.064321}. }
\label{DD2dr}
\end{figure}

To estimate the sensitivity of $\sigma_{\Delta N}$ with variations of $\Delta r_{np}$ and $L$, cross sections were calculated using theoretical density distributions obtained with RMF models, as the DD2 interaction  developed in Ref. \cite{TypelPRC81.015803}   The slope parameter $L$ was systematically varied to optimize the isovector parameters reproducing nuclear properties such as masses and radii \cite{TypelPRC89.064321}. The same procedure was used for the DD interaction \cite{TypelPRC71.064301}. The left panel in Fig. \ref{DD2dr} displays the predicted neutron-skin thicknesses for tin isotopes. In these calculations, different interactions were used ranging from $L$ values of 25\,MeV (DD2$^{--}$) to 100\,MeV (DD2$^{+++}$) which accordingly also predict  different values of $\Delta r_{np}$ ranging from 0.15 to 0.34 fm in $^{132}$Sn. This results in a corresponding variation of $\sigma_{R}$ from about 2550 to 2610\,mb, i.e.,  a 2.5\% change. The most sensitive quantity to $\Delta r_{np}$ is $\sigma_{\Delta N}$ is shown in the right frame of Fig.\ref{DD2dr}. A variation within 460 and 540\,mb is clearly visible for $^{132}$Sn, i.e., a cross section change of almost 20\%. One concludes that  $\sigma_{\Delta N}$ has a larger potential to constrain $L$ and is less sensitive to reaction theory uncertainties.

\begin{figure}[t]
\includegraphics[width=0.9\columnwidth]{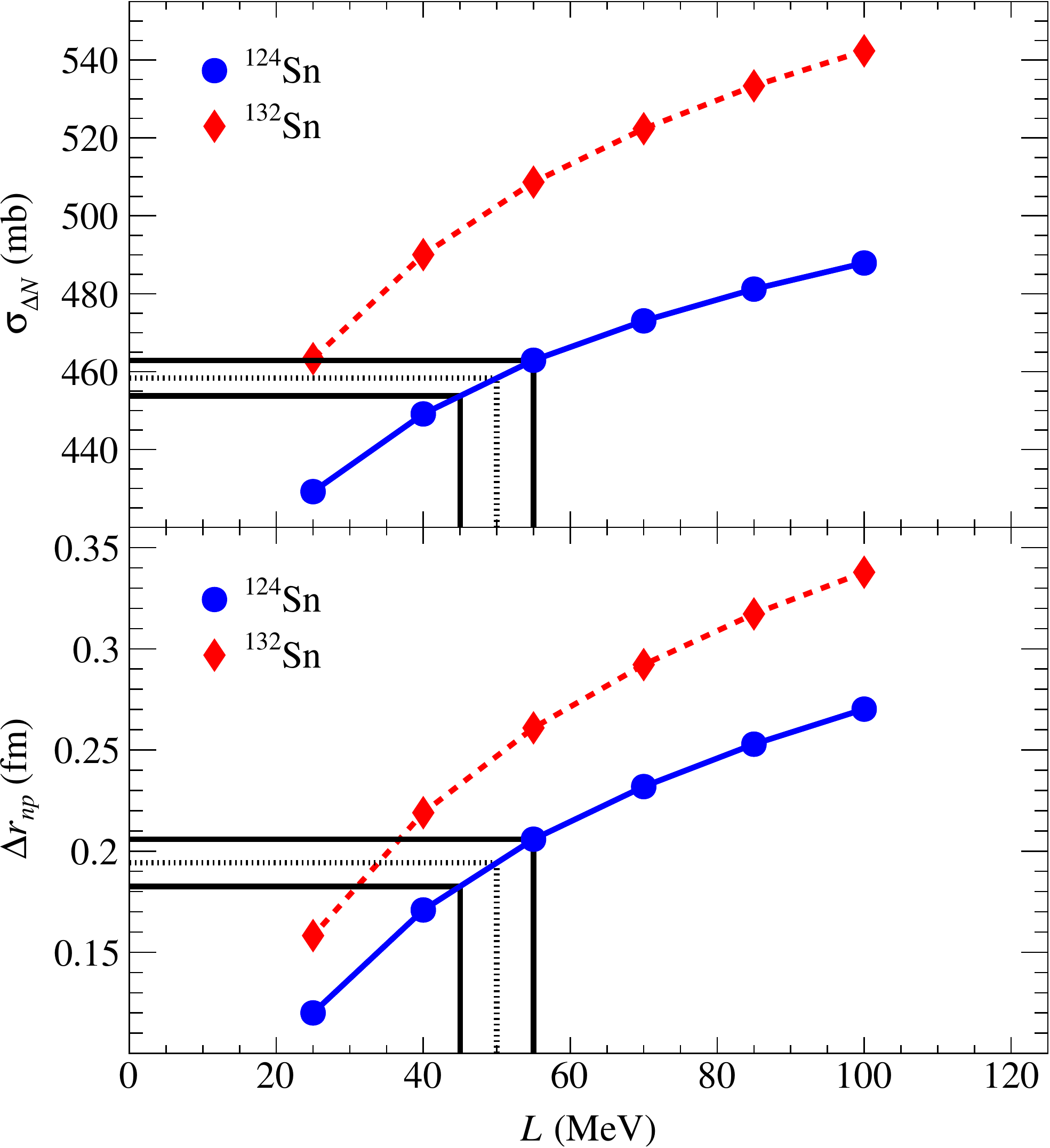}
\caption{Relation between $\sigma_{\Delta N}$ (top) and $\Delta r_{np}$ (bottom) with the slope parameter $L$ based on RMF calculations for $^{124}$Sn and $^{132}$Sn. The lines indicate the sensitivity of the observables with $L$ for a variation of  $L$ within 10 MeV.}
\label{XNandDR}
\end{figure}

Fig. \ref{XNandDR} shows the correlation of the value of $L$ obtained using the DD2 interaction and $\Delta r_{np}$ for $^{124}$Sn and $^{132}$Sn. A variation of $L$ by $\pm 5$ MeV changes the calculated neutron skin in $^{124}$Sn by about $\pm 0.01$ fm. The same variation in $L$ yields a modification of $\sigma_{\Delta N}$ by about $\pm 5$ mb, i.e., of only $\pm 1$\%. Therefore,  a determination of $\sigma_{\Delta N}$ within a 1\% accuracy in a combination of  experiment and theory can be reached to constrain $L$ via a  comparison with DFT. The scatter of results for various relativistic and non-relativistic models predicting a given $L$ on $\sigma_{\Delta N}$ is expected to be similar to the case of the scatter of $\Delta r_{np}$ analyzed in Ref. \cite{RocaPRL106.252501}, namely, about 10 MeV in the determination of $L$.  The dependence of the cross section on the slope parameter $L$ is steeper in the case of the more neutron-rich nucleus $^{132}$Sn, thus providing an even higher sensitivity. 

As we stressed before, nuclear fragmentation in high-energy collisions  is often modeled via two completely disconnected assumptions: (a) the production of  primary fragments via multi-nucleon knockout in binary nucleon-nucleon collisions (as we described previously), followed by (b) the production of secondary fragments due to nucleon evaporation from the energy deposited in the primary fragments. The second step, using the Hauser-Feshbach theory of  compound-nucleus decay, is highly model dependent. The method discussed in Ref. \cite{AumannPRL119.262501} does not need to consider the nuclear evaporation step, because the total neutron and charge removal cross sections include the completeness of the sum over all possible decay channels. Also, proton or charged-particle evaporation is negligible in the cases of $^{124}$Sn and heavier tin isotopes considered in Ref. \cite{AumannPRL119.262501}. For a typical case, one obtains $\sigma_{\Delta N}=485.6$~mb for the production of primary fragments in the reaction of  580\,MeV/nucleon $^{124}$Sn incident on $^{12}$C. The same cross section calculated after the CN evaporation stage, with standard parameters used in the Hauser-Feshbach formalism, yields  $\sigma_{\Delta N}=483.4$~mb. That is, less than 0.5\% of the neutron-removal cross section is moved to the charge-changing cross section after the primary production stage. For more neutron-rich tin isotopes, this contribution will be even smaller. Modifications of the input parameters in Hauser-Feshbach calculations are by no means enough to increase this effect appreciably. 

In addition to fragmentation processes induced by nucleon-nucleon collisions, the projectile can emit a nucleon after an inelastic excitation to collective states in the continuum of the projectile, such as giant resonances. For heavy neutron-rich nuclei, this process adds almost exclusively to the neutron-removal channel and to the total interaction cross section. The later is defined as the sum of the two processes, $\sigma_{I}=\sigma_{R}+\sigma_{inel}$. As we discuss later  $\sigma_{inel}$ contains a nuclear and an electromagnetic contribution as well as their interference. The contribution of $\sigma_{inel}$ will be shown to b e about 1\% for $^{12}$C + $^{12}$C at  500-1000 MeV/nucleon, but it can attain 100\,mb for $^{132}$Sn+$^{12}$C. This corresponds to 4\% or 20\% of $\sigma_{I}$ or $\sigma_{\Delta N}$, respectively.  Since it is the neutron-removal cross section which provides most sensitivity to the neutron skin, this additional channel has to be known within less than 5\% to achieve the required constraint on $L$. This seems impossible to achieve with reaction theory. But it is possible using state-of-the-art kinematically complete experimental measurements to separate the nuclear excitation contribution and to determine its cross section. In fact, the angular distributions of the neutrons are very different for the two processes: evaporated neutrons (typically with energies of 2\,MeV in the projectile rest frame) are kinematically boosted to the forward direction at high bombarding energies and can be well detected with nearly the beam velocity and a typical angular distribution covering the angular range of $0$ to $5^{}\circ$, while neutrons stemming from binary nucleon-nucleon collisions display a broader angular distribution scattered within 0 and $90^{}\circ$ and a maximum around $45^{}\circ$. The expected overlap region between the two processes is thus negligible. 

Base on the discussion above the primary process of binary nucleon-nucleon collisions remains the only significant hurdle to relate $\sigma_{\Delta N}$  with $\Delta r_{np}$ or $L$. This step was investigated in Ref. \cite{AumannPRL119.262501} for  the symmetric system $^{12}$C + $^{12}$C, compared to the available experimental information on $\sigma_R$, using the eikonal scattering theory as described previously. In this model, the known free nucleon-nucleon cross sections and the nuclear densities are only input.  The  total reaction cross section as a function of the laboratory energy are displayed by black triangles in the lower frame of Fig.\,\ref{xsvse}. The calculated cross sections are larger than the experimental data for energies above 200 MeV/nucleon. In-medium effects are expected to be responsible for the deviations at high beam energies. A large fraction of this effect can be attributed to Pauli blocking, calculated as in Ref.\,\cite{BertulaniConti10}, which when included yield  the red diamonds. However, the high-energy data point at about 950\,MeV/nucleon is still overestimated, although by only  2\%. Below 400 MeV/nucleon, the data start to deviate strongly from the calculation, probably due to the failure of the method below these energies. In Ref.\,\cite{TakechiPRC.79.061601}, the effect of Fermi motion was shown to be important at low energies and to increase the cross sections. In the most relevant energy region ($400-1200$ MeV/nucleon) there are only three data points, and no other data in the important region of 400 to 800 MeV/nucleon, where the cross section shows an increase with bombarding energy. Deviations from the eikonal approximation, such as in-medium and higher-order effects certainly depend on the beam energy, and high-precision data covering this energy region with less than $1\%$ accuracy are thus of utmost urgency for a more stringent test of reaction cross sections and further developments in reaction theory. They will also be used for a better the quantification of the uncertainties in neutron-changing ceros sections. 

\begin{figure}[t]
\includegraphics[width=0.99\columnwidth]{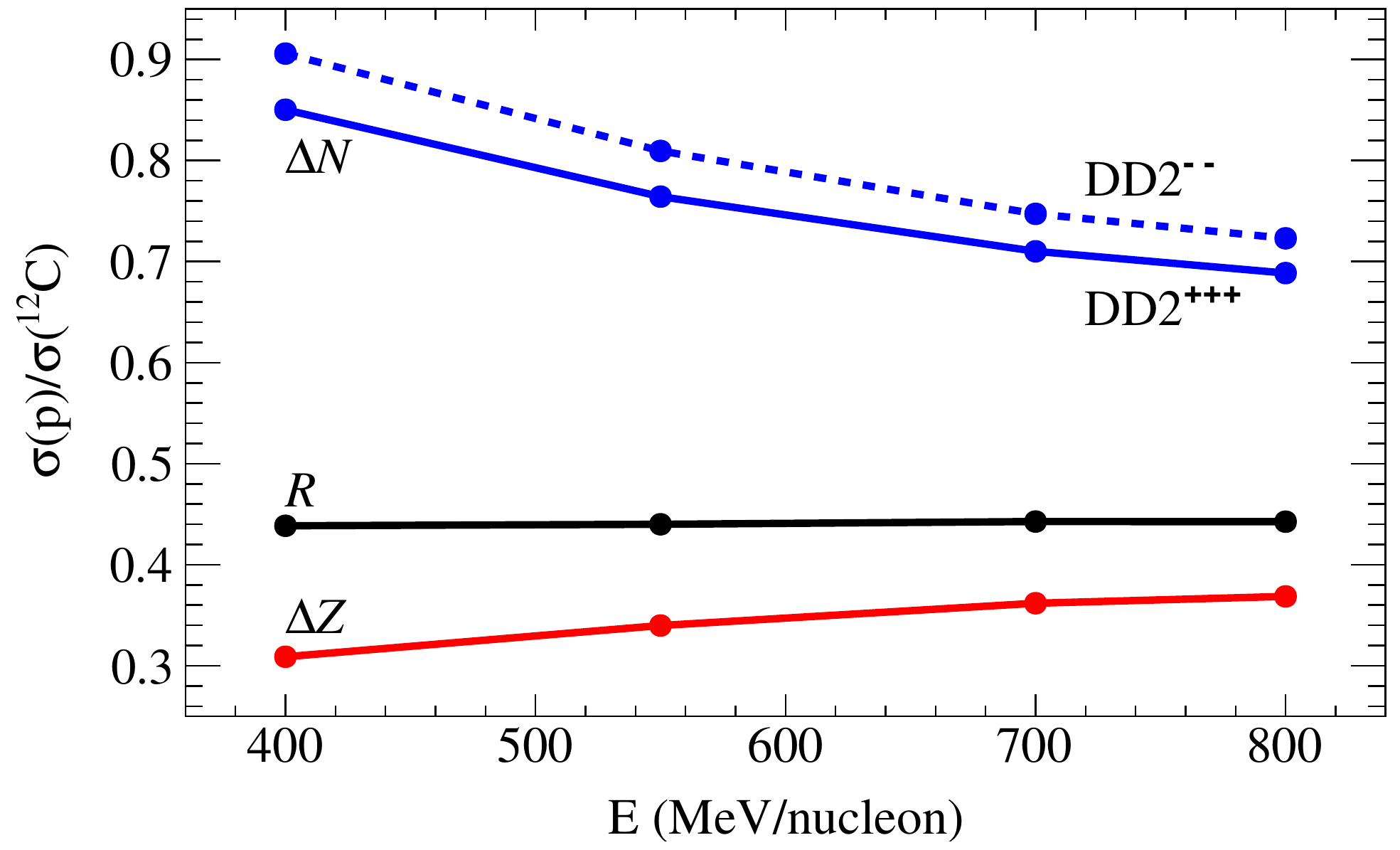}
\caption{Bombarding energy dependence of the ratios of $\sigma_R$, $\sigma_{\Delta Z}$ and $\sigma_{\Delta N}$ cross sections for $^{134}$Sn projectiles incident on proton and $^{12}$C targets.}
\label{XSRatio}
\end{figure}

Further sensitivity on the neutron skin can be obtained by changing the target. The $np$ and $pp$ cross sections have a very different energy dependence as seen in Fig.\,\ref{xsvse}. One thus expects that a corresponding change in the ratio of neutron-removal to charge-changing cross sections exists as a function of bombarding energy. This should be most pronounced in the case of a proton target as the proton probes the neutron skin exclusively via $pn$ collisions, whereas charge-changing reactions  are exclusively related to $pp$ collisions. Other additional subtle effects such as the proton passing across the nucleus without knocking out another nucleon adds a difference to the use of $^{12}$C targets that probes only the nuclear surface. This  becomes evident in Fig.\,\ref{XSRatio} where we show the ratios of $\sigma_R$, $\sigma_{\Delta Z}$ and $\sigma_{\Delta N}$ for $^{134}$Sn bombarding proton targets compared with those on $^{12}$C targets. While we do not see an energy dependence for  $\sigma_R({\rm p})/\sigma_R(^{12}{\rm C})$ ratio, the charge-changing $\sigma_{\Delta Z}({\rm p})/\sigma_{\Delta Z}(^{12}{\rm C})$ and $\sigma_{\Delta N}({\rm proton})/\sigma_{\Delta N}(^{12}{\rm C})$ ratios clearly show an energy dependence. This energy dependence  and the fact that the target ratio is significantly larger for $\sigma_{\Delta N}$ is due to the strong energy dependence of the $pp$ cross section (as shown in Fig.\,\ref{xsvse}) leading  to a substantial proton survival probability when proton targets are used around 400\,MeV/nucleon and therefore yielding a larger $\sigma_{\Delta N}$. This effect becomes visibly smaller at  800 MeV/nucleon and above. Therefore, the energy dependence of the  $\sigma_{\Delta N}$ target ratio provides  an additional sensitivity test of the reaction theory, if experimentally obtained with accuracy. Both ratios  for $\sigma_R$ and $\sigma_{\Delta Z}$ display negligible dependence on the neutron skin whereas the ratio for $\sigma_{\Delta N}$ displays a much stronger dependence on $\Delta r_{np}$ as evidenced by using the DD2$^{+++}$ and DD2$^{--}$ RMF interactions. Since the $rms$ radii of the charge distribution in the nuclei are known, the charge-changing cross sections obtained with proton and carbon targets by varying the bombarding energy will serve as a crucial test on the accuracy of the calculated cross sections. 

\subsubsection{Coulomb excitation followed by neutron emission}

\begin{figure}[t]
\centerline{
\includegraphics[width=1.\columnwidth]{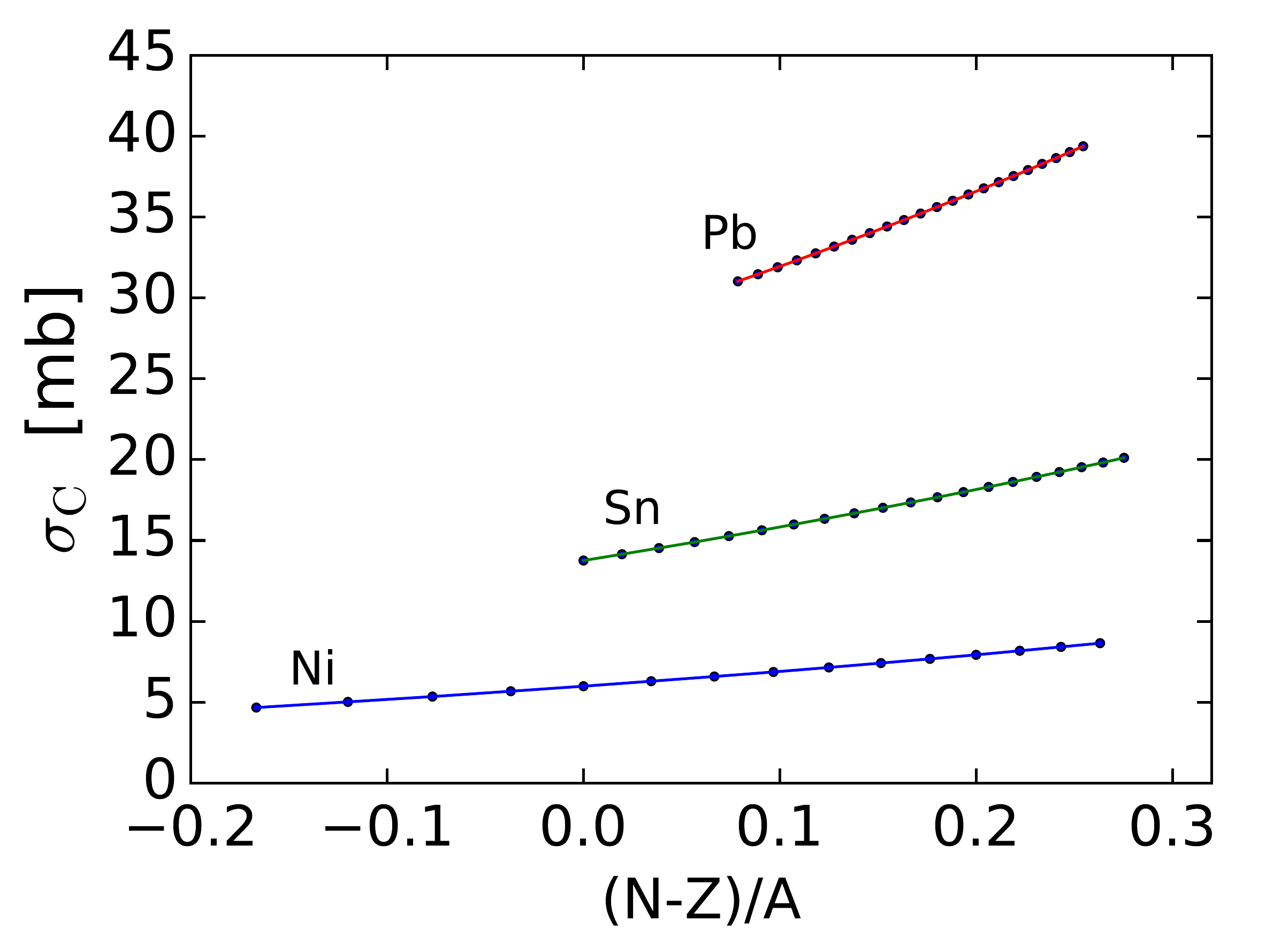}}
\caption{Cross sections in milibarns for the Coulomb excitation of the IVGDR in nickel, tin and lead projectiles incident on carbon targets at 1 GeV/nucleon, as a function of the asymmetry coefficient $\delta=(N-Z)/A$ of the projectile.}
\label{coulexni}
\end{figure}

We now extend the discussion of ``small'' corrections to the cross sections  $\sigma_{\Delta N}$ and $\sigma_{I}$. We first discuss  Coulomb excitation of giant resonances followed by neutron emission. In  Figure \ref{coulexni} we plot the results for the excitation cross sections of the IVGDR in nickel, tin and lead projectiles incident on carbon targets at 1 GeV/nucleon. The dependence on the asymmetry coefficient $\delta=(N-Z)/A$ of the projectile is shown. The Coulomb cross section shows little dependence on the neutron skin. The neutron skin only enters at small impact parameters entering the imaginary phase $\chi_{OL}(b)$  in Eq. \eqref{g}. The cross section depends on the asymmetry coefficient $\delta$, mainly through the isotopic dependence of the photo-nuclear cross section as shown in Eq. \eqref{gdr}. The mass dependence of the resonance centroid energy also has an influence on the virtual photon numbers $n_{E1}$ because of their fast decrease with the excitation energy.

As displayed in Fig. \ref{coulexni}, the Coulomb cross sections increase linearly with $(N-Z)/A$. They are very small for nickel and tin isotopes and negligible compared to the neutron changing and interaction cross sections. But the cross sections for lead projectiles are not negligible for carbon targets. The Coulomb cross sections are proportional to the square of the target charge yielding for proton targets cross sections that are smaller than those for carbon targets by a factor $30-40$.  By means of a  comparison of experimental results obtained with carbon and proton targets one can easily separate the contributions of Coulomb cross sections to the fragmentation \cite{AumannPRC.51.416}. There is a large variation of the Coulomb cross sections with bombarding energy \cite{BERTULANI1988299} which can also help disentangling their contribution from nuclear excitation.

\subsubsection{Nuclear excitation followed by neutron emission}

\begin{figure}[t]
\centerline{
\includegraphics[width=0.95\columnwidth]{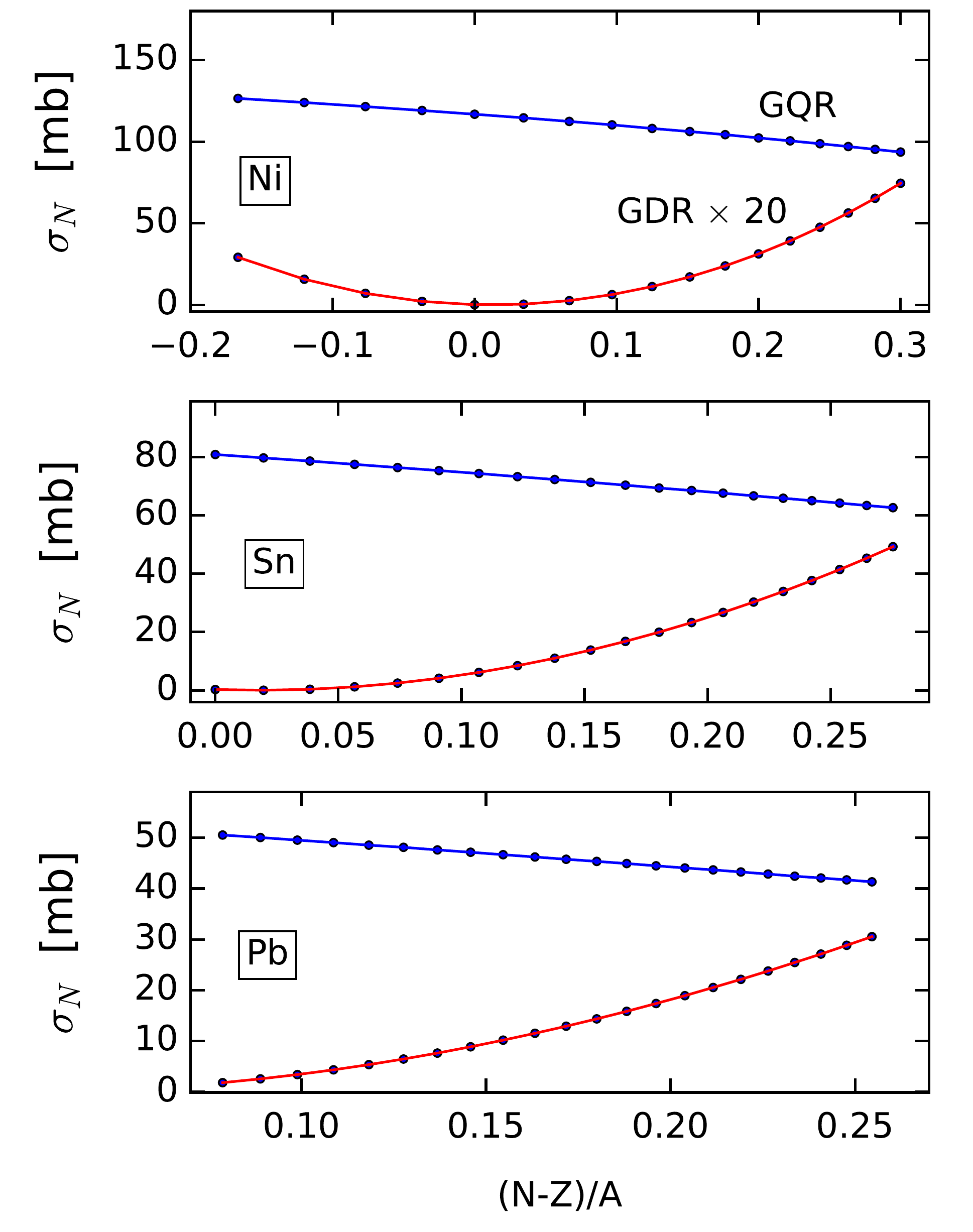}}
\caption{Nuclear excitation cross sections of ISGQR (GQR) and IVGDR (GDR) resonances in nickel, tin and lead projectiles bombarding carbon targets at 1 GeV/nucleon. The dependence on the asymmetry coefficient $(N-Z)/A$ of the projectile is shown. The upper curves in each frame display the excitation of ISGQR and the lower ones are calculations for the excitation of IVGDR multiplied by 20.}
\label{nucgdr}
\end{figure}

Using the deformed potential model, described in section \ref{necm} to obtain the cross sections for nuclear excitations, we show in Figure \ref{nucgdr} the cross sections for nuclear excitation of ISGQR (GQR) and IVGDR (GDR) resonances in nickel, tin and lead projectiles incident on carbon targets at 1 GeV/nucleon. The dependence on  the asymmetry coefficient $\delta=(N-Z)/A$ of the projectile is displayed. The upper curves in each frame are calculations  for the ISGQR and the lower curves are for the excitation of IVGDR multiplied by a factor of 20. For $^{208}$Pb  projectiles the cross sections are of the order of 43 mb (1.11) mb, for the ISGQR (IVGDR).

The IVGDR cross sections are negligible for $N=Z$ with a negligible neutron skin. Light nickel isotopes exhibit a non-zero proton skin  and  a reversing trend of the IVGDR excitation cross section around $\delta=0$ is observed. The cross sections for IVGDR resonances are bemire than a factor 20 smaller than for ISGQR resonances. Hence, they are unimportant for the purposes of extracting neutron skins at such bombarding energies. Nonetheless, the method has been used previously at lower energies,  below 100 MeV/nucleon, by measuring differential cross sections that can display marked differences between angular distributions for $L=1$ and $L=2$. The energy dependence of the cross sections at these energies has also been used as a tool \cite{KRASZNAHORKAY1994521}.

The ISGQR cross sections decrease along an isotopic chain as the neutron numbers increase. This can be understood as due to the decrease of the deformation parameter $\delta_2$ with the increase of the ISGQR centroid energy with mass number,  as inferred from Eq. \eqref{deform1}.

Larger theoretical uncertainties in treating nuclear excitations in high energy collisions exist as compared to Coulomb excitation. Whereas the Coulomb interaction is well known,  optical potentials used in the deformed potential model,  Eq.  \eqref{deform1}, are not so constrained. Little can be done to improve these models with the state of the art knowledge of high energy nuclear reactions.  The deformed potential and the Tassie model \cite{Tassie1956} are should be taken as rough approximations for nuclear reactions at high energies. More theoretical efforts are certainly needed.

The deformed potential model used in  collisions with a proton target yields cross sections that similar to those shown in Fig. \ref{nucgdr}.  Use different targets does not display appreciable changes in the nuclear excitation of giant resonances.  A noticeable variation of the Coulomb excitation is possible. By varying the bombarding energies in the range 100-1000 MeV/nucleon will not help either because the cross sections for nuclear excitation remain practically unchanged. Thus, the  50 mb to 100 mb of nuclear excitation cross sections mainly contributing to the one-neutron decay channel will be hard to control systematically without adding other observables to the angle integrated cross sections.   But, as discussed previously \cite{AumannPRL119.262501}, simulations show that nuclear excitation events can be separated using the angular distribution of neutrons.

\subsubsection{Neutron changing and interaction cross sections}

\begin{figure}[t]
\centerline{
\includegraphics[width=0.95\columnwidth]{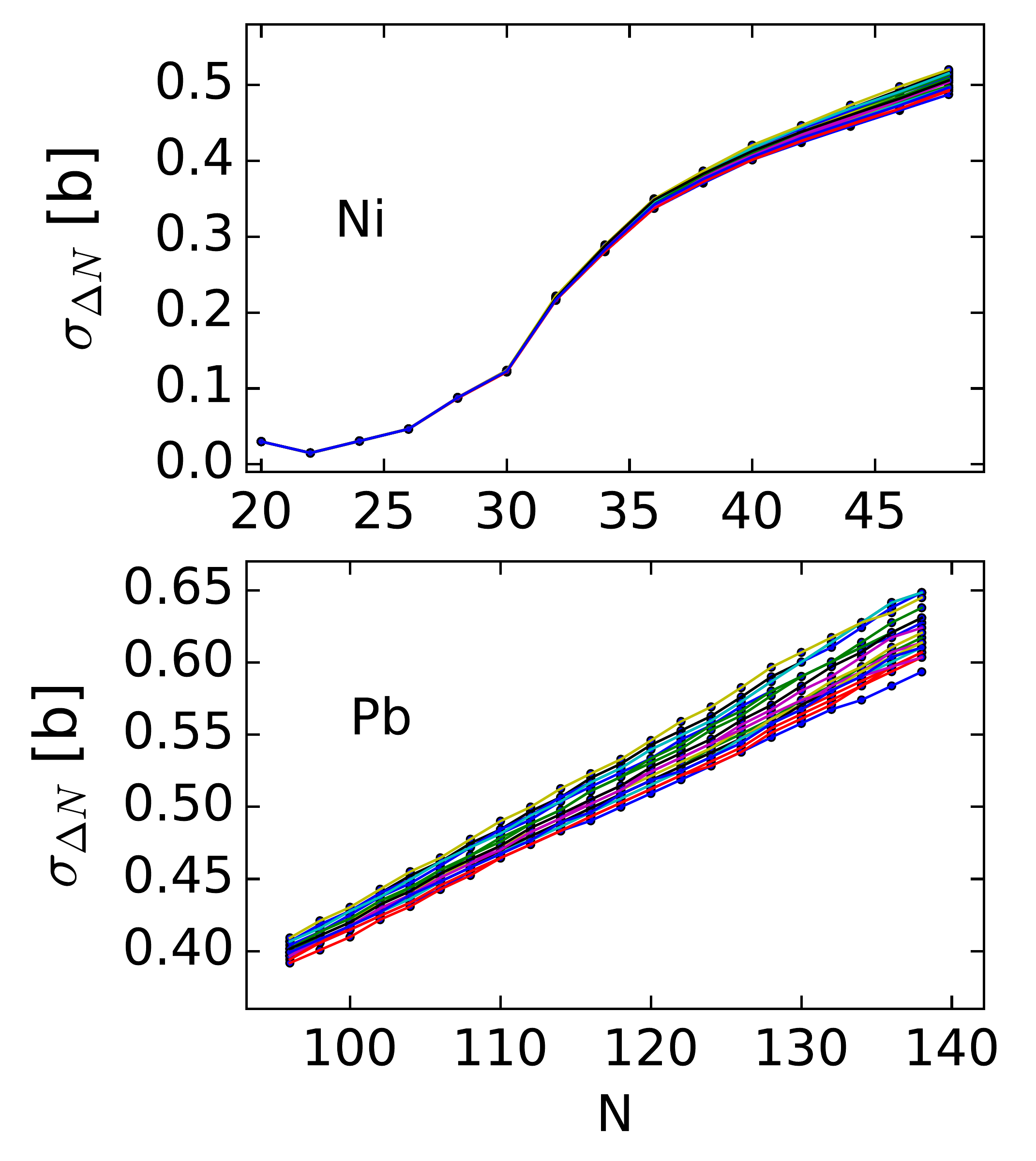}}
\caption{Neutron-changing cross sections in barns following the model of section \ref{necm}, for nickel (upper frame) and lead (lower frame) isotopes and  the 23 Skyrme interactions  \cite{Bertulani.PRC.100.015802}. The dependence on the neutron number is shown. The lines are drawn to guide the eyes and each one of them represents a prediction with one of the Skyrme interactions along an isotopic chain.}
\label{sdeltn1}
\end{figure}

Figure \ref{sdeltn1} displays neutron-changing  cross sections, following the model described in section \ref{necm}, for nickel (upper frame) and lead (lower frame) isotopes and for several Skyrme interactions, as a function of the neutron number. Nickel isotopes display a very small dependence on the neutron number with  a given Skyrme interaction.  The nickel radius is not much larger that for carbon and the geometric variation along the isotopic chain with a given Skyrme interaction do not lead to a sizable cross section variation. For $^{64}$Ni, the heaviest stable nickel isotope,  the cross sections range from $337$ to $350$ mb, yielding  a 4\% sensitivity on the choice of the Skyrme interaction. However, for  lead the cross sections show a very strong dependence on the neutron number with a linear dependence with the neutron number for a given Skyrme interaction. This seems to be is a robust result that can be employed in the experimental analysis. For $^{208}$Pb,  the heaviest stable lead isotope, the cross sections range from  $537$  to $576$ mb,  an approximate 7\% variation with the choice of the Skyrme interaction. Therefore, it seems that neutron-changing cross sections can constrain the several Skyrme models by comparison with calculations. The linear relation between $\sigma_{\Delta N}$ and the neutron number is also a feature worthwhile to explore.  

\begin{figure}[t]
\centerline{
\includegraphics[width=0.95\columnwidth]{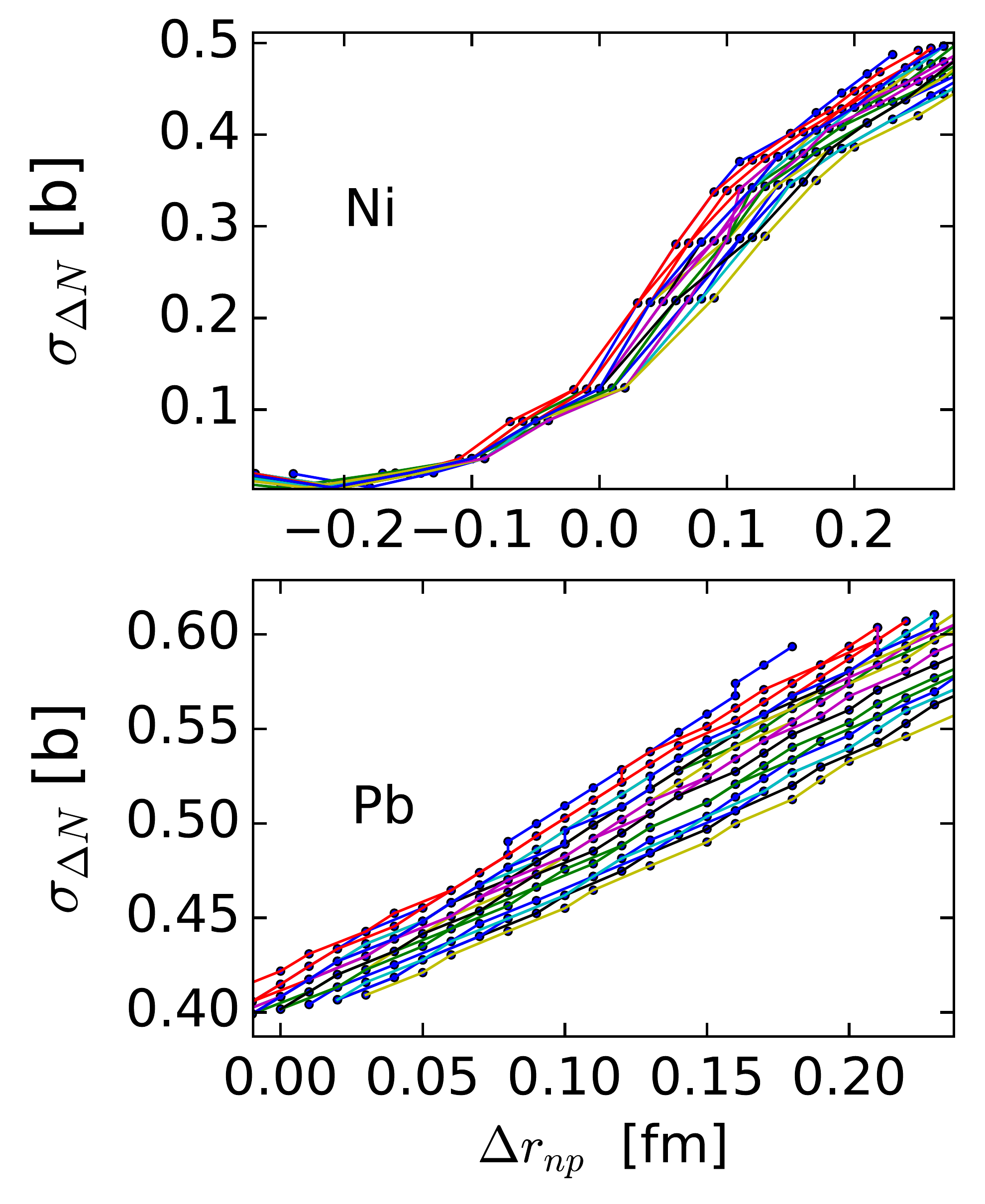}}
\caption{Same  as in Figure \eqref{sdeltn1} but plotted as a function of the neutron skin $\Delta r_{np}$ for different isotopes and Skyrme interactions. The lines are drawn to simply guide the eyes and are the prediction for each Skyrme interaction  along an isotopic chain.}
\label{sdeltn2}
\end{figure}

The same calculations are displayed  in Figure \ref{sdeltn2} now as a function of the neutron skin in the various isotopes.  A row of vertical points  correspond to different isotopes and each curve along an isotopic chain is obtained with a single Skyrme interaction. Because $\Delta r_{np}$ and the neutron number show a strong correlation (see Figure \ref{leadiso}),  no additional information is gained as compared to Figure \ref{sdeltn1}. But such dependencies can be used  to deduce the accuracy needed to obtain a given value of $\Delta r_{np}$. For example, a neutron skin of 0.15 fm in Ni and Pb isotopes, yield cross sections within the range of $0.32$ to $0.42$ b and $0.47$  to $0.52$ b, respectively. They  correspond to sensitivities of the neutron skin with the choice of Skyrme interactions of 20\% for nickel and 10\% for lead isotopes.  Plots like Figure \ref{sdeltn2} are a  combination of  the theoretical predictions shown in Figures \ref{leadiso} and \ref{sdeltn1}. They are useful if a large number of projectile isotopes can be tested with experiment.

\begin{figure}[t]
\centerline{
\includegraphics[width=0.95\columnwidth]{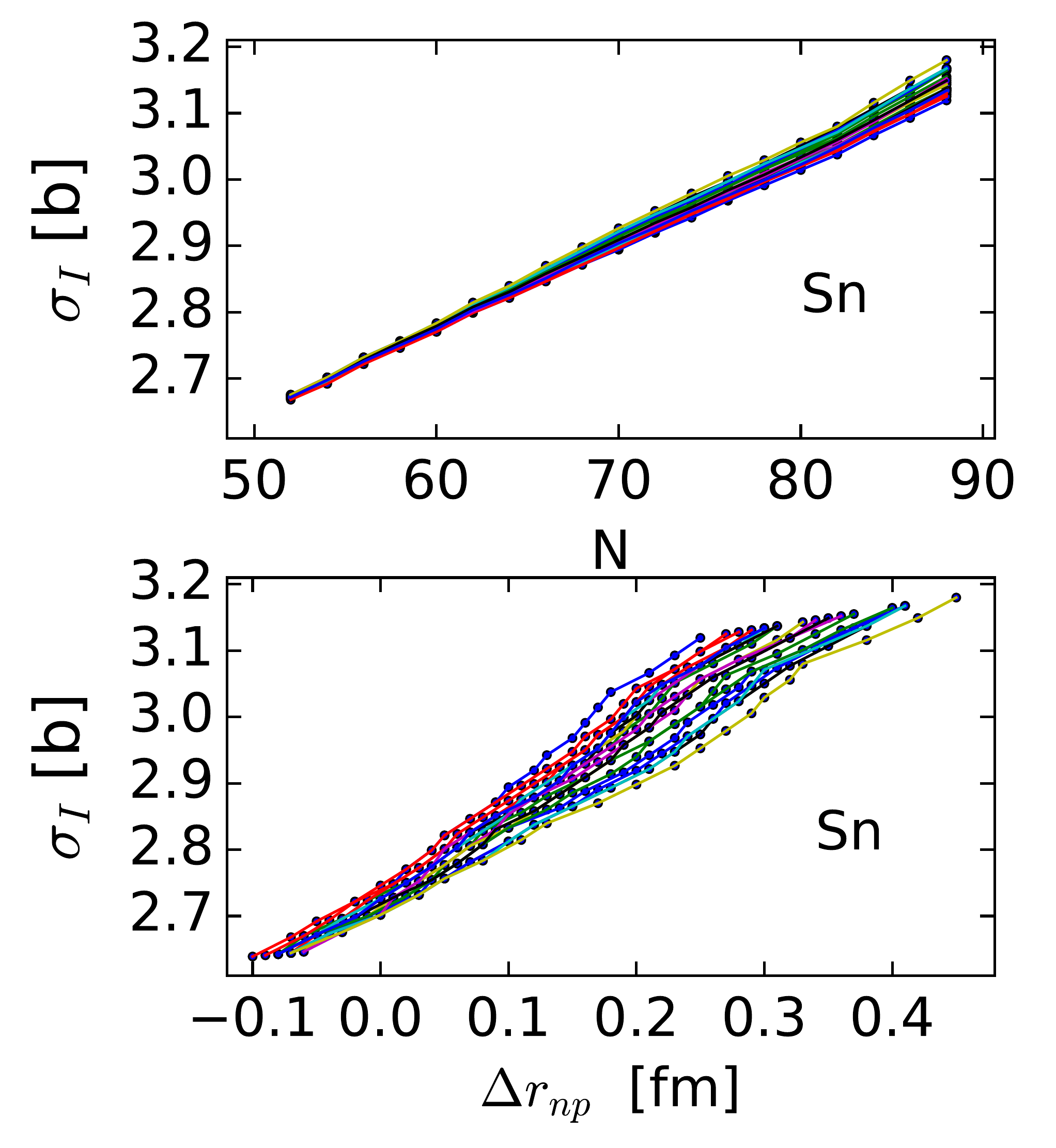}}
\caption{Dependence on several Skyrme interactions of the total interaction cross sections for tin isotopes bombarding carbon targets at 1 GeV/nucleon, using Eq. \eqref{sigma}. The results in the upper frame are displayed as a function of the neutron number $N$, while the lower frame displays the same data in terms of the neutron skin $\Delta r_{np}$.}
\label{sdeltn3}
\end{figure}

 Fig. \ref{sdeltn3} shows the total interaction cross sections for tin isotopes bombarding carbon targets at 1 GeV/nucleon. Calculations follow Eq. \eqref{sigma}, and use the same Skyrme interactions as above. The upper frame displays the calculations as a function of the neutron number $N$, while the lower frame shows the same data in terms of the neutron skin $\Delta r_{np}$. The cross sections change negligibly by varying the Skyrme interaction used for a given isotope. The reason is that for a given isotope all interactions yield essentially the same total matter density. Also, similar values for neutron skins are obtained for different isotopes using two or more Skyrme interactions. This is clear from the lower frame of Figure \ref{sdeltn3} where a much larger change of $\sigma_{I}$ with $\Delta r_{np}$ is observed. Hence, a systematic study of measurements of neutron-changing and interaction cross sections will be useful to test theoretical predictions of nuclear densities.

\begin{figure}[t]
\centerline{
\includegraphics[width=0.95\columnwidth]{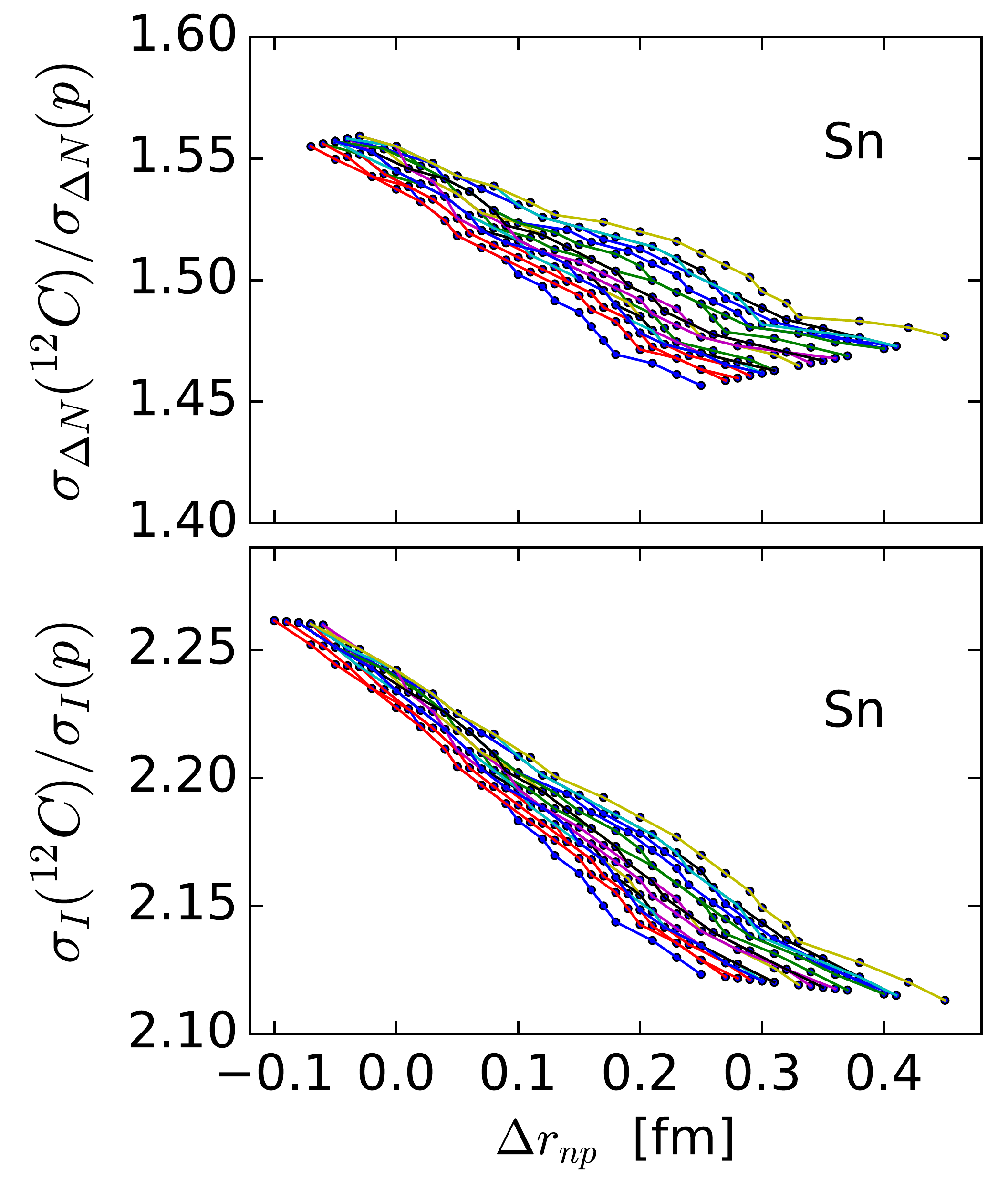}}
\caption{{\it Top frame:} Ratio of neutron changing cross sections, $\sigma_{\Delta N}$,  for 1 GeV/nucleon tin isotopes  obtained with carbon and  proton targets. The dependence on the neutron skin, $\Delta r_{np}$, is shown. The set of points along each curve correspond to a single Skyrme interaction. {\it Bottom frame:} Same ratio as in defined in the upper frame, but this time for the total interaction cross sections, $\sigma_{I}$. }
\label{sdeltn4}
\end{figure}

As a final remark, we show in Fig. \ref{sdeltn4} the ratio of neutron changing cross sections, $\sigma_{\Delta N}$,  for 1 GeV/nucleon tin isotopes  obtained with carbon and  proton targets. The dependence on the neutron skin, $\Delta r_{np}$, is shown. The set of points along each curve correspond to a single Skyrme interaction. In the lower frame we show  the same ratio, but this time for the total interaction cross sections, $\sigma_{I}$.  It is visible in the figure that the cross sections with proton targets have a steeper variation with the neutron skin than those obtained with carbon targets. This feature is better seen in the ratio of interaction cross sections. Thus, by using both carbon and proton targets allows for a better constraint on the proper Skyrme interaction that reproduces experimental data. Both neutron changing and total interaction cross sections will  help constraining these interactions and the EoS of symmetric and asymmetric nuclear matter.

\subsection{Polarized protons and the neutron skin}

The spin-orbit interaction is one of the most celebrated findings in our quest to understand the nature of nuclei. In 1936, Inglis \cite{InglisPR50.783} investigated the microscopic origin of the nuclear spin-orbit interaction by adding to nuclear forces a copycat of the atomic spin-orbit coupling, also known as the relativistic  Thomas effect \cite{Thomas1926}. Years later, the phenomenological inclusion of the spin-orbit interaction in the nuclear shell model, together with the Pauli principle, allowed  an astonishingly simple explanation of the magic numbers appearing in nuclear energy spectra. The interaction was assumed to be much larger than the relativistic effect proposed by Inglis. This achievement had a large impact in our understanding of nuclear systems and lead to a Nobel prize for  Mayer and Jensen in 1950 \cite{MayerPR78.16,Haxel1950}. The microscopic origins of the nucleon-nucleus spin-orbit force are now explained in terms of a quantum field description of $\sigma$ and  $\omega$ meson exchange.

\begin{figure}[t]
\begin{center}
\includegraphics[
width=3.in]{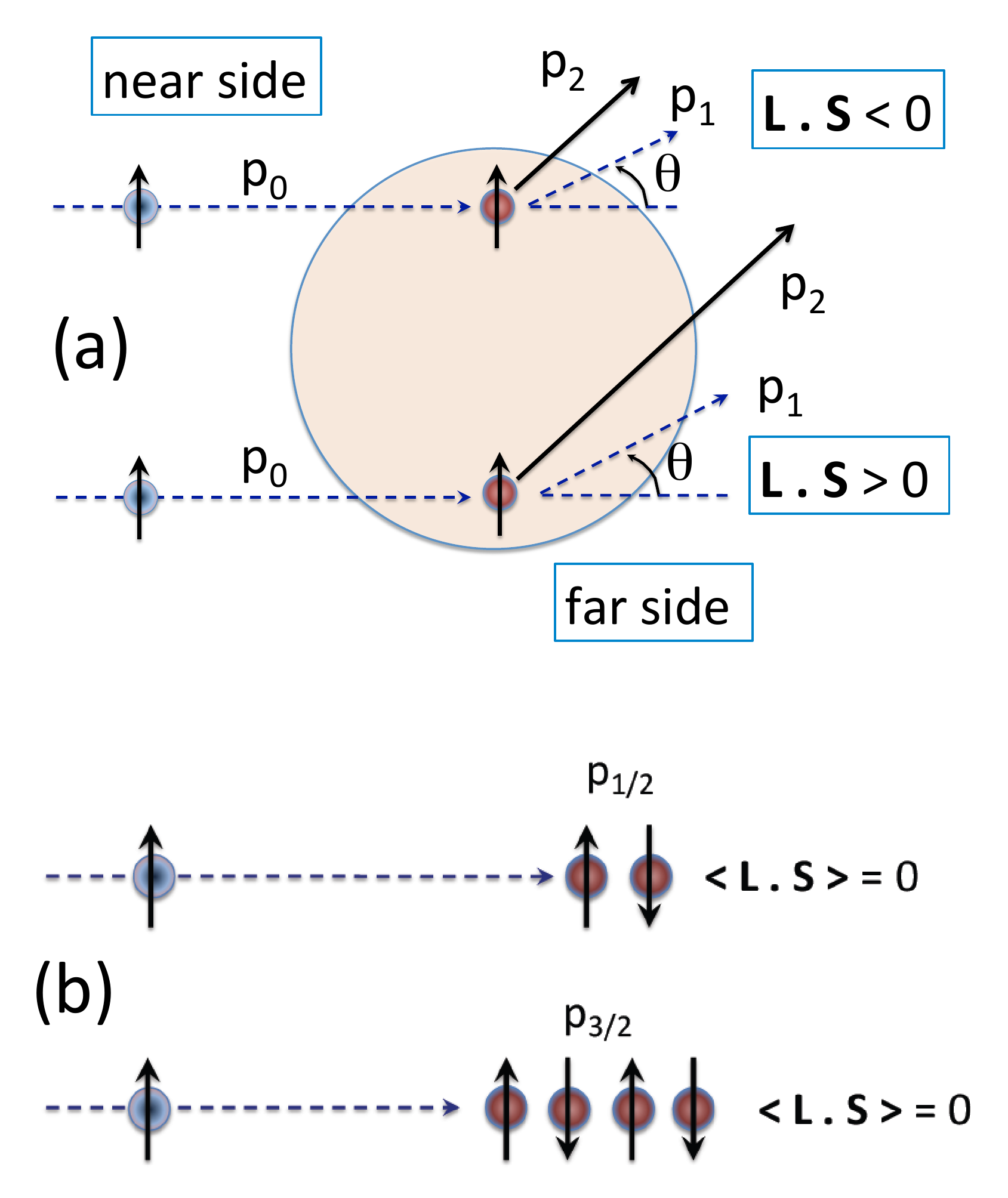}
\end{center}
\caption{{\it (a):}  a proton with spin up knocks out a nucleon (proton or neutron) with spin up. The  near and the far side scattering have opposite signs in the spin-orbit part to the optical potential. Near and far side scattering also yields a shorter  or a longer scattering path within the nucleus, changing the absorption of the scattered wave and its interference. {\it (b):} averaging the collisions of the incoming proton with nucleons within closed subshell tends to keep the initial proton polarization. But a net depolarization of the incoming proton will occur due to the spin-dependent part of the NN-interaction. This depolarization effect  will increase as the number of nucleons in the closed subshell also increases. The final polarization of the scattered proton will be sensitive to a combined effect of the  interference between the near and far side paths due to their different  absorption attenuation, the spin-orbit parts of the optical potential,  and the number of nucleons in the subshell \cite{MARIS1958577,JACOB1976517,MARIS1979461}.}
\label{marisf}
\end{figure}

The  spin-orbit interaction is of fundamental importance to explain basic phenomena observed in atomic and nuclear collisions. In nuclear physics, the simplest of all collisional cases, namely, elastic collision differential cross sections, display interference of polarized protons scattering through the near side and the far side of the nucleus. This interference pattern can be explained in terms of the opposite signs of the  ${\bf s}\cdot{\bf L}$ spin-orbit term due to the angular momentum flip (see, e.g., Ref. \cite{Bertulani:2004}) in changing from the near to the far side.   Evidently, other types of direct collisions using polarized protons are also influenced by the strength of the spin-orbit force and serve as a probe of its modification in the nuclear medium.  For example, one has speculated modifications of nucleon and meson masses and sizes, and also of meson-nucleon coupling constants in nuclear medium, motivated by  strong relativistic nuclear fields in the medium, deconfinement of quarks, and also  partial chiral symmetry restoration \cite{Miklukho2013,BrownPRL66.2720,FurnstahlPRC46.1507,HATSUDA199227,Walecka1986}.  A density dependence of the nucleon-nucleon interaction is obviously expected, modifying the expectations for nucleon induced reactions based on bare interactions. 

High-energy radioactive beams, and in particular quasifree (p,2p) and (p,pn) reactions have resurfaced as standard experimental tools to investigate nuclear spectroscopy. New  and more efficient detectors have allowed accurate experiments using inverse kinematics with hydrogen targets and have also opened new  possibilities for studies of the single-particle structure and nucleon-nucleon correlations in nuclei  as the neutron-to-proton ratio in the projectile increases. The detection of all outgoing particles has provided kinematically complete measurements of reactions studied at the GSI/Germany, RIKEN/Japan, and other nuclear physics laboratories. First experiments using (p,2p) and (p,pn) with newly developed experimental techniques have already been reported  with success \cite{AUMANN20073,Kobayashi:2008,Aum13,Pan16}.  These experiments have been concentrated on  using quasi-free scattering (QFS) in inverse kinematics as a tool to assess the shell-evolution in neutron-rich nuclei. Problems such as quenching of spectroscopy factors and single-particle properties of neutron-rich nuclei have been explored.  Recent theoretical work on (p,2p) reactions have also been reported \cite{Aum13,Oga15,Mor15,Cra16}.

The induced polarization due to a combination of absorption and spin is  known as the  ``Maris effect" \cite{MARIS1958577,JACOB1976517,MARIS1979461}. The idea is rather simple and invokes a combination of absorption and spin-orbit interaction. Suppose that the primary polarized proton is detected on the large-angle side of the momentum transfer $q$. Proton initial momenta directed toward the large-angle side of $q$, correspond to spin-up protons with $j = l - 1/2$ on the near side and to $j = l + 1/2$ on the far side. Initially polarized nucleons knocked out from the near side  will undergo less attenuation on their way out than those from the far side. Therefore they are less polarized,   $P_N < 0$, for $j - 1/2$. The reverse is true, and $P_N > 0$, for $j = l + 1/2$. The resulting net polarization of the knocked out nucleons when summed over their subshells would vanish for a closed-shell nucleus if the subshell momentum distributions were identical and if the NN interaction were spin-independent.  But they are not and will cause differences in $P_N^{(near)}$ and $P_N^{(far)}$. Therefore, one expect that due to absorption and the spin-orbit part of the optical potential,  Maris polarization is  approximately twice as large for $1p_{1/2}$ as for $1p_{3/2}$ and also opposite in sign.   The net polarization of the knocked-out nucleon can be observed using polarized proton targets and exploiting the  difference between the (spin-up)-(spin-up) and (spin-down)-(spin-up) cross sections in triplet and singlet scattering, respectively  \cite{MARIS1958577,JACOB1976517,MARIS1979461}. 

(p,2p) reactions are thought to be simpler than the elastic nuclear scattering mentioned above. This statement is based on the argument that in elastic scattering one deals with the scattering amplitudes of all nucleons in the nucleus, whereas (p,2p) reactions involves the scattering amplitude of a single nucleon in the nucleus. Absorption in this case is used as a benefit to enhance the effective polarization (see Fig. {marisf}).  

Polarized protons can also be used to probe the density dependence of the nucleon-nucleon (NN) interaction. A reduction of the analyzing power, $A_y$, due to density dependence  has indeed been discussed in Refs. \cite{Serot:1986,PhysRevC.33.2059}. An approximate 40\% reduction of $A_y$ has been predicted in the case of the $^{12}$C(p,2p) reaction, where the averaged density within $^{12}$C is about 50\% of the saturation density. In Ref. \cite{PhysRevC.51.2646} it was also suggested that the reduction of meson masses and coupling constants in dense nuclear matter will cause modifications of spin observables in quasifree reactions, explaining  why the $A_y$ are reduced by about 40\% when the matter density is about 50\% of the saturation density. These expectations have been verified experimentally \cite{HatanakaPRL78.1014}.

Before we assess the importance of Maris polarization to probe asymmetric nuclear matter using neutron-rich projectiles, we discuss how well existing experimental data can be understood with our calculations based on a standard theory of quasifree reactions. The triple differential cross sections for QFS in the Distorted Wave Impulse Approximation (DWIA) is given by \cite{Jacob:1973}  
\begin{eqnarray}
{d^3\sigma \over  d\Omega_1 d\Omega_2dT_1} &=&C^2S\cdot K_F \nonumber \\
&\times& \left| \left< \chi_{\sigma_2 {\bf k}_{p_2}}^{(-)} \chi_{\sigma_1 {\bf k}_{1}}^{(-)}          
\left| \tau_{pN}\right| \chi_{\sigma_0 {\bf k}_0}^{(+)}\psi_{jlm}\right>\right|^2 , \label{tripleX}
\end{eqnarray}
where $K_F$ is a kinematic factor,  $p_0$ ($p_1$) denotes the incoming (outgoing) proton, $p_2$ the knocked-out nucleon, $T_{2}$ its  energy, $C^2S$ is the spectroscopic factor associated with the single-particle properties of $p_2$ in the nucleus and $\psi_{jlm}$ is the nucleon wavefunction, which in the na\"\i ve single-particle model is labelled by the $jlm$  quantum numbers. $\chi_{\sigma {\bf k}_{p}}$ denote distorted scattering waves for the  reaction channel with spin $\sigma$ and momentum $p$). The DWIA matrix element includes the scattering waves for the incoming and outgoing nucleons, and the information on their spins and momenta, ($\sigma {\bf k}$), as well and the t-matrix for the nucleon-nucleon scattering. In first-order perturbation theory this t-matrix is directly proportional to the NN interaction.  For unpolarized protons, Eq. \eqref{tripleX} has to be averaged over initial spin orientations  besides a sum over final spin orientations. This formalism has been used previously in several  calculations and a good description of experimental data has been achieved with proper choices of optical potential and the nucleon-nucleon interaction (see, e.g., Refs. \cite{Cha77,ChantRooss83}). Ref. \cite{Aum13} reported that momentum distributions of the residual nuclei obtained in quasi-free scattering are well described using the eikonal wave functions $\chi_{{\bf k}_i}$ in Eq. \eqref{tripleX}. The Maris effect  has mostly been studied using the partial wave method  \cite{MARIS1958577,JACOB1976517,MARIS1979461,PhysRevC.33.2059,PhysRevC.51.2646,HatanakaPRL78.1014,Jacob:1973,Cha77,ChantRooss83}. 

\begin{figure}[t]
\begin{center}
\includegraphics[
width=3.5in]{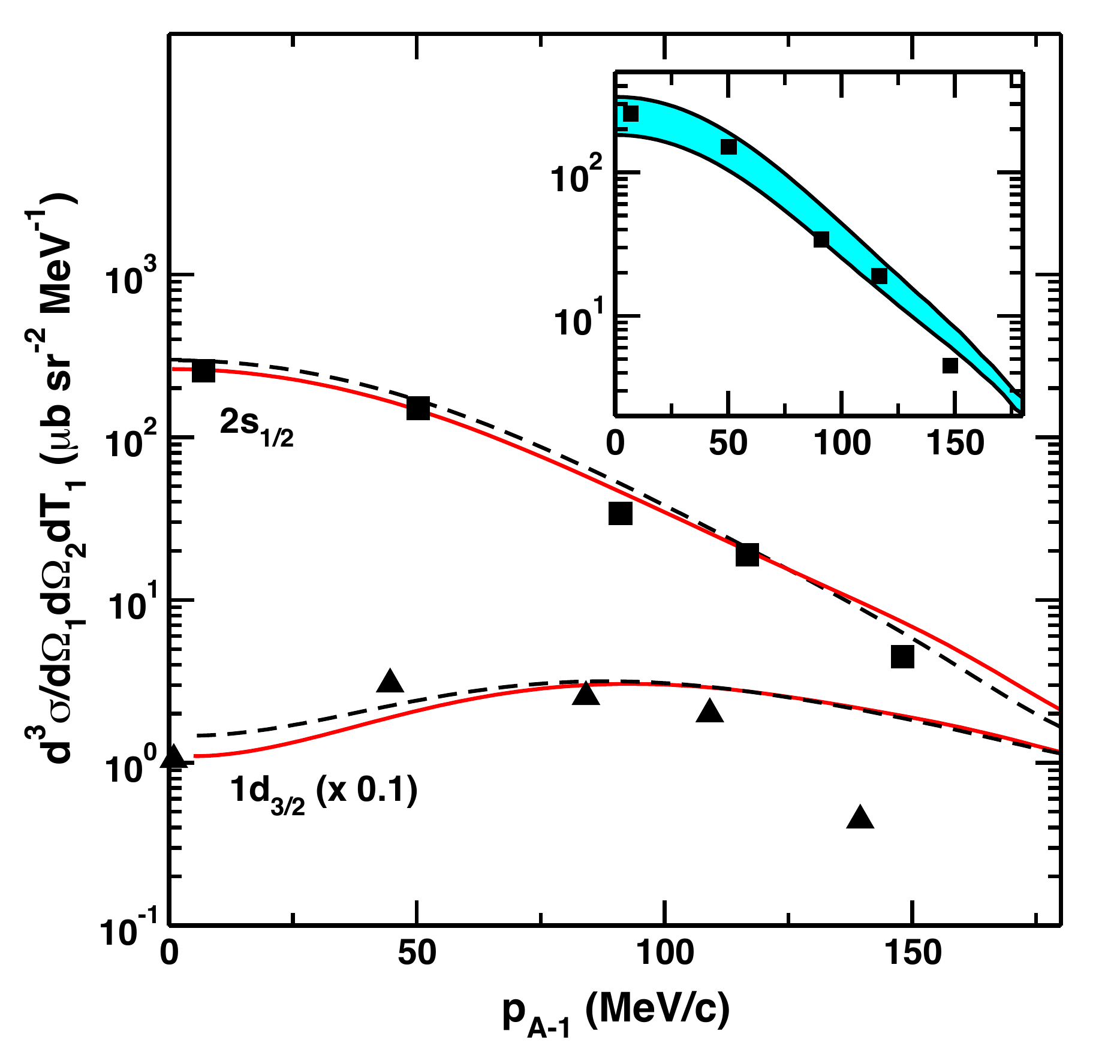}
\end{center}
\caption{Cross sections for $^{40}$Ca(p,2p)$^{39}$K and $E_p= 148$ MeV, in terms of the recoil momentum, $p_{A-1}$ of the the residual nucleus. The proton knockout are assumed to be from the $1d_{3/2}$ and $2s_{1/2}$ orbitals in $^{40}$Ca. The cross sections are integrated over the energy of removed proton and given in units of $\mu$b sr$^{-2}$ MeV$^{-1}$. The data are from  Ref. \cite{RoosPRL40.1439}. The big panel displays a dashed (solid) line which include (do not) the spin-orbit interaction. The inset shows the modification of the calculation for the $1s_{1/2}$ state with the inclusion of several NN-interactions. The shaded lines include a broad range of results obtained with the different NN-interactions.}
\label{polar1}
\end{figure}

Figure \ref{polar1}, taken from Ref. \cite{SHUBHCHINTAK201830}  shows the cross sections for $^{40}$Ca(p,2p)$^{39}$K and $E_p= 148$ MeV, as a function of the recoil momentum, $p_{A-1}$ of the the residual nucleus. The proton knockout is assumed to be from the $1d_{3/2}$ and $2s_{1/2}$ orbitals in $^{40}$Ca.   The cross sections are integrated over the energy of knocked-out proton and given in units of $\mu$b sr$^{-2}$ MeV$^{-1}$. The experimental data is from  Ref. \cite{RoosPRL40.1439}. In the big panel the dashed (solid) lines include (do not) the spin-orbit interaction. The optical potential of Ref. \cite{Nadasen:1981} was used together with NN-interaction from Ref. \cite{reid93}. In agreement with  Refs. \cite{RoosPRL40.1439,ChantPRL43.495},  the spin-orbit effect is found to be rather small for unpolarized protons.  The inset shows the comparison with the experimental data for the $1s_{1/2}$ state as the NN-interaction is changed.  The shaded area includes results for seven NN-interactions taken from Refs. \cite{YamaguchiPTP.70.459,PhysRevC.31.488,PhysRevC.38.51,KellyPRC39.2120,RayPRC41.2816,reid93,Amos:00}. It is clear that the proper choice of the interaction has a greater impact on the results for unpolarized protons than the spin-orbit interaction. The same conclusion applies for the proton removal from the 1$d_{3/2}$ orbital.  Not shown for simplicity are sources of uncertainty in the numerical results arising with the adoption of different global optical potentials.  They also yield a broad range of results, as with the case of the  NN interactions. 

We now turn to the effects of the density dependence on the cross sections and analyzing power, 
\begin{equation}
A_y={{d\sigma(\uparrow)-d\sigma(\downarrow)}\over {d\sigma(\uparrow)+d\sigma(\downarrow)}},
\end{equation} 
which requires the detection of knocked out nucleons by incoming polarized protons with opposite polarizations. To describe the analyzing power one needs to properly account for the spin variables in the transition matrix of Eq. \eqref{tripleX}. This procedure has been described in details in Refs. \cite{MARIS1958577,JACOB1976517,MARIS1979461,PhysRevC.33.2059,PhysRevC.51.2646,HatanakaPRL78.1014,Jacob:1973,Cha77,ChantRooss83}. The density dependence of the interaction has been assumed to be of the form proposed in Ref. \cite{PhysRevC.33.2059}, namely, one assumes that the NN t-matrix is modified because the nucleon mass in the nuclear medium, $m^*(r)$ changes locally according to   
\begin{equation}
m_N^*(r)=\left[ 1-0.44 {\rho(r)\over \rho_0}\right]m_N, \label{mN*}
\end{equation} 
where $\rho(r)$ is the density at radius $r$, $\rho_0$ is the nuclear saturation density of $0.17$ fm$^{-3}$, and the factor $-0.44$ stems from the  relativistic mean field theory \cite{Serot:1986}. This effect is obtained from a Schr\"odinger equivalent form of the Dirac description of the scattering waves $\chi_i$. The spinor parts of these waves are then incorporated into the NN t-matrix, with the nucleon mass replaced by $m^*_N$ within the nucleon spinors \cite{PhysRevC.33.2059}. This effect is somewhat reminiscent of the modification of meson masses and coupling constant in the Rho-Brown scaling conjecture \cite{Brown:81} so that the masses of the mesons giving rise to the interaction are modified according to $m_\sigma^*/m_\sigma = m_\rho^*/m_\rho=m_\omega^*/m_\omega = \xi$ and $g_{\sigma N}^*/g_{\sigma_N}= g_{\omega N}^*/g_\omega N=\chi$. It has been applied previously in Ref. \cite{PhysRevC.51.2646} to study nuclear medium effects in quasi-free scattering, with the parameters $\xi$ and $\chi$ varying within the range $0.6 - 0.9$.

\begin{figure}[t]
\begin{center}
\includegraphics[
width=3.5in]{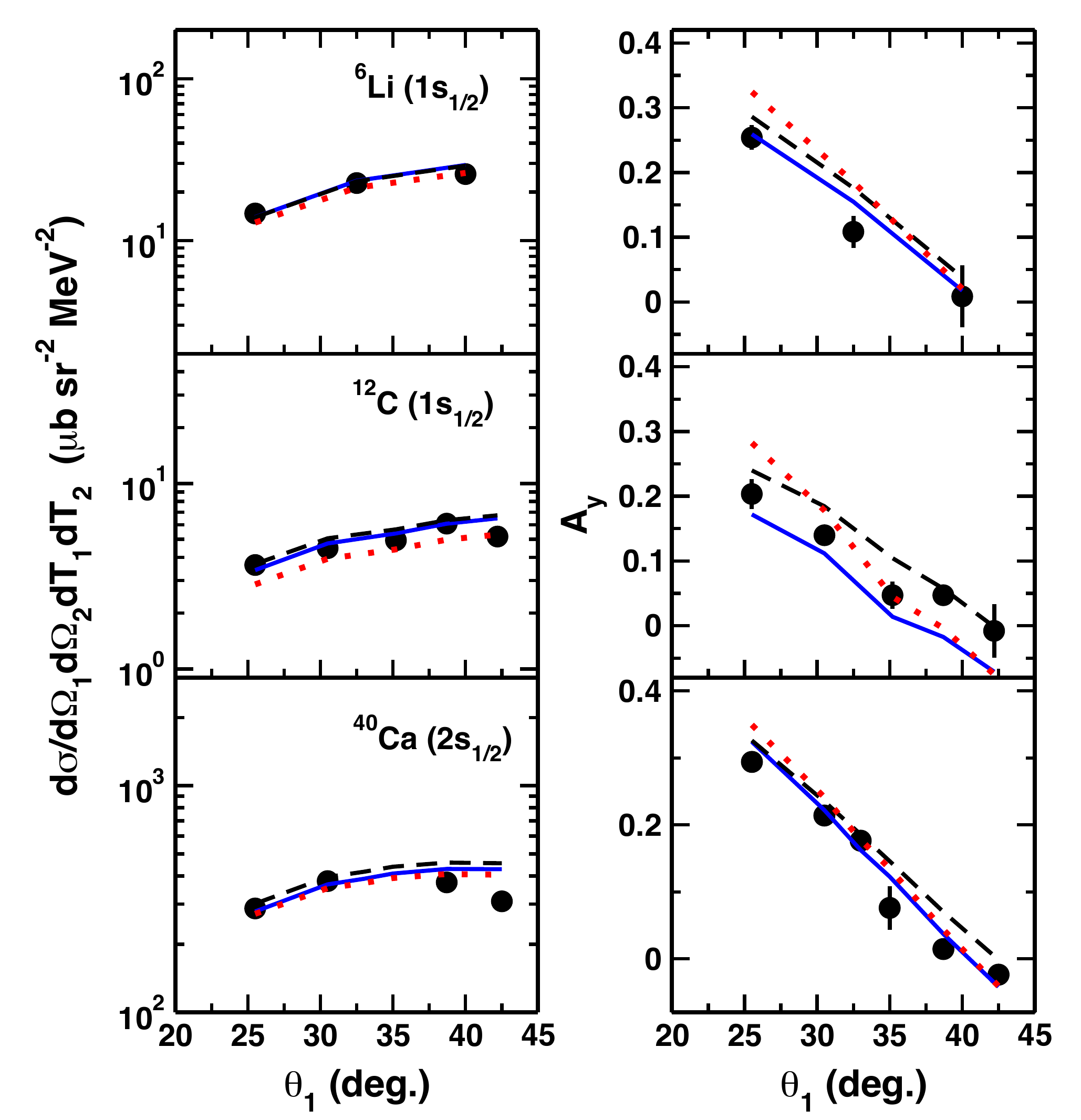}
\end{center}
\caption{Cross sections and analyzing powers for (p,2p) reactions on $^6$Li, $^{12}$C and $^{40}$Ca at 392 MeV. The solid lines contain the spin-orbit part of the optical potential, the dashed lines are results without the spin-orbit part, and the dotted-lines contain spin-orbit but neglect the density dependence in the NN t-matrix. The data are from Ref. \cite{HatanakaPRL78.1014}.}
\label{polar2}
\end{figure}

In Figure \ref{polar2} we show the results of Ref. \cite{SHUBHCHINTAK201830} for (p,2p) reactions on $^6$Li, $^{12}$C and $^{40}$Ca at 392 MeV, based on Eq. \eqref{mN*} in the model proposed by Horowitz and Iqbal \cite{PhysRevC.33.2059}. The data are from Ref. \cite{HatanakaPRL78.1014}. The NN interaction from Ref. \cite{HorowitzPRC31.1340} and the Dirac phenomenological optical potential  from Ref. \cite{Cooper:1993} was used, where the inclusion of the modification in Eq. \eqref{mN*} is straightforward. The solid lines contain the spin-orbit part of the optical potential and the calculations are normalized to the data  for  $d^3\sigma / d\Omega_1 d\Omega_2 dT_1 $. Due to the nature of the data analysis \cite{HatanakaPRL78.1014}, no attempt was made to identify  them as spectroscopic factors which are also irrelevant for the calculation of $A_y$. The dashed lines display the results without the spin-orbit part, and the dotted-lines contain spin-orbit but neglect the density dependence in the NN t-matrix. The usage of s-shell protons is chosen  because the interpretation is rather simplified since Maris polarization (discussed below) should be small, although the knocked out nucleon can still acquire a non-zero angular momentum with respect to the (A-1) residue after the collision. In fact, the target protons are unpolarized and the scattering asymmetry should be nearly equal to the asymmetry for the scattering of free protons. But we observe that  the spin-orbit effect still plays a role even for nucleons knocked from  s-waves because of the non-negligible angular momentum transfer in the collision. 

\begin{figure}[t]
\begin{center}
\includegraphics[
width=3.5in]{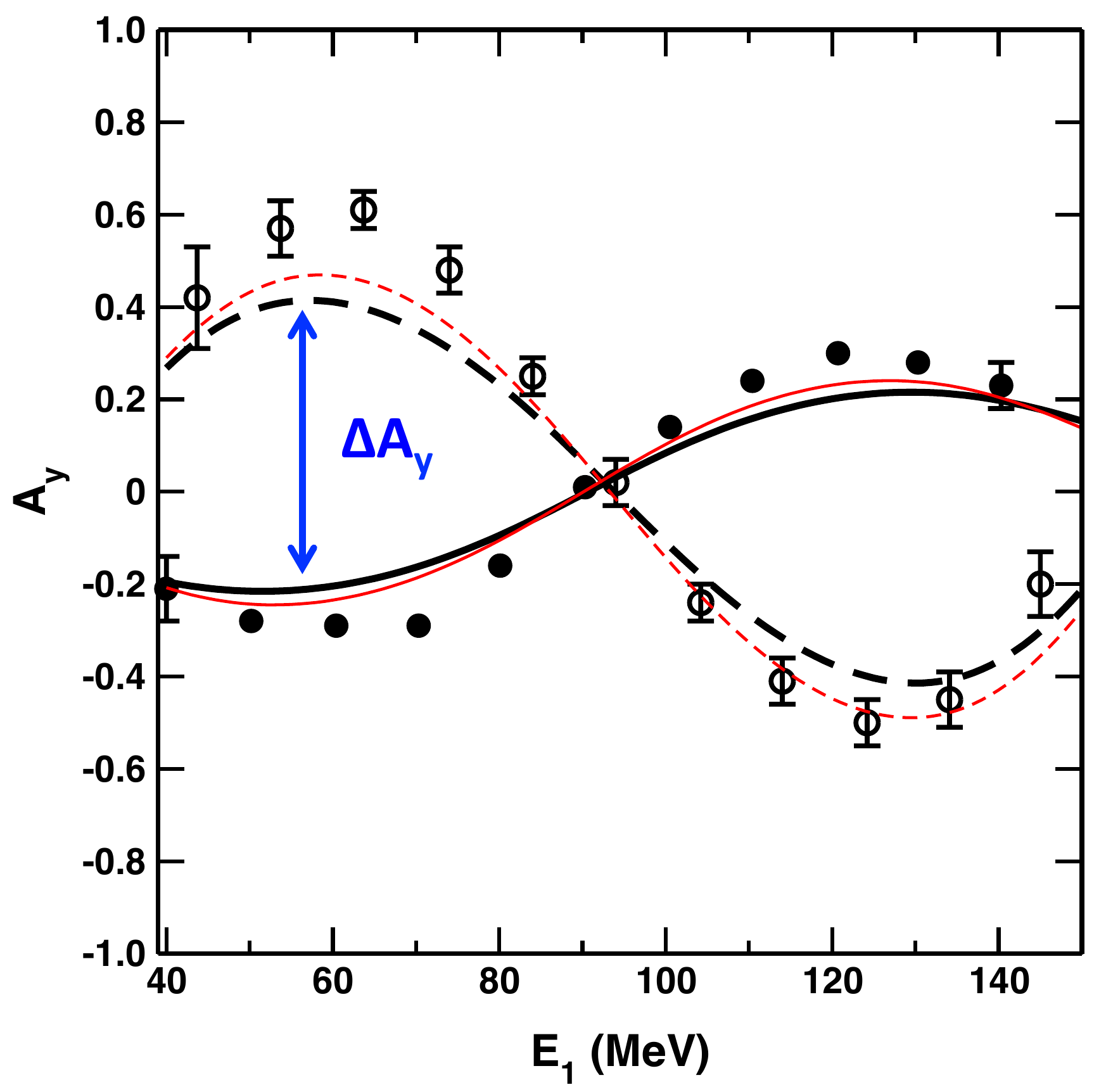}
\end{center}
\caption{Analyzing powers for the $1p_{3/2}$ and $1p_{1/2}$ states  in $^{16}$O(p,2p) reaction at 200 MeV as a function of
the kinetic energy of the ejected proton. Thin (thicker) lines include (do not include) the medium modification of the NN interaction. One proton is measured at 30$^\circ$ and the other at $-30^\circ$. The open circles  are  for $1p_{1/2}$ and the solid circles  for $1p_{3/2}$. The thiner (thicker) lines include (do not include) the medium modification of the NN interaction.}
\label{polar3}
\end{figure}

Maris polarization is a combined action of absorption and spin-orbit force  for a nucleon knocked out a non-zero angular momentum orbital, such as $p_{1/2}$ and $p_{3/2}$ nucleons. The ejected nucleon  is polarized before the collision and an average over the spin tends to  wash out this polarization, except that the interaction with the incoming polarized proton  has a strong spin dependence. This leads to a net effective polarization which depends on the sub-shell where the nucleon is ejected from. But, as shown in Ref. \cite{MARIS1979461}, the effective polarization is not far from being proportional to $A_y$.  Therefore, Maris polarization is also directly visible in analyzing power data. This is best seen if $A_y$ is displayed for fixed angles of the outgoing nucleons and scanning the energy of the ejected nucleon, as seen in Figure \ref{polar3}.  The data are from Ref. \cite{KITCHING1980423}.  Both nucleons are measured at 30$^\circ$. The open circles are for $1p_{1/2}$ and the solid ones for $1p_{3/2}$. Thiner (thicker) lines include (do not include) the medium modification of the NN interaction.  

Maris polarization in neutron-rich nuclei and its dependence on the neutron number was explored in Ref. \cite{SHUBHCHINTAK201830} by studying the tin isotopic chain which is well described with standard mean field theories. The density dependence of the interaction was included following the prescription in Eq. \eqref{mN*} with nuclear densities obtained from HFB calculations and the BSk2 Skyrme interaction, described in Ref. \cite{SAMYN2002142}. One needs single-particle energies as well as wavefunctions of the ejected nucleon. It is possible, but complicated and not necessarily reliable, to extract these quantities from the HFB mean field method. A simpler approach was adopted to determine these quantities from a global Woods-Saxon potential model. Protons are ejected from the $2p_{1/2}$ and $2p_{3/2}$ states were assumed \cite{SHUBHCHINTAK201830}.

The calculated analyzing powers for the tin isotopes have a similar feature as that displayed in Fig. \ref{polar3}.   The magnitude of the Maris polarization will be quantified in terms of the  difference between the first maximum of the $2p_{1/2}$ state and the first minimum of the $2p_{3/2}$ state, denoted by 
\begin{equation}
\Delta A_y= (A_y^{p_{1/2}})_{max} - (A_y^{p_{3/2}})_{min}.
\end{equation} 

\begin{figure}[t]
\begin{center}
\includegraphics[
width=3.5in]{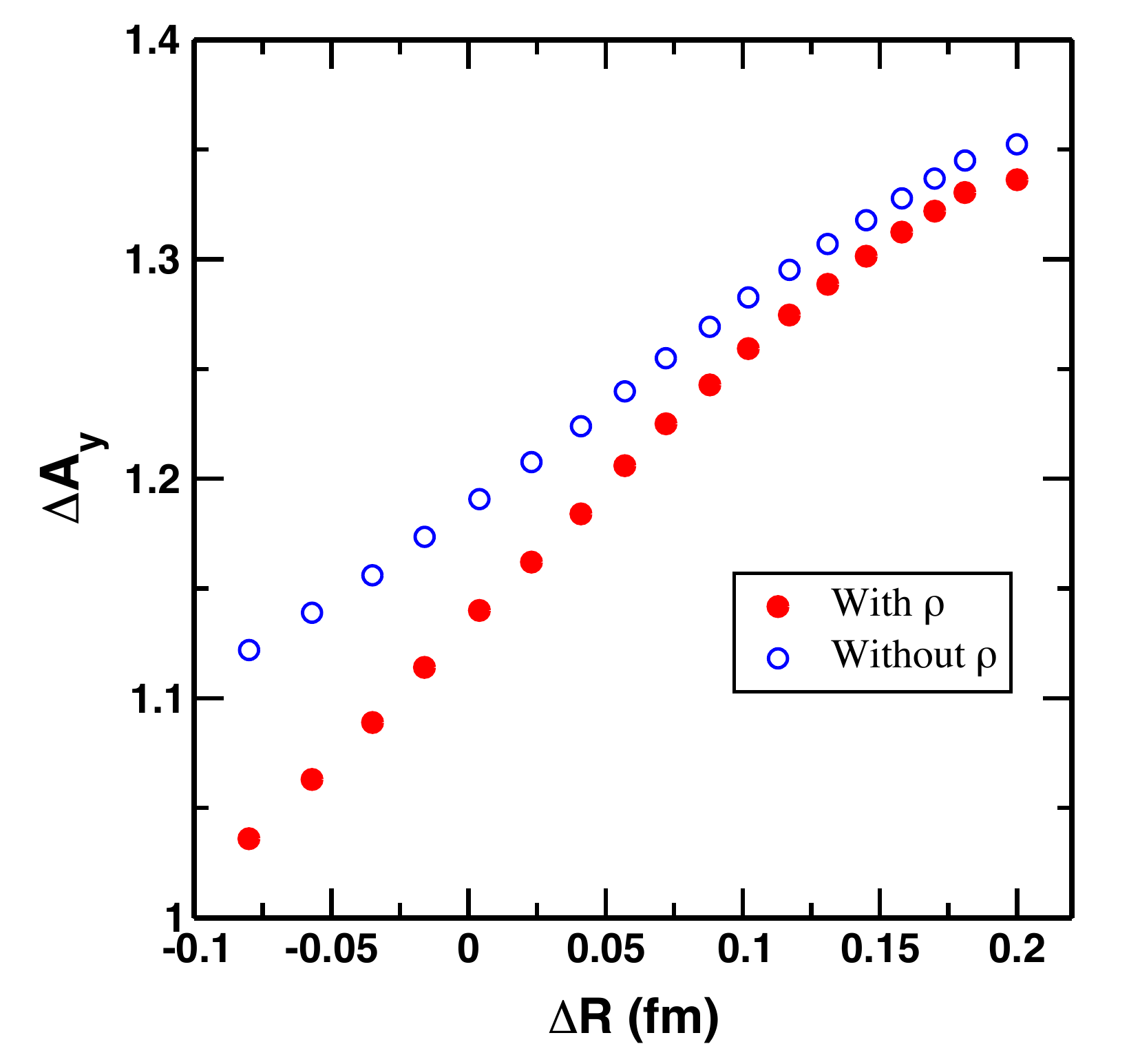}
\end{center}
\caption{Difference between the polarization maxima and minima for the $2p_{1/2}$ and $2p_{3/2}$ subshells in tin isotopes for (p,2p) reactions at 200 MeV as a function of the neutron skin. The solid (open) circles include (do not include) density dependence of the NN interaction. We assume that protons are detected at $\theta = 35^\circ$ and  $\theta = -35^\circ$, respectively.}
\label{polar4}
\end{figure}

In Figure \ref{polar4} we plot $\Delta A_y$ in tin isotopes for (p,2p) reactions at 200 MeV as a function of the neutron skin obtained from the calculated rms radii for the HFB neutron and proton densities. The protons are assumed to be detected at $\theta = 35^\circ$ and $\theta = -35^\circ$, respectively. The solid (open) circles include (do not include) the density dependence of the NN interaction.  It is evident that adding more neutrons to the system increases the magnitude of the Maris polarization. The polarization increases becomes larger than 30\% along this isotopic chain. The dependence with the neutron skin comes out nearly linear, although deviations from the linear behavior appears at large neutron numbers. The inclusion of the density dependence of the NN interaction decreases $\Delta A_y$ for small neutron numbers and small skins.  This difference is stronger for isospin symmetric nuclei than for asymmetric ones.

 As the nuclear size increases, the asymmetric behavior in analyzing power measurements due to the combination of spin-orbit and absorption effects increase accordingly. We have shown that the determination of neutron skins with such measurements can be done if a separate information on the nuclear charge density is known. An unequivocal determination of the neutron skin requires that the $A_y$ measurements also explore the choices of nuclear interactions and account of medium effects. The choice of NN interactions, some of them also including medium modifications due to Pauli blocking and many-body effects, using e.g., a G-matrix approach, can be tested for a large number of experimental data already available. 

The Maris polarization effect is an useful tool to investigate single-particle properties in nuclei and their evolution  in neutron rich isotopes. Its sensitiveness to the strength of the spin-orbit interaction, medium modification of nucleon masses, and nuclear absorption allows for new applications in the studies carried out with secondary radioactive beams.  Because experiments can now be carried out with a much larger accuracy than in the past, new techniques are increasingly being developed to extend our knowledge of the nuclear physics of neutron-rich nuclei. We have shown that the effective polarization of knocked out protons in (p,2p) reactions can be added to the new techniques to study the nuclear size measurements.   indeed, the determination of neutron skins in nuclei is one of the major research efforts due to its relation to neutron stars and their equation of state \cite{bertulani2012neutron}.

\subsection{Charge-exchange reactions}\label{sec:cex_th}

The rapid neutron capture process (r-process) is responsible for about half of all elements heavier than iron. Despite that, the nuclear physics properties of the nuclei involved in the r-process are not well known and its astrophysical site has not been clearly identified. Supernova explosions and neutron star mergers are possible stellar sites candidates for the r-process. It is uncertain what of these mechanisms are more effective \cite{Benoit:APJ2017,Nishimura:APJ2017}. The neutrino-driven wind model within core-collapse supernovae is a promising candidate for the r-process and it could explain the observation  that the abundances of r-nuclei in old halo-stars are similar to those in our solar system \cite{Truran:2002}.  Electron-capture on neutron-rich nuclei with mass number  $A \sim 45 - 120$  is also thought to become important as the density increases during core-collapse supernovae \cite{HixPRL91.201102}.  Electron capture can also occur on excited states which are energetically not allowed with atomic electrons on earth \cite{LANGANKE199819}.  Electron capture by nuclei in $pf$-shell plays a pivotal role in the deleptonization of a massive star before the core-collapse \cite{BETHE1979487}. During the silicon burning, supernova collapse proceeds via to a competition of  gravity and the weak interaction, with electron captures on nuclei and on protons and the $\beta$-decay processes playing crucial roles. In this scenario, weak-interaction phenomena become important when nuclei with masses $A \sim 45 - 120$ become more abundant in the supernova core. Weak interactions change the value of $Y_e$ and electron capture dominates, the $Y_e$ value is successively reduced from its initial value  $\sim 0.5$. Electron capture  yields more neutron-rich and the abundance of heavier nuclei, because nuclei with decreasing $Z/A$ ratios are more bound with increasing nuclear mass. For  densities $\rho \le 10^{11}$ g/cm$^3$, the weak-interaction processes are dominated by Gamow-Teller and sometimes by Fermi transitions.  Ref. \cite{Fuller:1980ApJS,Fuller:1980ApJS,FFN:1985ApJ} reported systematic estimates of electron capture rates in stellar environments. However, their calculations are only based on the centroid of the Gamow-Teller response.  $B(GT)$-distributions have also been obtained using modern shell-model calculations \cite{CAURIER1999439,BaumerPRC68.031303,LANGANKE2000481,LANGANKE20011}. Some notable deviations from the previous rates reported in  \cite{Fuller:1980ApJS,Fuller:1980ApJS,FFN:1985ApJ} have emerged, e.g., $Y_e$ increases to about 0.445 instead of the value of 0.43 found previously \cite{Fuller:1980ApJS,Fuller:1980ApJS,FFN:1985ApJ}.

\begin{figure}[t]
\begin{center}
\includegraphics[
width=3.5in]{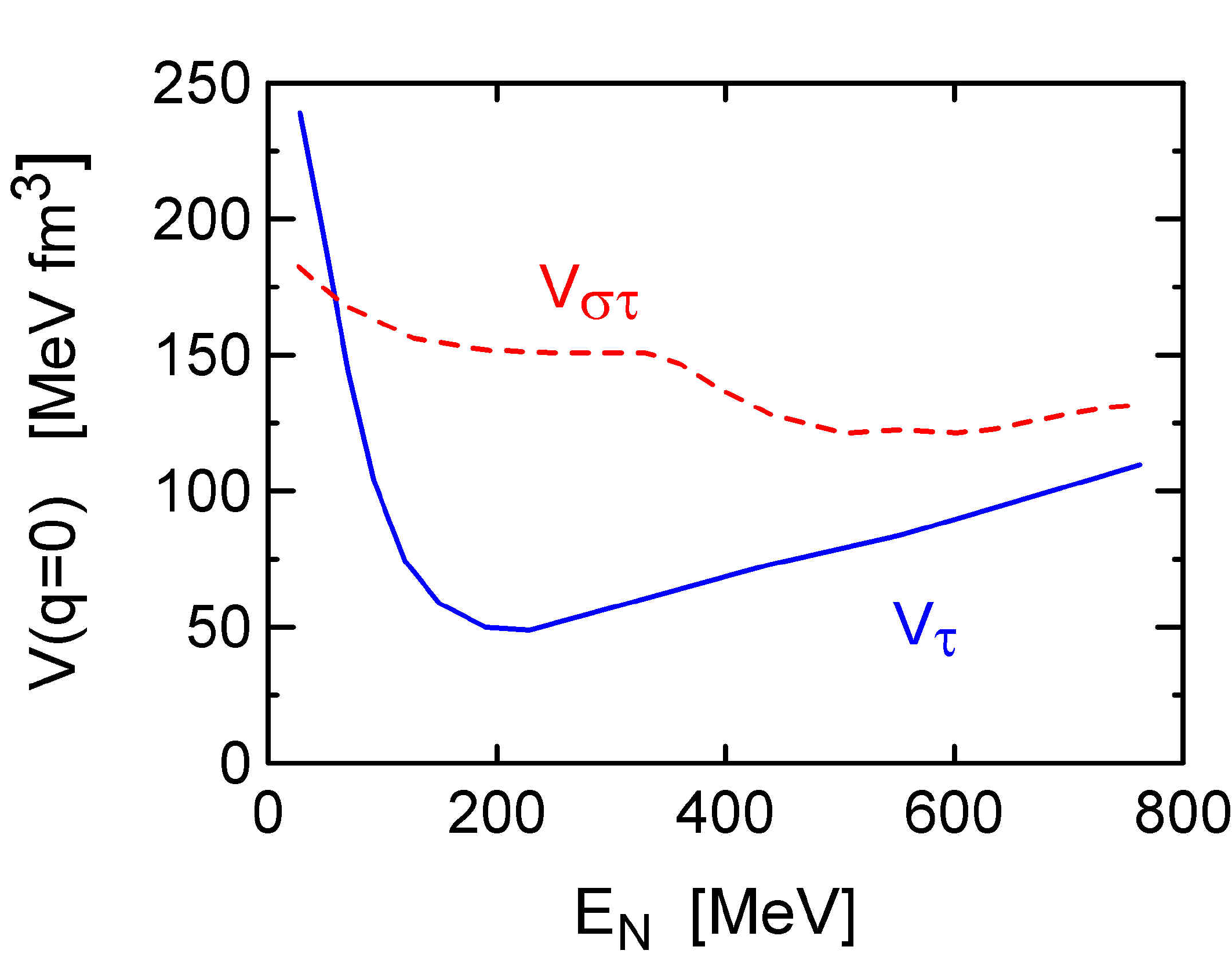}
\end{center}
\caption{Nucleon-nucleon potential (in momentum space) at forward angles. The picture shows the separate contributions from the spin-isospin, $\sigma \tau$, and the isospin, $\tau$, part of the interaction as a function of the laboratory energy.}
\label{vnnf}
\end{figure}

One cannot access these reaction rates directly in laboratory experiments. The medium nuclear mass range, accurate shell-model calculations are difficult and theoretical methods employing mean-field techniques have been introduced which include large uncertainties.  Theoretical calculations must be tested against experiment. Another even more difficult problem arises because laboratory-based experiments do not reproduce the conditions (density and temperature) present in stellar environments \cite{HixPRL91.201102,LangankePinedoRMP.75.819}. Thus the numerous electron capture reactions occurring in stars need coordinated efforts involving theory and experiments.

\subsubsection{Fermi and Gamow-Teller transitions}

Charge exchange induced reactions is often used to obtain values of Gamow-Teller, $B(GT)$,  and Fermi, $B(F)$,  matrix elements which
cannot be extracted from $\beta$-decay experiments \cite{TADDEUCCI1987125}. This method relies on the similarity in spin-isospin reaction operators in charge-exchange reactions and $\beta$-decay operators. In fact, it can be shown from first principles that, in the DWBA approximation,  the cross section for charge-exchange at small momentum transfer $q$ is closely proportional to $B(GT)$ and  $B(F)$ \cite{Bertulani:NPA1993:554:493}, 
\begin{equation}
{d\sigma\over d\Omega}(\theta=0^\circ)=\left( \mu \over 2\pi \hbar\right)^2 {k_f \over k_i} N_D|J_{\sigma\tau}|^2 \left[ B(GT) + {\cal C}B(F) \right], \label{tadeucci}
\end{equation}
where  $\mu$ is the reduced mass, $k_i(k_f)$  is the reactants relative momentum,  $N_D$ is a distortion factor
(which accounts for initial and final state interactions), $J_{\sigma\tau}$ is the Fourier transform of the GT part of the effective nucleon-nucleon interaction, ${\cal C} = \left| J_\tau/J_{\sigma\tau}\right|^2$, and  $B(\alpha=F,GT)$ is the reduced transition probability for non-spin-flip ($\tau_k$ is the isospin operator), 
$$B(F)= {1\over 2J_i+1}| \langle f ||\sum_k  \tau_k^{(\pm)} || i \rangle |^2,$$ 
and spin-flip  ($\sigma_k$ is the spin operator),
$$B(GT)= {1\over 2J_i+1}| \langle f ||\sum_k \sigma_k \tau_k^{(\pm)} || i \rangle |^2,$$ 
transitions. The condition that the momentum transfer is small, $q\sim 0$, is assumed to be valid for very small scattering angles, so that  $\theta  \ll 1/kR$, with $R$ being the nuclear radius and $k$ is the projectile wavenumber. 

At high energies, the charge-exchange reactions proceed via the exchange of charged pions and rho mesons which carry spin and isospin. Fig. \ref{vnnf} shows the nucleon-nucleon potential (in momentum space) at forward angles and the separate contributions from the spin-isospin, $\sigma \tau$, and the isospin, $\tau$, part of the interaction. One sees that, at $E\sim 100-300$ MeV, the $\sigma \tau$ contribution is larger than the  $\tau$ one. This hints to a favored energy region for studies of the Gamow-Teller matrix elements needed for astrophysics. At such energies, one expects
\begin{equation}
{d\sigma\over dq}(q=0)\sim KN_D|J_{\sigma\tau}|^2 B(GT) , \label{tadeucci2}
\end{equation}
where $K$ is a kinematical constant. When used to obtain $B(GT)$ values from experiments, Eq. \ref{tadeucci2} is purely empirical. It has been used indiscriminately in the analysis of charge exchange reactions, although it has been also shown that it fails in few cases. It lacks a solid theoretical basis and should be used with caution to reach the accuracy needed for the electron capture, beta-decay, or neutrino scattering response functions \cite{BERTULANI1997237}.  

Eq. (\ref{tadeucci}) can be derived easily in the plane-wave Born-approximation when the  charge-exchange matrix element becomes \cite{Bertulani:NPA1993:554:493}
\begin{equation}
{\mathcal M}_{exch}({\bf q})=\left<\Psi_a^{(f)} ({\bf r}_a)\Psi^{(f)}_b({\bf    r}_b) \left| e^{-i{\bf q}\cdot {\bf r}_a }  V_{exch}({\bf q})e^{i{\bf q}\cdot {\bf r}_b} \right|  \Psi_a^{(i)} ({\bf r}_a)\Psi^{(i)}_b({\bf r}_b) \right>, \label{Mexch}
\end{equation}
where ${\bf q}$ is the momentum transfer, $\Psi_{a,b}^{(i,f)}$ are the intrinsic wavefunctions of nuclei $a$ and $b$ for the initial and final states, ${\bf
  r}_{a,b}$ are the nucleon coordinates within $a$ and $b$, and $v_{exch}$ is the part of the nucleon-nucleon interaction responsible for charge exchange, which contains spin and isospin operators.  For forward scattering,  low-momentum transfers,  
${\bf q}\sim 0$, and small  reaction q-values, the matrix element \eqref{Mexch} becomes
\begin{equation}
{\mathcal M}_{exch}({\bf q} \sim 0) \sim V_{exch}^{(0)} ({\bf q}\sim 0) \, {\mathcal M}_a(F,GT)\, {\mathcal M}_b(F,GT)
\, ,
\label{q1}
\end{equation}
where $v_{exch}^{(0)} $ is the spinless part of the interaction, and 
$${\mathcal M}_{exch}(F,GT)= \left<\Psi_{a,b}^{(f)}\vert\vert (1 \ {\rm or} \ \sigma ) \tau \vert\vert   \Psi_{a,b}^{(i)}\right>$$  
are Fermi or Gamow-Teller (GT) matrix  elements for the nuclear transition. The result above emerges by using  eikonal scattering waves for the nuclei. In conclusion, extracting  Fermi or Gamow-Teller transition strength from experimental measurements of charge-exchange reactions depends on the validity of the low-momentum transfer assumption in collisions at high energies.  

\subsubsection{Double-charge-exchange and double-beta-decay}

The validity  of one-step processes in Eq. (\ref{tadeucci})  was proven to be a rather good assumption for $(p,n)$ reactions with a few exceptions. In  heavy-ion charge-exchange reactions this assumption might not be so good as shown in Refs. \cite{LenskePRL62.1457,Bertulani:NPA1993:554:493}. In Ref. \cite{LenskePRL62.1457}  multi-step processes involving the physical
exchange of a proton and a neutron were shown to still play an important role up to bombarding energies of 100 MeV/nucleon. Ref. \cite{BERTULANI1997237} explored the isospin terms of the effective interaction to show that deviations from Eq. (\ref{tadeucci})  are common under many circumstances. For those important GT transitions whose strengths are only a small fraction of the sum rule, a direct relation between $\sigma($p,\ n$)$ and $B(GT)$ values may cease to exist. Discrepancies have also been observed \cite{WatsonPRL55.1369} for reactions involving some odd-A nuclei including $^{13}$C, $^{15}$N, $^{35}$Cl, and $^{39}$K and for charge exchange using heavy ions \cite{SteinerPRL76.26}. 

\begin{figure}[t]
\begin{center}
\includegraphics[
width=3.5in]{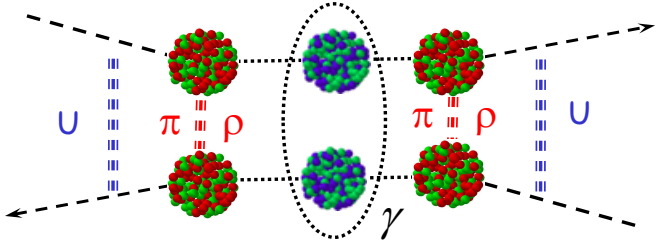}
\end{center}
\caption{Schematic view of a double-charge exchange reaction, involving a two-step process induced by the nucleon-nucleon interaction. The potential $U$ is responsible for the elastic scattering of the incoming and outgoing nuclei.}
\label{dcx}
\end{figure}

Double-charge exchange reactions, as shown schematically in Figure \ref{dcx}, could in principle be used to extract matrix elements for double beta decay in nuclei for  a number of nuclei where such decays are energetically allowed. The reaction mechanism using DWBA would involve the calculation of the amplitude
\begin{equation}
{\cal M}({\bf k},{\bf k}') = \sum_{\gamma,{\bf k''}} C_{\gamma}\left<\chi_{\bf k'}^{(-)}\left| V_{exch}{1 \over E_{\bf k}-\epsilon_{\gamma,{\bf k''}}-T-V_{exch}}V_{exch}  \right| \chi_{\bf k}^{+}\right> ,
\label{Tdcx}
\end{equation}
where $\chi_{\bf k}$ is the the distorted scattering wave due to an optical potential $U$ in the initial and final channels, $\bf k$, $\bf k'$ are the initial and final momenta of the scattering nuclei, $\bf k''$ is the momentum of the intermediate state $\gamma$ with energy $\epsilon_{\gamma,{\bf k''}}$ and $T$ is the kinetic energy. $C_{\gamma}$ are the spectroscopic amplitudes of the intermediate states. By using the Glauber scattering theory one can include the interaction $U$ in all orders. Using the assumptions of forward scattering,  and following the same approximations as used in Eq. (\ref{q1}) one can show again that a proportionality arises between double charge-exchange reactions and double beta-decay processes. The typical value of the cross section of a single step charge exchange relation is a few millibarns, while the a double charge exchange cross section is expected to be of the order of microbarns or less \cite{Bertulani:NPA1993:554:493}.

Usually, double beta decay are ground state to ground state transitions. It can be accompanied by two neutrino emission,  or by no emission of neutrinos. The latter process puts constraints on particle physics models beyond the standard one, such as the breaking of the lepton number conservation symmetry. In such case, the neutrino is a Majorana particle, e.g., its own anti-particle. To study neutrinoless double beta decay one needs to know the mass of the neutrino and the nuclear transition matrix element. Double beta decays emitting two neutrinos have been observed \cite{ElliottPRL59.2020} but the observation of neutrinoless double beta decay remains elusive.

Usually, the Fermi type operator does not contribute appreciably to double beta-decay with neutrinos emitted,  since the ground state of the final nucleus is not the double isobaric analog of the initial state. Therefore, the important transitions are those of double Gamow-Teller type. In the case of neutrinoless beta-decay one still expects that Gamow-Teller are larger than Fermi transitions  \cite{ZHENG1990343}.

The problem still remains if one can control the contributions of the matrix elements for intermediate states entering Eq. (\ref{Tdcx}). Perhaps, by measuring transitions to a large number of intermediate states in one step charge-exchange reactions, such as in (p,n) reactions, one in principle can determine the incoherent sum for the double charge-exchange transition. An obvious problem is that one is not sure if the same intermediate states excited in (p,n) and (n, p) experiments are involved in double-charge exchange. These intermediate states might also contribute very weakly in one-step reactions and very strongly in double charge-exchange, and vice-versa. Therefore, it seems that the best way to access information on the matrix elements needed for double beta-decay is to measure double charge exchange reactions directly. This idea is now being adopted by a few experimental groups (see, e.g., \cite{Matsubara2013,Kisamori:PRL.116.052501,Cappuzzello2018}) not only focused on double charge-exchange reactions related to neutrinoless double beta-decay but also to populate exotic nuclear structures. Recent theoretical studies on the relation between nuclear reactions and the neutrino induced matrix elements have also emerged (see, e.g., \cite{LenskePRC98.044620,ShiumizuPRL120.142502}).

\begin{figure}[t]
\begin{center}
{\includegraphics[width=9cm]{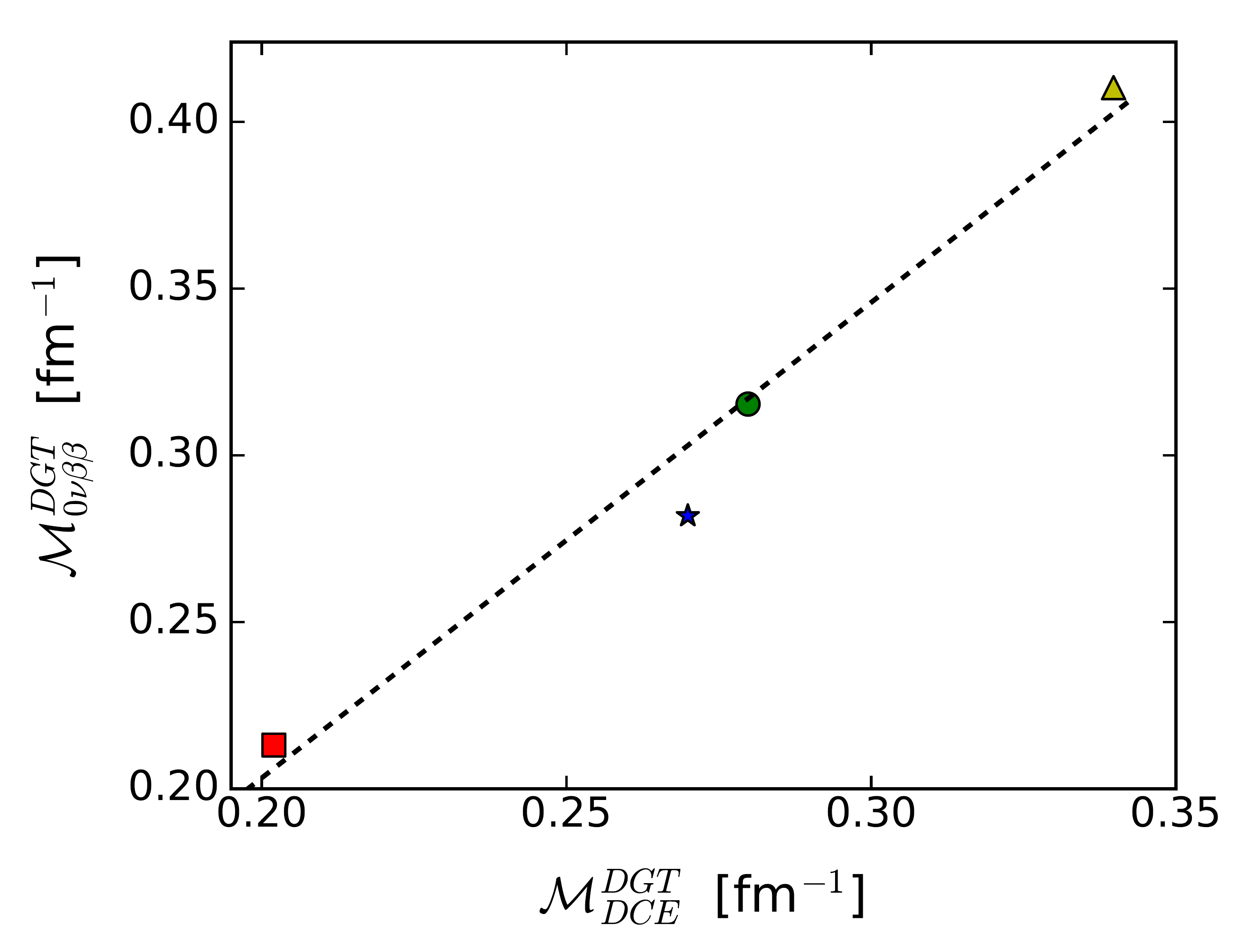}}
\end{center}
\caption{\label{dchexch}
 Correlation between calculated double charge-exchange (DCE) nuclear matrix elements (NME) for Gamow-Teller (GT) transitions 
and neutrino less double beta-decay (0$\nu\beta\beta$) \cite{SantopintoPRC98.061601}. The calculations have been done  for $^{116}$Cd $\rightarrow$ $^{116}$Sn, $^{128}$Te $\rightarrow$ $^{128}$Xe, $^{82}$Se $\rightarrow$ $^{82}$Kr, and $^{76}$Ge $\rightarrow$ $^{76}$Se, respectively.  
}
\end{figure}

In Fig. \ref{dchexch} we show a correlation between calculated double charge-exchange (DCE) nuclear matrix elements (NME) for Gamow-Teller (GT) transitions  and neutrino less double beta-decay (0$\nu\beta\beta$) obtained in Ref. \cite{SantopintoPRC98.061601}. The calculations have been done  for $^{116}$Cd $\rightarrow$ $^{116}$Sn, $^{128}$Te $\rightarrow$ $^{128}$Xe, $^{82}$Se $\rightarrow$ $^{82}$Kr, and $^{76}$Ge $\rightarrow$ $^{76}$Se, respectively.  The linear correlation is explained in terms of a simple reaction theory and if it holds for cases of interest, it opens the possibility of constraining neutrinoless double beta-decay NMEs in terms of the experimental data on DCE at forward angles. For a recent review on the use of charge-exchange reactions as a probe of nuclear $\beta$-decay, see Ref. \cite{LENSKE2019103716}.

\subsection{Central collisions}

Finally, we would like to briefly discuss the role of central collisions at energies of few hundreds of MeV/nucleon as a means to access information on the nuclear EoS (see Fig. \ref{dcx}). For a  gas of particles, with number density $n$, if their interaction distance is small compared to their mean separation distance $a$, then $na^{2}\ll1 $. In this situation, the particles interact only when they collide with the average distance travelled by a particle between two collisions being known known as its mean free path $\lambda$. Since the interaction cross section between particles is $\sigma \sim a^2$, one has for the mean free path $\lambda={1}/{n\sigma}$, and for a dilute gas  $\lambda\gg a$. If the gas is dilute, the probability of
three-body collisions is much lower than that for two-body collisions and they
can be neglected.

\begin{figure}[t]
\begin{center}
\includegraphics[
width=3.5in]{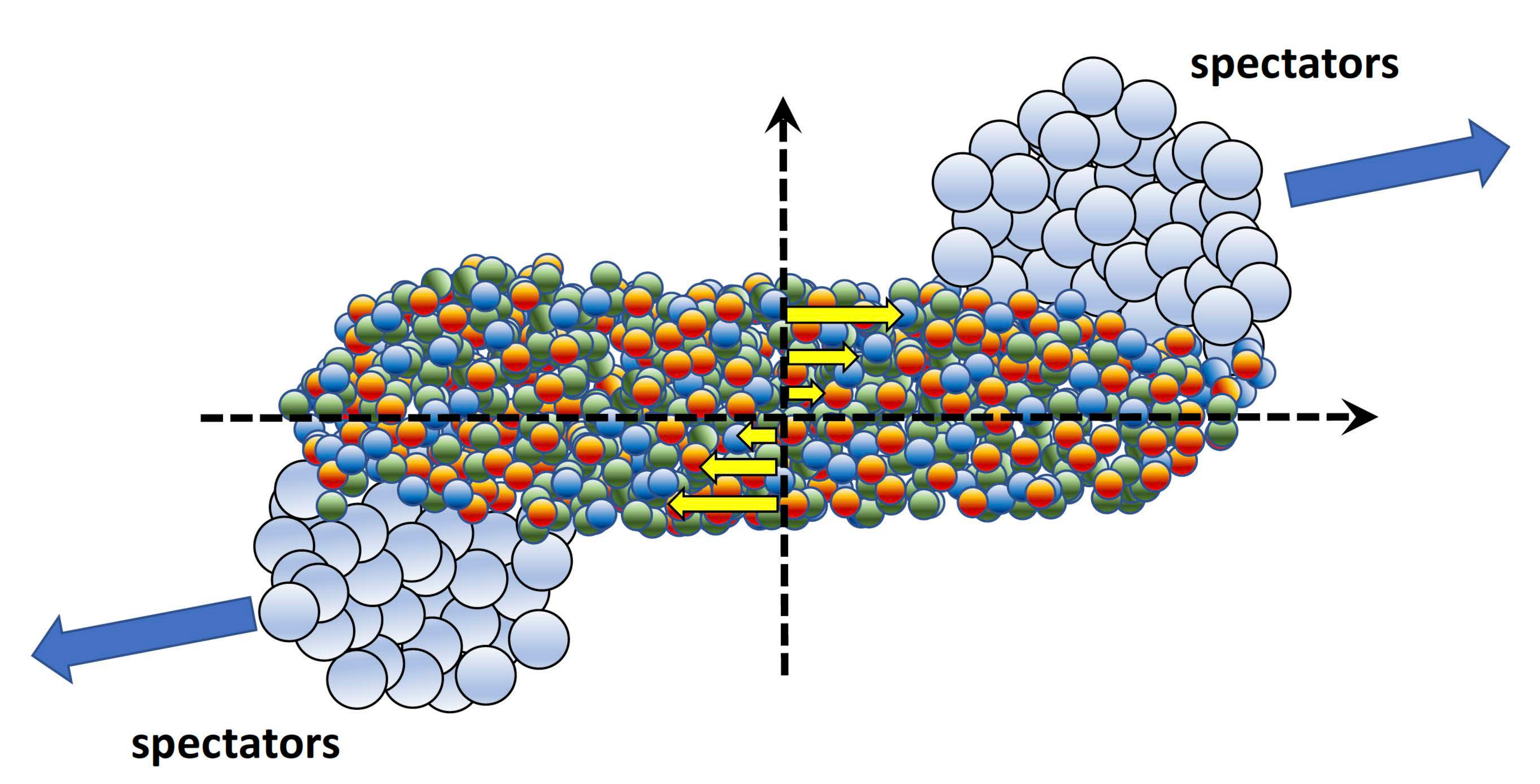}
\end{center}
\caption{Schematic view of a near central collision between two nuclei at high energies. The incoherent and coherent nucleon-nucleon collisions produce numerous particles, and at extreme relativistic energies available at RHIC/Brookhaven and CERN/Switzerland, it can de-confine the quarks and gluons (colored objects) within the nucleons as predicted by Quantum-Chromodynamics. Identifying the properties of the final  particles can lead to information about the nuclear EoS in the hadronic and de-confined phase.}
\label{centcoll}
\end{figure}

Assuming that these conditions are valid, several practical theoretical methods have been used to describe  nucleus-nucleus collisions from the microscopic point of view, i.e., using the collisions between the elementary particles composing the system. Most models are similar to the cascade model, where the elementary particles (nucleons, mesons, or quarks and gluons) move within a mean-field $U$ between collisions. Such models are often called transport models, for example solving the Boltzmann-Uehling-Uhlenbeck (BUU)  equation, 
\begin{eqnarray}  
&&\frac{\partial f}{\partial t}+\left(  \frac{\mathbf{p}}{m}+\mathbf{\nabla}_{\mathbf{p}}U\right)  \cdot\mathbf{\nabla}_{\mathbf{r}}f-\mathbf{\nabla}_{\mathbf{r}}U\cdot\mathbf{\nabla}_{\mathbf{r}}f = \nonumber \\
&& \int d^{3}p_{2}\int d\Omega\;\sigma\left(  \Omega\right)  \left\vert \mathbf{v}_{1}-\mathbf{v}_{2}\right\vert \nonumber\\
&  \times&\left\{  f_{1}^{\prime}f_{2}^{\prime}\left[  1-f_{1}\right]  \left[ 1-f_{2}\right]  -f_{1}f_{2}\left[  1-f_{1}^{\prime}\right]  \left[
1-f_{2}^{\prime}\right]  \right\}  ,\nonumber \\  \label{BE}%
\end{eqnarray}
where $\sigma$ is the elementary cross sections scattering the particles to within a solid angle $\Omega$ and the number of particles with a phase space volume is defined in terms of the distribution function $f$ as $f\left(  \mathbf{r,p},t\right)  d^{3}rd^{3}p$ where $\bf p$ and $\bf r$ are the coordinates of a particle and the labels 1 and 2 refer to the two colliding particles.   The  gain and loss (first and second) collisional terms on the right hand side  account for binary collisions between the particles and incorporates the Pauli principle through the $(1-f)$-terms to avoid scattering into occupied states \cite{Book:Ber04}. 

Eq. \eqref{BE} can be generalized to a covariant equations taking into account elements of relativity, although retardation and simultaneity are very difficult to handle. In such transport models the mean field $U$ and the elementary cross sections $\sigma$ are are correlated and one needs to use a self-consistent microscopic approach. In practice,  the simulations are often done with a phenomenological mean field and free nuclear cross sections. Skyrme-type interactions are often adopted with a momentum dependent part \cite{GalePRC35.1666}.
As in the case of mean field calculations of nuclear densities, mentioned in previous sections of this review, this procedure allows one to deduce the 
compressibility $K$ of nuclear matter, which refers to the second derivative of the compressional energy $E$ with as well as the symmetry energy $S$ related to the nuclear EoS.

\begin{figure}[t]
\begin{center}
{\includegraphics[width=8cm]{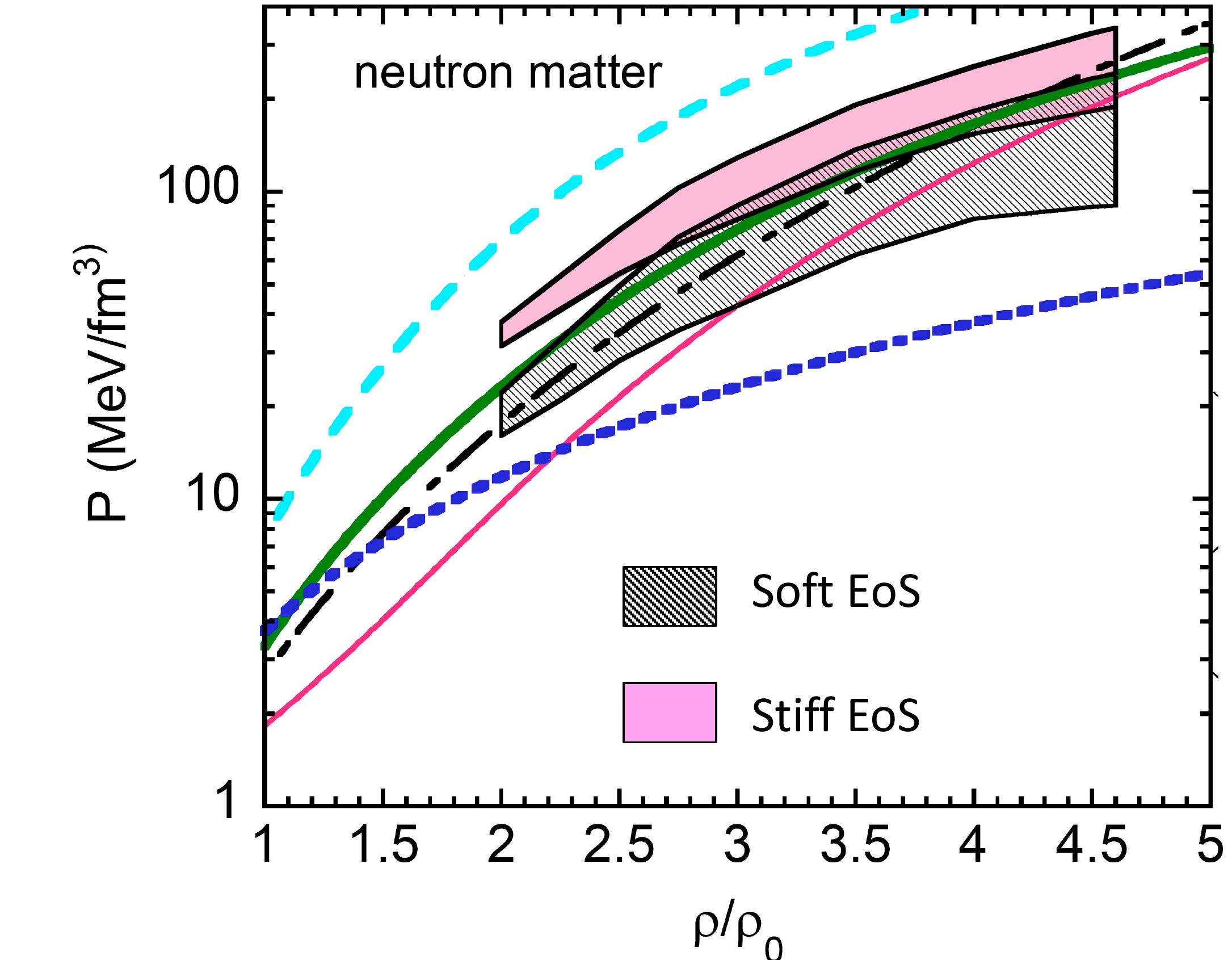}}
\end{center}
\caption{\label{lynch}
Pressure associated with neutron matter as a function of nucleon density. The various curves are different theoretical models. The shadow bands show  the numerous possibilities for the EoS based on the  experiment analyses using a soft (lower shadow region) or stiff (upper shadow region) EoS. (Adapted from \cite{Danielewicz1592}).  
}
\end{figure}

\begin{figure}[t]
\begin{center}
{\includegraphics[width=9cm]{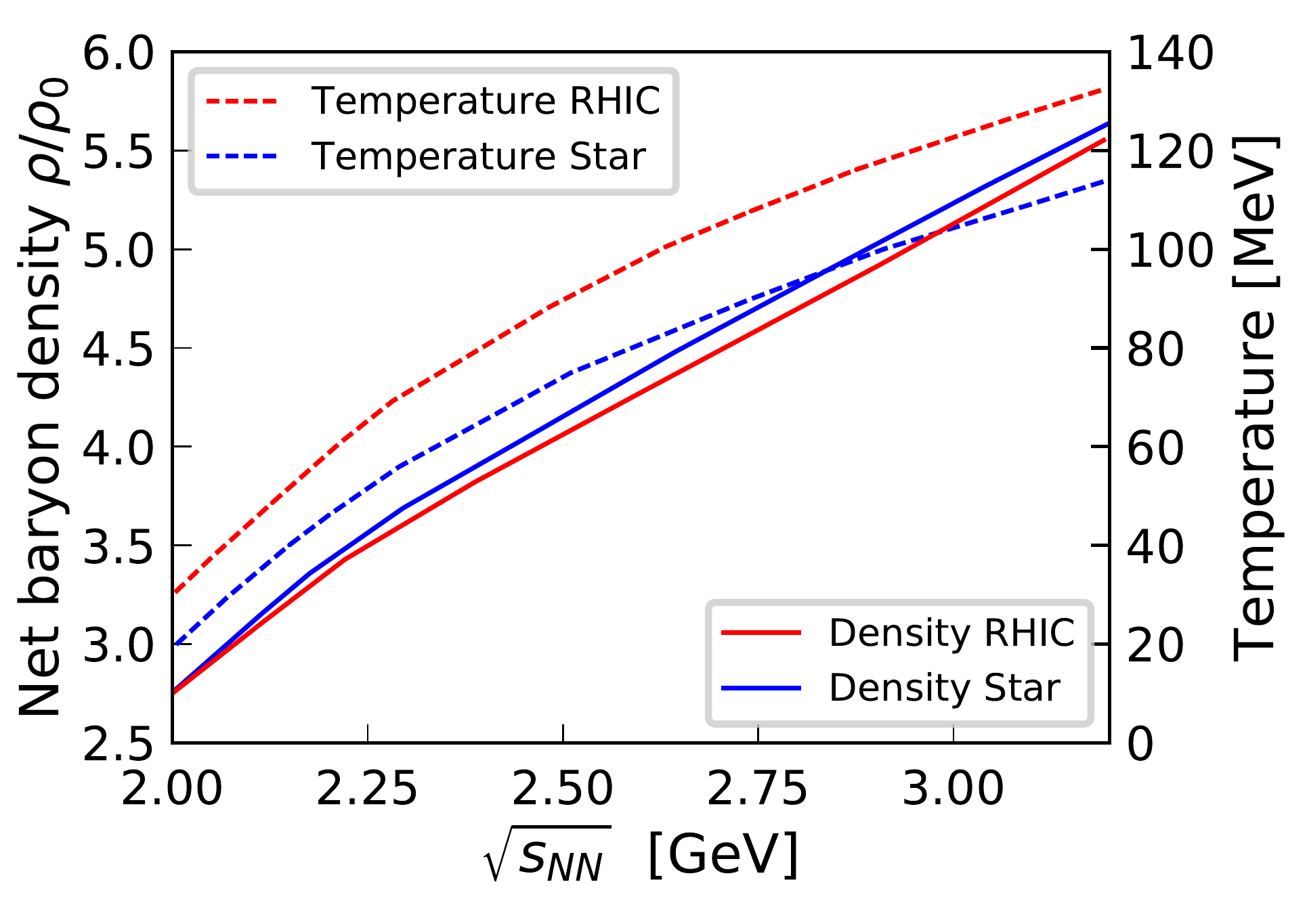}}
\end{center}
\caption{\label{rhicstar}
Largest values of baryon densities (solid lines) and temperatures (dashed lines) reached in relativistic heavy ion collisions (RHIC)  and in neutron star mergers as a function of the center of mass beam energy $\sqrt{s_{NN}} = 2 \gamma_{c.m.} m_N$.  The densities and temperatures were calculated using a quark-hadron chiral parity doublet model for the EoS which depend on the beam energy.  ] (Adapted from Ref. \cite{Hanauske_2017}).  
}
\end{figure}

After an initial compression  of the projectile and target densities, the elementary particle collisions start to thermalize matter in the collisional overlap region. The momentum distributions in this region are centered at zero momentum in the c.m. system, as shown schematically in Figure (\ref{lynch}).
The density in the overlap region rises above the saturation density,  driving the compressed matter to expand into transverse directions, a phenomenon known as collective flow. The detection of the matter distribution in arising form this compression phase,  with a comparison with transport theories, allows one to obtain the incompressibility of nuclear matter. Studies of the effect of the symmetry energy can be inferred from collisions involving neutron-rich nuclei. Other observables, such as particle production and their kinematic properties, also  help in the experimental analysis to deduce $K$ and $S$.   As an example, Ref. \cite{Danielewicz1592} determined that maximum pressures deduced from experiments  aimed at studying central nuclear collisions. Pressures in the range of  $P = 80$ to 130 MeV/fm$^3$ were deduced in collisions at 2 GeV/nucleon. In Pascal units, this corresponds to  $1.3 \times 10^{34}$ to $2.1 \times 10^{34}$ Pa). At 6 GeV/nucleon, the deduced pressures are even higher: $P = 210$ to 350 MeV/fm$^3$ (or $3.4 \times 10^{34}$ to $5.6 \times 10^{34}$ Pa), about  19 orders of magnitude larger than pressures within the core of the Sun and only comparable to pressures within neutron stars. The experimental analyses are consistent with  $K$ of Eq. \eqref{Kcomp} within the range $K=170-380$ MeV \cite{Danielewicz1592}. In Figure (\ref{lynch}) (adapted from Ref.  \cite{Danielewicz1592}), we show the pressure for neutron matter as a function of the density. The shadow bands represent the range  of possible theoretical EoS based on  soft and stiff mean-field potentials.

According to numerical simulations, in neutron star mergers, high temperatures $T\lesssim 100$ MeV can be reached \cite{Hanauske_2017}. In Fig. \ref{rhicstar} we show the largest values of baryon densities (solid lines) and temperatures (dashed lines) reached in relativistic heavy ion collisions (RHIC)  and in neutron stars as a function of the center of mass beam energy $\sqrt{s_{NN}} = 2 \gamma_{c.m.} m_N$.  Beam energies in the range $\sqrt{s_{NN}}=2.5 -3$ GeV have been considered. This is the energy region of the current SIS18 accelerator at the GSI/Darmstadt laboratory. The densities and temperatures were calculated using a quark-hadron chiral parity doublet model for the EoS which depend on the c.m. energy \cite{SteinheimerPRC84.045208}.   This EoS has different properties for different isospin content, and therefore the figure shows that the temperatures in RHIC are larger and densities slightly smaller at the same relative velocities. In nucleus-nucleus collisions at relativistic energies the isospin per baryon is of the order of -0.1, whereas in NS it is about -0.38. NS also have a  significantly different composition of strange particles compared to RHIC. The density compression in symmetric nuclear matter and in NS matter is very similar, but the temperature is quite different. This indicates that  the additional degrees of freedom, such as leptons in beta equilibrium and non-conserved strangeness, decrease the temperature at a given compression. Therefore, the study of neutron star mergers requires the use of a consistent and realistic temperature dependent EoS, probably incorporating quark degrees of freedom. Experiments carried out in the future FAIR/Darmstadt facility will be of crucial importance to study de isospin dependence of the EoS, with a connection to quarks and gluons degrees of freedom. For more details on this subject, see Ref. \cite{Hanauske_2017}.

\begin{figure}[t]
\begin{center}
{\includegraphics[width=9cm]{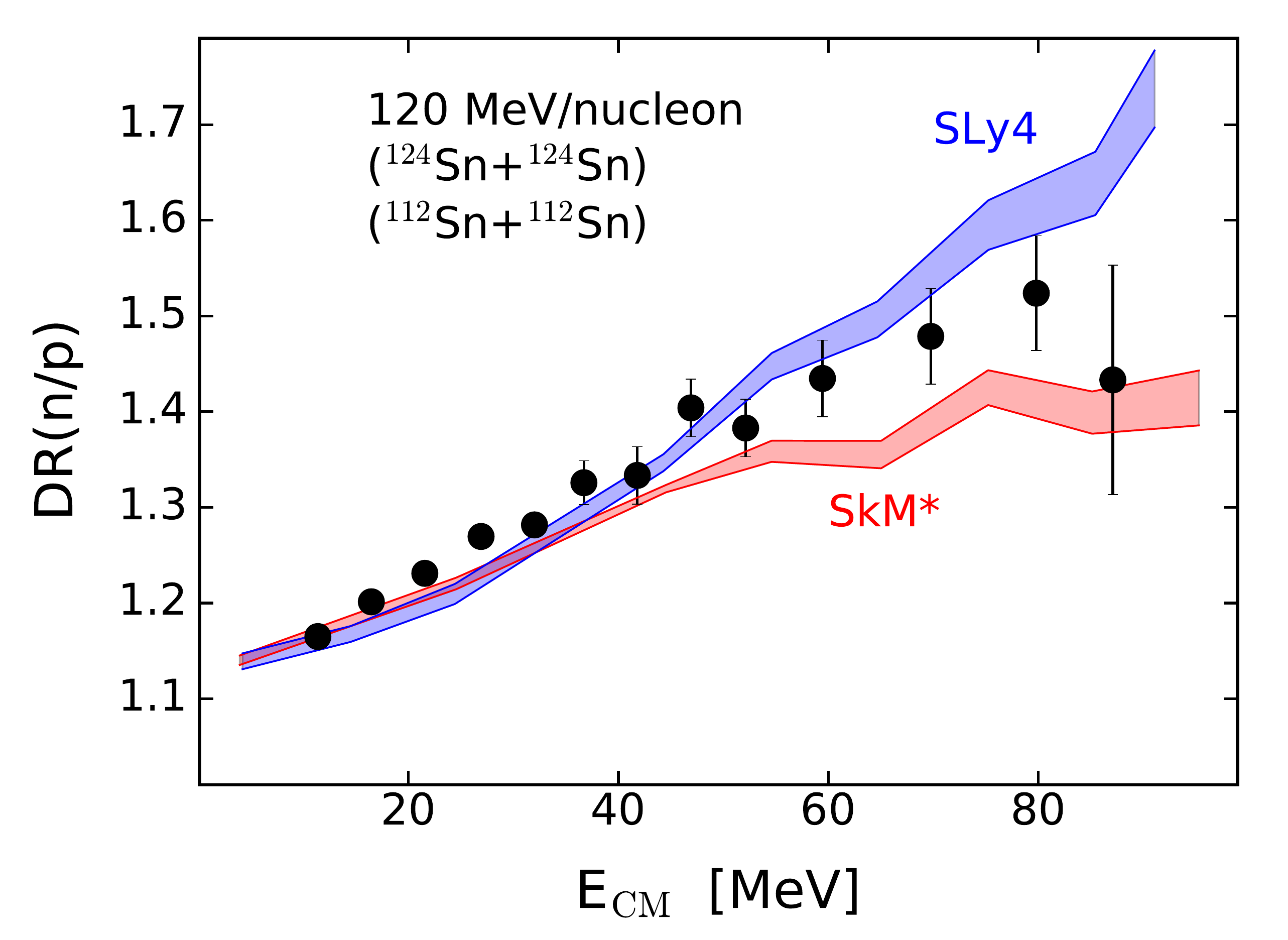}}
\end{center}
\caption{\label{tsang}
Double ratio (DR) of neutron over proton yields in collisions of $^{124}$Sn+$^{124}$Sn over $^{112}$Sn+$^{112}$Sn at  120 AMeV as a function of the center-of-mass energy of the emitted particles. The calculations with error bands use the Boltzmann-Uehling-Uhlenbeck transport equation with either the Skyrme functional SKM* or SLy4. (Adapted from
Ref. \cite{CouplandPRC94.011601}).  
}
\end{figure}

The use of radioactive beams has been crucial to understand the role of symmetry energy in central nuclear collisions.  Several indicators have been used experimentally. For example, the ``coalescence-invariant'', $DR(n/p)$ is one of such quantities \cite{CouplandPRC94.011601}, in which the ratio $R(n/p)$ neutrons and protons of all experimentally measured clusters with $A\le 4$ are measured for reactions with different combinations of projectile and target isotopes, e.g.,
\begin{equation}
DR({\rm n/p;^{124}Sn;^{112}Sn})={R({\rm n/p;^{124}Sn+^{124}Sn})\over R({\rm n/p;^{112}Sn+^{112}Sn})}.
\end{equation}
This ratio is plotted in Fig. \ref{tsang}, the double-ratio yields for n/p emission being compared to data from Ref. \cite{CouplandPRC94.011601} for Sn+Sn reactions at 120 MeV/nucleon. The double-ratio yields between two reactions of different neutron content can allow the extraction of nucleonic chemical potentials and the symmetry energy via isoscaling \cite{TsangPRL86.5023,TsangPRC64.041603}. Ratios of n/p yields in reactions with isospin partners should be sensitive to the symmetry dependence of the mean-field potential with density and also test their momentum dependence.

Ultra-relativistic central collisions, as depicted in Fig. \ref{centcoll}, are crucial to study on earth what happens with nuclear matter within the neutron star core for densities $\rho \gg \rho_{0}$ where a description of matter in terms of leptons and nucleons is not adequate. At such high densities,  hyperons and $\delta$ isobars may be produced and meson condensations may also occur. Ultimately, at extremely high densities,  a transition to a quark-gluon phase should occur \cite{WittenPRD30.272,BAYM1985181,glendenning1990}. But in contrast to relativistic nuclear collisions on earth, the cold quark-gluon phase within a neutron star poses additional experimental problems, difficult to assess with central collisions. If a cold quark-gluon phase exists inside a neutron star, its maximum mass can be substantially modified \cite{lattimer:2016:PREP}. Such studies have already been carried out for decades  at several worldwide facilities, such as RHIC/Brookhaven/USA or the CERN/ Switzerland laboratories. In a near future, the CBM (Compressed Baryonic Matter) experiment will be held at the FAIR (Facility for Antiproton and Ion Research) in Darmstadt/Germany to find out how matter changes at such high densities, extending previous experimental efforts on the study of compact stars.

\section{Conclusions}

In this review we have concentrated on the use of relativistic radioactive beams to study nuclear astrophysics with reactions in inverse kinematics. We have not covered se viral subjects associated with this topic, as they have been widely discussed in previous reviews, although we mentioned some of the challenges involved with them.
 
 Instead, we have focused the review on the needs to determined the equation of state of neutron stars and supernova explosions. We have shown that a great amount of  progress has been made in experimental nuclear physics over the past few decades. Most importantly, the application to solve long standing problems in astrophysics are remarkable. We have also discussed many aspects of theoretical nuclear physics, both for nuclear structure and nuclear reactions, which need improvement. This is particularly true in view of the construction of the  next-generation rare-isotope facilities  like GSI/FAIR in Germany and FRIB in the United States, RAON in Korea and major upgrades on the way at GANIL in France and TRIUMF in Canada.  

The RIBF facility in Japan,  will continue to provide a large number of exciting opportunities to advance the experimental research in nuclear astrophysics at all fronts. The nuclear equation of state (EoS), crucial for an understanding of neutron stars and supernova explosions, can already be probed with relativistic heavy-ion collisions induced by projectiles with neutron excess utilizing the present  GSI and RIBF facilities. 
The important scientific questions to be addressed  with relevance for astrophysics are: (a)  what are the paths to the formation and destruction of elements? (b) Do new types of radioactivity exist in stars? (c) Are new types of nuclear symmetry and spatial arrangements possible in stars? (d) What are the limits of nuclear existence and how do they depend on the environment? (e) How do the properties of nuclear matter change with variation of the nucleon density, temperature and proton-to-neutron ratio? (f) Where and how do thermal and quantum phase transitions occur with small nuclear systems? (g) What is responsible for  the shapes and symmetry properties of a nucleus? (h) How does quantum tunneling of composite particles, such as an alpha particle, occur during reactions and decay? (i) How do fundamental forces and symmetries change  in unusual stellar conditions? (j) How and where have the elements heavier than iron been formed? (k) How do rare isotopes influence the process of stellar explosions? (l) What kind of exotic nuclear structures exist and what role do they play in neutron stars? (m) Are quarks and gluons deconfined somewhere in the universe?

The answer to such questions could be born in small details of the physics that we already know but are difficult to solve theoretically or to be tested experimentally. We are not really sure where is the site of the r-process and how much neutron star mergers may contribute to it. A clearer understanding of nuclear structure and nuclear reactions is key to answer many of the questions discussed in this review. Recently we have been able to clarify some of the correlations existing in reaction networks involving unstable nuclei. The advent of new nuclear physics laboratories matched with efforts in nuclear theory have been of crucial importance to connect microscopic dynamics of nuclear systems and challenging questions in cosmology and stellar physics. Some reactions needed for astrophysics modeling  seem to be impossible to measure directly with present techniques leading to some of the nuclear physics problems relying heavily on theory. As nuclear physics is certainly one of the hardest problems to tackle in all science, we will need many more decades of dedicated scientific work.         

\section*{Acknowledgement} 

This project was supported by the BMBF via Project No. 05P15RDFN1, through the GSI-TU Darmstadt co- operation agreement, the DFG via SFB1245, and in part by U.S. DOE Grant No. DE- FG02-08ER41533 and U.S. NSF Grant No. 1415656. We thank HIC for FAIR for supporting visits (C. A. B.) to the TU Darmstadt.

\section*{References}


\end{document}